\pdfoutput=1

\documentclass[12pt,a4paper]{article}

\usepackage{ifthen}
\newboolean{pdflatex}
\setboolean{pdflatex}{true}

\newboolean{articletitles}
\setboolean{articletitles}{true}

\newboolean{uprightparticles}
\setboolean{uprightparticles}{false}

\newboolean{inbibliography}
\setboolean{inbibliography}{false}

\def\paperauthors{LHCb collaboration}
\def\paperasciititle{Amplitude analysis of the B+ -> pi+ pi+ pi- decay}
\def\papertitle{Amplitude analysis of the ${\decay{\Bp}{\pip\pip\pim}}$ decay}
\def\paperkeywords{{High Energy Physics}, {LHCb}}
\def\papercopyright{\the\year\ CERN for the benefit of the LHCb collaboration}
\def\paperlicence{CC-BY-4.0 licence}
\def\paperlicenceurl{https://creativecommons.org/licenses/by/4.0/}

\usepackage[top=1in, bottom=1.25in, left=1in, right=1in]{geometry}

\usepackage{microtype}
\usepackage{xspace}
\usepackage{caption}

\usepackage{graphicx}
\usepackage{color}
\usepackage{colortbl}
\graphicspath{{./figs/}}

\usepackage{amsmath}
\usepackage{amssymb}
\usepackage{amsfonts}
\usepackage{upgreek}

\usepackage{booktabs}

\usepackage{lineno}

\newcommand*\patchAmsMathEnvironmentForLineno[1]{%
\expandafter\let\csname old#1\expandafter\endcsname\csname #1\endcsname
\expandafter\let\csname oldend#1\expandafter\endcsname\csname
end#1\endcsname
 \renewenvironment{#1}%
   {\linenomath\csname old#1\endcsname}%
   {\csname oldend#1\endcsname\endlinenomath}%
}
\newcommand*\patchBothAmsMathEnvironmentsForLineno[1]{%
  \patchAmsMathEnvironmentForLineno{#1}%
  \patchAmsMathEnvironmentForLineno{#1*}%
}
\AtBeginDocument{%
\patchBothAmsMathEnvironmentsForLineno{equation}%
\patchBothAmsMathEnvironmentsForLineno{align}%
\patchBothAmsMathEnvironmentsForLineno{flalign}%
\patchBothAmsMathEnvironmentsForLineno{alignat}%
\patchBothAmsMathEnvironmentsForLineno{gather}%
\patchBothAmsMathEnvironmentsForLineno{multline}%
\patchBothAmsMathEnvironmentsForLineno{eqnarray}%
}

\usepackage{hyperxmp}

\usepackage[pdftex,
            pdfauthor={\paperauthors},
            pdftitle={\paperasciititle},
            pdfkeywords={\paperkeywords},
            pdfcopyright={Copyright (C) \papercopyright},
            pdflicenseurl={\paperlicenceurl}]{hyperref}

\usepackage[all]{hypcap}

\usepackage{xspace} 
\usepackage{upgreek}

\def\lhcb   {\mbox{LHCb}\xspace}

\def\MagUp {\mbox{\em Mag\kern -0.05em Up}\xspace}

\ifthenelse{\boolean{uprightparticles}}%
{

 \def\Pmu         {\ensuremath{\upmu}\xspace}

 \def\Ppi         {\ensuremath{\uppi}\xspace}

 \def\Ppsi        {\ensuremath{\uppsi}\xspace}

 \def\PDelta      {\ensuremath{\Delta}\xspace}                 
 \def\PXi         {\ensuremath{\Xi}\xspace}                 
 \def\PLambda     {\ensuremath{\Lambda}\xspace}                 
 \def\PSigma      {\ensuremath{\Sigma}\xspace}                 
 \def\POmega      {\ensuremath{\Omega}\xspace}                 
 \def\PUpsilon    {\ensuremath{\Upsilon}\xspace}

 \def\PB      {\ensuremath{\mathrm{B}}\xspace}                 
                  
 \def\PD      {\ensuremath{\mathrm{D}}\xspace}

 \def\PJ      {\ensuremath{\mathrm{J}}\xspace}                 
 \def\PK      {\ensuremath{\mathrm{K}}\xspace}

 \def\Pb      {\ensuremath{\mathrm{b}}\xspace}                 
 \def\Pc      {\ensuremath{\mathrm{c}}\xspace}

 \def\Ph      {\ensuremath{\mathrm{h}}\xspace}                 
 \def\Pi      {\ensuremath{\mathrm{i}}\xspace}

 \def\Pp      {\ensuremath{\mathrm{p}}\xspace}

 \def\thebaroffset{0.0em}
}
{

 \def\Pmu         {\ensuremath{\mu}\xspace}

 \def\Ppi         {\ensuremath{\pi}\xspace}

 \def\Ppsi        {\ensuremath{\psi}\xspace}                 
                  
 \mathchardef\PDelta="7101
 \mathchardef\PXi="7104
 \mathchardef\PLambda="7103
 \mathchardef\PSigma="7106
 \mathchardef\POmega="710A
 \mathchardef\PUpsilon="7107
                  
 \def\PB      {\ensuremath{B}\xspace}                 
                  
 \def\PD      {\ensuremath{D}\xspace}

 \def\PJ      {\ensuremath{J}\xspace}                 
 \def\PK      {\ensuremath{K}\xspace}

 \def\Pb      {\ensuremath{b}\xspace}                 
 \def\Pc      {\ensuremath{c}\xspace}

 \def\Ph      {\ensuremath{h}\xspace}                 
 \def\Pi      {\ensuremath{i}\xspace}

 \def\Pp      {\ensuremath{p}\xspace}

 \def\thebaroffset{0.18em}
}
\newcommand{\offsetoverline}[2][\thebaroffset]{\kern #1\overline{\kern -#1 #2}}%

\makeatletter
\ifcase \@ptsize \relax% 10pt
  \newcommand{\miniscule}{\@setfontsize\miniscule{4}{5}}% \tiny: 5/6
\or% 11pt
  \newcommand{\miniscule}{\@setfontsize\miniscule{5}{6}}% \tiny: 6/7
\or% 12pt
  \newcommand{\miniscule}{\@setfontsize\miniscule{5}{6}}% \tiny: 6/7
\fi
\makeatother

\DeclareRobustCommand{\optbar}[1]{\shortstack{{\miniscule (\rule[.5ex]{1.25em}{.18mm})}
  \\ [-.7ex] $#1$}}

\def\mumu       {{\ensuremath{\Pmu^+\Pmu^-}}\xspace}

\def\cquark    {{\ensuremath{\Pc}}\xspace}

\def\bquark    {{\ensuremath{\Pb}}\xspace}

\def\hadron {{\ensuremath{\Ph}}\xspace}
\def\pion   {{\ensuremath{\Ppi}}\xspace}
\def\piz    {{\ensuremath{\pion^0}}\xspace}
\def\pip    {{\ensuremath{\pion^+}}\xspace}
\def\pim    {{\ensuremath{\pion^-}}\xspace}
\def\pipm   {{\ensuremath{\pion^\pm}}\xspace}
\def\pimp   {{\ensuremath{\pion^\mp}}\xspace}

\def\kaon    {{\ensuremath{\PK}}\xspace}
\def\Kbar    {{\ensuremath{\offsetoverline{\PK}}}\xspace}

\def\KorKbar {\kern \thebaroffset\optbar{\kern -\thebaroffset \PK}{}\xspace}

\def\Kp      {{\ensuremath{\kaon^+}}\xspace}
\def\Km      {{\ensuremath{\kaon^-}}\xspace}

\def\Dbar    {{\ensuremath{\offsetoverline{\PD}}}\xspace}
\def\D       {{\ensuremath{\PD}}\xspace}

\def\DorDbar {\kern \thebaroffset\optbar{\kern -\thebaroffset \PD}\xspace}
\def\Dz      {{\ensuremath{\D^0}}\xspace}
\def\Dzb     {{\ensuremath{\Dbar{}^0}}\xspace}

\def\Dstarp  {{\ensuremath{\D^{*+}}}\xspace}

\def\B       {{\ensuremath{\PB}}\xspace}

\def\BorBbar {\kern \thebaroffset\optbar{\kern -\thebaroffset \PB}\xspace}
\def\Bz      {{\ensuremath{\B^0}}\xspace}

\def\Bu      {{\ensuremath{\B^+}}\xspace}
\def\Bub     {{\ensuremath{\B^-}}\xspace}
\def\Bp      {{\ensuremath{\Bu}}\xspace}
\def\Bm      {{\ensuremath{\Bub}}\xspace}
\def\Bpm     {{\ensuremath{\B^\pm}}\xspace}

\def\jpsi     {{\ensuremath{{\PJ\mskip -3mu/\mskip -2mu\Ppsi\mskip 2mu}}}\xspace}

\def\Y#1S{\ensuremath{\PUpsilon{(#1S)}}\xspace}

\def\proton      {{\ensuremath{\Pp}}\xspace}

\def\LorLbar     {\kern \thebaroffset\optbar{\kern -\thebaroffset \PLambda}\xspace}

\newcommand{\decay}[2]{\mbox{\ensuremath{#1\!\to #2}}\xspace}         % {\Pa}{\Pb \Pc}

\def\to                 {\ensuremath{\rightarrow}\xspace}

\def\CP                {{\ensuremath{C\!P}}\xspace}

\def\AT#1     {\ensuremath{A_{\mathrm{T}}^{#1}}\xspace}           % 2

\def\C#1      {\ensuremath{\mathcal{C}_{#1}}\xspace}                       % 9
\def\Cp#1     {\ensuremath{\mathcal{C}_{#1}^{'}}\xspace}                    % 7
\def\Ceff#1   {\ensuremath{\mathcal{C}_{#1}^{\mathrm{(eff)}}}\xspace}        % 9  
\def\Cpeff#1  {\ensuremath{\mathcal{C}_{#1}^{'\mathrm{(eff)}}}\xspace}       % 7
\def\Ope#1    {\ensuremath{\mathcal{O}_{#1}}\xspace}                       % 2
\def\Opep#1   {\ensuremath{\mathcal{O}_{#1}^{'}}\xspace}                    % 7

\newcommand{\nospaceunit}[1]{\ensuremath{\text{#1}}}       
\newcommand{\aunit}[1]{\ensuremath{\text{\,#1}}}       
\newcommand{\tev}{\aunit{Te\kern -0.1em V}\xspace}
\newcommand{\gev}{\aunit{Ge\kern -0.1em V}\xspace}
\newcommand{\mev}{\aunit{Me\kern -0.1em V}\xspace}
\newcommand{\kev}{\aunit{ke\kern -0.1em V}\xspace}
\newcommand{\ev}{\aunit{e\kern -0.1em V}\xspace}
\newcommand{\mevc}{\ensuremath{\aunit{Me\kern -0.1em V\!/}c}\xspace}
\newcommand{\gevc}{\ensuremath{\aunit{Ge\kern -0.1em V\!/}c}\xspace}
\newcommand{\mevcc}{\ensuremath{\aunit{Me\kern -0.1em V\!/}c^2}\xspace}
\newcommand{\gevcc}{\ensuremath{\aunit{Ge\kern -0.1em V\!/}c^2}\xspace}
\newcommand{\gevgevcccc}{\ensuremath{\gev^2/c^4}\xspace} % for q^2

\def\mm   {\aunit{mm}\xspace}

\def\mum  {\ensuremath{\,\upmu\nospaceunit{m}}\xspace}

\def\fm   {\aunit{fm}\xspace}

\def\fb   {\ensuremath{\aunit{fb}}\xspace}
\def\invfb   {\ensuremath{\fb^{-1}}\xspace}

\def\gsim{{~\raise.15em\hbox{$>$}\kern-.85em
          \lower.35em\hbox{$\sim$}~}\xspace}
\def\lsim{{~\raise.15em\hbox{$<$}\kern-.85em
          \lower.35em\hbox{$\sim$}~}\xspace}

\def\pt         {\ensuremath{p_{\mathrm{T}}}\xspace}

\def\ptot       {\ensuremath{p}\xspace}

\def\mrad{\aunit{mrad}}

\def\evtgen     {\mbox{\textsc{EvtGen}}\xspace}

\def\geant      {\mbox{\textsc{Geant4}}\xspace}

\def\photos     {\mbox{\textsc{Photos}}\xspace}

\def\pythia     {\mbox{\textsc{Pythia}}\xspace}

\xspace

\def\tell1  {TELL1\xspace}
\def\ukl1   {UKL1\xspace}

\newcommand{\ie}{\mbox{\itshape i.e.}\xspace}

\usepackage{siunitx}

\usepackage{cite}
\usepackage{mciteplus}

\usepackage{longtable}
\usepackage{rotating}
\usepackage{pdflscape}
\usepackage[section]{placeins}

\begin{document}

\renewcommand{\thefootnote}{\fnsymbol{footnote}}
\setcounter{footnote}{1}
\begin{titlepage}
\pagenumbering{roman}

\vspace*{-1.5cm}
\centerline{\large EUROPEAN ORGANIZATION FOR NUCLEAR RESEARCH (CERN)}
\vspace*{1.5cm}
\noindent
\begin{tabular*}{\linewidth}{lc@{\extracolsep{\fill}}r@{\extracolsep{0pt}}}
\ifthenelse{\boolean{pdflatex}}
{\vspace*{-1.5cm}\mbox{\!\!\!\includegraphics[width=.14\textwidth]{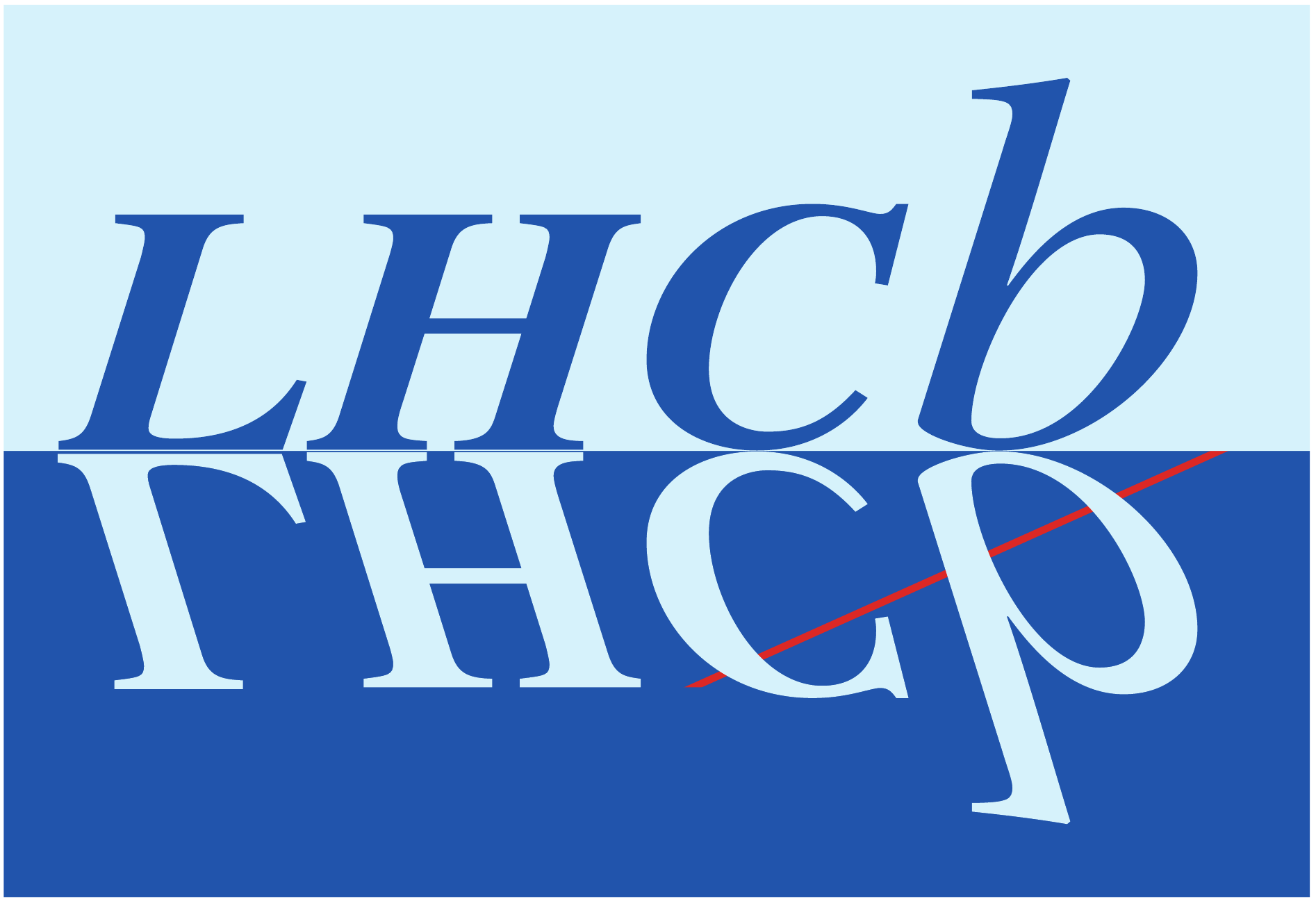}} & &}%
{\vspace*{-1.2cm}\mbox{\!\!\!\includegraphics[width=.12\textwidth]{lhcb-logo.eps}} & &}%
\\
 & & CERN-EP-2019-157 \\
 & & LHCb-PAPER-2019-017 \\
 & & 23 January 2020 \\
 & & \\
\end{tabular*}

\vspace*{2.5cm}

{\normalfont\bfseries\boldmath\huge
\begin{center}
  \papertitle
\end{center}
}

\vspace*{1.0cm}

\begin{center}
\paperauthors\footnote{Authors are listed at the end of this paper.}
\end{center}

\vspace{\fill}

\begin{abstract}
  \noindent
  The results of an amplitude analysis of the charmless three-body decay \decay{\Bp}{\pip\pip\pim}, in which \CP-violation effects are taken into account, are reported.
  The analysis is based on a data sample corresponding to an integrated luminosity of $3 \invfb$ of $\proton\proton$ collisions recorded with the LHCb detector.
  The most challenging aspect of the analysis is the description of the behaviour of the $\pi^+ \pi^-$ S-wave contribution, which is achieved by using three complementary approaches based on the isobar model, the K-matrix formalism, and a quasi-model-independent procedure.
  Additional resonant contributions for all three methods are described using a common isobar model, and include the $\rho(770)^0$, $\omega(782)$ and $\rho(1450)^0$ resonances in the $\pip\pim$ P-wave, the $f_2(1270)$ resonance in the $\pip\pim$ D-wave, and the $\rho_3(1690)^0$ resonance in the $\pip\pim$ F-wave.
  Significant \CP-violation effects are observed in both S- and D-waves, as well as in the interference between the S- and P-waves.
  The results from all three approaches agree and provide new insight into the dynamics and the origin of \CP-violation effects in \decay{\Bp}{\pip\pip\pim} decays.
\end{abstract}

\vspace*{1.5cm}

\begin{center}
  Published in Phys.~Rev.~D 101 (2020) 012006
\end{center}

\vspace{\fill}

{\footnotesize
\centerline{\copyright~\papercopyright. \href{\paperlicenceurl}{\paperlicence}.}}
\vspace*{2mm}

\end{titlepage}

\newpage
\setcounter{page}{2}
\mbox{~}

\cleardoublepage

\renewcommand{\thefootnote}{\arabic{footnote}}
\setcounter{footnote}{0}

\pagestyle{plain}
\setcounter{page}{1}
\pagenumbering{arabic}

\section{Introduction}
\label{sec:Introduction}

In the Standard Model (SM), \CP\ violation originates from a single irreducible complex phase in the Cabibbo--Kobayashi--Maskawa (CKM) matrix~\cite{Cabibbo:1963yz,Kobayashi:1973fv}.
Thus far, all measurements of \CP\ violation in particle decays are consistent with this explanation.
Nevertheless, the degree of \CP\ violation permitted in the SM is inconsistent with the macroscopic matter-antimatter asymmetry observed in the Universe~\cite{Shaposhnikov:1991cu}, motivating further studies and searches for sources of \CP\ violation beyond the SM.

For the manifestation of \CP\ violation in decay, at least two interfering amplitudes with different strong and weak phases are required.
In the SM, weak phases are associated with the complex elements of the CKM matrix and have opposite sign between charge-conjugate processes, while strong phases are associated with hadronic final-state effects and do not change sign under \CP\ conjugation.
In decays of \bquark hadrons to charmless hadronic final states, contributions from both tree and loop (so-called ``penguin'') diagrams, which can provide the relative weak phase that is necessary for \CP violation to manifest, are possible with comparable magnitudes.
Indeed, significant \CP asymmetries have been observed in both $\Bz \to \Kp\pim$~\cite{Lees:2012mma,Duh:2012ie,LHCb-PAPER-2013-018,Aaltonen:2014vra} and $\Bz \to \pip\pim$ decays~\cite{Lees:2012mma,Adachi:2013mae,LHCb-PAPER-2018-006}.
In multibody decays, variation across the phase space of strong phases, caused by hadronic resonances, allows for further enhancement of \CP\ violation effects and a richer phenomenology compared to two-body decays.
Large \CP asymmetries localised in regions of phase space of charmless three-body $B$-meson decays have been observed in model-independent analyses~\cite{LHCb-PAPER-2013-027,LHCb-PAPER-2013-051,LHCb-PAPER-2014-044,Hsu:2017kir}, but until recently there has been no description of these effects with an accurate model of the contributing resonances.
An amplitude analysis of \decay{\Bp}{\Kp\Kp\pim} decays~\cite{LHCb-PAPER-2018-051} has shown that $\pi\pi \leftrightarrow \kaon\Kbar$ rescattering plays an important role in the observed \CP\ violation, and it is anticipated that similar effects will occur in other charmless three-body $B$-meson decays.

This paper documents an analysis of the \decay{\Bp}{\pip\pip\pim} decay amplitude in the two-dimensional phase space known as the Dalitz plot~\cite{Dalitz:1953cp, Fabri:1954zz}.
The inclusion of charge-conjugate processes is implied, except where asymmetries are discussed.
Previous studies of this decay mode indicate that the amplitude contains a sizable $\rho(770)^0$ component~\cite{Jessop:2000bv,Gordon:2002yt,Aubert:2005sk,Aubert:2009av}. The amplitude analysis performed by the BaBar
collaboration~\cite{Aubert:2009av} additionally observed a large S-wave contribution, however, measurements of \CP-violating quantities
were limited by statistical precision.

Phenomenological studies~\cite{Xu:2013rua, Bhattacharya:2013boa, Bhattacharya:2014eca, Krankl:2015fha, Klein:2017xti, Li:2018lbd} have focussed on investigating the localised \CP asymmetries seen in the model-independent analysis of \decay{\Bp}{\pip\pip\pim} decays~\cite{LHCb-PAPER-2014-044}, with some works indicating the potential importance of the $\rho$--$\omega$ mixing effect between the $\rho(770)^0$ and $\omega(782)$ resonances~\cite{Guo:2000uc, Wang:2015ula, Cheng:2016shb}, and the interference between the $\rho(770)^0$ resonance and the broad S-wave contribution~\cite{Dedonder:2010fg,Zhang:2013oqa,Nogueira:2015tsa,Bediaga:2015mia}.
Furthermore, the relative pattern of \CP\ asymmetries between \decay{\Bp}{\hadron^+\hadron^{+}\hadron^{-}} decays, where $\hadron$ is a
kaon or pion, could be indicative of \CP\ violation induced by $\pi\pi \leftrightarrow \kaon\Kbar$ rescattering~\cite{PhysRevD.71.014030, Dedonder:2010fg, Nogueira:2015tsa, IgnacioCPT}.

The present analysis is performed on data corresponding to 3\invfb collected by the LHCb experiment, of which $1\invfb$ was collected in 2011 with a \proton\proton collision centre-of-mass energy of $7\tev$ and $2\invfb$ was collected in 2012 with a centre-of-mass energy of $8\tev$.
A model of the Dalitz plot distribution is constructed in terms of the intermediate resonant and nonresonant structures.
Due to its magnitude and potential importance to the observed \CP violation in this decay, particular attention is given to the $\pip\pim$ S-wave contribution, which is known to consist of numerous overlapping
resonances and open decay channels~\cite{Pelaez:2015qba}.
Three different state-of-the-art approaches to the modelling of the S-wave are used to ensure that any inaccuracies in the description of this part of the amplitude do not impact the interpretation of the physical quantities reported.

This paper is organised as follows: Section~\ref{sec:Detector} gives a brief description of the LHCb detector
and the event reconstruction and simulation software; the signal candidate selection procedure is described in
Section~\ref{sec:Selection}; Section~\ref{sec:Mass_fit} describes the procedure for estimating the signal and
background yields that enter into the amplitude fit; Section~\ref{sec:DPformalism} outlines the formalism used
for the construction of the amplitude models, as well as a description of the mass lineshapes used to
parameterise the intermediate structures; Section~\ref{sec:systematics} describes the systematic uncertainties associated with the analysis procedure; Section~\ref{sec:results} documents the physics parameters of interest
obtained from the amplitude models and presents projections of the fit models on the selected data; these results are
then discussed in Section~\ref{sec:discussion}; and a summary of the work as a whole can be found in
Section~\ref{sec:conclusions}.
A shorter description of the analysis, more focussed on the first observations of different sources of \CP-violation effects, can be found in a companion article~\cite{LHCb-PAPER-2019-018}.

\section{Detector and simulation}
\label{sec:Detector}

The \lhcb detector~\cite{Alves:2008zz,LHCb-DP-2014-002} is a single-arm forward
spectrometer covering the \mbox{pseudorapidity} range $2<\eta <5$,
designed for the study of particles containing \bquark or \cquark
quarks. The detector includes a high-precision tracking system
consisting of a silicon-strip vertex detector surrounding the $pp$
interaction region~\cite{LHCb-DP-2014-001}, a large-area silicon-strip detector located
upstream of a dipole magnet with a bending power of about
$4{\mathrm{\,Tm}}$, and three stations of silicon-strip detectors and straw
drift tubes~\cite{LHCb-DP-2013-003} placed downstream of the magnet.
The tracking system provides a measurement of the momentum \ptot of charged particles with
relative uncertainty that varies from 0.5\% at low momentum to 1.0\% at 200\gevc.
The minimum distance of a track to a primary vertex (PV), or impact parameter (IP),
is measured with a resolution of $(15+29/\pt)\mum$,
where \pt is the component of the momentum transverse to the beam (in\,\gevc).
Different types of charged hadrons are distinguished using information
from two ring-imaging Cherenkov detectors~\cite{LHCb-DP-2012-003}.
Photons, electrons and hadrons are identified by a calorimeter system consisting of
scintillating-pad and preshower detectors, an electromagnetic
and a hadronic calorimeter. Muons are identified by a
system composed of alternating layers of iron and multiwire
proportional chambers~\cite{LHCb-DP-2012-002}.
The magnetic field deflects oppositely charged particles in opposite directions and this can lead to detection asymmetries. Periodically reversing the magnetic field polarity throughout the data-taking reduces this effect to a negligible level.
Approximately $60\%$ of 2011 data and $52\%$ of 2012 data was collected in the ``down'' polarity configuration, and the rest in the ``up'' configuration.

The online event selection is performed by a trigger~\cite{LHCb-DP-2012-004}
which consists of a hardware stage followed by a software stage. The hardware stage is based on information from the calorimeter and muon systems in which events are required to contain a muon with high \pt, or a hadron, photon or electron with high transverse energy in the calorimeters.
The software trigger requires a two- or three-track secondary vertex with significant displacement from all primary $pp$ interaction vertices. 
All charged particles with \mbox{$\pt>500\,(300)\mevc$}  are reconstructed, for data collected in 2011\,(2012), in events where at least one charged particle has transverse momentum $\pt > 1.7\,(1.6)\gevc$ and is inconsistent with originating from a PV. 
A multivariate algorithm~\cite{BBDT} is used for the identification of secondary vertices consistent with the decay of a \bquark hadron.

Simulated data samples are used to investigate backgrounds from other
\bquark-hadron decays and also to study the detection and reconstruction efficiency of the signal.
In the simulation, $pp$ collisions are generated using
\pythia~\cite{Sjostrand:2006za,*Sjostrand:2007gs} with a specific \lhcb
configuration~\cite{LHCb-PROC-2010-056}.
Decays of unstable particles are described by \evtgen~\cite{Lange:2001uf},
in which final-state radiation is generated using \photos~\cite{Golonka:2005pn}.
The interaction of the generated particles with the detector and its
response are implemented using the \geant toolkit~\cite{Allison:2006ve,
*Agostinelli:2002hh} as described in Ref.~\cite{LHCb-PROC-2011-006}.

\section{Selection}
\label{sec:Selection}

The selection of signal candidates follows closely the procedure used in the model-independent analysis of the same data sample~\cite{LHCb-PAPER-2014-044}.
Signal \Bp candidates are formed from three good-quality tracks
that are consistent with originating from the same secondary vertex (SV), which must also be at least $3\mm$ away from any PV.
The reconstructed \Bp candidate is associated with the PV that is most consistent with its flight direction.
A requirement is also imposed on the angle between the \Bp momentum and the vector between the PV and SV, that must be less than approximately $6 \mrad$.

To reject random associations of tracks (combinatorial background), a boosted decision-tree classifier~\cite{Breiman} is trained
to discriminate between simulated signal candidates and candidates in collision data residing in a region where this background dominates, \mbox{$5.4 < m(\pip\pip\pim) < 5.8\gevcc$}.
The variables that enter this classifier are \Bp and decay product kinematic properties, quantities based on the quality of the reconstructed tracks and decay vertices, as well as the \Bp displacement from the PV.
The requirement on the output of this classifier is optimised to maximise the expected approximate signal
significance, $N_s/\sqrt{N_s + N_b}$, where $N_s$ is the expected signal yield within $40 \mevcc$ of the
known \Bp mass~\cite{PDG2018}, and $N_b$ is the corresponding combinatorial background level within the same region.

To suppress backgrounds that arise when any number of kaons are misidentified as pions,
requirements are placed on the particle-identification information associated with each final-state track.
Furthermore, tracks associated with a hit in the muon system are also removed, as are tracks that are outside the fiducial region of the particle-identification system.

A veto in both combinations of the opposite-sign dipion mass, where \mbox{$1.740 < m(\pip\pim) < 1.894 \gevcc$}, is
applied to remove \decay{\Bp}{(\Dzb \rightarrow \pip\pim)\pip} decays,
along with partially reconstructed decays involving intermediate \Dzb mesons and decays of \Dzb mesons where one or more kaons are misidentified as pions.
Approximately $2\%$ of events contain multiple \Bp decay candidates following the aforementioned selection procedure, of which one is chosen at random.

The Dalitz-plot variables are calculated following a kinematic mass constraint, fixing the \Bp candidate mass to the known value to improve resolution and to ensure that all decays remain within the Dalitz-plot boundary.
In consequence, the experimental resolution in the region with the narrowest resonance considered in this analysis, the $\omega(782)$ state, is better than $3 \mevcc$, with a corresponding full width at half maximum of better than $7 \mevcc$.
This is smaller than the $\omega(782)$ width, and therefore effects related to the finite resolution in the Dalitz plot are not considered further.

\section{\boldmath \Bp\ candidate invariant-mass fit}
\label{sec:Mass_fit}

An extended, unbinned, maximum-likelihood fit is performed to the $m(\pip\pip\pim)$ invariant-mass spectrum to extract yields and charge asymmetries of the \decay{\Bp}{\pip\pip\pim} signal and various contributing backgrounds.
The fit is performed to candidates in the range $5.080 < m(\pip\pip\pim) < 5.580 \gevcc$, and its results are used to obtain signal and background yields in the signal region, $5.249 < m(\pip\pip\pim) < 5.317 \gevcc$, in which the subsequent Dalitz-plot fit is performed.
All shape parameters of the probability density functions (PDFs) comprising the fit model are shared between \Bp and \Bm candidates and only the yields are permitted to vary between these categories, which are fitted simultaneously.
The data are also subdivided by data-taking year, and whether the hardware trigger decision is due to hadronic calorimeter deposits associated with the signal candidate, or due to other particles in the event, to permit correction for possible differences in efficiency between subsamples (see Section~\ref{sec:efficiencies}).

The shape of the \decay{\Bp}{\pip\pip\pim} signal decay is parameterised by the sum of a core Gaussian with two
Crystal Ball functions~\cite{Skwarnicki:1986xj}, with tails on opposite sides of the peak in order to describe the asymmetric tails of the distribution
due to detector resolution and final-state radiation. The tail mean and width parameters of the Crystal Ball functions are determined from simulation relative to the core mean and width, which are
left free in the fit to collision data to account for small differences between simulation and data.
All remaining parameters, apart from the total yield, are obtained from a fit to simulated events.

Partially reconstructed
backgrounds, which predominantly arise from four-body \mbox{\B-meson} decays where a charged hadron or neutral particle is not reconstructed, are modelled with an ARGUS function~\cite{Albrecht:1990cs} convolved with a Gaussian resolution function.
The smooth combinatorial background is modelled with a falling exponential function.
The only significant source of cross-feed background, where one or more kaons are misidentified as pions, is the \decay{\Bp}{\Kp\pip\pim} decay.
To obtain an accurate model for this background, simulated \decay{\Bp}{\Kp\pip\pim} decays are weighted according to the amplitude model obtained by the BaBar collaboration~\cite{Aubert:2008bj}, also accounting for the probability to be reconstructed as a \decay{\Bp}{\pip\pip\pim} candidate.
A model for \decay{\Bp}{\Kp\pip\pim} decays based on a similar-sized data sample has also been obtained by the Belle collaboration~\cite{Garmash:2005rv}; the details of the model used do not impact this analysis.
This shape is modelled in the invariant-mass fit as a sum of a Gaussian with
two Crystal Ball functions, where all parameters are determined from a fit to the weighted simulation.
Furthermore, the yield of this component is constrained to the \decay{\Bp}{\pip\pip\pim} signal yield
multiplied by the product of the relative branching fractions of these decays and the inverse of the
relative overall reconstruction and selection efficiencies, which are described in
Section~\ref{sec:efficiencies}.

\begin{figure}[tb]
\centering
\includegraphics[width=0.67\linewidth]{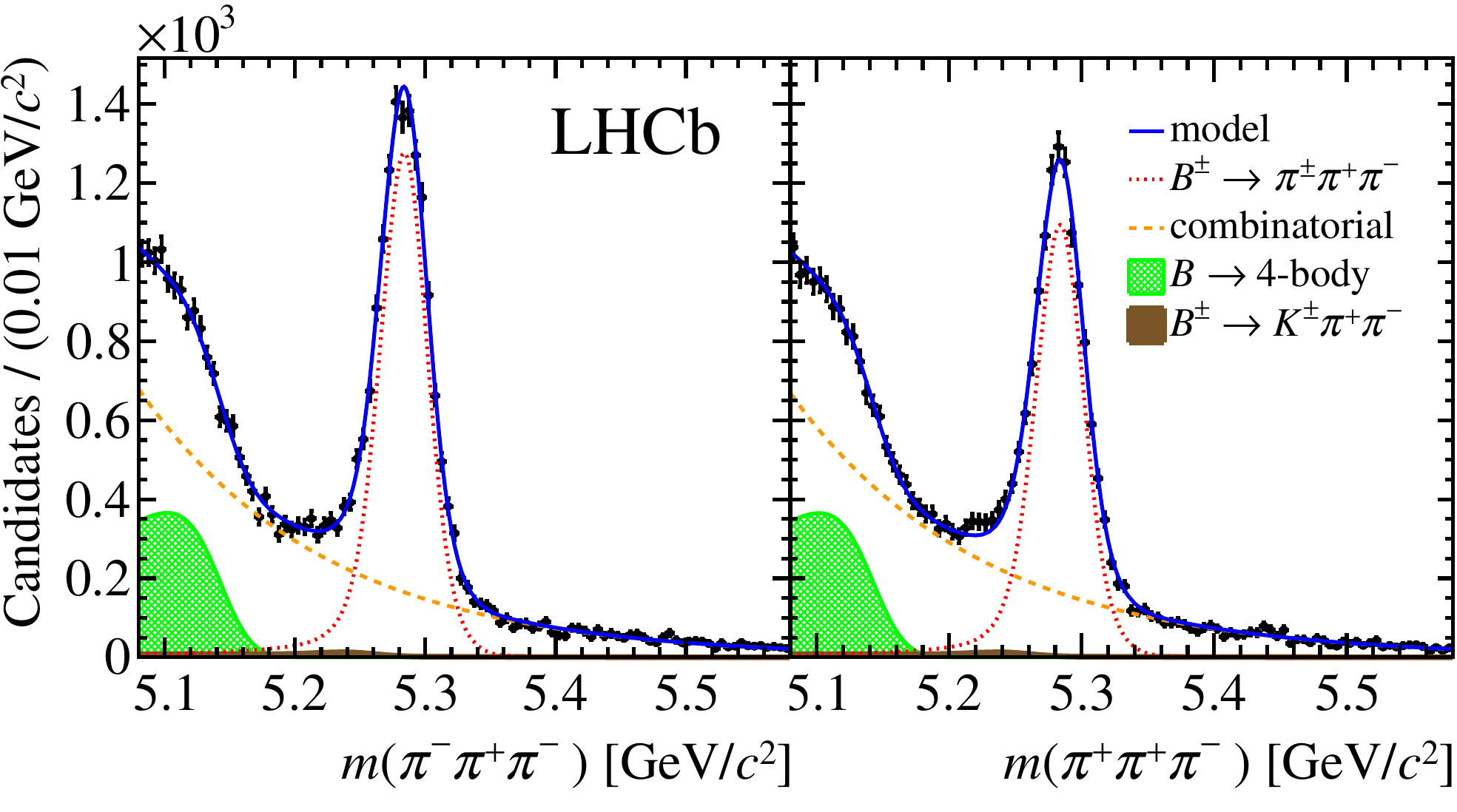}
\put(-265,140){(a)}
\put(-130,140){(b)}

\caption{
  Invariant-mass fit model for (a) \Bm\ and (b) \Bp\ candidates reconstructed in the \pimp\pip\pim final state for the combined 2011 and 2012 data taking samples.
  Points with error bars represent the data while the components comprising the model are listed in the plot legend.
}
\label{fig:massFit}
\end{figure}

The mass fit results are shown in Fig.~\ref{fig:massFit},
while Table~\ref{tab:massFit} quotes the component yields and phase-space-integrated raw detection asymmetries in the \Bp\ signal region, which are subsequently used in the Dalitz-plot fit.
The quoted uncertainties account for systematic effects evaluated with the procedures outlined in Section~\ref{sec:systematics}.

\begin{table}[ht]
  \centering
  \caption{Component yields and phase-space-integrated raw detection asymmetries in the \Bp\ signal region, calculated from the results of the invariant-mass fit. The uncertainties include both statistical and systematic effects.}
  \label{tab:massFit}
  \begin{tabular}
    {@{\hspace{0.5cm}}l@{\hspace{0.25cm}}  @{\hspace{0.25cm}}r@{\hspace{0.5cm}}}
    \hline \hline
    Parameter & Value \\ \hline
    Signal yield & $20\,600 \pm 1\,600$\\
    Combinatorial background yield & $4\,400 \pm 1\,600$\\
    \decay{\Bp}{\Kp\pip\pim} background yield & $143 \pm \phantom{0\,0}11$\\ \hline
    Combinatorial background asymmetry & $+0.005 \pm 0.010$\\
    \decay{\Bp}{\Kp\pip\pim} background asymmetry & $\phantom{+}0.000 \pm 0.008$\\    \hline \hline
  \end{tabular}
\end{table}

\section{Dalitz-plot model}
\label{sec:DPformalism}

The \decay{\Bp}{\pi^+_1\pi^+_2\pi^-_3} decay amplitude can be expressed fully in terms of the invariant-mass-squared of two pairs
of the decay products $m^2_{13}$ and $m^2_{23}$.
As no resonances are expected to decay to $\pip\pip$, the squared invariant-masses of the two combinations of oppositely charged pions are used as these two pairs.
Due to Bose symmetry, the amplitude is invariant under exchange of the two like-sign pions, $A(m_{13}^2, m_{23}^2) \equiv A(m_{23}^2, m_{13}^2)$, meaning that the assignments of $\pi^+_1$ and $\pi^+_2$ are arbitrary.

\begin{figure}[tb]
  \centering
  \includegraphics[width=1.0\linewidth]{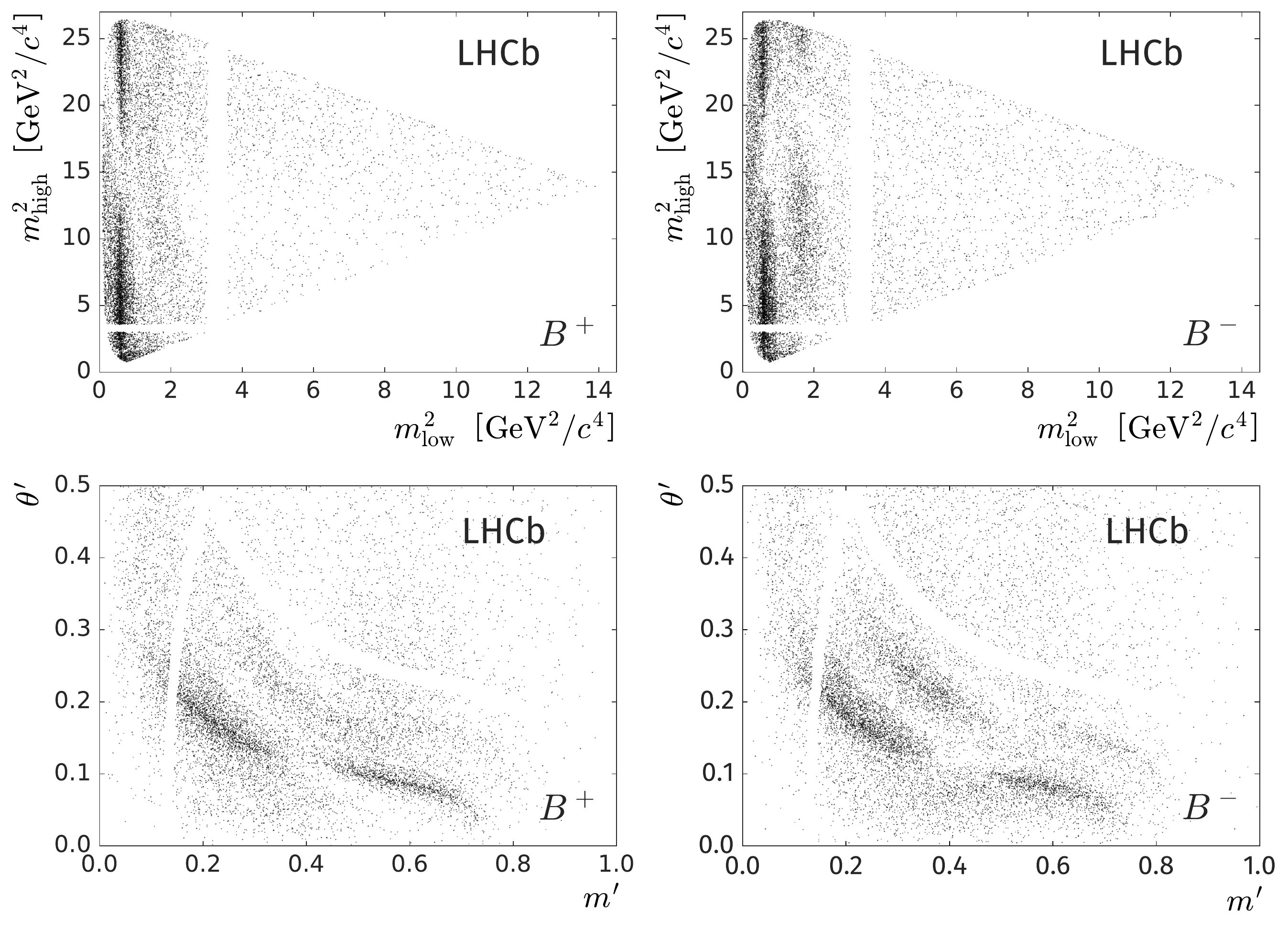}
  \put(-255,298){(a)}
  \put(-30,298){(b)}
  \put(-255,130){(c)}
  \put(-30,130){(d)}

  \caption{
    Conventional Dalitz-plot distributions for (a)~\Bp and (b)~\Bm, and square Dalitz-plot (defined in Section~\ref{sec:sqDP}) distributions for (c)~\Bp and (d)~\Bm candidate decays to $\pipm\pip\pim$.
    Depleted regions are due to the \Dzb veto.
  }
  \label{fig:dataDP}
\end{figure}
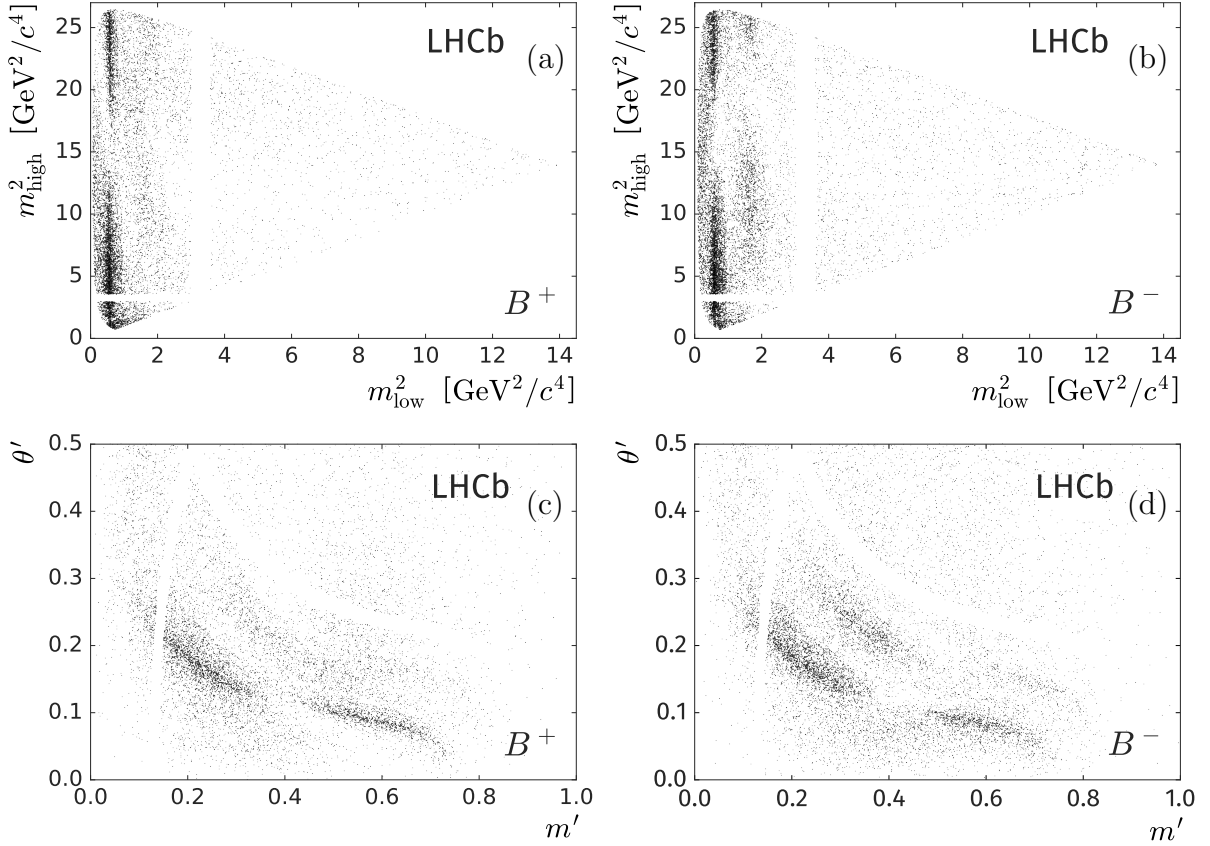

Due to this symmetry, a natural ``folding'' occurs in the Dalitz plot about the axis $m_{13}^2 = m_{23}^2$.
Since the majority of the resonant structure is expected to be at low mass
$m(\pip\pim) < 2\gevcc$, the data and its projections are presented with the two axes being the squares of the low-mass $m_{\rm low}$
and high-mass $m_{\rm high}$ combinations of the opposite-sign pion pairs, for visualisation purposes.
Plots of this kind are therefore similar to those found in other analyses with Dalitz plots that are expected to
contain resonances along only one axis. In this case, structure resulting predominantly from the mass lineshape appears
in $m_{\rm low}$, while $m_{\rm high}$ is influenced by the angular
momentum eigenfunctions. The Dalitz-plot distributions of the selected candidates can be seen in Fig.~\ref{fig:dataDP}(a) and~(b).

\subsection{Amplitude analysis formalism}
\label{sec:DPformalism:amplitude}

In general, the isobar model is used to define the total amplitude for \Bp decays as a coherent sum over $N$ components,
each described by a function $F_j$ that parameterises the intermediate resonant or nonresonant processes:
\begin{equation}
\label{eqn:isobar1}
A^+(m_{13}^2, m_{23}^2) = \sum_j^N A^+_j(m_{13}^2, m_{23}^2) = \sum_j^N c^+_j F_j(m_{13}^2, m_{23}^2)\,,
\end{equation}
and similarly for \Bm:
\begin{equation}
\label{eqn:isobar2}
A^-(m_{13}^2, m_{23}^2) = \sum_j^N A^-_j(m_{13}^2, m_{23}^2) = \sum_j^N c^-_j F_j(m_{13}^2, m_{23}^2)\,,
\end{equation}
where the complex coefficients $c_j$ represent the relative contribution of component $j$.
These are expressed in the ``Cartesian'' \CP-violating convention,
\begin{equation}
  \label{eq:cartesian}
  c_j^{\pm} = (x \pm \delta x) + i(y \pm \delta y)\,,
\end{equation}
for $c_j^+$ ($c_j^-$) coefficients corresponding to \Bp (\Bm) decays.
The function $F$ contains only strong dynamics and, for a resonant or nonresonant contribution to the $m_{13}$ spectrum, is parameterised as
\begin{equation}
F(m_{13}^2, m_{23}^2) \propto R(m_{13}) \cdot T(\vec{p}, \vec{q}) \cdot X(|\vec{p}\,|r_{\rm BW}^P) \cdot X(|\vec{q}\,|r_{\rm BW}^R)\,,
\end{equation}
where $R$ describes the mass lineshape, $T$ describes the angular dependence, and $X$ are Blatt--Weisskopf barrier factors~\cite{blatt-weisskopf} depending on a radius parameter $r_{\rm BW}$.
Here and in the following, the momentum of one of the $m_{13}$ decay products is denoted as $\vec{q}$ and the momentum of the third pion ($\pi_2$) as $\vec{p}$, where both momenta are evaluated in the rest frame of the dipion system.

Using the Zemach tensor formalism~\cite{Zemach:1963bc,Zemach:1968zz}, the angular probability distribution terms
$T(\vec{p},\vec{q})$ are given by
\begin{eqnarray}
\label{eq:ZTFactors}
L = 0 \ : \ T(\vec{p},\vec{q}) & = & \phantom{-}\,1\,, \nonumber \\[4pt]
L = 1 \ : \ T(\vec{p},\vec{q}) & = & -\,2\,\vec{p}\cdot\vec{q}\,,\nonumber \\[4pt]
L = 2 \ : \ T(\vec{p},\vec{q}) & = & \phantom{-}\,\frac{4}{3} \left[3(\vec{p}\cdot\vec{q}\,)^2 - (|\vec{p}\,||\vec{q}\,|)^2\right]\,, \nonumber \\[4pt]
L = 3 \ : \ T(\vec{p},\vec{q}) & = & -\,\frac{24}{15} \left[5(\vec{p}\cdot\vec{q}\,)^3 - 3(\vec{p}\cdot\vec{q}\,)(|\vec{p}\,||\vec{q}\,|)^2\right]\,.
\end{eqnarray}
These are related to the Legendre polynomials $P_L(\cos \theta_{\rm hel})$, where the helicity angle $\theta_{\rm hel}$
is the angle between $\vec{p}$ and $\vec{q}$ and provide a good visual indicator of the spin of the intermediate state.

The Blatt--Weisskopf barrier factors account for the finite size of the decaying hadron and
correct for the unphysical increase in the amplitude above the angular momentum barrier introduced by the form of
angular momentum distributions given in Eq.~\eqref{eq:ZTFactors}.
They are expressed in terms of $z=|\vec{q}\,|r^P_{\rm BW}$ for the \Bp decay and $z=|\vec{p}\,|r^R_{\rm BW}$ for the intermediate state,
\begin{eqnarray}
\label{eq:BWFormFactors}
L = 0 \ : \ X(z) & = & 1\,, \nonumber \\[4pt]
L = 1 \ : \ X(z) & = & \sqrt{\frac{1 + z_0^2}{1 + z^2}}\,, \nonumber \\[4pt]
L = 2 \ : \ X(z) & = & \sqrt{\frac{z_0^4 + 3z_0^2 + 9}{z^4 + 3z^2 + 9}}\,, \nonumber \\[4pt]
L = 3 \ : \ X(z) & = & \sqrt{\frac{z_0^6 + 6z_0^4 + 45z_0^2 + 225}{z^6 + 6z^4 + 45z^2 + 225}}\,,
\end{eqnarray}
where $L$ is the relative angular momentum between the $\Bp$ meson and the resonance, which is equal to the spin of the
resonance since the $\Bp$ meson is spinless. The variable $z_0$ is equal to the value of $z$ when the mass of the propagator
is equal to the mass of the resonance. Unless otherwise stated, the barrier radius of the \Bp meson and the intermediate resonance is taken to be
$r_{\rm BW}^P = r_{\rm BW}^R = 4.0 \gev^{-1}$ ($\approx 0.8 \fm$), for all resonances.
(To simplify expressions, natural units with $c = \hbar = 1$ are used throughout Sections~\ref{sec:DPformalism:amplitude}--\ref{sec:DPformalism:S-wave}.)

\subsubsection{Square Dalitz plot}
\label{sec:sqDP}

Since resonances tend to populate the edges of the conventional Dalitz plot in charmless \B decays, it is useful to define the so-called ``square'' Dalitz plot~\cite{Aubert:2005sk}, which provides improved resolution in these critical regions when using uniform binning, for example when modelling efficiency effects.  Furthermore, the mapping to a square space aligns the bin boundaries to the kinematic boundaries of the phase space. As such, all efficiencies and backgrounds described in Sections~\ref{sec:efficiencies} and~\ref{sec:backgrounds} are determined as a function of the square Dalitz-plot variables, however the data and fit projections in Section~\ref{sec:results} are displayed as a function of the squares of the invariant-mass pairs.

The square Dalitz plot is defined in terms of $m^\prime$ and $\theta^\prime$
\begin{equation}
m^\prime \equiv \frac{1}{\pi} \arccos \left( 2 \, \frac{m(\pip\pip) - m(\pip\pip)^{\rm min}}{m(\pip\pip)^{\rm max} - m(\pip\pip)^{\rm min}} - 1 \right), \ \ \ \ \theta^\prime \equiv \frac{1}{\pi} \theta(\pip\pip),
\end{equation}
where $m(\pip\pip)^{\rm max} = m_{B^+} - m_{\pim}$ and $m(\pip\pip)^{\rm min} = 2 m_{\pip}$ represent the kinematic limits
permitted in the \decay{\Bp}{\pip\pip\pim} decay and $\theta(\pip\pip)$ is the angle between \pip and \pim in the $\pip\pip$ rest frame.
The Bose symmetry of the final state requires that distributions are symmetric with respect to $\theta^\prime = 0.5$. The square Dalitz-plot distributions of the selected candidates can be seen in Fig.~\ref{fig:dataDP}(c) and~(d).

\subsection{Mass lineshapes}

Resonant contributions are mostly described by the relativistic Breit--Wigner lineshape
\begin{equation}
R(m) =\frac{1}{(m_0^2 - m^2) - i m_0 \Gamma(m)} \,,
\label{BW}
\end{equation}
with a mass-dependent decay width
\begin{equation}
\Gamma(m) = \Gamma_0 \left( \frac{q}{q_0} \right)^{2L + 1} \left( \frac{m_0}{m} \right) X^2(qr^{R}_{\rm BW}) \,,
\end{equation}
where $q_0$ is the value of $q = |\vec{q}\,|$ when the invariant mass, $m$, is equal to the pole mass, $m_0$, of the resonance. Here, the nominal resonance width is given by $\Gamma_0$.

For the broad $\rho(770)^0$ resonance, an analytic dispersive term is included to ensure unitarity far from the pole mass, known as the Gounaris--Sakurai model~\cite{GS}. It takes the form
\begin{equation}
\label{eq:GS}
R(m) = \frac{1+D\,\Gamma_0/m_0}
                {(m_0^2 - m^2) + f(m) - i\, m_0 \Gamma(m)} \,,
\end{equation}
with an additional mass dependence
\begin{equation}
\label{eq:GSfm}
f(m) = \Gamma_0 \,\frac{m_0^2}{q_0^3}\,
       \left[\;
             q^2 \left[h(m)-h(m_0)\right] +
             \left(\,m_0^2-m^2\,\right)\,q^2_0\,
             \frac{{\rm d}h}{{\rm d}m^2}\bigg|_{m_0}
       \;\right] \,,
\end{equation}
where
\begin{equation}
\label{eq:GShm}
h(m) = \frac{2}{\pi}\,\frac{q}{m}\,
       \log\left(\frac{m+2q}{2m_\pi}\right) \,,
\end{equation}
and
\begin{equation}
\label{eq:GSdh}
\frac{{\rm d}h}{{\rm d}m^2}\bigg|_{m_0} =
h(m_0)\left[(8q_0^2)^{-1}-(2m_0^2)^{-1}\right] \,+\, (2\pi m_0^2)^{-1} \,.
\end{equation}
The constant parameter $D$ is given by
\begin{equation}
\label{eq:GSd}
D = \frac{3}{\pi}\frac{m_\pi^2}{q_0^2}\,
    \log\left(\frac{m_0+2q_0}{2m_\pi}\right)
    + \frac{m_0}{2\pi\,q_0}
    - \frac{m_\pi^2 m_0}{\pi\,q_0^3} \,.
\end{equation}

Isospin-violating $\omega(782)$ decays to two charged pions can occur via $\omega(782)$ mixing with the $\rho(770)^0$ state.
To account for such effects, a model is constructed that directly parameterises the interference between these two contributions
following Ref.~\cite{Rensing:259802},
\begin{equation}
\label{eqn:rho-omegamixing}
R_{\rho\omega}(m) = R_{\rho}(m) \left[ \frac{1 + R_{\omega}(m) \,\Delta\, |\zeta| \exp(i \phi_{\zeta})}{1 - \Delta^2 \, R_{\rho}(m) \, R_{\omega}(m)} \right]\,,
\end{equation}
where $R_{\rho}(m)$ is the Gounaris--Sakurai $\rho(770)^0$ lineshape, $R_{\omega}(m)$ is the relativistic Breit--Wigner
$\omega(782)$ lineshape, $|\zeta|$ and $\phi_{\zeta}$ are free parameters of the fit that denote the respective magnitude and phase of the production amplitude of $\omega(782)$ with respect to that for the $\rho(770)^0$ state,
and $\Delta \equiv \delta \left(m_{\rho} + m_{\omega}\right)$, where $\delta$ governs the electromagnetic mixing of $\rho(770)^0$ and $\omega(782)$ and $m_{\rho}$ and $m_{\omega}$ represent the known particle masses~\cite{PDG2018}.
This is equivalent to the parameterisation described in Ref.~\cite{Akhmetshin:2001ig} if the small $\Delta^2$ term in the denominator is ignored.
The value for $\delta$ is fixed in the fit to $\delta = 0.00215 \pm 0.00035 \gev$~\cite{Rensing:259802}.

\subsection{S-wave models}
\label{sec:DPformalism:S-wave}

The S-wave ($L = 0$) component of the \decay{\Bp}{\pip\pip\pim} amplitude is both large in
magnitude and contains many overlapping resonances and decay channel thresholds, \ie, where increasing two-body
invariant mass opens additional decay channels that were previously inaccessible, thereby modulating the observed intensity.
This analysis includes three distinct treatments of the \mbox{S-wave} component in \decay{\Bp}{\pip\pip\pim} in an attempt to
better understand its behaviour. They also increase confidence that parameters reported for the non-S-wave contributions
are robust and provide additional information for further study.

As such, three sets of results are presented here, corresponding to the cases where the $\pip\pim$ S-wave is described by:
(i) a coherent sum of specific resonant contributions (isobar);
(ii) a monolithic, unitarity-preserving model informed by historical scattering data \mbox{(K-matrix);}
and (iii) a quasi-model-independent binned approach (QMI).
All approaches contain identical contributions to higher partial waves, where $L > 0$.

\subsubsection{Isobar model}
\label{sec:Isobar}
The isobar model S-wave amplitude is represented by the coherent sum of contributions from the $\sigma$, or $f_0(500)$, meson and
a $\pi \pi \leftrightarrow \kaon\Kbar$ rescattering amplitude within the mass range \mbox{$1.0 < m(\pip\pim) < 1.5\gev$}.
The $\sigma$ meson is represented as a simple pole~\cite{cpole, Pelaez:2015qba}, parameterised as
\begin{equation}
A_{\sigma}(m) = \frac{1}{s_\sigma - m^2}\,,
\end{equation}
with $s_\sigma$ the square of the pole position  $\sqrt{s_\sigma} =m_\sigma -\,i\,\Gamma_\sigma$, extracted from the fit.

The concept of $\pi\pi \leftrightarrow \kaon\Kbar$ rescattering was originally developed inside the context of two-body interactions. For three-body decays,
rescattering means that a pair of mesons produced in one channel will appear in the final state of a coupled channel.
Therefore, a model is used that describes the source of the rescattering~\cite{IgnacioCPT, Nogueira:2015tsa},
\begin{equation}
  \label{eq:isobar:Asource}
A_{\rm{source}}(m) = [1 + (m/\Delta^2_{\pi\pi})]^{-1}[1 + (m/\Delta^2_{KK})]^{-1} \, ,
\end{equation}
where $\Delta^2_{\pi\pi}=\Delta^2_{KK} = 1\gev$.
The total rescattering amplitude in the three-body \Bp decay is then
\begin{equation}
A_{\rm{scatt}}(m) = A_{\rm{source}}(m)\,f_{\rm{rescatt}}(m)\,.
\end{equation}
The amplitude $f_{\rm{rescatt}}(m)= \sqrt{1-\eta(m)^2}e^{2i\delta(m)}$ is described in terms of the inelasticity, $\eta(m)$, and a phase shift, $\delta(m)$.
Functional forms of these are used that combine constraints from unitarity and analyticity with dispersion relation techniques~\cite{pelaez2005}.
The inelasticity is described by
\begin{equation}
\eta(m)=1-\left(\epsilon_1\dfrac{k_2(m)}{m}+\epsilon_2\dfrac{k_2^2(m)}{m^2}\right)\,\dfrac{M^{\prime 2}-m^2}{m^2}\,,
\end{equation}
with
\begin{equation}
k_2(m)=\frac{1}{2}\sqrt{m^2-4m^2_K}\,,
\end{equation}
where $m_K = 0.494\gev$ is the charged kaon mass, $\epsilon_1 = 2.4$,  $\epsilon_2 = -5.5$,
and $M^\prime = 1.5\gev$.
The phase shift is given by
\begin{equation}
\cot\delta(m)=c_0\,\dfrac{(m^2-M^2_s)(M^2_{f}-m^2)}{M^2_{f} m}\,\dfrac{|k_2(m)|}{k_2^2(m)} \,,
\end{equation}
where $M_f = 1.32\gev$, $c_0 = 1.3$  and $M_s = 0.92\gev$.
Except for $m_K$, values of all parameters are taken from Ref.~\cite{pelaez2005}.

\subsubsection{K-matrix model}
\label{sec:Kmatrix}

The coherent sum of resonant contributions modelled with Breit--Wigner lineshapes can be used to describe the dynamics of three-body decays when the quasi-two-body resonances are relatively narrow and isolated.
However, when there are broad, overlapping resonances (with the same isospin and spin-parity quantum numbers) or structures that are near open decay channels, this
model does not satisfy $S$-matrix unitarity, thereby violating the conservation of quantum mechanical probability current.

Assuming that the dynamics is dominated by two-body processes (\ie\ that the \mbox{S-wave} does not interact with other decay products in the final state),
then two-body unitarity is naturally conserved within the K-matrix approach~\cite{Chung:1995dx}.
This approach was originally developed for two-body scattering~\cite{Dalitz:1960du} and the study of resonances in nuclear reactions~\cite{Wigner:1946zz,Wigner:1947zz}, but was extended
to describe resonance production and $n$-body decays in a more general way~\cite{Aitchison:1972ay}.
This section provides a brief introduction to the K-matrix approach as applied in this analysis; for more detail
see Ref.~\cite{Back:2017zqt}.

From unitarity conservation, the form-factor for two-body production is related to the scattering amplitude
for the same channel, when including all coupled channels. In this way, the K-matrix model describes the amplitude, $F_u$,
of a channel $u$ in terms of the initial $\hat{P}$-vector preparation of channel states $v$,
that has the same form as $\hat{K}$, ``scattering'' into the final state $u$ via the propagator term $(\hat{I} - i \hat{K} \hat{\rho})^{-1}$,
\begin{equation}
\label{eq:KMatProd}
F_u = \sum_{v=1}^{n} [(\hat{I} - i \hat{K} \hat{\rho})^{-1}]_{uv} \, \hat{P}_{v} \,.
\end{equation}
The diagonal matrix $\hat{\rho}$ accounts for phase space, where the element for the two-body channel $u$ is given by~\cite{PDG2018}
\begin{equation}
\label{eq:rhoterm}
\hat{\rho}_{uu} = \sqrt{\left(1 - \frac{(m_{1} + m_{2})^2}{s}\right)
\left(1 - \frac{(m_{1} - m_{2})^2}{s}\right)} \,,
\end{equation}
where $s = m_{\pi^+\pi^-}^2$ and $m_{1}$ and $m_{2}$ are the rest masses of the two decay products.
This expression is analytically continued by setting $\hat{\rho}_{uu}$ to be $i|\hat{\rho}_{uu}|$ when the channel is below
its mass threshold (provided it does not cross into another channel). For the coupled multi-meson final states,
the corresponding expression can be found in Ref.~\cite{Back:2017zqt}.

The scattering matrix, $\hat{K}$, can be parameterised as a combination of the sum
of $N_p$ poles with real bare masses $m_{\alpha}$, together with nonresonant ``slowly-varying'' parts
(SVPs). These slowly-varying parts are so-called as they have a $1/s$ dependence, and incorporate real coupling constants,
$f^{\rm{scatt}}_{uv}$~\cite{Anisovich:2002ij}. The scattering matrix is symmetric in $u$ and $v$,
\begin{equation}
\label{eq:KMatTerms}
\hat{K}_{uv}(s) = \left( \sum_{\alpha=1}^{N_p}{\frac{g^{\alpha}_u g^{\alpha}_v}{m^2_{\alpha} - s}} +
f^{\rm{scatt}}_{uv} \frac{m^2_0 + s^{\rm{scatt}}_0}{s + s^{\rm{scatt}}_0} \right) f_{\!A0}(s) \,,
\end{equation}
where $g_u^{\alpha}$ and $g_v^{\alpha}$ denote the real coupling constants of the pole $m_{\alpha}$ to the
channels $u$ and $v$, respectively. The factor
\begin{equation}
\label{eq:adler}
f_{\!A0}(s) = \left(\frac{1 \gev^2 - s_{\!A0}}{s - s_{\!A0}}\right) \left(s - \frac{1}{2} m_{\pi}^2\right)
\end{equation}
is the Adler zero term, which suppresses the false kinematic singularity due to left-hand cuts when $s$ goes
below the $\pi\pi$ production threshold~\cite{Adler:1965ga}. The parameters $m_0^2$, $s^{\rm{scatt}}_0$ and $s_{\!A0}$ are real constants.
Note that the masses $m_{\alpha}$ are those of the poles, or the so-called \emph{bare} states of the system, which do not correspond to the masses and widths of \emph{resonances} -- mixtures of bare states.

Extension to three-body decays is achieved by fitting for the complex coefficients $\beta_{\alpha}$ and $f^{\rm{prod}}_{v}$ of the
production pole and SVP terms in the production vector, $\hat{P}$,
\begin{equation}
\label{eq:Pvector}
  \hat{P}_v = \sum_{\alpha=1}^{N_p} \frac{\beta_\alpha g_v^{\alpha}}{m_\alpha^2 - s} + \frac{m_0^2 + s_0^{\rm{prod}}}{s + s_0^{\rm{prod}}} f^{\rm{prod}}_{v}\, ,
\end{equation}
where these coefficients are different for \Bp and \Bm decays to allow for \CP violation.
The parameter $s_0^{\rm{prod}}$ is dependent on the production environment and is taken from Ref.~\cite{Link:2003gb}.

Using the above expressions, the amplitude for each production pole $\alpha$ to the $\pi\pi$ channel (denoted by the subscript $u=1$) is given by
\begin{equation}
\label{eq:prodPole}
A_{\alpha}(s) = \frac{\beta_{\alpha}}{m^2_{\alpha} - s} \sum_{v=1}^{n} \, [(\hat{I} - i \hat{K} \hat{\rho})^{-1}]_{1v} \, g^{\alpha}_v \,,
\end{equation}
where the sum is over the intermediate channels, $v$, while the SVP production amplitudes are separated out for each individual channel as
\begin{equation}
\label{eq:prodSVP}
A_{{\rm{SVP}},v}(s) = \frac{m_0^2 + s_0^{\rm{prod}}}{s + s_0^{\rm{prod}}} \,
[(\hat{I} - i \hat{K} \hat{\rho})^{-1}]_{1v} \, f^{\rm{prod}}_{v} \,.
\end{equation}
All of these contributions are then summed to give the total S-wave amplitude
\begin{equation}
\label{eq:prodPoleSVP}
F_1 = \sum_{\alpha=1}^{N_p} A_{\alpha} + \sum_{v=1}^{n}A_{{\rm{SVP}},v} \,.
\end{equation}

The $\hat{K}$ matrix elements in Eq.~\eqref{eq:KMatTerms} are completely defined
using the values quoted in Ref.~\cite{Aubert:2008bd} and given in Table~\ref{table:kmatrixParameters}
(for five channels $n=5$ and five poles $N_p=5$),
which are obtained from a global analysis of $\pi\pi$ scattering data~\cite{Anisovich:2002ij}.

\renewcommand{\arraystretch}{1.35}
\begin{table}[!bt]
\centering
\caption{\small
  K-matrix parameters quoted in Ref.~\cite{Aubert:2008bd}, which are obtained from a global analysis of $\pi\pi$ scattering data~\cite{Anisovich:2002ij}.
  Only $f_{1v}$ parameters are listed here, since only the dipion final state is relevant to the analysis. Masses $m_\alpha$ and couplings $g_u^{\alpha}$ are given in $\!\gev$, while units of $\!\gev^2$ for $s$-related quantities are implied; $s^{\rm{prod}}_0$ is taken from Ref.~\cite{Link:2003gb}.}
  \label{table:kmatrixParameters}
  \begin{tabular}
          {@{\hspace{0.5cm}}c@{\hspace{0.25cm}} @{\hspace{0.25cm}}c@{\hspace{0.25cm}}  @{\hspace{0.25cm}}c@{\hspace{0.25cm}}  @{\hspace{0.25cm}}c@{\hspace{0.25cm}}  @{\hspace{0.25cm}}c@{\hspace{0.25cm}}  @{\hspace{0.25cm}}c@{\hspace{0.25cm}}  @{\hspace{0.25cm}}c@{\hspace{0.5cm}}}
\hline\hline

$\alpha$ & $m_{\alpha}$ & $g^{\alpha}_1 [\pi\pi]$ & $g^{\alpha}_2 [\kaon\Kbar]$
& $g^{\alpha}_3 [4\pi]$ & $g^{\alpha}_4 [\eta\eta]$ & $g^{\alpha}_5 [\eta\eta']$ \\
\hline
$1$ & $0.65100$ & $0.22889$ &           $-0.55377$  &  \phantom{$-$}$0.00000$ &            $-0.39899$ &            $-0.34639$ \\
$2$ & $1.20360$ & $0.94128$ & \phantom{$-$}$0.55095$  &  \phantom{$-$}$0.00000$ &  \phantom{$-$}$0.39065$ &  \phantom{$-$}$0.31503$ \\
$3$ & $1.55817$ & $0.36856$ & \phantom{$-$}$0.23888$  &  \phantom{$-$}$0.55639$ &  \phantom{$-$}$0.18340$ &  \phantom{$-$}$0.18681$ \\
$4$ & $1.21000$ & $0.33650$ & \phantom{$-$}$0.40907$  &  \phantom{$-$}$0.85679$ &  \phantom{$-$}$0.19906$ &            $-0.00984$ \\
$5$ & $1.82206$ & $0.18171$ &           $-0.17558$  &            $-0.79658$ &            $-0.00355$ &  \phantom{$-$}$0.22358$ \\[15pt]

& $s_0^{\rm{scatt}}$ & $f_{11}^{\rm{scatt}}$ & $f_{12}^{\rm{scatt}}$ & $f_{13}^{\rm{scatt}}$ & $f_{14}^{\rm{scatt}}$ & $f_{15}^{\rm{scatt}}$ \\
\cmidrule{2-7}
& $3.92637$      & $0.23399$ & \phantom{$-$}$0.15044$ & $-0.20545$ & \phantom{$-$}$0.32825$ & \phantom{$-$}$0.35412$
\\[6pt]
& $s_0^{\rm{prod}}$ & $m_0^2$ & $s_{\!A0}$ & & \\
\cmidrule{2-4}
& $3.0$          & $1.0$  & $-0.15$ & &\\

\hline\hline
\end{tabular}
\end{table}

\subsubsection{Quasi-model-independent analysis}
\label{sec:qmi}

In the quasi-model-independent (QMI) analysis, the amplitude for the $\pi\pi$ S-wave is described by individual magnitudes and phases within each of $17$ bins in $m(\pip\pim)$.
The QMI approach exploits the fact that the S-wave amplitude is constant in $\cos \theta_{\rm hel}$ (Eq.~\eqref{eq:ZTFactors}) to disentangle this component from other contributions to the Dalitz plot, assuming the higher-order waves to be well modelled by the isobar approach.

The bins are defined identically for \Bp and \Bm: $13$ below the charm veto and $4$ above.
The binning scheme is chosen {\it ad hoc} as optimisation requires {\it a priori} knowledge of the S-wave and total \decay{\Bp}{\pip\pip\pim} amplitude model.
The bin boundaries are chosen adaptively, by requiring initially roughly equal numbers of candidates in each region and using the isobar model subsequently to tune the bins in order to reduce intrinsic bias in the method's ability to reproduce a known S-wave.

Similar QMI approaches have previously been performed in amplitude analyses of various decays of $B$ and $D$ mesons to study the $K\pi$~\cite{Aitala:2005yh,Bonvicini:2008jw,Link:2009ng,delAmoSanchez:2010fd}, $\pi\pi$~\cite{Aubert:2008ao} and $D\pi$~\cite{LHCb-PAPER-2016-026} S-waves.
In contrast to these previous approaches, no interpolation between bins is performed in this analysis --- the amplitudes are constant within each bin.
This choice is made as interpolation is based on a premise of smoothness, which is appropriate for goals such as confirmation of a new resonance.
However, interpolation is not optimal in the description of the $\pi\pi$ S-wave due to the opening of various decay channels that become kinematically allowed as the mass increases, and which cause sharp changes in the amplitude on scales less than the bin width.
A key difference between this and the other two S-wave approaches is that the QMI absorbs the average contribution from potential interactions with the other decay products in the final state.
These final state interactions may be quantified by comparing with similar dipion S-wave distributions obtained by rescattering experiments.

\subsection{Measurement quantities}
The primary outputs of the Dalitz-plot fit are the complex isobar coefficients $c_j^\pm$ defined in Eq.~\eqref{eqn:isobar1}.
However, since these depend on the choice of phase convention, amplitude formalism and normalisation, they are not
straightforwardly comparable between analyses and have limited physical meaning. Instead, it is useful to compare fit fractions,
defined as the integral of the absolute value of the amplitude squared for each intermediate component, $j$, divided by that
of the coherent matrix-element squared for all intermediate contributions,
\begin{equation}
  {\cal F}^\pm_j = \frac{ \int_{\rm DP} |A^\pm_j(m_{13}^2, m_{23}^2)|^2~{\rm d}m_{13}^2~{\rm d}m_{23}^2 }
{\int_{\rm DP} |A^{\pm}(m_{13}^2, m_{23}^2)|^2~{\rm d}m_{13}^2~{\rm d}m_{23}^2}\,.
\label{equation:fitFracs}
\end{equation}
These fit fractions will not sum to unity if there is net constructive or destructive interference. The
interference fit fractions are given by
\begin{equation}
    {\cal I}^\pm_{i\,>\,j} = \frac{\int_{\rm DP} 2{\rm Re} [A_i^\pm(m_{13}^2, m_{23}^2)A_j^{\pm *}(m_{13}^2, m_{23}^2)]~{\rm d}m_{13}^2~{\rm d}m_{23}^2 }{\int_{\rm DP} |A^{\pm}(m_{13}^2, m_{23}^2)|^2~{\rm d}m_{13}^2~{\rm d}m_{23}^2}\,,
    \label{equation:intFracs}
\end{equation}
where the sum of the fit fractions and interference fit fractions must be unity.
The fit fractions are defined for the \Bp and \Bm decay amplitudes separately, however a measure of the
\CP-averaged contribution can be obtained from the \CP-conserving fit fraction
\begin{equation}
{\cal F}_{j} = \frac{ \int_{\rm DP} \left( |A^+_j(m_{13}^2, m_{23}^2)|^2 + |A^-_j(m_{13}^2, m_{23}^2)|^2\right)~{\rm d}m_{13}^2~{\rm d}m_{23}^2 }{\int_{\rm DP} \left( |A^+(m_{13}^2, m_{23}^2)|^2 + |A^-(m_{13}^2, m_{23}^2)|^2\right)~{\rm d}m_{13}^2~{\rm d}m_{23}^2}\,.
\end{equation}

Since the $\rho(770)^0$ and $\omega(782)$ components cannot be completely decoupled in the mixing amplitude, their effective lineshapes are defined for the purpose of calculating fit fractions.
This is achieved by separating Eq.~\eqref{eqn:rho-omegamixing} into the two respective terms with a common denominator.

Another important physical quantity is the quasi-two-body parameter of \CP violation in decay associated with a particular intermediate contribution
\begin{equation}
\label{eq:cpAsy}
\mathcal{A}_{\CP}^j = \frac{|c^-_j|^2 - |c^+_j|^2}{|c^-_j|^2 + |c^+_j|^2}\,.
\end{equation}
The \CP asymmetry associated with the S-wave cannot be determined using Eq.~\eqref{eq:cpAsy} since this component involves several contributions.
Instead, it is determined as the asymmetry of the relevant \Bm\ and \Bp\ decay rates,
\begin{equation}
    \mathcal{A}_{\CP}^{\rm S} = \frac{\int_{\rm DP} |A^-_{\rm S}(m_{13}^2, m_{23}^2)|^2~{\rm d}m_{13}^2~{\rm d}m_{23}^2 - \int_{\rm DP} |A^+_{\rm S}(m_{13}^2, m_{23}^2)|^2~{\rm d}m_{13}^2~{\rm d}m_{23}^2}{\int_{\rm DP} |A^-_{\rm S}(m_{13}^2, m_{23}^2)|^2~{\rm d}m_{13}^2~{\rm d}m_{23}^2 + \int_{\rm DP} |A^+_{\rm S}(m_{13}^2, m_{23}^2)|^2~{\rm d}m_{13}^2~{\rm d}m_{23}^2} \, ,
\end{equation}
where $A^\pm_{\rm S}$ is the coherent sum of all contributions to the S-wave.

\subsection{Efficiency model}
\label{sec:efficiencies}

The efficiency of selecting a signal decay is parameterised in the two-dimensional square Dalitz plot and determined separately for \Bp and \Bm decays.
Non-uniformities in these distributions arise as a result of the detector geometry, reconstruction and trigger algorithms, particle identification selections and other background rejection requirements such as that imposed on the boosted decision-tree classifier to discriminate against combinatorial background.
The efficiency map is primarily obtained using simulated decays, however effects arising from the hardware trigger and particle identification efficiency are determined using data calibration samples.

The hardware-trigger efficiency correction is calculated using pions from \decay{\Dz}{\Km\pip} decays, arising from promptly produced \decay{\Dstarp}{\Dz(\Km\pip)\pip} decays, and affects two disjoint subsets of the selected candidates: those where the trigger requirements were satisfied by hadronic calorimeter deposits as a result of the signal decay and those where the requirements were satisfied only by deposits from the rest of the event.
In the first case, the probability to satisfy the trigger requirements is calculated using calibration data as a function of the transverse energy of each final-state particle of a given species, the dipole-magnet polarity, and the hadronic calorimeter region.
In the second subset, a smaller correction is applied following the same procedure in order to account for the requirement that these tracks did not fire the hardware trigger.
These corrections are combined according to the relative abundance of each category in data.

\begin{figure}[tb]
\centering
\includegraphics[width=0.49\linewidth]{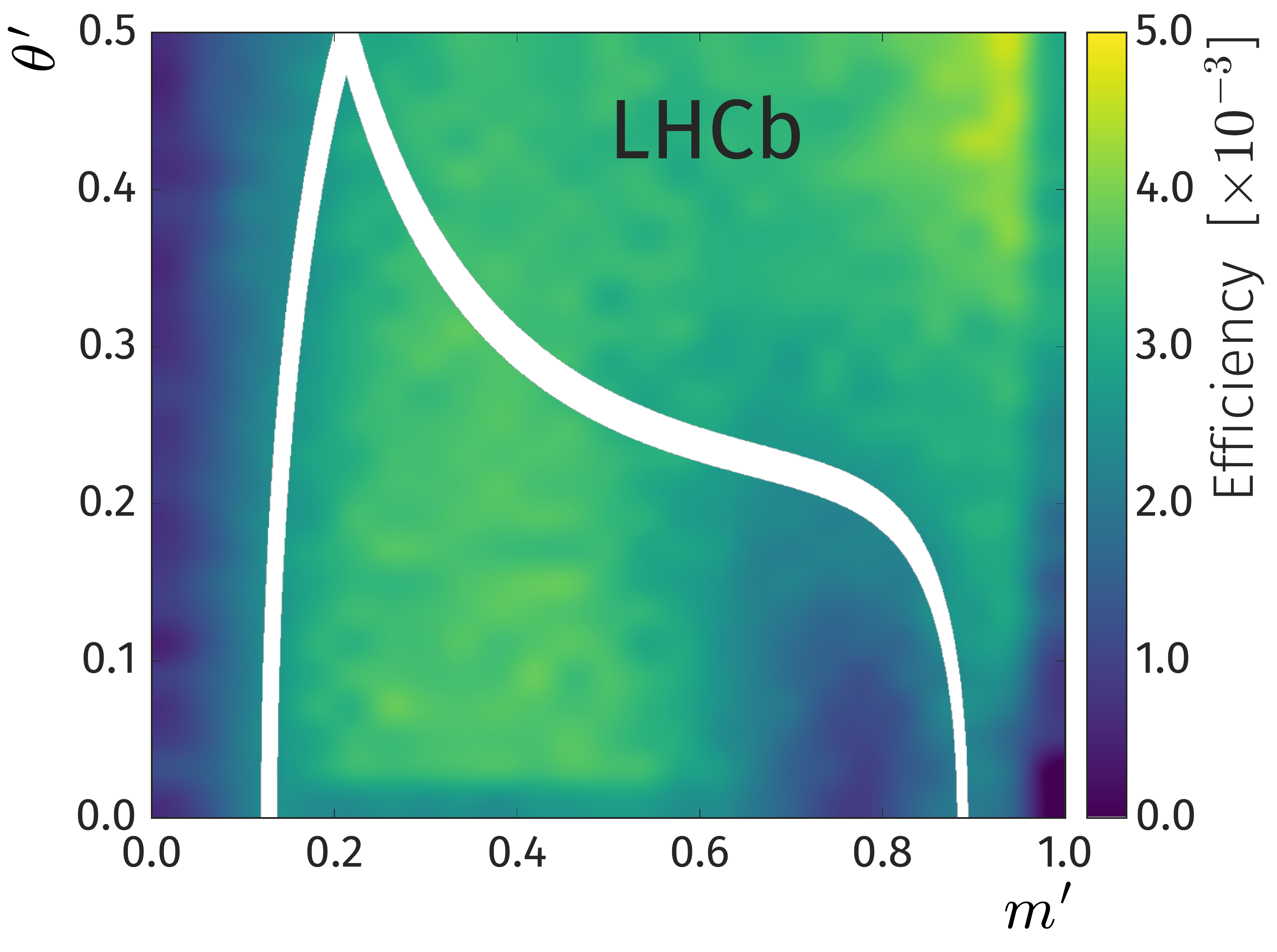}
\includegraphics[width=0.49\linewidth]{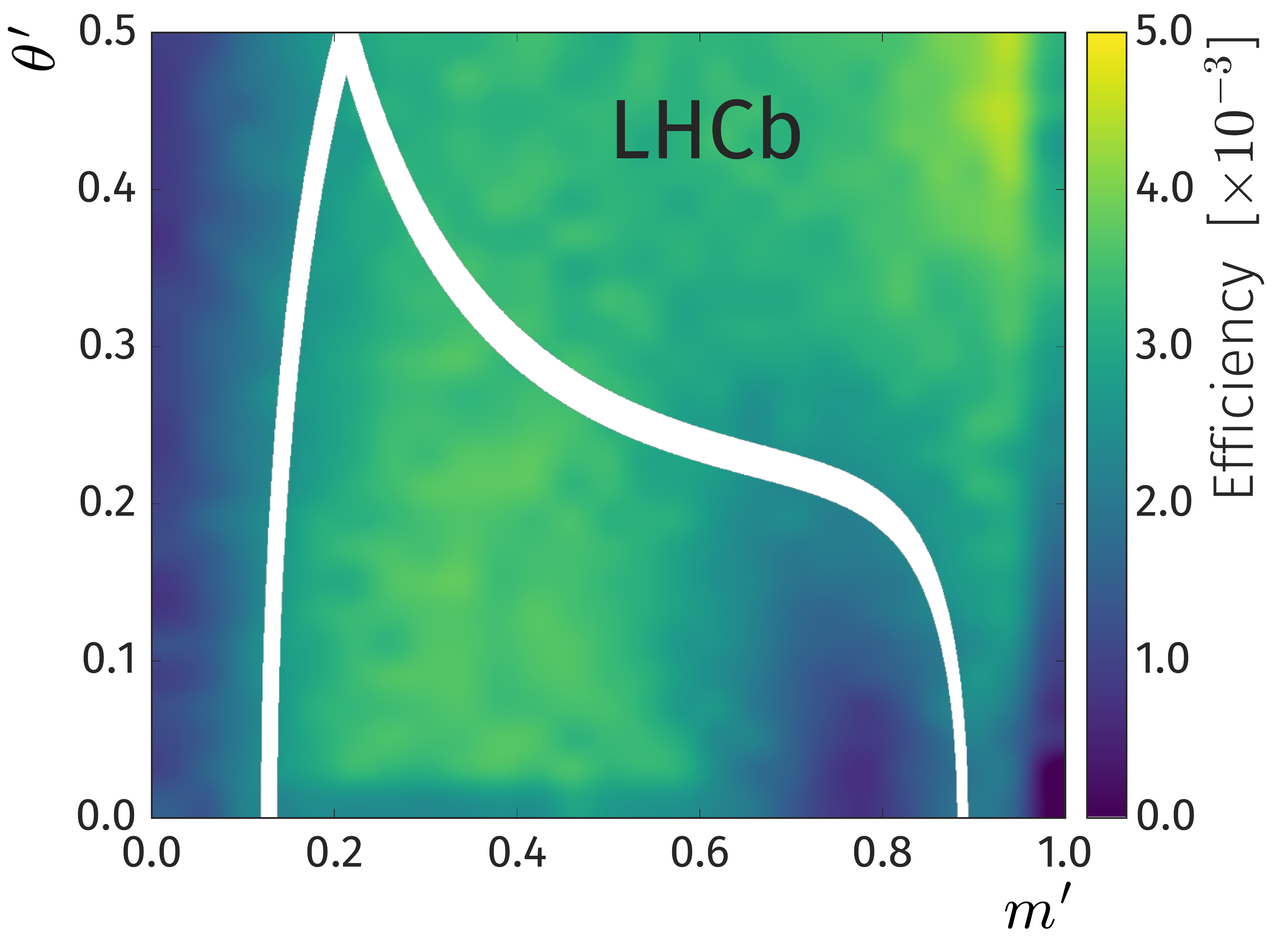}
\caption{Square Dalitz-plot distributions for the (left) \Bp and (right) \Bm signal efficiency models, smoothed using a two-dimensional cubic spline. Depleted regions are due to the \Dzb veto.}
\label{fig:effSDP}
\end{figure}

The particle identification efficiency is calculated from calibration data also corresponding to the
\decay{\Dstarp}{\Dz(\Km\pip)\pip} decay, where pions and kaons can be identified without the use
of the LHCb particle identification system~\cite{LHCb-PUB-2016-021}. The particle identification efficiencies
for the background-subtracted pions and kaons are parameterised in terms of their total and transverse momentum,
and the number of tracks in the event.
This efficiency is assumed to factorise with respect to the final-state tracks and therefore the efficiency for each track is multiplied to form the overall efficiency.

The effect of an asymmetry between the production rates of $\Bm$ and $\Bp$ mesons is indistinguishable in this analysis from a global detection efficiency asymmetry.
Therefore, the \Bp production asymmetry, as measured within the LHCb acceptance~\cite{LHCb-PAPER-2016-062}, is taken into
account by introducing a global asymmetry of approximately $-0.6\%$ into the efficiency maps. This is obtained as an average of the measured production asymmetries, weighted by the relative integrated luminosity obtained in 2011 and 2012.
The overall efficiency, as a function of square Dalitz-plot position, can be seen in Fig.~\ref{fig:effSDP} for \Bp and \Bm decays separately.
These histograms are smoothed by a two-dimensional cubic spline to mitigate effects of discontinuity at the bin edges, with bins abutting kinematic boundaries repeated to ensure good behaviour at the edge of the phase space.
The signal PDF for \Bp or \Bm decays is then given by
\begin{equation}
  {\cal P}^{\pm}_{\rm sig}(m^2_{13}, m^2_{23}) = \frac{\epsilon^{\pm}(m^\prime, \theta^\prime) |A^{\pm}(m^2_{13}, m^2_{23})|^2}{\int_{\rm DP} \epsilon^{\pm}(m^\prime, \theta^\prime) |A^{\pm}(m^2_{13}, m^2_{23})|^2~{\rm d}m^2_{13} {\rm d}m^2_{23}},
\end{equation}
where $\epsilon^\pm$ represents the Dalitz-plot dependent efficiency for the $B^\pm$ decay.

\subsection{Background model}
\label{sec:backgrounds}

The dominant source of background in the signal region is combinatorial in nature. In the Dalitz-plot fit, the distribution of this background is modelled separately for \Bp and \Bm decays using square Dalitz-plot histograms of upper sideband data, from the region $5.35 < m(\pip\pip\pim) < 5.68\gevcc$, with a uniform $16\times16$ binning in $m^\prime$ and $\theta^\prime$. In this region, a feature is observed that can be identified as arising from real \decay{\Bz}{\pip\pim} decays combined with a random track from the rest of the event. However, this background does not enter the signal region due to kinematics.
As such, this feature is modelled in the $m(\pip\pim)$ spectrum using a Gaussian function located at the known \Bz mass, and events are subtracted from the combinatorial background histograms accordingly.

\begin{figure}[tb]
\centering
\includegraphics[width=0.49\linewidth]{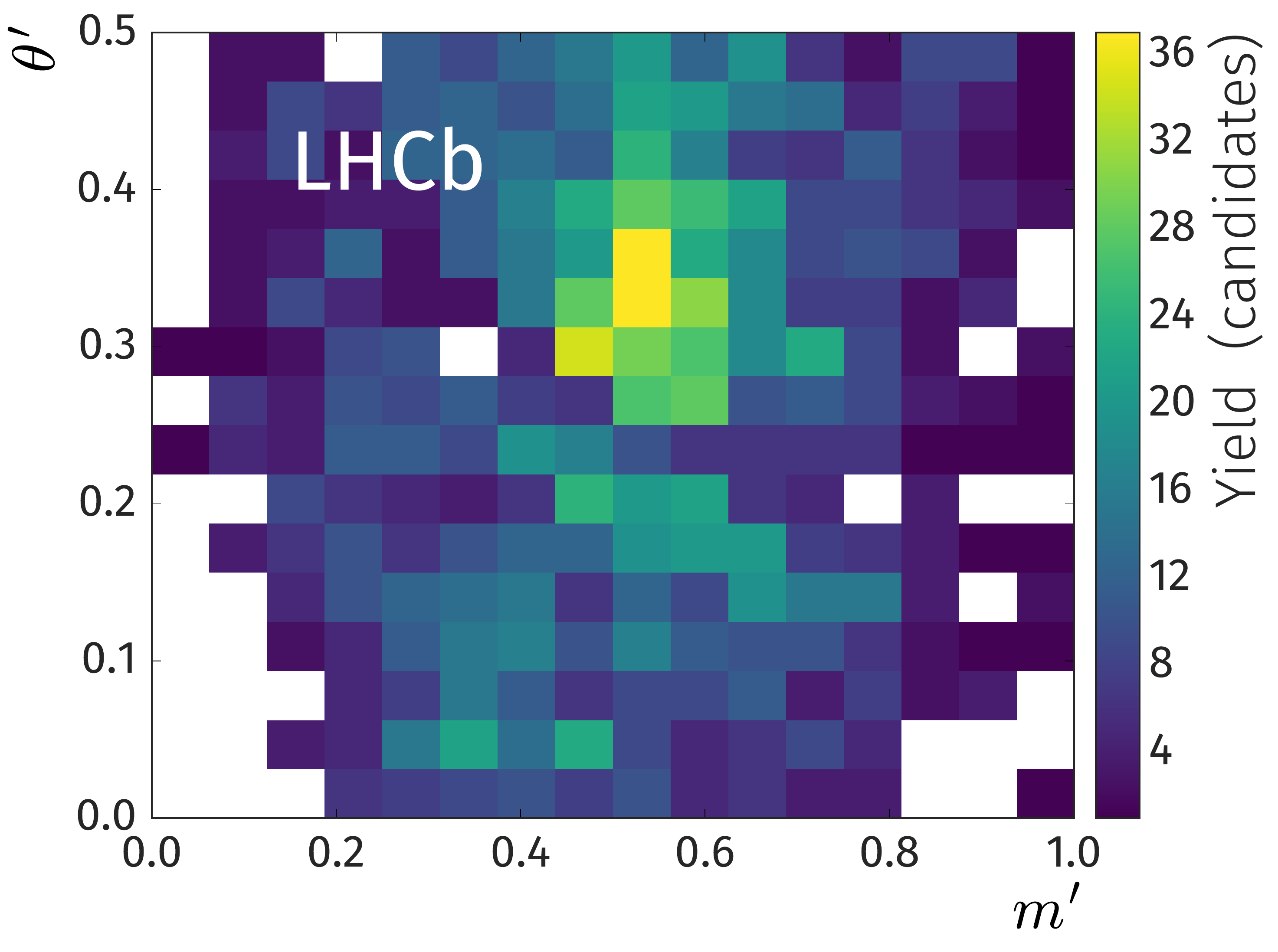}
\includegraphics[width=0.49\linewidth]{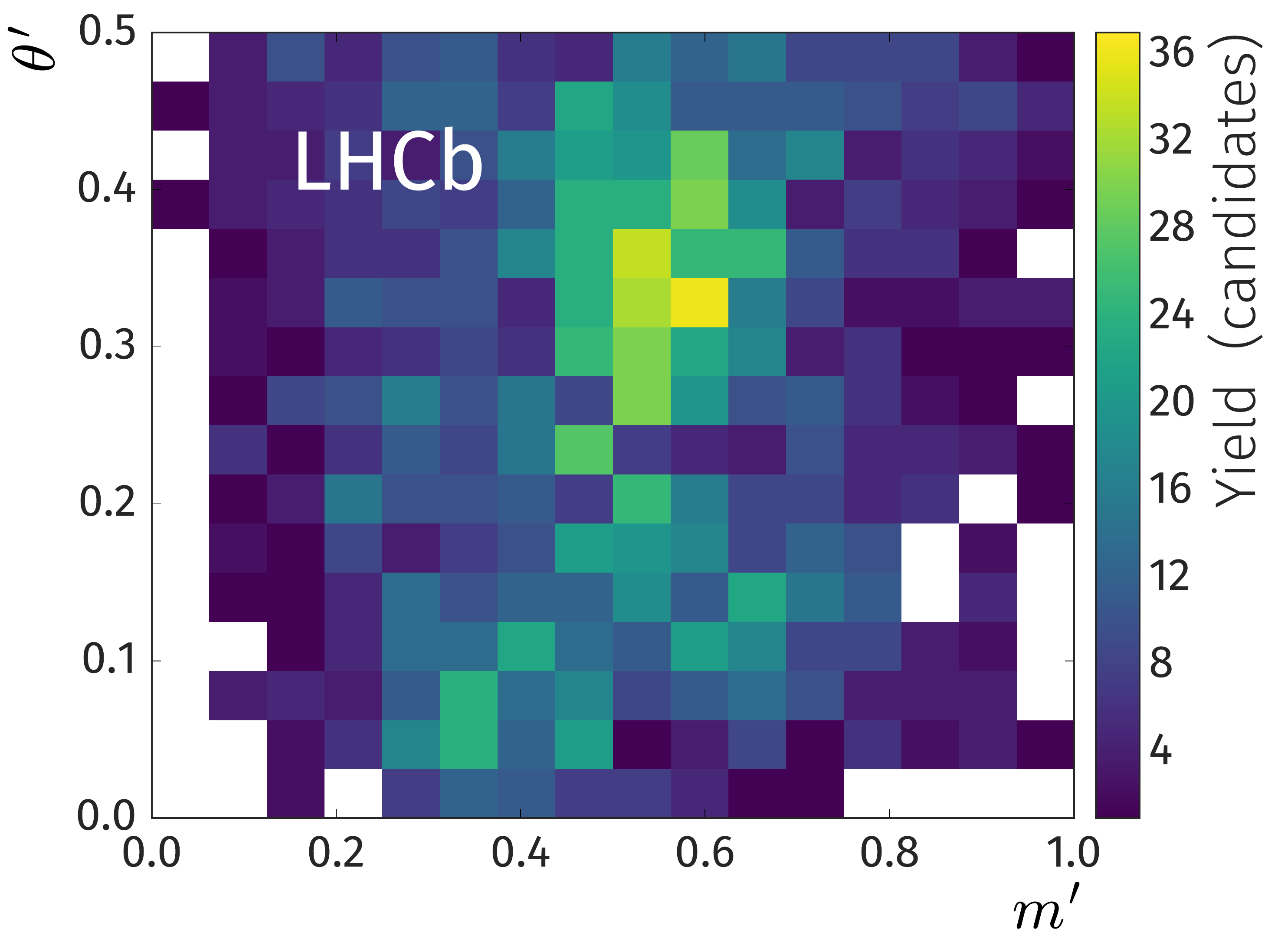}
\caption{Square Dalitz-plot distributions for the (left) \Bp and (right) \Bm combinatorial background models,
scaled to represent their respective yields in the signal region.}
\label{fig:combinatorialBkg}
\end{figure}

The corresponding combinatorial background distributions can be seen in Fig.~\ref{fig:combinatorialBkg}.
For use in the fit, these histograms are smoothed using a two-dimensional cubic spline to mitigate effects of discontinuity at the bin edges.
In the Dalitz-plot fit, the charge asymmetry in the combinatorial background yield is fixed to that obtained in the \Bp invariant-mass fit described in Section~\ref{sec:Mass_fit}.

\begin{figure}[tb]
\centering
\includegraphics[width=0.65\linewidth]{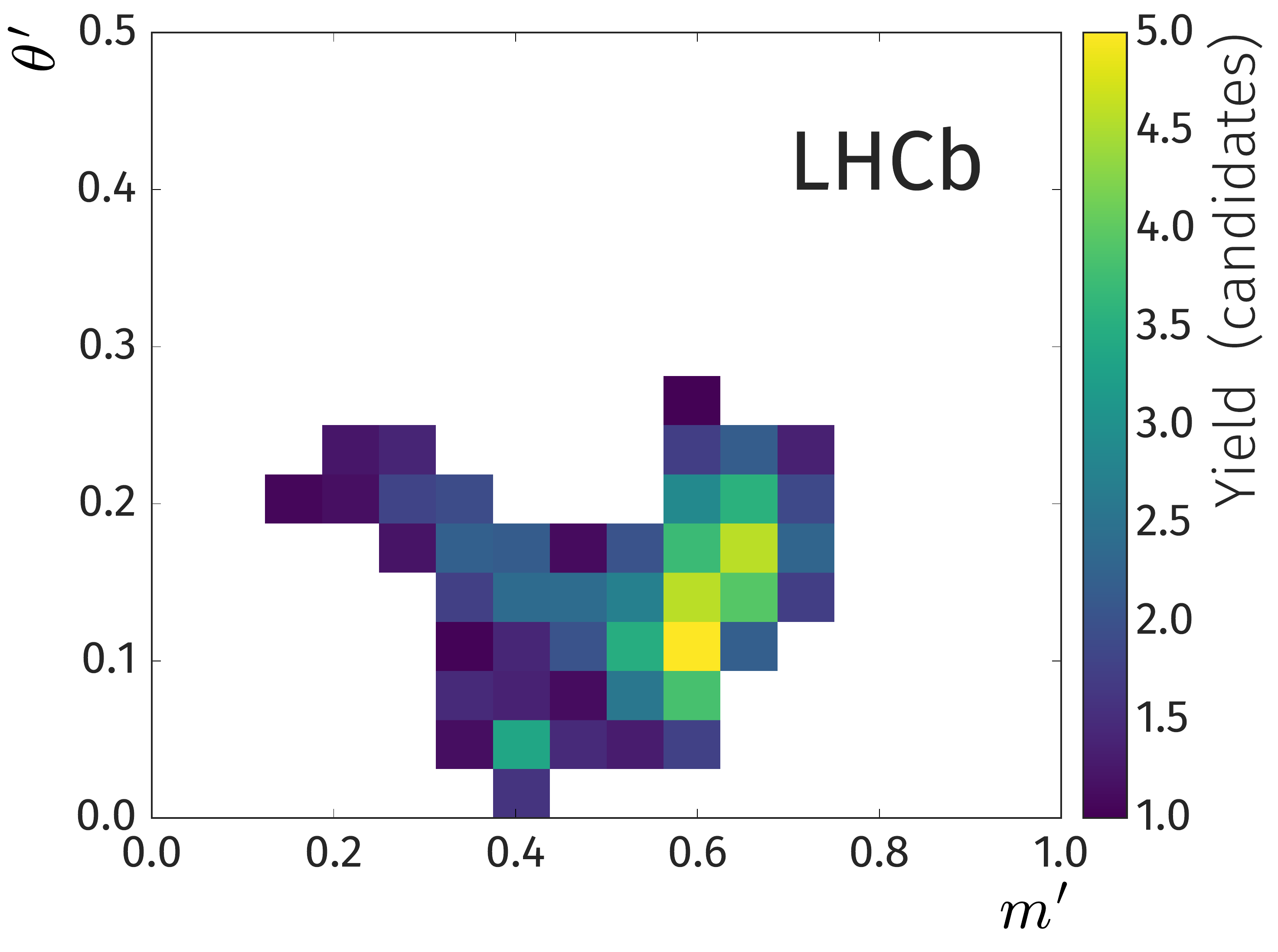}
\caption{Square Dalitz-plot distribution for the misidentified \decay{\Bp}{\Kp\pip\pim} background model, scaled to
represent its yield in the signal region.}
\label{fig:crossFeedBkg}
\end{figure}

A secondary source of background arises from misidentified \decay{\Bp}{\Kp\pip\pim} decays.
This background is modelled using simulated \decay{\Bp}{\Kp\pip\pim} events, reconstructed under the \decay{\Bp}{\pip\pip\pim} decay hypothesis.
To account for the phase-space distribution of this background, the events are weighted according to the amplitude model obtained by the BaBar collaboration~\cite{Aubert:2008bj}.
Similarly to the combinatorial background, this contribution is described in terms of a uniform $16\times16$ binned square Dalitz-plot histogram, smoothed with a two-dimensional cubic spline.
The corresponding distribution, without the smoothing, can be seen in Fig.~\ref{fig:crossFeedBkg}.

\subsection{Fit procedure}

Of the three approaches to the S-wave, the isobar and K-matrix fits are performed using the \verb|Laura++| Dalitz-plot
fitter package v3.5~\cite{Back:2017zqt}, which interfaces to the \verb|MINUIT| function minimisation
algorithm~\cite{James:1975dr,Brun:1997pa}.
In contrast, the QMI approach relies on
the \verb|Mint2|~\cite{jeremysbrain} amplitude-analysis interface to \verb|Minuit2|~\cite{Brun:1997pa}.
The fundamental difference between these amplitude-analysis software packages is in the handling of the normalisation.
The former approximates the definite integral by employing a Gaussian quadrature approach, while the latter invokes a Monte Carlo technique.
Additionally, due to the size of its parameter space, the QMI greatly benefits from the use of GPU-accelerated solutions.

In all cases, the combined $\rho$--$\omega$ mixing component is set to be the ``reference'' amplitude.
In practice, this means that the average magnitude of the \Bp and \Bm coefficients for this component is set to unity (in terms of Eq.~\eqref{eq:cartesian}, $x=1$), while the $\delta x$ parameter is left free to vary to allow for \CP violation.
Since there is no sensitivity to the phase difference between the \Bp and \Bm amplitudes, the imaginary part of the $\rho$--$\omega$ component is set to zero for both \Bp and \Bm ($y = \delta y = 0$), which means that all other contributions to the model are measured relative to this component.

The extended likelihood function that is optimised is of the form
\begin{equation}
\mathcal{L} = e^{-N} \prod_{i=1}^{N_{\rm cand}} \left[ \sum_k N_k \mathcal{P}_k^i \right]
\end{equation}
where $N_k$ is the yield for the candidate category $k$ (given in Table~\ref{tab:massFit}), $N$ is equal to $\sum_k N_k$, $N_{\rm cand}$ is the
total number of candidates, and $\mathcal{P}_k^i$ is the probability density function for candidates in
category $k$ in terms of the Dalitz-plot coordinates.
The optimal values of the fitted parameters are found by minimising twice the negative log-likelihood, $-2\log \mathcal{L}$.

Since Dalitz-plot analyses involve multidimensional parameter spaces, depending upon the initial parameter values the results may correspond to a local, rather than global, minimum of the $-2\log \mathcal{L}$ function.
To attempt to find the global minimum, a large number of fits are performed where the initial values of the complex isobar coefficients $c_j$ are randomised.
The fit result with the smallest $-2\log \mathcal{L}$ value out of this ensemble is then taken to be the nominal result for each S-wave method, and solutions near to this are also inspected (see Appendix~\ref{sec:kMatrixSecondary} for the K-matrix model fit).

\subsection{Model selection}
\label{sec:modelSel}

The inclusion or exclusion of various resonant contributions to the amplitude is studied using the isobar and K-matrix S-wave approaches.
This is not practical with the QMI approach as the large S-wave parameter space requires a detailed search for the global minimum given each model hypothesis.
Starting with resonant contributions identified during previous analyses of the \decay{\Bp}{\pip\pip\pim} decay~\cite{Jessop:2000bv,Gordon:2002yt,Aubert:2005sk,Aubert:2009av}, additional resonances are examined iteratively, in the order that maximises the change in log-likelihood between the current and proposed model with respect to the data.
This procedure is terminated when the log-likelihood gain from including any contribution not yet in the model is less than 10.
Only resonances that have been observed by two or more experiments and have been seen to decay to two charged pions are considered initially.
Scalar and vector nonresonant contributions, and possible virtual excited $\B^{*}$ states~\cite{LHCb-PAPER-2014-036}, are then investigated as possible improvements to the model, however none are found to have a significant contribution.

After this initial iterative procedure, a second step is performed that involves {\it ad hoc} trials of alternative mass lineshapes for resonances already included in the model, and investigation of additional, more speculative, states.
These form the basis of several important systematic uncertainties listed in Section~\ref{sec:systematics} and are further discussed in Section~\ref{sec:discussion}.
Lastly, tests are performed for ``latent'' resonant contributions up to spin $4$, where a resonance is inserted as a relativistic Breit--Wigner shape with a width of $0.025\gev$, $0.050\gev$, $0.100\gev$, or $0.150\gev$, in mass steps of $0.2\gev$.
No significant evidence of any resonant structure not captured by the previously established model is observed.

The goodness of fit is assessed by comparing the fit model with the data in square Dalitz-plot bins and determining an associated $\chi^2$ value (see Appendix~\ref{sec:isobarTables} for the distribution for the isobar model fit, Appendix~\ref{sec:kMatrixTables} for the distribution for the K-matrix model fit, and Appendix~\ref{sec:qmiTables} for the distribution for the QMI model fit).
The binning is chosen through an adaptive procedure that requires an approximately constant number of candidates from the data sample in each bin.
For various values of the required number of candidates per bin, the ratio of the $\chi^2$ to the number of bins is approximately 1.5 accounting for statistical uncertainties only.
Given the impact of the systematic uncertainties on the results, as shown in Section~\ref{sec:results}, the agreement of the fit models with the data is reasonable.
Smaller $\chi^2$ values are obtained for the S-wave models with larger numbers of free parameters, such that all three approaches have comparable goodness-of-fit overall.
The distribution in the square Dalitz plot of bins that contribute significantly to the $\chi^2$ does not reveal any clear source of mismodelling. Nevertheless, a discrepancy between all of the models and the data is observed in the region around the $f_2(1270)$ resonance, which is investigated further in Section~\ref{sec:f2Discussion}.

Resonant contributions with spin greater than zero that were identified through the model selection procedure are common to all three approaches and are listed in Table~\ref{tab:resonances}.
Furthermore, the mass and width of the dominant $\rho(770)^0$ contribution are left free to vary, which results in a significantly better fit quality.

\renewcommand{\arraystretch}{1.35}
\begin{table}[ht]
  \centering
  \caption{
    Non-S-wave resonances and their default lineshapes as identified by the model selection procedure.
    These are common to all S-wave approaches.
  }
  \label{tab:resonances}

  \begin{tabular}
    {@{\hspace{0.5cm}}l@{\hspace{0.25cm}}  @{\hspace{0.25cm}}c@{\hspace{0.25cm}}  @{\hspace{0.25cm}}r@{\hspace{0.5cm}}}
    \hline \hline
    Resonance & Spin & Mass lineshape\\
    \hline
    $\rho(770)^0$ & $1$ & Gounaris--Sakurai ($\rho$--$\omega$ mixing) \\
    $\omega(782)$ & $1$ & Relativistic Breit--Wigner ($\rho$--$\omega$ mixing) \\
    $f_2(1270)$ & $2$ & Relativistic Breit--Wigner \\
    $\rho(1450)^0$ & $1$ & Relativistic Breit--Wigner \\
    $\rho_3(1690)^0$ & $3$ & Relativistic Breit--Wigner \\
    \hline \hline
  \end{tabular}%
\end{table}

\section{Systematic uncertainties}
\label{sec:systematics}

Sources of systematic uncertainty are separated into two categories: those that arise from experimental effects and those from the inherent lack of knowledge on the amplitude models.
The experimental systematic uncertainties comprise those from the uncertainty on the signal and background yields, the phase-space-dependent efficiency description, the combinatorial and \decay{\Bp}{\Kp\pip\pim} background models in the Dalitz plot, and the intrinsic fit bias.
Model systematic uncertainties comprise those introduced by the uncertainty on the known resonance masses and widths, the radius parameter of the \mbox{\it ad hoc} \mbox{Blatt--Weisskopf} barrier factors and from potential additional resonant contributions to the amplitude.
Furthermore, this latter category also includes sources of uncertainty that are specific to each S-wave approach.
The effects in each category are considered to be uncorrelated and are therefore combined in quadrature to obtain the total systematic uncertainty.

The uncertainties on the signal yield and the background yields and asymmetries, given in Table~\ref{tab:massFit}, comprise a statistical component as well as systematic effects due to the invariant-mass fit procedure.
The uncertainty arising from assumptions regarding the signal parameterisation is found by replacing the model with two Crystal Ball functions with a common mean and width, but independent tail parameters.
Similarly, the model for the combinatorial background is replaced with a first-order polynomial.
The uncertainty on the cross-feed \decay{\Bp}{\Kp\pip\pim} background shape in the three-body invariant-mass fit is negligible, however the yield of this component is varied by three times the nominal uncertainty on the expectation from the simulation to account for possible inaccuracies in the constraint.
Additionally, effects associated with allowing different relative signal and partially reconstructed background yields in the data subcategories separated by source of hardware trigger decision are investigated by constraining them to be common in both subcategories.
The combined statistical and systematic uncertainties on the signal and background yields and the background asymmetries are then propagated to the Dalitz-plot fit, where those variations causing the largest upward and downward deviations with respect to the nominal yield values are taken to assign the systematic uncertainty relating to the three-body invariant-mass fit.

To account for the statistical uncertainty on the efficiency description, an ensemble of efficiency maps is created by sampling bin-by-bin from the baseline description, according to uncorrelated Gaussian distributions with means corresponding to the central value of the nominal efficiency in each bin, and widths corresponding to the uncertainty.
The standard deviation of the distribution of resulting Dalitz-plot fit parameters obtained when using this ensemble is then taken to be the associated systematic uncertainty.
To account for potential biases in the method used to correct the hardware trigger efficiency, an alternative method using \decay{\Bz}{\jpsi(\mumu)\Kp\pim} decays, requiring a positive trigger decision on the muons from the \jpsi decay, is used to apply corrections to the simulation~\cite{LHCB-PAPER-2017-010, LHCB-PAPER-2018-045}.
The effect on the baseline results is assigned as an uncertainty.

Additionally, to account for potential variation of the efficiency within a nominal square Dalitz-plot bin, the efficiency map is constructed using a finer binning scheme, and the total deviation of the results is taken as the systematic uncertainty.
The effect arising from the uncertainty on the measured \Bp production asymmetry is also considered, but is found to be negligible.

The statistical uncertainty on the combinatorial background distribution is propagated to the Dalitz-plot fit results in a procedure similar to that for the efficiency map.
Uncertainty associated with the Dalitz-plot model of the \decay{\Bp}{\Kp\pip\pim} decay is also assigned.
This is calculated by fluctuating the parameters obtained in the \decay{\Bp}{\Kp\pip\pim} fit according to their uncertainties~\cite{Aubert:2008bj}, taking into account the reported correlations on the statistical uncertainties.
The standard deviation in the variation of the subsequent Dalitz-plot fit results is taken to be the systematic uncertainty due to this effect.

Systematic uncertainties related to possible intrinsic fit bias are investigated using an ensemble of pseudoexperiments.
Differences between the input and fitted values from the ensemble for the fit parameters are generally found to be small.
Systematic uncertainties are assigned as the sum in quadrature of the difference between the input and output values and the uncertainty on the mean of the output value determined from a fit to the ensemble.

Sources of model uncertainty independent of the S-wave approach are those arising from the uncertainties on the masses and widths of resonances in the baseline model, the Blatt--Weisskopf barrier factors and contributions from additional resonances.
The systematic uncertainty due to resonance masses and widths are again assigned with an ensemble technique, where the parameter values, excluding those that appear in the isobar S-wave model, are fluctuated according to the uncertainties listed in the Particle Data Group tables~\cite{PDG2018}.
Where appropriate, these are taken to be those from combinations only considering decays to \pip\pim.
The uncertainty arising from the lack of knowledge of the radius parameter of the Blatt--Weisskopf barrier factors is estimated by modifying the value of this between $3$ and $5 \gev^{-1}$, with the maximum deviation of the fit parameters taken to be the systematic uncertainty.

To account for mismodelling in the $f_2(1270)$ region, discussed in Section~\ref{sec:discussion}, an additional systematic uncertainty is assigned as the maximal variation in fit parameters when either an additional spin-$2$ component with mass and width parameters determined by the fit is included into the model, or when the $f_2(1270)$ resonance mass and width are permitted to vary in the fit.
Furthermore, a possible contribution from the $\rho(1700)^0$ resonance cannot be excluded.
Using perturbative QCD calculations, the branching fraction of the \decay{\Bp}{\rho(1700)^0\pip} decay has been calculated to be around $3\times 10^{-7}$~\cite{Li:2017mao}, which is plausibly within the sensitivity of this analysis.
Therefore a systematic uncertainty is assigned as the deviation of the fit parameters with respect to the nominal values, when the $\rho(1700)^0$ contribution is included.

For fits related to the isobar approach, the nominal rescattering parametrisation relies on a source term with two components as given in Eq.~\eqref{eq:isobar:Asource}.
The fits have little sensitivity to the values chosen for the $\Delta^2_{\pi\pi}$ and $\Delta^2_{KK}$ parameters, so the robustness of this parametrisation is investigated by using instead a source term with only one component, $A_{\rm{source}} = [1 + (m/\Delta^2_{KK})]^{-1}$, and the difference in the results obtained assigned as a systematic uncertainty.
The parameters of the $\sigma$ contribution to the S-wave are also varied within the uncertainties on the world-average mass and width, and the effect on the results taken as a systematic uncertainty.

For fits using the K-matrix approach, both the fourth $\hat{P}$-vector pole and the fourth slowly varying part result in a negligible change to the total likelihood when removed, and therefore a systematic uncertainty is assigned that corresponds to the maximum deviation of the parameters, with respect to the nominal values, when these components are removed from the K-matrix model.
Furthermore, in the baseline fit the $s_0^{\rm prod}$ parameter appearing in the slowly varying parts of Eq.~\eqref{eq:prodSVP} is fixed to a value of $-3 \gevgevcccc$.
However as this comprises part of the production component of the K-matrix, this is not fixed by scattering data and can depend on the production environment.
As such, this value is varied between $-1$ and $-5 \gevgevcccc$ based on the likelihood profile and the maximum deviation from the nominal fit results
taken to be the systematic uncertainty due to this effect.

For the fits involving the QMI approach, an additional bias may arise from the intrinsic ability of the approach to reproduce the underlying analytic S-wave.
Causes of such a bias can include the definition of the binning scheme, the extent to which the S-wave interferes with other partial waves in a particular bin, and the approximation of an analytic lineshape by a constant amplitude in each bin.
This systematic uncertainty is evaluated reusing the ensemble of pseudoexperiments generated for estimating the K-matrix fit bias, fitting them with the QMI model, and determining the difference between the obtained and true bin-averaged values of the S-wave amplitude.
The QMI intrinsic bias is by far the dominant systematic uncertainty on the S-wave magnitude and phase motion.
Previous quasi-model-independent partial wave analyses have not recognised such an effect as a possible source of bias; an important conclusion of this study is that the associated systematic uncertainty must be accounted for in analyses in which quantitative results from binned partial-wave amplitude models are obtained.

The systematic uncertainties for the \CP-averaged fit fractions and quasi-two-body \CP asymmetries are summarised in Tables~\ref{tab:syst:iso:ff} and~\ref{tab:syst:iso:acp} for the isobar approach, Tables~\ref{tab:syst:km:ff} and~\ref{tab:syst:km:acp} for the K-matrix approach, and Tables~\ref{tab:syst:qmi:ff} and~\ref{tab:syst:qmi:acp} for the QMI approach.
In general the largest sources of systematic uncertainty are due to variations in the model, which tend to dominate the total uncertainties for the \CP-averaged fit fractions while the \CP asymmetries for well established resonances are somewhat more robust against these effects.
In particular, the inclusion of an additional tensor or vector resonance, \ie the $f_2(1430)$ or $\rho(1700)^0$ states, can have a large effect on parameters associated with other resonances, particularly when they are in the same partial wave.
With larger data samples it may be possible to clarify the contributions from these amplitudes and thereby reduce these uncertainties.
Intrinsic fit bias is also an important source of uncertainty for several measurements, in particular those using the QMI description of the S-wave.

\sisetup{round-mode=places,round-precision=1}
\begin{table}[ht]
  \centering
  \caption{Systematic uncertainties on the \CP-averaged fit fractions, given in units of $10^{-2}$, for the isobar method. Uncertainties are given both for the total S-wave, and for the individual components due to the $\sigma$ pole and the rescattering amplitude. For comparison, the statistical uncertainties are also listed at the bottom.}
  \label{tab:syst:iso:ff}
  \resizebox{\textwidth}{!}{
    \begin{tabular}
      {@{\hspace{0.5cm}}l@{\hspace{0.25cm}}  @{\hspace{0.25cm}}r@{\hspace{0.25cm}}  @{\hspace{0.25cm}}r@{\hspace{0.25cm}}  @{\hspace{0.25cm}}r@{\hspace{0.25cm}}  @{\hspace{0.25cm}}r@{\hspace{0.25cm}}  @{\hspace{0.25cm}}r@{\hspace{0.25cm}}  @{\hspace{0.25cm}}r@{\hspace{0.25cm}}  @{\hspace{0.25cm}}r@{\hspace{0.25cm}}  @{\hspace{0.25cm}}r@{\hspace{0.25cm}}  @{\hspace{0.25cm}}r@{\hspace{0.5cm}}}
      \hline \hline

      Category & \multicolumn{1}{c}{$\rho(770)^0$} & \multicolumn{1}{c}{$\omega(782)$} & \multicolumn{1}{c}{$f_2(1270)$} & \multicolumn{1}{c}{$\rho(1450)^0$} & \multicolumn{1}{c}{$\rho_3(1690)^0$} & \multicolumn{1}{c}{S-wave} & \multicolumn{1}{c}{Rescattering} & \multicolumn{1}{c}{$\sigma$}\\ \hline
      $B$ mass fit                       & $0.23$ & $0.01$ & $0.68$ & $0.07$ & $0.03$ & $0.40$ & $0.16~~$ & $0.02$\\
      Efficiency& \\
      \hspace{10pt}Simulation sample size & $0.10$ & $<0.01$ & $0.06$ & $0.05$ & $0.01$ & $0.08$ & $0.02~~$ & $0.09$\\
      \hspace{10pt}Binning               & $0.07$ & $<0.01$ & $0.03$ & $0.08$ & $0.02$ & $0.09$ & $0.01~~$ & $0.08$\\
      \hspace{10pt}L0 Trigger            & $0.02$ & $<0.01$ & $<0.01$ & $<0.01$ & $<0.01$ & $0.02$ & $<0.01~~$ & $0.02$ \\
      Combinatorial bkgd                 & $0.26$ & $<0.01$ & $0.14$ & $0.15$ & $0.03$ & $0.28$ & $0.04~~$ & $0.31$ \\
      $B^+ \to K^+ \pip \pim$            & $0.01$ & $<0.01$ & $<0.01$ & $<0.01$ & $<0.01$ & $0.01$ & $<0.01~~$ & $0.01$\\
      Fit bias                           & $0.03 $ & $<0.01 $ & $0.01 $ & $0.01 $ & $0.01 $ & $0.04 $ & $<0.01~~ $ & $0.04$\\
      \hline
      Total experimental                 & $\num{0.37}\phantom{0}$ & $\num{0.01}\phantom{0}$ & $\num{0.69}\phantom{0}$ & $\num{0.19}\phantom{0}$ & $\num{0.05}\phantom{0}$ & $\num{0.5}\phantom{0}$ & $\num{0.17}\phantom{0}~~$ & $\num{0.34}\phantom{0}$ \\
      \hline
      Amplitude model & \\
      \hspace{10pt}Resonance properties       & $0.63$ & $0.01$ & $0.17$ & $0.39$ & $0.05$ & $0.29$ & $0.01~~$ & $0.41$\\
      \hspace{10pt}Barrier factors            & $0.82$ & $0.01$ & $0.18$ & $0.40$ & $0.01$ & $0.05$ & $0.04~~$ & $0.17$\\
      Alternative lineshapes & \\
      \hspace{10pt}$f_2(1270)$                & $0.23$ & $<0.01$ & $0.68$ & $0.07$ & $0.03$ & $0.40$ & $0.16~~$ & $0.02$\\
      \hspace{10pt}$f_2(1430)$                & $0.40$ & $<0.01$ & $0.88$ & $0.25$ & $0.10$ & $0.90$ & $0.21~~$ & $0.66$\\
      \hspace{10pt}$\rho(1700)^0$             & $0.88$ & $0.02$ & $0.09$ & $1.28$ & $0.01$ & $0.01$ & $<0.01~~$ & $<0.01$\\
      Isobar specifics & \\
      \hspace{10pt}$\sigma$ from PDG          & $2.00$ & $0.03$ & $0.69$ & $1.18$ & $0.32$ & $3.40$ & $0.35~~$ & $4.90$\\
      \hspace{10pt}Rescattering & $0.01$ & $<0.01$ & $0.19$ & $0.03$ & $<0.01$ & $0.11$ & $0.07~~$ & $0.24$\\
      \hline
      Total model                             & $\num{2.46}\phantom{0}$ & $\num{0.04}\phantom{0}$ & $\num{1.35}\phantom{0}$ & $\num{1.85}\phantom{0}$ & $\num{0.34}\phantom{0}$ & $\num{3.55}\phantom{0}$ & $\num{0.45}\phantom{0}~~$ & $\num{4.97}\phantom{0}$\\
      \hline
      Statistical uncertainty & $\num{0.60}\phantom{0}$ & $\num{0.03}\phantom{0}$ & $\num{0.28}\phantom{0}$ & $\num{0.29}\phantom{0}$ & $\num{0.08}\phantom{0}$ & $\num{0.49}\phantom{0}$ & $\num{0.15}\phantom{0}~~$ & $\num{0.48}\phantom{0}$\\

      \hline \hline
    \end{tabular}
  }
\end{table}
\sisetup{round-mode=off}

\sisetup{round-mode=places,round-precision=1}
\begin{table}[ht]
  \centering
  \caption{Systematic uncertainties on ${\cal A}_{\CP}$ values, given in units of $10^{-2}$, for the isobar method. Uncertainties are given both for the total S-wave, and for the individual components due to the $\sigma$ pole and the rescattering amplitude. For comparison, the statistical uncertainties are also listed at the bottom.}
  \label{tab:syst:iso:acp}
  \resizebox{\textwidth}{!}{
    \begin{tabular}
      {@{\hspace{0.5cm}}l@{\hspace{0.25cm}}  @{\hspace{0.25cm}}r@{\hspace{0.25cm}}  @{\hspace{0.25cm}}r@{\hspace{0.25cm}}  @{\hspace{0.25cm}}r@{\hspace{0.25cm}}  @{\hspace{0.25cm}}r@{\hspace{0.25cm}}  @{\hspace{0.25cm}}r@{\hspace{0.25cm}}  @{\hspace{0.25cm}}r@{\hspace{0.25cm}}  @{\hspace{0.25cm}}r@{\hspace{0.25cm}}  @{\hspace{0.25cm}}r@{\hspace{0.25cm}}}
      \hline \hline

      Category & \multicolumn{1}{c}{$\rho(770)^0$} & \multicolumn{1}{c}{$\omega(782)$} & \multicolumn{1}{c}{$f_2(1270)$} & \multicolumn{1}{c}{$\rho(1450)^0$} & \multicolumn{1}{c}{$\rho_3(1690)^0$} & \multicolumn{1}{c}{S-wave} & \multicolumn{1}{c}{Rescattering} & \multicolumn{1}{c}{$\sigma$}\\ \hline
      $B$ mass fit                       & $0.12$ & $0.10$ & $0.89$ & $0.40$ & $4.19$ & $0.58$ & $4.20\quad$ & $0.54$\\
      Efficiency& \\
      \hspace{10pt}Simulation sample size & $0.34$ & $0.71$ & $0.61$ & $0.92$ & $1.24$ & $0.36$ & $1.00\quad$ & $0.35$\\
      \hspace{10pt}Binning               & $0.27$ & $0.87$ & $0.23$ & $1.19$ & $0.52$ & $0.28$ & $1.43\quad$ & $0.22$\\
      \hspace{10pt}L0 Trigger            & $0.02$ & $0.37$ & $0.17$ & $0.31$ & $0.28$ & $0.14$ & $0.32\quad$ & $0.19$\\
      Combinatorial bkgd                 & $0.40$ & $0.50$ & $1.02$ & $3.06$ & $5.75$ & $0.75$ & $3.16\quad$ & $0.75$\\
      $B^+ \to K^+ \pip \pim$            & $<0.01$ & $0.01$ & $0.02$ & $0.03$ & $0.05$ & $0.01$ & $0.04\quad$ & $0.01$\\
      Fit bias                           & $0.05 $ & $0.35 $ & $0.25 $ & $1.10 $ & $2.95 $ & $0.04 $ & $0.96\quad$ & $0.09$\\
      \hline
      Total experimental                 & $\num{0.60}\phantom{0}$ & $\num{1.33}\phantom{0}$ & $\num{1.49}\phantom{0}$ & $\num{3.61}\phantom{0}$ & $\num{7.75}\phantom{0}$ & $\num{0.99}\phantom{0}$ & $\num{5.54}\phantom{0}\quad$ & $\num{0.99}\phantom{0}$ \\
      \hline
      Amplitude model & \\
      \hspace{10pt}Resonance properties       & $0.20$ & $0.53$ & $0.55$ & $2.66$  & $5.58$  & $0.41$ & $1.58\quad$ & $0.29$\\
      \hspace{10pt}Barrier factors            & $0.18$ & $0.95$ & $0.80$ & $3.84$  & $1.56$  & $1.27$ & $0.34\quad$ & $1.25$\\
      Alternative lineshapes & \\
      \hspace{10pt}$f_2(1270)$                & $0.11$ & $0.10$ & $0.82$ & $0.30$  & $4.05$  & $0.49$ & $4.07\quad$ & $0.45$\\
      \hspace{10pt}$f_2(1430)$                & $0.02$ & $0.04$ & $2.84$ & $1.76$  & $12.05$ & $0.98$ & $6.39\quad$ & $1.05$\\
      \hspace{10pt}$\rho(1700)^0$             & $1.49$ & $0.81$ & $0.75$ & $27.78$ & $4.57$  & $0.73$ & $6.32\quad$ & $0.66$\\
      Isobar specifics & \\
      \hspace{10pt}$\sigma$ from PDG          & $0.01$ & $3.26$ & $2.97$ & $21.83$ & $19.04$ & $0.11$ & $12.9\quad$ & $0.53$ \\
      \hspace{10pt}Rescattering & $0.02$ & $0.14$ & $0.81$ & $0.19$  & $1.97$  & $0.29$ & $1.24\quad$ & $0.17$ \\
      \hline
      Total model                             & $\num{1.52}\phantom{0}$ & $\num{3.54}\phantom{0}$ & $\num{4.44}\phantom{0}$ & $\num{35.68}\phantom{0}$ & $\num{24.13}\phantom{0}$ & $\num{1.90}\phantom{0}$ & $\num{16.37}\phantom{0}\quad$ & $\num{1.92}\phantom{0}$ \\
      \hline
      Statistical uncertainty & $\num{1.07}\phantom{0}$ & $\num{6.51}\phantom{0}$ & $\num{6.10}\phantom{0}$ & $\num{3.25}\phantom{0}$ & $\num{11.36}\phantom{0}$ & $\num{1.79}\phantom{0}$ & $\num{8.59}\phantom{0}\quad$ & $\num{1.73}\phantom{0}$\\
      \hline \hline
    \end{tabular}
  }
\end{table}
\sisetup{round-mode=off}

\sisetup{round-mode=places,round-precision=1}
\begin{table}[ht]
  \centering
  \caption{Systematic uncertainties on the \CP-averaged fit fractions, given in units of $10^{-2}$, for the K-matrix method. For comparison, the statistical uncertainties are also listed at the bottom.}
  \label{tab:syst:km:ff}
  \resizebox{\textwidth}{!}{
    \begin{tabular}
      {@{\hspace{0.5cm}}l@{\hspace{0.25cm}}  @{\hspace{0.25cm}}r@{\hspace{0.25cm}}  @{\hspace{0.25cm}}r@{\hspace{0.25cm}}  @{\hspace{0.25cm}}r@{\hspace{0.25cm}}  @{\hspace{0.25cm}}r@{\hspace{0.25cm}}  @{\hspace{0.25cm}}r@{\hspace{0.25cm}}  @{\hspace{0.25cm}}r@{\hspace{0.25cm}}  @{\hspace{0.25cm}}r@{\hspace{0.5cm}}}
      \hline \hline

      Category & \multicolumn{1}{c}{$\rho(770)^0$} & \multicolumn{1}{c}{$\omega(782)$} & \multicolumn{1}{c}{$f_2(1270)$} & \multicolumn{1}{c}{$\rho(1450)^0$} & \multicolumn{1}{c}{$\rho_3(1690)^0$} & \multicolumn{1}{c}{S-wave} \\ \hline
      $B$ mass fit & $ 1.31 $ & $ 0.01 $ & $ 0.51 $ & $ 0.65 $ & $ 0.04 $ & $ 2.53 $ \\
      Efficiency & \\
      \hspace{10pt}Simulation sample size & $ 0.13 $ & $< 0.01 $ & $ 0.07 $ & $ 0.09 $ & $ 0.02 $ & $ 0.09 $ \\
      \hspace{10pt}Binning & $ 0.46 $ & $< 0.01 $ & $ 0.08 $ & $ 0.32 $ & $ 0.07 $ & $ 0.33 $ \\
      \hspace{10pt}L0 trigger & $ 0.02 $ & $< 0.01 $ & $ 0.01 $ & $ 0.04 $ & $< 0.01 $ & $ 0.03 $  \\
      Combinatorial bkgd & $ 0.41 $ & $ 0.01 $ & $ 0.15 $ & $ 0.35 $ & $ 0.08 $ & $ 0.24 $ \\
      $B^+ \to K^+ \pip \pim$ & $ 0.01 $ & $< 0.01 $ & $< 0.01 $ & $ 0.02 $ & $< 0.01 $ & $ 0.01 $ \\
      Fit bias & $ 0.06 $ & $< 0.01 $ & $ 0.05 $ & $ 0.09 $ & $ 0.04 $ & $ 0.06 $ \\
      \hline
      Total experimental & $\num{1.45}\phantom{0}$ & $0.01$ & $\num{0.55}\phantom{0}$ & $\num{0.81}\phantom{0}$ & $\num{0.13}\phantom{0}$ & $\num{2.56}\phantom{0}$ \\
      \hline
      Amplitude model & \\
      \hspace{10pt}Resonance properties & $ 1.02 $ & $ 0.01 $ & $ 0.18 $ & $ 1.41 $ & $ 0.09 $ & $ 0.32 $ \\
      \hspace{10pt}Barrier factors & $ 0.24 $ & $< 0.01 $ & $ 0.34 $ & $ 0.19 $ & $ 0.06 $ & $ 0.57 $ \\
      Alternative lineshapes & \\
      \hspace{10pt}$f_2(1270)$ & $ 0.29 $ & $ 0.01 $ & $ 0.62 $ & $ 0.60 $ & $ 0.03 $ & $ 0.05 $ \\
      \hspace{10pt}$f_2(1430)$ & $ 2.30 $ & $< 0.01 $ & $ 2.24 $ & $ 4.17 $ & $ 0.36 $ & $ 0.01 $ \\
      \hspace{10pt}$\rho(1700)^0$ & $ 1.66 $ & $ 0.01 $ & $ 0.08 $ & $ 0.55 $ & $ 0.02 $ & $ 0.97 $ \\
      K-matrix specifics & \\
      \hspace{10pt}$s^0_{\rm prod}$ & $ 0.63 $ & $< 0.01 $ & $ 0.06 $ & $ 0.21 $ & $ 0.03 $ & $ 0.48 $ \\
      \hspace{10pt}K-matrix components & $ 0.48 $ & $ 0.01 $ & $ 0.04 $ & $ 0.36 $ & $ 0.01 $ & $ 0.57 $ \\
      \hline
      Total model & $\num{3.14}\phantom{0}$ & $0.02$ & $\num{2.36}\phantom{0}$ & $\num{4.50}\phantom{0}$ & $\num{0.38}\phantom{0}$ & $\num{1.39}\phantom{0}$ \\
      \hline
      Statistical uncertainty & $\num{0.8}\phantom{0}$ & $0.04$ & $\num{0.4}\phantom{0}$ & $\num{0.7}\phantom{0}$ & $\num{0.1}\phantom{0}$ & $\num{0.6}\phantom{0}$ \\
      \hline \hline
    \end{tabular}
  }
\end{table}
\sisetup{round-mode=off}

\sisetup{round-mode=places,round-precision=1}
\begin{table}[ht]
  \centering
  \caption{Systematic uncertainties on ${\cal A}_{\CP}$ values, given in units of $10^{-2}$, for the K-matrix method. For comparison, the statistical uncertainties are also listed at the bottom.}
  \label{tab:syst:km:acp}
  \resizebox{\textwidth}{!}{
    \begin{tabular}
      {@{\hspace{0.50cm}}l@{\hspace{0.25cm}}  @{\hspace{0.25cm}}r@{\hspace{0.25cm}}  @{\hspace{0.25cm}}r@{\hspace{0.25cm}}  @{\hspace{0.25cm}}r@{\hspace{0.25cm}}  @{\hspace{0.25cm}}r@{\hspace{0.25cm}}  @{\hspace{0.25cm}}r@{\hspace{0.25cm}}  @{\hspace{0.25cm}}r@{\hspace{0.50cm}}}
      \hline \hline

      Category & \multicolumn{1}{c}{$\rho(770)^0$} & \multicolumn{1}{c}{$\omega(782)$} & \multicolumn{1}{c}{$f_2(1270)$} & \multicolumn{1}{c}{$\rho(1450)^0$} & \multicolumn{1}{c}{$\rho_3(1690)^0$} & \multicolumn{1}{c}{S-wave} \\ \hline
      $B$ mass fit & $ 1.97 $ & $ 0.12 $ & $ 1.42 $ & $ 9.74 $ & $ 5.77 $ & $ 1.03 $ \\
      Efficiency & \\
      \hspace{10pt}Simulation sample size & $ 0.22 $ & $ 0.88 $ & $ 0.73 $ & $ 0.97 $ & $ 1.34 $ & $ 0.42 $ \\
      \hspace{10pt}Binning & $ 1.53 $ & $ 5.48 $ & $ 0.15 $ & $ 2.89 $ & $ 1.72 $ & $ 1.54 $ \\
      \hspace{10pt}L0 trigger & $ 0.15 $ & $ 0.59 $ & $ 0.19 $ & $ 0.32 $ & $ 0.30 $ & $ 0.02 $ \\
      Combinatorial bkgd & $ 0.61 $ & $ 0.60 $ & $ 1.31 $ & $ 3.45 $ & $ 5.82 $ & $ 0.93 $ \\
      $B^+ \to K^+ \pip \pim$ & $ 0.01 $ & $ 0.03 $ & $ 0.03 $ & $ 0.04 $ & $ 0.12 $ & $ 0.03 $ \\
      Fit bias & $ 0.02 $ & $ 0.04 $ & $ 0.24 $ & $ 0.85 $ & $ 0.40 $ & $ 0.36 $ \\
      \hline
      Total experimental & $\num{ 2.60 }\phantom{0}$ & $\num{ 5.60 }\phantom{0}$ & $\num{ 2.09 }\phantom{0}$ & $\num{ 10.81 }\phantom{0}$ & $\num{ 8.49 }\phantom{0}$ & $\num{ 2.13 }\phantom{0}$ \\
      \hline
      Amplitude model & \\
      \hspace{10pt}Resonance properties & $ 0.62 $ & $ 0.91 $ & $ 1.08 $ & $ 4.35 $ & $ 5.34 $ & $ 1.27 $ \\
      \hspace{10pt}Barrier factors & $ 1.97 $ & $ 3.54 $ & $ 0.04 $ & $ 12.53 $ & $ 2.79 $ & $ 3.50 $ \\
      Alternative lineshapes & \\
      \hspace{10pt}$f_2(1270)$ & $ 0.58 $ & $ 0.56 $ & $ 0.48 $ & $ 2.96 $ & $ 4.41 $ & $ 1.13 $ \\
      \hspace{10pt}$f_2(1430)$ & $ 3.04 $ & $ 1.69 $ & $ 8.78 $ & $ 41.78 $ & $ 33.96 $ & $ 4.77 $ \\
      \hspace{10pt}$\rho(1700)^0$ & $ 3.38 $ & $ 1.17 $ & $ 0.39 $ & $ 8.82 $ & $ 8.80 $ & $ 1.60 $ \\
      K-matrix specifics & \\
      \hspace{10pt}$s^0_{\rm prod}$ & $ 2.08 $ & $ 4.42 $ & $ 0.20 $ & $ 3.42 $ & $ 0.98 $ & $ 2.41 $ \\
      \hspace{10pt}K-matrix components & $ 2.11 $ & $ 5.31 $ & $ 0.01 $ & $ 8.11 $ & $ 0.21 $ & $ 1.03 $ \\
      \hline
      Total model & $\num{5.84}\phantom{0}$ & $\num{8.10}\phantom{0}$ & $\num{8.87}\phantom{0}$ & $\num{45.67}\phantom{0}$ & $\num{35.88}\phantom{0}$ & $\num{6.88}\phantom{0}$ \\
      \hline
      Statistical uncertainty & $\num{1.5}\phantom{0}$ & $\num{8.4}\phantom{0}$ & $\num{4.3}\phantom{0}$ & $\num{8.4}\phantom{0}$ & $\num{11.8}\phantom{0}$ & $\num{2.6}\phantom{0}$ \\
      \hline \hline
    \end{tabular}
  }
\end{table}
\sisetup{round-mode=off}

\sisetup{round-mode=places,round-precision=1}
\begin{table}[ht]
  \centering
  \caption{Systematic uncertainties on the \CP-averaged fit fractions, given in units of $10^{-2}$, for the QMI method. For comparison, the statistical uncertainties are also listed at the bottom.}
  \label{tab:syst:qmi:ff}
  \resizebox{\textwidth}{!}{
    \begin{tabular}
      {@{\hspace{0.5cm}}l@{\hspace{0.25cm}}  @{\hspace{0.25cm}}r@{\hspace{0.25cm}}  @{\hspace{0.25cm}}r@{\hspace{0.25cm}}  @{\hspace{0.25cm}}r@{\hspace{0.25cm}}  @{\hspace{0.25cm}}r@{\hspace{0.25cm}}  @{\hspace{0.25cm}}r@{\hspace{0.25cm}}  @{\hspace{0.25cm}}r@{\hspace{0.25cm}}  @{\hspace{0.25cm}}r@{\hspace{0.5cm}}}
      \hline \hline
      Category & \multicolumn{1}{c}{$\rho(770)^0$} & \multicolumn{1}{c}{$\omega(782)$} & \multicolumn{1}{c}{$f_2(1270)$} & \multicolumn{1}{c}{$\rho(1450)^0$} & \multicolumn{1}{c}{$\rho_3(1690)^0$} & \multicolumn{1}{c}{S-wave} \\ \hline
      $B$ mass fit & $1.03$ & $0.03$ & $0.29$ & $0.42$ & $0.01$ & $1.34$ \\
      Efficiency& \\
      \hspace{10pt}Simulation sample size & $0.15$ & $0.01$ & $0.22$ & $0.12$ & $0.02$ & $0.25$ \\
      \hspace{10pt}Binning & $0.03$ & $0.01$ & $0.03$ & $0.21$ & $0.04$ & $0.01$ \\
      \hspace{10pt}L0 trigger & $0.04$ & $<0.01$ & $0.01$ & $0.01$ & $<0.01$ & $0.04$ \\
      Combinatorial bkgd & $0.60$ & $0.01$ & $0.19$ & $0.67$ & $0.06$ & $0.62$ \\
      $B^+ \to K^+ \pip \pim$ & $0.03$ & $<0.01$ & $0.01$ & $0.02$ & $<0.01$ & $0.03$ \\
      Fit bias & $1.06$ & $0.10$ & $0.46$ & $0.61$ & $0.14$ & $0.68$ \\
      \hline
      Total experimental & $\num{1.72}\phantom{0}$ & $\num{0.12}\phantom{0}$ & $\num{0.65}\phantom{0}$ & $\num{1.26}\phantom{0}$ & $\num{0.16}\phantom{0}$ & $\num{1.77}\phantom{0}$ \\
      \hline
      Amplitude model & \\
      \hspace{10pt}Resonance properties & $0.63$ & $0.04$ & $0.21$ & $0.73$ & $0.03$ & $0.18$ \\
      \hspace{10pt}Barrier factors & $0.95$ & $0.05$ & $0.58$ & $0.80$ & $0.04$ & $0.78$\\
      Alternative lineshapes & \\
      \hspace{10pt}$f_2(1270)$ & $0.10$ & $0.04$ & $0.30$ & $0.34$ & $0.04$ & $0.36$ \\
      \hspace{10pt}$f_2(1430)$ & $0.28$ & $0.01$ & $3.83$ & $0.49$ & $0.04$ & $0.36$ \\
      \hspace{10pt}$\rho(1700)^0$ & $0.24$ & $0.01$ & $0.07$ & $0.52$ & $<0.01$ & $0.45$ \\
      QMI specifics & \\
      \hspace{10pt}QMI bias & $0.89$ & $0.11$ & $0.32$ & $3.65$ & $0.47$ & $0.93$ \\
      \hline
      Total model & $\num{1.35}\phantom{0}$ & $\num{0.12}\phantom{0}$ & $\num{3.90}\phantom{0}$ & $\num{3.82}\phantom{0}$ & $\num{0.48}\phantom{0}$ & $\num{1.39}\phantom{0}$ \\
      \hline
      Statistical uncertainty & $\num{0.61}\phantom{0}$ & $\num{0.09}\phantom{0}$ & $\num{0.38}\phantom{0}$ & $\num{0.42}\phantom{0}$ & $\num{0.13}\phantom{0}$ & $\num{0.56}\phantom{0}$ \\
      \hline \hline
    \end{tabular}
  }
\end{table}
\sisetup{round-mode=off}

\sisetup{round-mode=places,round-precision=1}
\begin{table}[ht]
  \centering
  \caption{Systematic uncertainties on ${\cal A}_{\CP}$ values, given in units of $10^{-2}$, for the QMI method. For comparison, the statistical uncertainties are also listed at the bottom.}
  \label{tab:syst:qmi:acp}
  \resizebox{\textwidth}{!}{
    \begin{tabular}
      {@{\hspace{0.50cm}}l@{\hspace{0.25cm}}  @{\hspace{0.25cm}}r@{\hspace{0.25cm}}  @{\hspace{0.25cm}}r@{\hspace{0.25cm}}  @{\hspace{0.25cm}}r@{\hspace{0.25cm}}  @{\hspace{0.25cm}}r@{\hspace{0.25cm}}  @{\hspace{0.25cm}}r@{\hspace{0.25cm}}  @{\hspace{0.25cm}}r@{\hspace{0.50cm}}}
      \hline \hline
      Category & \multicolumn{1}{c}{$\rho(770)^0$} & \multicolumn{1}{c}{$\omega(782)$} & \multicolumn{1}{c}{$f_2(1270)$} & \multicolumn{1}{c}{$\rho(1450)^0$} & \multicolumn{1}{c}{$\rho_3(1690)^0$} & \multicolumn{1}{c}{S-wave} \\ \hline
      $B$ mass fit & $0.40$ & $1.02$ & $0.23$ & $0.92$ & $0.31$ & $0.04$ \\
      Efficiency& \\
      \hspace{10pt}Simulation sample size & $0.54$ & $1.59$ & $2.29$ & $1.19$ & $0.67$ & $0.46$ \\
      \hspace{10pt}Binning & $0.26$ & $1.46$ & $0.25$ & $1.31$ & $0.87$ & $0.24$ \\
      \hspace{10pt}L0 trigger & $0.15$ & $0.75$ & $0.14$ & $0.07$ & $0.12$ & $0.04$ \\
      Combinatorial bkgd & $0.91$ & $3.05$ & $1.96$ & $10.99$ & $2.88$ & $2.72$ \\
      $B^+ \to K^+ \pip \pim$ & $0.01$ & $0.04$ & $0.11$ & $0.33$ & $0.30$ & $0.07$ \\
      Fit bias & $1.92$ & $13.45$ & $5.14$ & $8.24$ & $7.07$ & $2.86$ \\
      \hline
      Total experimental & $\num{2.29}\phantom{0}$ & $\num{14.20}\phantom{0}$ & $\num{6.04}\phantom{0}$ & $\num{14.25}\phantom{0}$ & $\num{8.00}\phantom{0}$ & $\num{4.17}\phantom{0}$ \\
      \hline
      Amplitude model & \\
      \hspace{10pt}Resonance properties & $0.47$ & $2.31$ & $0.88$ & $3.23$ & $2.06$ & $1.26$\\
      \hspace{10pt}Barrier factors & $0.17$ & $3.39$ & $1.99$ & $12.01$ & $3.03$ & $5.12$ \\
      Alternative lineshapes & \\
      \hspace{10pt}$f_2(1270)$ & $0.02$ & $0.68$ & $0.70$ & $0.98$ & $0.32$ & $0.67$ \\
      \hspace{10pt}$f_2(1430)$ & $0.51$ & $0.72$ & $0.08$ & $2.96$ & $1.52$ & $0.67$ \\
      \hspace{10pt}$\rho(1700)^0$ & $0.63$ & $2.37$ & $0.97$ & $4.09$ & $0.29$ & $1.39$ \\
      QMI specifics & \\
      \hspace{10pt}QMI bias & $1.35$ & $5.56$ & $4.70$ & $29.40$ & $37.89$ & $4.40$ \\
      \hline
      Total model & $\num{1.58}\phantom{0}$ & $\num{7.00}\phantom{0}$ & $\num{5.24}\phantom{0}$ & $\num{32.17}\phantom{0}$ & $\num{38.05}\phantom{0}$ & $\num{6.96}\phantom{0}$ \\
      \hline
      Statistical uncertainty & $\num{1.27}\phantom{0}$ & $\num{15.44}\phantom{0}$ & $\num{3.63}\phantom{0}$ & $\num{5.55}\phantom{0}$ & $\num{17.01}\phantom{0}$ & $\num{1.52}\phantom{0}$ \\
      \hline \hline
    \end{tabular}
  }
\end{table}
\sisetup{round-mode=off}

\section{Results}
\label{sec:results}

Numerical results for the fit fractions and quasi-two-body \CP asymmetries are given in this section;
Appendices~\ref{sec:isobarTables}--\ref{sec:qmiTables} also provide the correlation matrices for the \CP-averaged fit fractions and \CP asymmetries, while Appendix~\ref{sec:extraApp} presents the the combined values for the three S-wave approaches.
The complex coefficients are fitted in terms of Cartesian parameters as shown in Eq.~\eqref{eq:cartesian}, but it can also be convenient to interpret them in terms of magnitudes and phases.
A comparison of the phases of the non-S-wave contributions between the three approaches can be found in Appendix~\ref{sec:phaseComp}.
The fitted complex coefficients are
recorded for completeness in Appendices~\ref{sec:isobarTables}, \ref{sec:kMatrixTables} and \ref{sec:qmiTables} for the isobar, K-matrix and QMI approaches, respectively. Throughout the tables in this section, separate parameters for the $\rho(770)^0$ and $\omega(782)$ resonances are presented, which are extracted from the combined component described by Equation~\ref{eqn:rho-omegamixing}.
In general, the statistical uncertainty is lowest for the model with the fewest parameters (isobar), and highest for the model with the largest number of parameters (QMI), as expected.

\begin{figure}[t]
\centering

\includegraphics[width=1.0\linewidth]{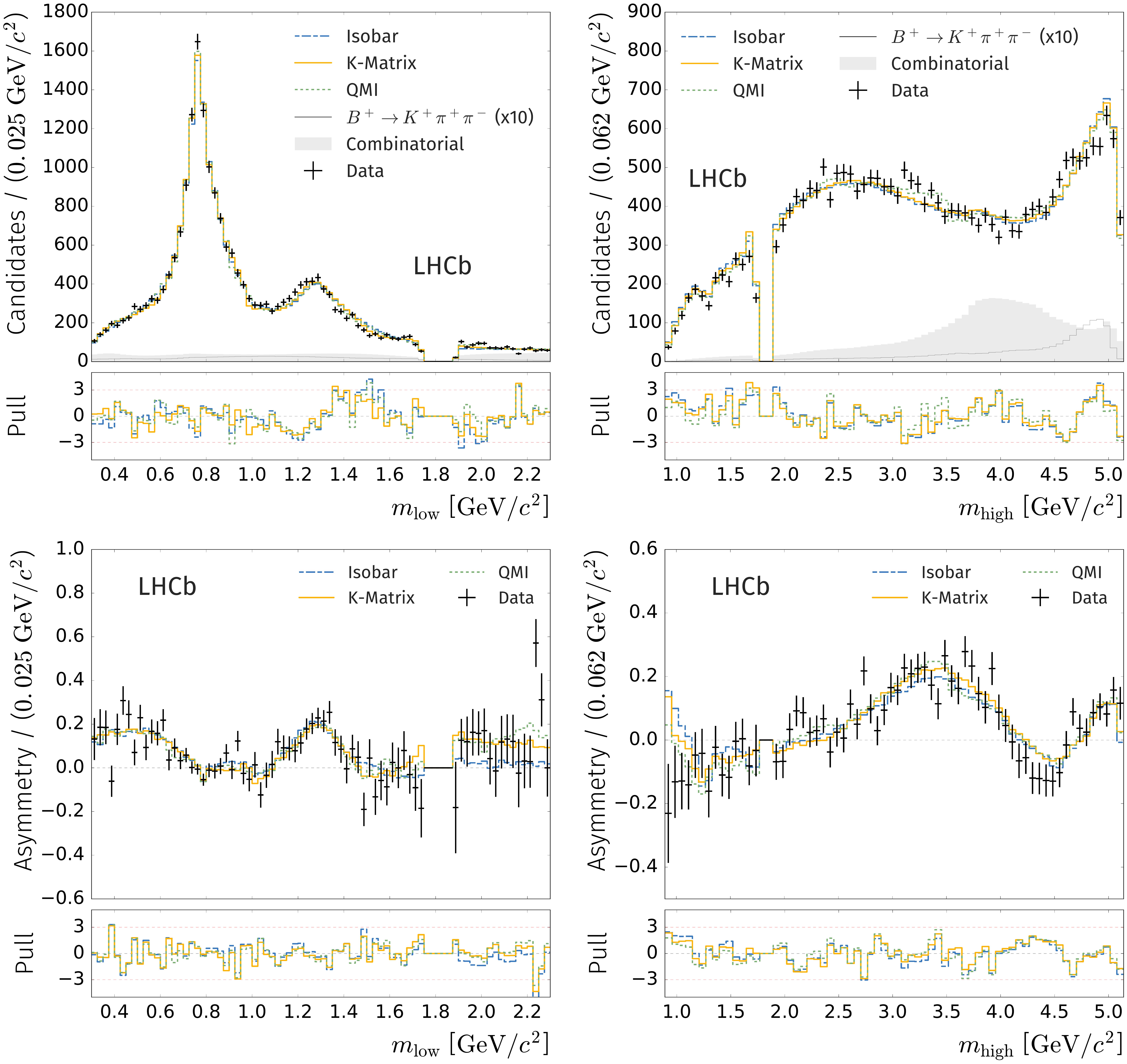}
\put(-405,330){(a)}
\put(-178,330){(b)}
\put(-407,74){(c)}
\put(-178,74){(d)}

\caption{Fit projections of each model (a)~in the low $m_{\rm low}$ region and (b)~in the full range of $m_{\rm high}$, with the corresponding asymmetries shown beneath in (c) and (d). The normalised residual or pull distribution, defined as the difference between the bin value less the fit value over the uncertainty on the number of events in that bin, is shown below each fit projection.}
\label{fig:projAll}
\end{figure}

\begin{figure}[t]
\centering

\includegraphics[width=1.0\linewidth]{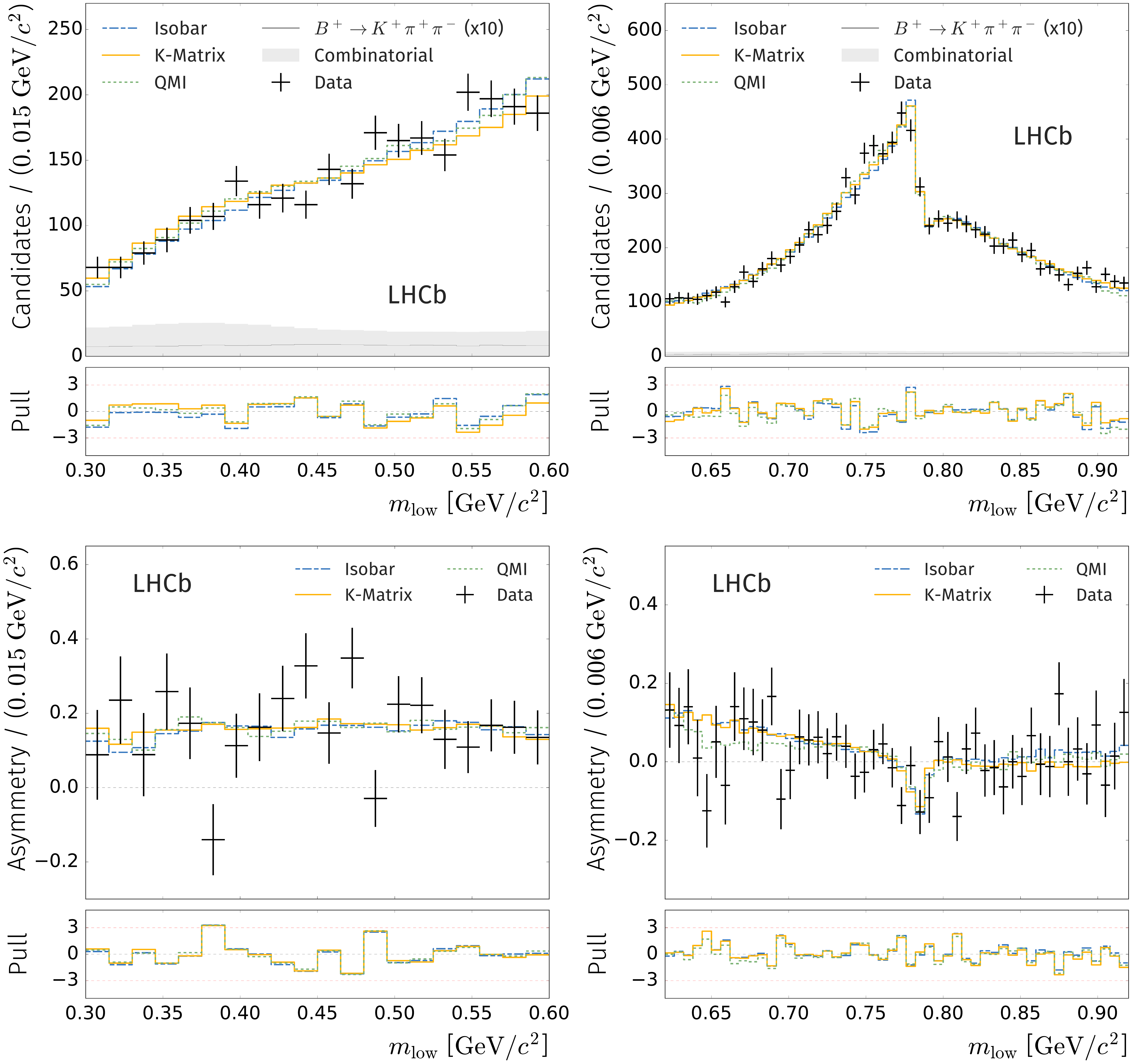}
\put(-411,294){(a)}
\put(-183,294){(b)}
\put(-410,73){(c)}
\put(-183,73){(d)}

\caption{Fit projections of each model on $m_{\rm low}$ (a)~in the region below the $\rho(770)^0$ resonance and (b)~in the $\rho(770)^0$ region, with the corresponding asymmetries shown beneath in (c) and (d). The pull distribution is shown below each fit projection.}
\label{fig:projLowRho}
\end{figure}

\begin{figure}[t]
\centering

\includegraphics[width=1.0\linewidth]{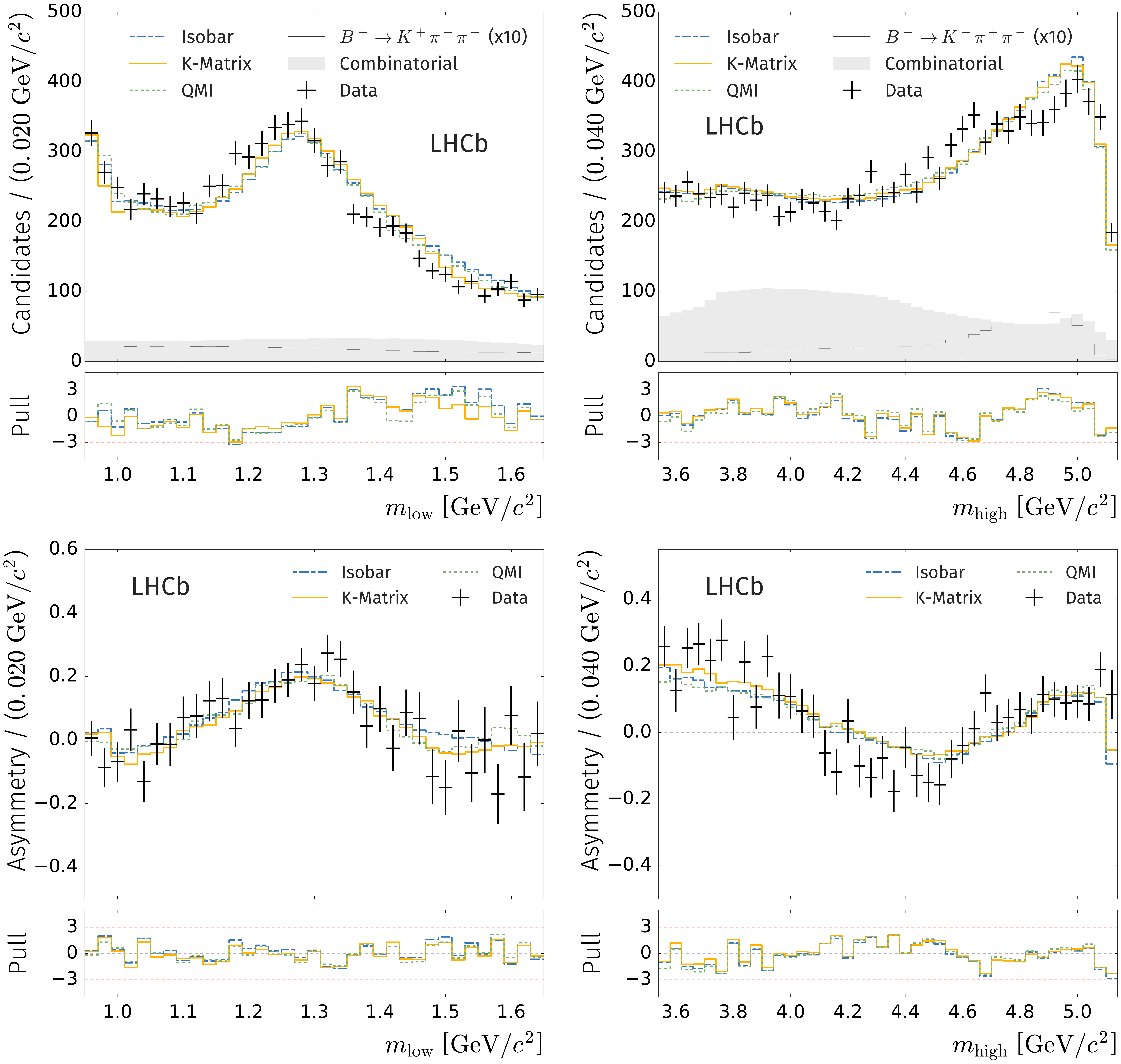}
\put(-411,294){(a)}
\put(-183,294){(b)}
\put(-410,73){(c)}
\put(-183,73){(d)}

\caption{Fit projections of each model on $m_{\rm low}$ (a)~in the region around the $f_2(1270)$ resonance and (b)~in the high $m_{\rm high}$ region, with the corresponding asymmetries shown beneath in (c) and (d). The pull distribution is shown below each fit projection.}
\label{fig:projf2High}
\end{figure}

\begin{figure}[t]
\centering

\includegraphics[width=1.0\linewidth]{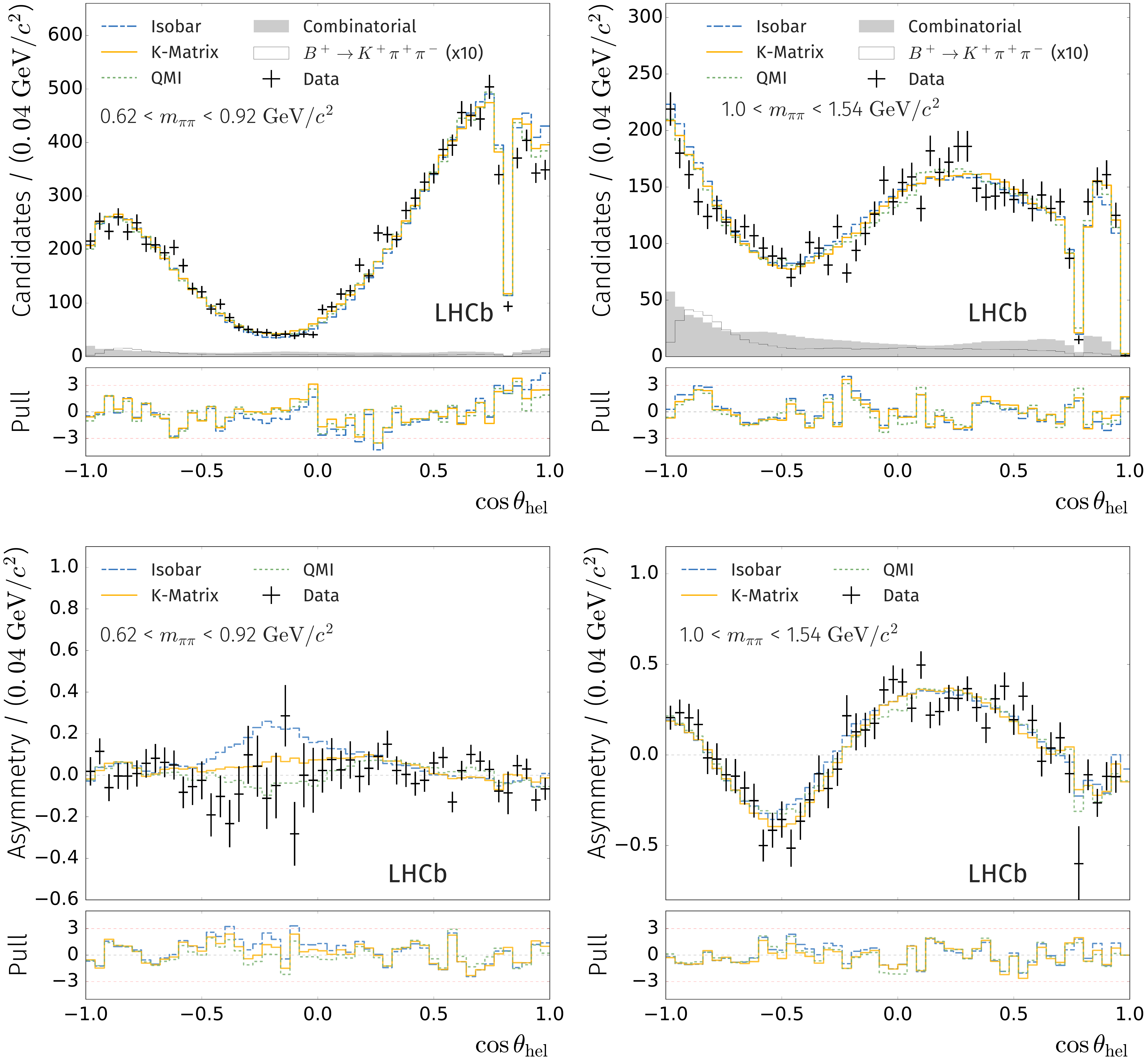}
\put(-411,299){(a)}
\put(-183,299){(b)}
\put(-410,73){(c)}
\put(-183,73){(d)}

\caption{Fit projections of each model on $\cos\,\theta_{\rm hel}$ (a)~in the region around the $\rho(770)^0$ resonance and (b)~in the $f_2(1270)$ region, with the corresponding asymmetries shown beneath in (c) and (d). The pull distribution is shown below each fit projection.}
\label{fig:projHelRhof2}
\end{figure}

\begin{figure}[t]
\centering

\includegraphics[width=1.0\linewidth]{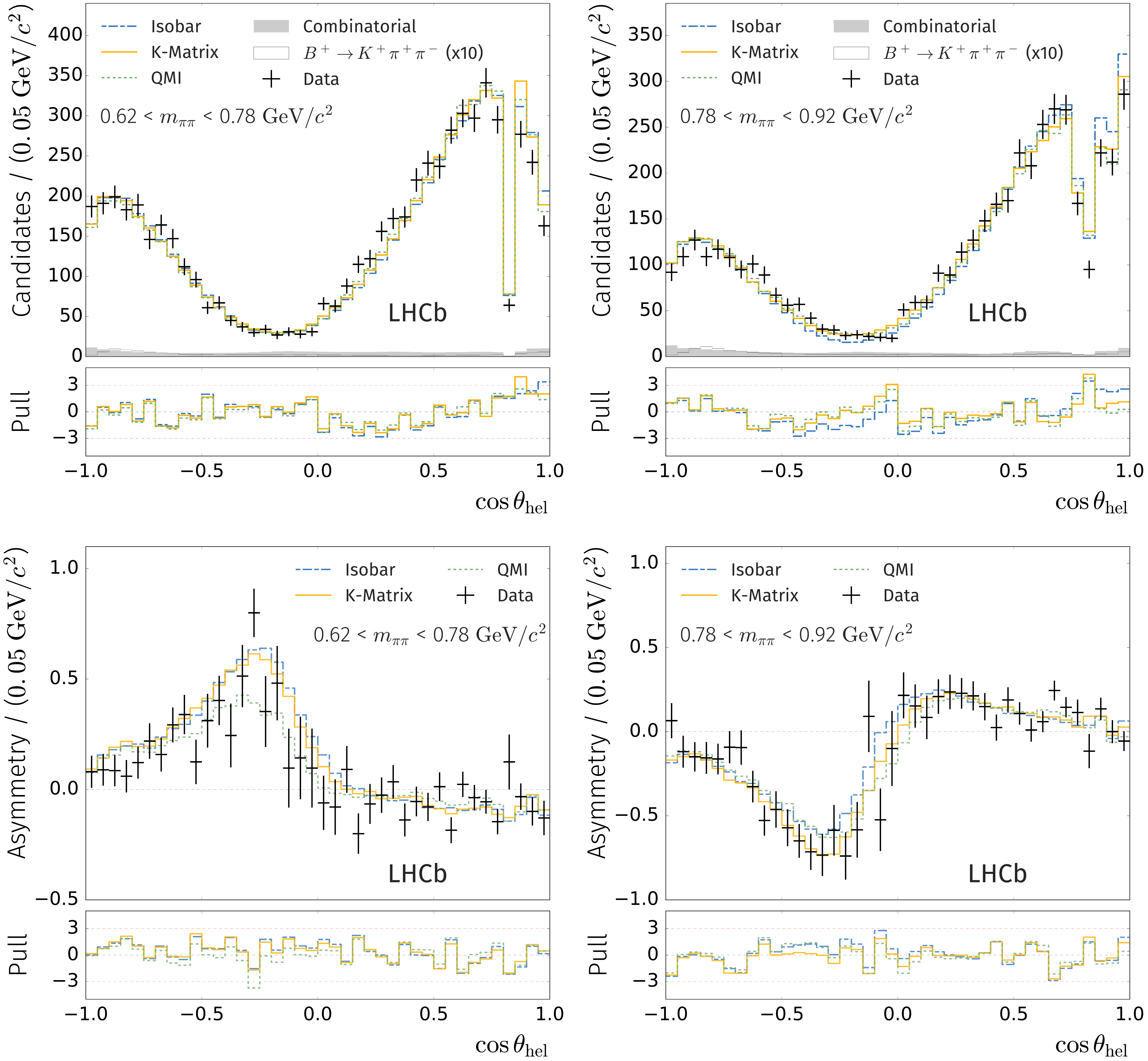}
\put(-411,296){(a)}
\put(-183,296){(b)}
\put(-410,74){(c)}
\put(-183,74){(d)}

\caption{Fit projections of each model on $\cos\,\theta_{\rm hel}$ in the regions (a)~below and (b)~above the $\rho(770)^0$ resonance pole, with the corresponding asymmetries shown beneath in (c) and (d). The pull distribution is shown below each fit projection.}
\label{fig:projHelRhoUL}
\end{figure}

\begin{figure}[t]
\centering

\includegraphics[width=0.5\linewidth]{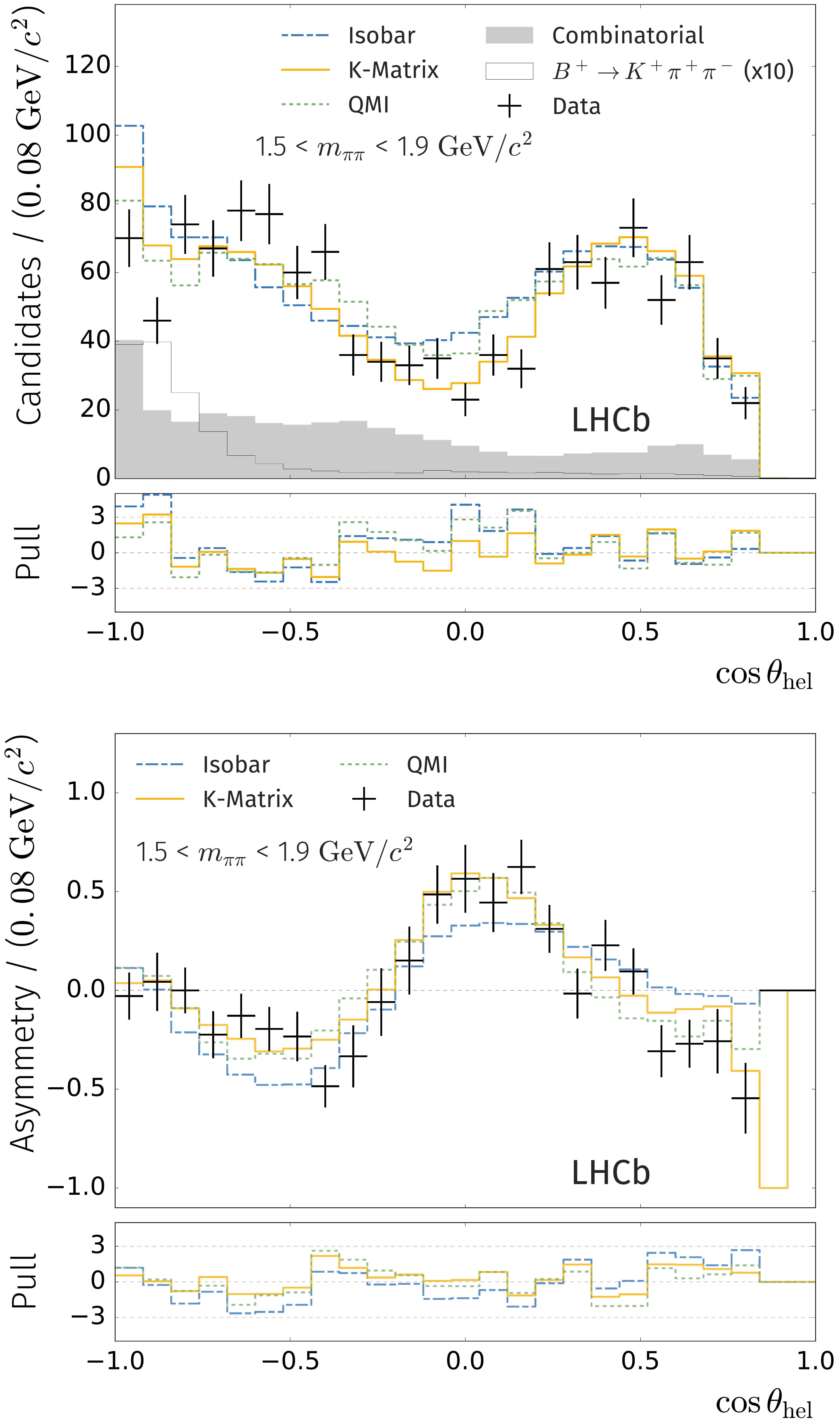}
\put(-180,285){(a)}
\put(-180,64){(b)}

\caption{Fit projections of each model (a)~on $\cos\,\theta_{\rm hel}$ in the $\rho_3(1690)$ region, with (b)~the corresponding asymmetry shown beneath. The pull distribution is shown below each fit projection.}
\label{fig:projHelRho3}
\end{figure}

\begin{figure}[t]
\centering
\includegraphics[width=0.49\linewidth]{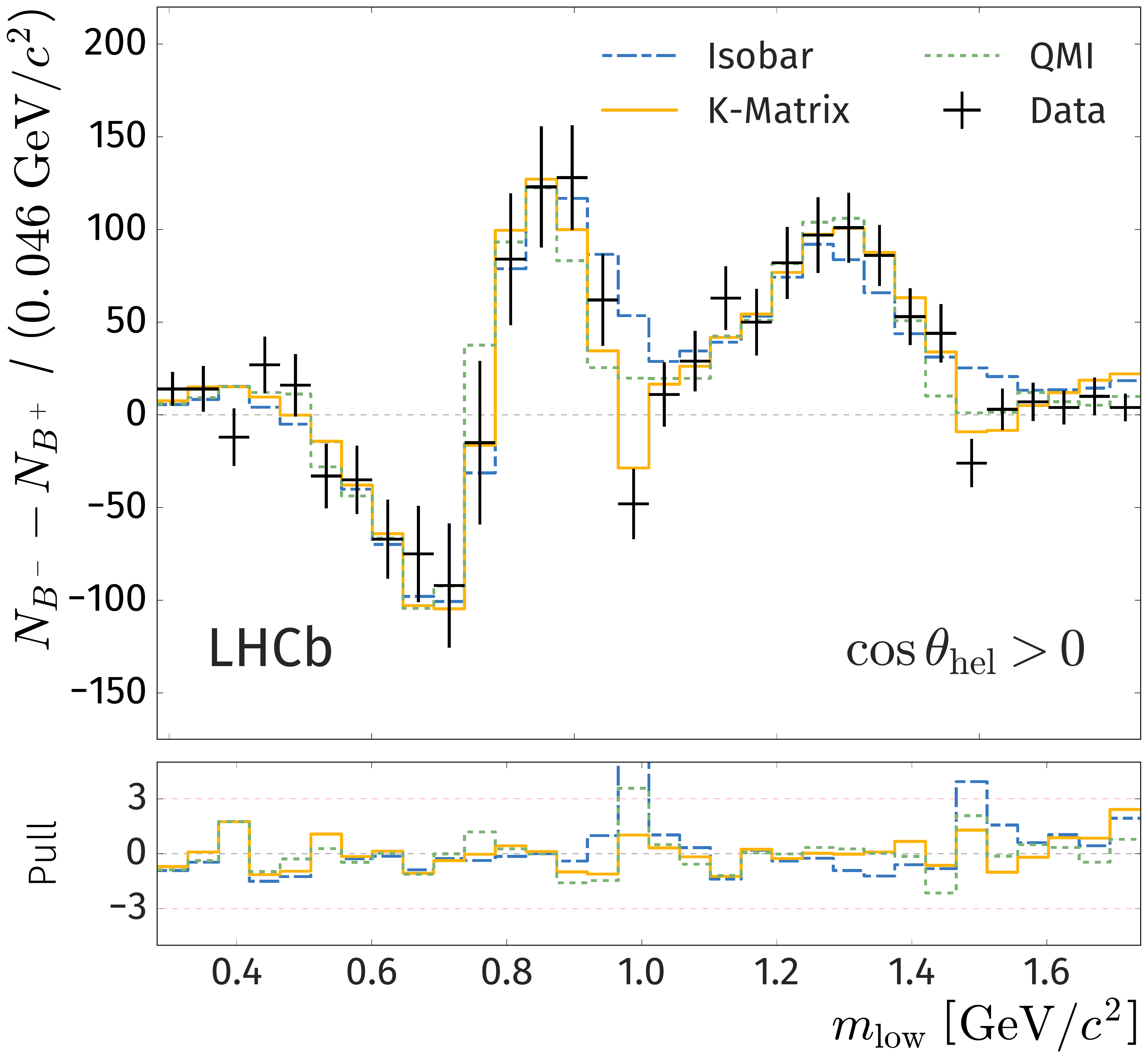}
\includegraphics[width=0.49\linewidth]{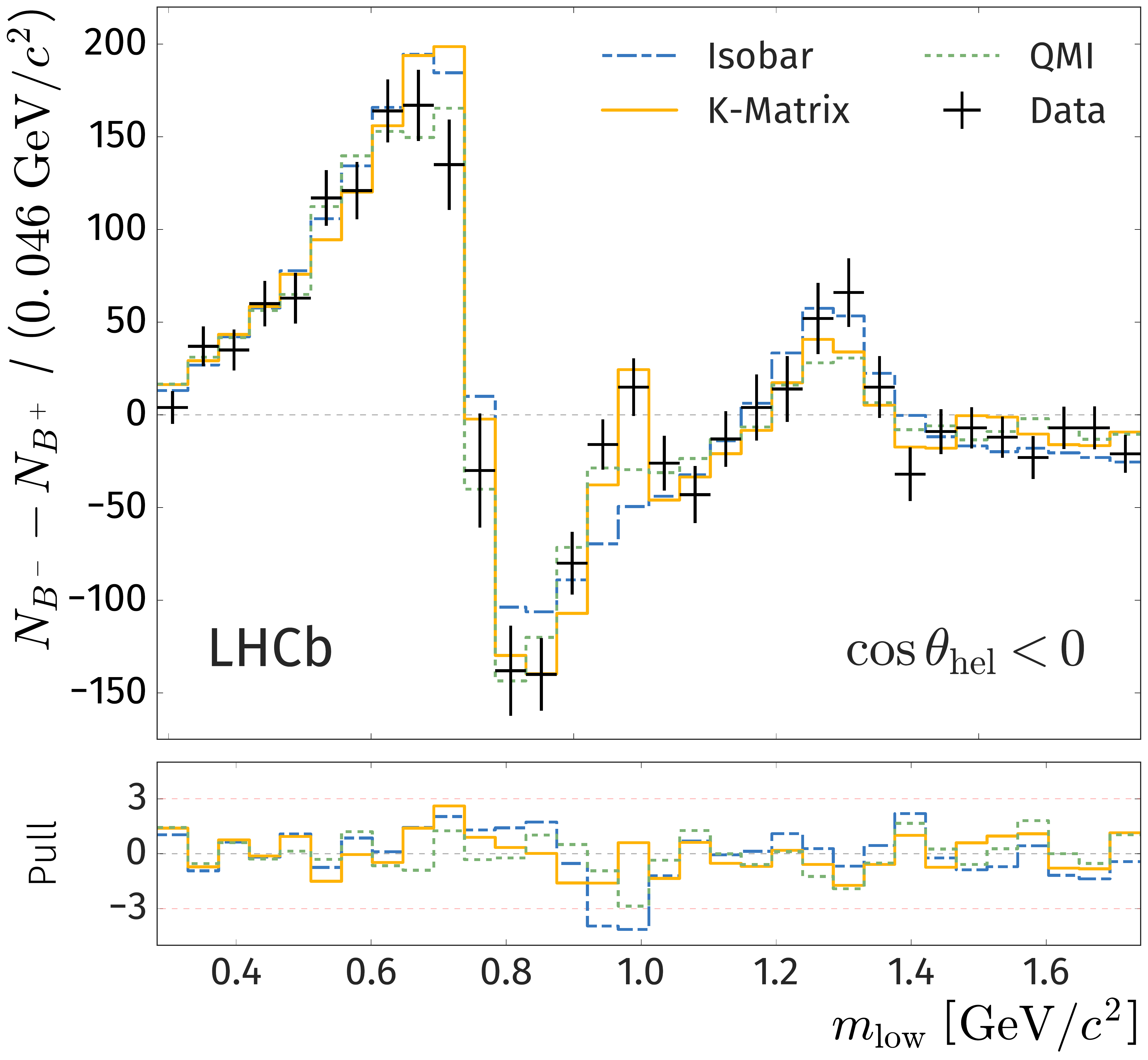}
\put(-252,92){(a)}
\put(-27,92){(b)}

\caption{Raw difference in the number of $\Bm$ and $\Bp$ candidates in the low $m_{\rm low}$ region, for (a)~positive, and (b)~negative cosine of the helicity angle. The pull distribution is shown below each fit projection.}
\label{fig:projHelCut}
\end{figure}

A comparison of the data and all three fit models, projected onto $m_{\rm low}$ for a large dipion mass range, along with the asymmetries between $\Bm$ and $\Bp$ decays, can be seen in Fig.~\ref{fig:projAll}. In these and subsequent figures, the difference between the data and model expectation, divided by the uncertainty on this quantity (the ``pull'') is also shown below in the same binning scheme.
Projections focussing on the low $m_{\rm low}$ and the $\rho(770)^0$ regions are shown in Fig.~\ref{fig:projLowRho}, while the $f_2(1270)$ and high $m_{\rm high}$ regions are displayed in Fig.~\ref{fig:projf2High}.
The fit result projected on the helicity angle in each of the $\rho(770)^0$ and $f_2(1270)$ regions, shown in Fig.~\ref{fig:projHelRhof2}, is also given separately above and below the $\rho(770)^0$ pole in Fig.~\ref{fig:projHelRhoUL}.
The projection on the helicity angle in the vicinity of the $\rho_3(1690)^0$ resonance is shown in Fig.~\ref{fig:projHelRho3}.
Figure~\ref{fig:projHelCut} shows the raw difference in the number of $\Bm$ and $\Bp$ candidates in the low $m_{\rm low}$ region for negative and positive helicity angle cosines.
Additional projections separating the contributions for various components of the amplitude model are shown in Appendices~\ref{sec:isobarProj}, \ref{sec:kMatrixProj} and~\ref{sec:qmiProj}.

\subsection{Fit fractions}

The \CP-averaged fit fractions are given in Table~\ref{tab:ff} for all three S-wave approaches.
The fit fractions and interference fit fractions, separated by \Bpm charge for each S-wave approach, are given in Tables~\ref{tab:iso:ffp}--\ref{tab:qmi:ffm}.
In all cases, statistical uncertainties are calculated using $68\%$ confidence intervals obtained from the results of fits performed to data sets sampled from the nominal fit models.
Throughout this paper, if three uncertainties are listed, they are separated into statistical, systematic and amplitude model sources, whereas if only two are listed, the systematic and models sources have been combined in quadrature for brevity.
The total sums of fit fractions for the \Bp and \Bm amplitudes can be found in Table~\ref{tab:ffSum}.

\renewcommand{\arraystretch}{1.35}
\begin{table}[ht]
  \centering
  \caption{The \CP-separated sum of fit fractions in units of $10^{-2}$, for each approach, where the first uncertainty is statistical, the second the experimental systematic and the third is the model systematic.
}
  \label{tab:ffSum}
  % \resizebox{\textwidth}{!}{
    \begin{tabular}
      {@{\hspace{0.5cm}}l@{\hspace{0.25cm}}  @{\hspace{0.25cm}}r@{\hspace{0.25cm}}  @{\hspace{0.25cm}}r@{\hspace{0.25cm}}}
      \hline \hline

      S-wave approach & \multicolumn{1}{c}{\Bp} & \multicolumn{1}{c}{\Bm} \\
      \hline
      Isobar & $93.7 \pm 2.6 \pm 1.5 \pm 4.5$ & $100.7 \pm 2.7 \pm 1.7 \pm 6.0$ \\
      K-Matrix & $99.2 \pm 1.8 \pm 4.1 \pm 5.7$ & $108.3 \pm 1.7 \pm 3.3 \pm 9.3$ \\
      QMI & $92.2 \pm 1.2 \pm 7.7 \pm 3.2$ & $108.0 \pm 1.7 \pm 3.7 \pm 6.3$ \\

      \hline \hline
    \end{tabular}
  % }
\end{table}

\renewcommand{\arraystretch}{1.35}
\begin{table}[ht]
  \centering
  \caption{The \CP-averaged fit fractions in units of $10^{-2}$, for each approach, where the first uncertainty is statistical, the second the experimental systematic and the third is the model systematic.
}
  \label{tab:ff}
  \resizebox{\textwidth}{!}{
    \begin{tabular}
      {@{\hspace{0.5cm}}l@{\hspace{0.25cm}}  @{\hspace{0.25cm}}r@{\hspace{0.25cm}}  @{\hspace{0.25cm}}r@{\hspace{0.25cm}}  @{\hspace{0.25cm}}r@{\hspace{0.5cm}}}
      \hline \hline

      Component & \multicolumn{1}{c}{Isobar} & \multicolumn{1}{c}{K-matrix} & \multicolumn{1}{c}{QMI} \\
      \hline
      $\rho(770)^0$ & $55.5\phantom{0} \pm 0.6\phantom{0} \pm 0.4\phantom{0} \pm 2.5\phantom{0}$ &
      $ 56.5\phantom{0} \pm 0.7\phantom{0} \pm 1.5\phantom{0} \pm 3.1\phantom{0}$ &
      $54.8\phantom{0} \pm 1.0\phantom{0} \pm 1.9\phantom{0} \pm 1.0\phantom{0}$ \\

      $\omega(782)$ & $0.50 \pm 0.03 \pm 0.01 \pm 0.04$ &
      $ 0.47 \pm 0.04 \pm 0.01 \pm 0.03$ &
      $\phantom{0}0.57 \pm 0.10 \pm 0.12 \pm 0.12$ \\

      $f_2(1270)$ & $9.0\phantom{0} \pm 0.3\phantom{0} \pm 0.7\phantom{0} \pm 1.4\phantom{0}$ &
      $ 9.3\phantom{0} \pm 0.4\phantom{0} \pm 0.6\phantom{0} \pm 2.4\phantom{0}$ &
      $9.6\phantom{0} \pm 0.4\phantom{0} \pm 0.7\phantom{0} \pm 3.9\phantom{0}$ \\

      $\rho(1450)^0$ & $5.2\phantom{0} \pm 0.3\phantom{0} \pm 0.2\phantom{0} \pm 1.9\phantom{0}$ &
      $ 10.5\phantom{0} \pm 0.7\phantom{0} \pm 0.8\phantom{0} \pm 4.5\phantom{0}$ &
      $\phantom{0}7.4\phantom{0} \pm 0.5\phantom{0} \pm 3.9\phantom{0} \pm 1.1\phantom{0}$ \\

      $\rho_3(1690)^0$ & $0.5\phantom{0} \pm 0.1\phantom{0} \pm 0.1\phantom{0} \pm 0.3\phantom{0}$ &
      $ 1.5\phantom{0} \pm 0.1\phantom{0} \pm 0.1\phantom{0} \pm 0.4\phantom{0}$ &
      $\phantom{0}1.0\phantom{0} \pm 0.1\phantom{0} \pm 0.5\phantom{0} \pm 0.1\phantom{0}$ \\

      S-wave & $25.4\phantom{0} \pm 0.5\phantom{0} \pm 0.5\phantom{0} \pm 3.6\phantom{0}$ &
      $ 25.7\phantom{0} \pm 0.6\phantom{0} \pm 2.6\phantom{0} \pm 1.4\phantom{0}$ &
      $26.8\phantom{0} \pm 0.7\phantom{0} \pm 2.0\phantom{0} \pm 1.0\phantom{0}$ \\

      \hline \hline
    \end{tabular}
  }
\end{table}

\begin{table}[ht]
  \centering
  \caption{Fit (diagonal) and interference (off-diagonal) fractions for \Bp decay in units of $10^{-2}$, between amplitude components in the isobar approach. The first uncertainty is statistical and the second the quadratic sum of systematic and model sources.}
  \label{tab:iso:ffp}
  \resizebox{\textwidth}{!}{
    \begin{tabular}
      {@{\hspace{0.5cm}}c@{\hspace{0.25cm}} @{\hspace{0.25cm}}|r@{\hspace{0.25cm}}  @{\hspace{0.25cm}}r@{\hspace{0.25cm}}  @{\hspace{0.25cm}}r@{\hspace{0.25cm}}  @{\hspace{0.25cm}}r@{\hspace{0.25cm}}  @{\hspace{0.25cm}}r@{\hspace{0.25cm}}  @{\hspace{0.25cm}}r@{\hspace{0.25cm}}  @{\hspace{0.25cm}}r@{\hspace{0.5cm}}}
      \hline \hline

                                         & \multicolumn{1}{c}{$\rho(770)^0$--$\omega(782)$}     & \multicolumn{1}{c}{$f_2(1270)$}              & \multicolumn{1}{c}{$\rho(1450)^0$}             & \multicolumn{1}{c}{$\rho_3(1690)^0$}         & \multicolumn{1}{c}{rescattering}             & \multicolumn{1}{c}{$\sigma$} \\ \hline
      $\rho(770)^0$--$\omega(782)$       & $ 57.9 \pm 0.8 \pm 1.6$          & $ -1.8 \pm 0.1 \pm 0.3$  & $+8.3 \pm 0.6 \pm 4.1$     & $+0.8 \pm 0.1 \pm 0.1$  & $ -0.7 \pm 0.1 \pm 0.1$  & $+1.3 \pm 0.2 \pm 0.4$\\
      $f_2(1270)$                         &                                & $ 5.1 \pm 0.4 \pm 1.1$    & $ -0.4 \pm 0.1 \pm 0.3$    & $ -0.2 \pm 0.0 \pm 0.0$  & $+0.2 \pm 0.0 \pm 0.1$  & $-0.2 \pm 0.1 \pm 0.3$\\
      $\rho(1450)^0$                     &                                 &                    & $ 6.2 \pm 0.5 \pm 1.1$            & $+0.1 \pm 0.0 \pm 0.1$   & $ -0.1 \pm 0.0 \pm 0.1$  & $-2.7 \pm 0.1 \pm 0.4$\\
      $\rho_3(1690)^0$                   &                                 &                    &                                   & $ 1.0 \pm 0.2 \pm 0.2$   & $ 0.0 \pm 0.0 \pm 0.0$   & $+0.3 \pm 0.1 \pm 0.1$\\
      rescattering                       &                                 &                    &                                  &                          & $ 0.8 \pm 0.1 \pm 0.2$   & $0.0 \pm 0.2 \pm 0.6$\\
      $\sigma$                           &                                 &                    &                                  &                           &                         & $22.2 \pm 0.6 \pm 0.9$\\

      \hline \hline
    \end{tabular}
  }
\end{table}

\begin{table}[ht]
  \centering
  \caption{Fit (diagonal) and interference (off-diagonal) fractions for \Bm decay in units of $10^{-2}$, between amplitude components in the isobar approach. The first uncertainty is statistical and the second the quadratic sum of systematic and model sources.}
  \label{tab:iso:ffm}
  \resizebox{\textwidth}{!}{
    \begin{tabular}
      {@{\hspace{0.5cm}}c@{\hspace{0.25cm}} @{\hspace{0.25cm}}|r@{\hspace{0.25cm}}  @{\hspace{0.25cm}}r@{\hspace{0.25cm}}  @{\hspace{0.25cm}}r@{\hspace{0.25cm}}  @{\hspace{0.25cm}}r@{\hspace{0.25cm}}  @{\hspace{0.25cm}}r@{\hspace{0.25cm}}  @{\hspace{0.25cm}}r@{\hspace{0.25cm}}  @{\hspace{0.25cm}}r@{\hspace{0.5cm}}}
      \hline \hline
      & \multicolumn{1}{c}{$\rho(770)^0$--$\omega(782)$}     
      & \multicolumn{1}{c}{$f_2(1270)$}              
      & \multicolumn{1}{c}{$\rho(1450)^0$}             
      & \multicolumn{1}{c}{$\rho_3(1690)^0$}         
      & \multicolumn{1}{c}{rescattering}             
      & \multicolumn{1}{c}{$\sigma$} \\ \hline
      $\rho(770)^0$--$\omega(782)$     & $ 53.3 \pm 0.8 \pm 2.1$          & $ -1.4 \pm 0.1 \pm 0.6$ & $+1.8 \pm 0.6 \pm 1.4$   & $+0.2 \pm 0.1 \pm 0.2$     & $ -0.5 \pm 0.1 \pm 0.1$    & $+5.7 \pm 0.4 \pm 0.3$\\
      $f_2(1270)$                      &                                  & $ 12.6 \pm 0.4 \pm 1.6$ & $ -0.9 \pm 0.1 \pm 0.3$    & $ 0.0 \pm 0.0 \pm 0.0$    & $+0.5 \pm 0.0 \pm 0.1$    & $ -1.1 \pm 0.1 \pm 0.6$\\
      $\rho(1450)^0$                   &                                  &                      & $ 4.3 \pm 0.4 \pm 2.9$   & $ 0.0 \pm 0.0 \pm 0.0$     & $ -0.2 \pm 0.1 \pm 0.1$      & $ -2.1 \pm 0.2 \pm 0.7$\\
      $\rho_3(1690)^0$                 &                                  &                      &                       & $ 0.1 \pm 0.0 \pm 0.1$      & $ 0.0 \pm 0.0 \pm 0.0$         & $+0.2 \pm 0.1 \pm 0.1$\\
      rescattering                     &                                 &                       &                       &                      & $ 2.0 \pm 0.2 \pm 0.6$        & $ -2.9 \pm 0.3 \pm 0.5$\\
      $\sigma$                         &                                  &                      &                      &                      &                             & $ 27.9 \pm 0.7 \pm 1.5$\\

      \hline \hline
    \end{tabular}
  }
\end{table}

\begin{table}[ht]
  \centering
  \caption{Fit (diagonal) and interference (off-diagonal) fractions for \Bp decay in units of $10^{-2}$, between amplitude components in the K-matrix approach. The first uncertainty is statistical and the second the quadratic sum of systematic and model sources.}
  \label{tab:km:ffp}
  \resizebox{\textwidth}{!}{
    \begin{tabular}
      {@{\hspace{0.5cm}}c@{\hspace{0.25cm}} @{\hspace{0.25cm}}|r@{\hspace{0.25cm}}  @{\hspace{0.25cm}}r@{\hspace{0.25cm}}  @{\hspace{0.25cm}}r@{\hspace{0.25cm}}  @{\hspace{0.25cm}}r@{\hspace{0.25cm}}  @{\hspace{0.25cm}}r@{\hspace{0.5cm}}}
      \hline \hline

      & \multicolumn{1}{c}{$\rho(770)^0$--$\omega(782)$}     & \multicolumn{1}{c}{$f_2(1270)$}              & \multicolumn{1}{c}{$\rho(1450)^0$}             & \multicolumn{1}{c}{$\rho_3(1690)^0$}         & \multicolumn{1}{c}{S-wave} \\ \hline
      $\rho(770)^0$--$\omega(782)$     & $  59.5 \pm 2.1 \pm 9.8$ & $ -1.7 \pm 0.1 \pm 0.5$ & $ +2.8 \pm 1.5 \pm 7.7$ & $ +1.0 \pm 0.1 \pm 0.3$ & $ +0.6 \pm 0.5 \pm 2.3$\\
      $f_2(1270)$ & & $  5.6 \pm 0.5 \pm 2.5$ & $ -0.3 \pm 0.1 \pm 0.4$ & $ -0.3 \pm 0.0 \pm 0.1$ & $  0.0 \pm 0.2 \pm 0.5 $\\
      $\rho(1450)^0$ & & & $  10.2 \pm 0.8 \pm 3.1$ & $ +0.4 \pm 0.1 \pm 0.4$ & $ -3.0 \pm 0.3 \pm 1.0$\\
      $\rho_3(1690)^0$ & & & & $  2.1 \pm 0.2 \pm 0.3$ & $  0.0 \pm 0.2 \pm 0.6$\\
      S-wave & & & & & $  23.1 \pm 0.9 \pm 3.7$\\

      \hline \hline
    \end{tabular}
  }
\end{table}

\begin{table}[ht]
  \centering
  \caption{Fit (diagonal) and interference (off-diagonal) fractions for \Bm decay in units of $10^{-2}$, between amplitude components in the K-matrix approach. The first uncertainty is statistical and the second the quadratic sum of systematic and model sources.}
  \label{tab:km:ffm}
  \resizebox{\textwidth}{!}{
    \begin{tabular}
      {@{\hspace{0.5cm}}c@{\hspace{0.25cm}} @{\hspace{0.25cm}}|r@{\hspace{0.25cm}}  @{\hspace{0.25cm}}r@{\hspace{0.25cm}}  @{\hspace{0.25cm}}r@{\hspace{0.25cm}}  @{\hspace{0.25cm}}r@{\hspace{0.25cm}}  @{\hspace{0.25cm}}r@{\hspace{0.5cm}}}
      \hline \hline
      & \multicolumn{1}{c}{$\rho(770)^0$--$\omega(782)$}     & \multicolumn{1}{c}{$f_2(1270)$}              & \multicolumn{1}{c}{$\rho(1450)^0$}             & \multicolumn{1}{c}{$\rho_3(1690)^0$}         & \multicolumn{1}{c}{S-wave} \\ \hline
      $\rho(770)^0$--$\omega(782)$     & $ 56.1 \pm 2.0 \pm 14.0$ & $ -0.2 \pm 0.2 \pm 0.9$ & $ -8.6 \pm 1.3 \pm 10.9$ & $+0.7 \pm 0.1 \pm 0.3$ & $+4.4 \pm 0.4 \pm 1.4$\\
      $f_2(1270)$ & & $ 12.5 \pm 0.6 \pm 2.4$ & $ -1.1 \pm 0.1 \pm \phantom{0}0.3$ & $ -0.2 \pm 0.0 \pm 0.2$ & $ -1.2 \pm 0.2 \pm 0.9 $\\
      $\rho(1450)^0$ & & & $ 10.8 \pm 1.0 \pm \phantom{0}7.4$ & $ -0.1 \pm 0.1 \pm 0.2$ & $ -3.0 \pm 0.3 \pm 1.4$\\
      $\rho_3(1690)^0$ & & & & $ 0.9 \pm 0.2 \pm 0.6$ & $+0.8 \pm 0.1 \pm 0.5$\\
      S-wave & & & & & $ 28.1 \pm 0.7 \pm 3.1$\\

      \hline \hline
    \end{tabular}
  }
\end{table}

\begin{table}[ht]
  \centering
  \caption{Fit (diagonal) and interference (off-diagonal) fractions for \Bp decay in units of $10^{-2}$, between amplitude components in the QMI approach. The first uncertainty is statistical and the second the quadratic sum of systematic and model sources.}
  \label{tab:qmi:ffp}
  \resizebox{\textwidth}{!}{
    \begin{tabular}
      {@{\hspace{0.5cm}}c@{\hspace{0.25cm}} @{\hspace{0.25cm}}|r@{\hspace{0.25cm}}  @{\hspace{0.25cm}}r@{\hspace{0.25cm}}  @{\hspace{0.25cm}}r@{\hspace{0.25cm}}  @{\hspace{0.25cm}}r@{\hspace{0.25cm}}  @{\hspace{0.25cm}}r@{\hspace{0.5cm}}}
      \hline \hline
      & \multicolumn{1}{c}{$\rho(770)^0$--$\omega(782)$}     & \multicolumn{1}{c}{$f_2(1270)$}              & \multicolumn{1}{c}{$\rho(1450)^0$}             & \multicolumn{1}{c}{$\rho_3(1690)^0$}         & \multicolumn{1}{c}{S-wave} \\ \hline
      $\rho(770)^0$--$\omega(782)$ & $52.4 \pm 1.3 \pm 4.4$ & $-1.7 \pm 0.1 \pm 0.7$ & $+5.6 \pm 1.3 \pm 6.1$ & $+0.8 \pm 0.1 \pm 0.3$ & $-1.1 \pm 0.6 \pm 3.3$\\
      $f_2(1270)$ & & $6.0 \pm 0.5 \pm 3.5$ & $-0.4 \pm 0.0 \pm 0.4$ & $-0.4 \pm 0.0 \pm 0.1$ & $+0.2 \pm 0.1 \pm 0.5$\\
      $\rho(1450)^0$ & & & $8.5 \pm 0.8 \pm 3.4$ & $+0.1 \pm 0.0 \pm 0.4$ & $-2.3 \pm 0.6 \pm 2.0$\\
      $\rho_3(1690)^0$ & & & & $1.9 \pm 0.3 \pm 0.5$ & $-0.3 \pm 0.1 \pm 0.4$\\
      S-wave & & & & & $22.8 \pm 1.1 \pm 3.4$\\

      \hline \hline
    \end{tabular}
    }
\end{table}

\begin{table}[ht]
  \centering
  \caption{Fit (diagonal) and interference (off-diagonal) fractions for \Bm decay in units of $10^{-2}$, between amplitude components in the QMI approach. The first uncertainty is statistical and the second the quadratic sum of systematic and model sources.}
  \label{tab:qmi:ffm}
  \resizebox{\textwidth}{!}{
    \begin{tabular}
      {@{\hspace{0.5cm}}c@{\hspace{0.25cm}} @{\hspace{0.25cm}}|r@{\hspace{0.25cm}}  @{\hspace{0.25cm}}r@{\hspace{0.25cm}}  @{\hspace{0.25cm}}r@{\hspace{0.25cm}}  @{\hspace{0.25cm}}r@{\hspace{0.25cm}}  @{\hspace{0.25cm}}r@{\hspace{0.5cm}}}
      \hline \hline
      & \multicolumn{1}{c}{$\rho(770)^0$--$\omega(782)$}     & \multicolumn{1}{c}{$f_2(1270)$}              & \multicolumn{1}{c}{$\rho(1450)^0$}             & \multicolumn{1}{c}{$\rho_3(1690)^0$}         & \multicolumn{1}{c}{S-wave} \\ \hline
      $\rho(770)^0$--$\omega(782)$ & $57.2 \pm 1.3 \pm 4.8$ & $-0.6 \pm 0.2 \pm 0.9$ & $-3.8 \pm 1.3 \pm 5.1$ & $+0.1 \pm 0.1 \pm 0.1$ & $+6.0 \pm 0.7 \pm 2.3$\\
      $f_2(1270)$ & & $13.2 \pm 0.6 \pm 7.6$ & $-0.7 \pm 0.1 \pm 0.5$ & $-0.1 \pm 0.0 \pm 0.2$ & $+0.1 \pm 0.2 \pm 1.9$\\
      $\rho(1450)^0$ & & & $6.2 \pm 0.7 \pm 5.5$ & $ 0.0 \pm 0.0 \pm 0.1$ & $-2.9 \pm 0.4 \pm 2.3$\\
      $\rho_3(1690)^0$ & & & & $0.1 \pm 0.1 \pm 0.6$ & $ 0.0 \pm 0.1 \pm 0.4$\\
      S-wave & & & & & $30.8 \pm 0.8 \pm 2.7$\\

      \hline \hline
    \end{tabular}
    }
\end{table}

\subsection{\texorpdfstring{\boldmath{\CP} asymmetries}{CP asymmetries}}

Quasi-two-body \CP asymmetries associated to each component are shown in Table~\ref{tab:acp}.
A detailed discussion of these results is given in Section~\ref{sec:discussion}, however it should be stressed that \CP-violation effects can manifest in Dalitz-plot distributions through interference effects that leave values of the quasi-two-body \CP asymmetries consistent with zero, and indeed this occurs in \decay{\Bp}{\pip\pip\pim} decays.
The phase-space integrated \CP asymmetry is consistent, in all three models, with the value previously determined through model-independent analysis~\cite{LHCb-PAPER-2014-044}.

\renewcommand{\arraystretch}{1.35}
\begin{table}[ht]
  \centering
  \caption{Quasi-two-body \CP asymmetries in units of $10^{-2}$, for each approach. The first uncertainty is statistical, the second the experimental systematic and the third is the model systematic.}
  \label{tab:acp}
  \resizebox{\textwidth}{!}{
    \begin{tabular}
      {@{\hspace{0.5cm}}l@{\hspace{0.25cm}}  @{\hspace{0.25cm}}r@{\hspace{0.25cm}}  @{\hspace{0.25cm}}r@{\hspace{0.25cm}}  @{\hspace{0.25cm}}r@{\hspace{0.5cm}}}
      \hline \hline
      Component & \multicolumn{1}{c}{Isobar} & \multicolumn{1}{c}{K-matrix} & \multicolumn{1}{c}{QMI} \\
      \hline
      $\rho(770)^0$ & $+0.7 \pm \phantom{0}1.1 \pm \phantom{0}0.6 \pm \phantom{0}1.5$ &
      $ +4.2 \pm \phantom{0}1.5 \pm \phantom{0}2.6 \pm \phantom{0}5.8$ &
      $+4.4 \pm \phantom{0}1.7 \pm \phantom{0}2.3 \pm \phantom{0}1.6$ \\

      $\omega(782)$ & $-4.8 \pm \phantom{0}6.5\pm \phantom{0}1.3 \pm \phantom{0}3.5$ &
      $ -6.2 \pm \phantom{0}8.4 \pm \phantom{0}5.6 \pm \phantom{0}8.1$ &
      $-7.9 \pm 16.5 \pm 14.2 \pm \phantom{0}7.0$ \\

      $f_2(1270)$ & $+46.8 \pm \phantom{0}6.1 \pm \phantom{0}1.5 \pm \phantom{0}4.4$ &
      $ +42.8 \pm \phantom{0}4.1 \pm \phantom{0}2.1 \pm \phantom{0}8.9$ &
      $+37.6 \pm \phantom{0}4.4 \pm \phantom{0}6.0 \pm \phantom{0}5.2$ \\

      $\rho(1450)^0$ & $-12.9 \pm \phantom{0}3.3 \pm \phantom{0}3.6 \pm 35.7$ &
      $ +9.0 \pm \phantom{0}6.0 \pm 10.8 \pm 45.7$ &
      $-15.5 \pm \phantom{0}7.3 \pm 14.3 \pm 32.2$ \\

      $\rho_3(1690)^0$ & $-80.1 \pm 11.4 \pm 7.8 \pm 24.1$ &
      $ -35.7 \pm 10.8 \pm \phantom{0}8.5 \pm 35.9$ &
      $-93.2 \pm \phantom{0}6.8 \pm \phantom{0}8.0 \pm 38.1$ \\

      S-wave & $+14.4 \pm \phantom{0}1.8 \pm \phantom{0}1.0 \pm \phantom{0}1.9$ &
      $ +15.8 \pm \phantom{0}2.6 \pm\phantom{0} 2.1 \pm \phantom{0}6.9$ &
      $+15.0 \pm \phantom{0}2.7 \pm \phantom{0}4.2 \pm \phantom{0}7.0$ \\

      \hline \hline
    \end{tabular}
  }
\end{table}

\subsection{\texorpdfstring{\boldmath S-wave projections}{S-wave projections}}

\begin{figure}[ht]
  \centering
  \includegraphics[width=1.0\textwidth]{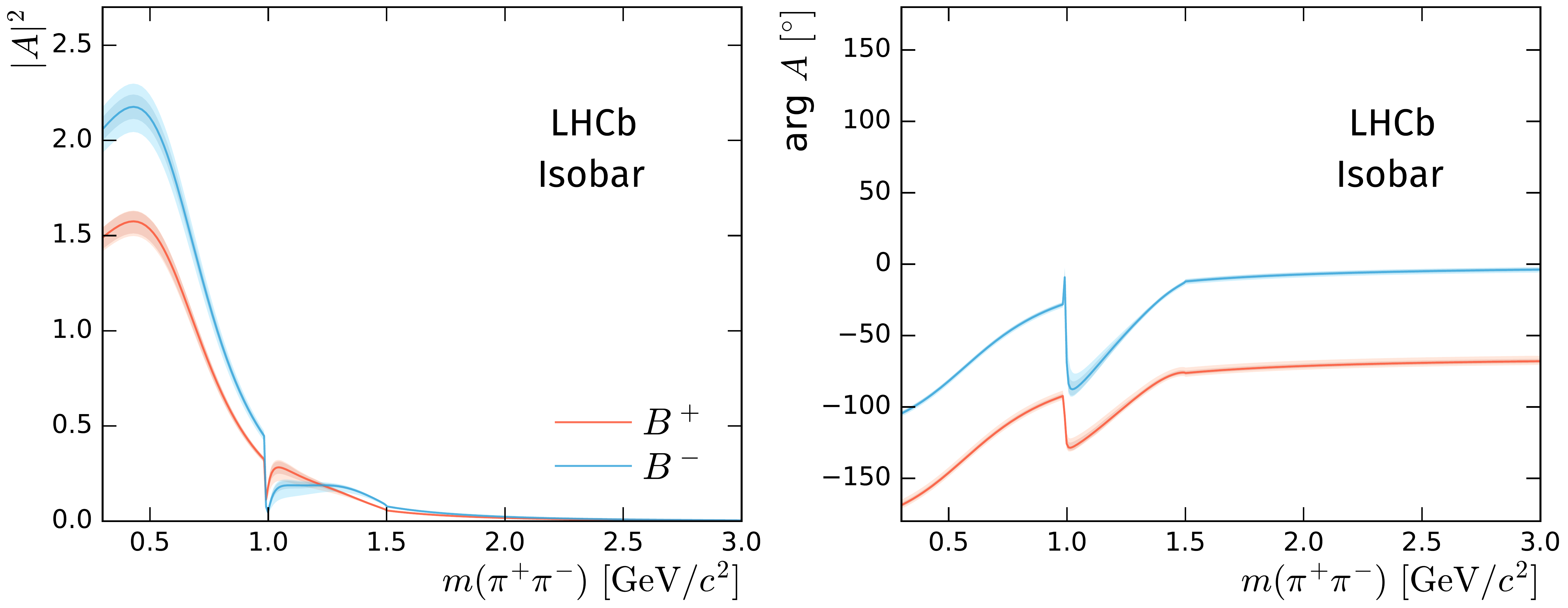}
  \put(-255,150){(a)}
  \put(-30,150){(b)}

  \includegraphics[width=1.0\textwidth]{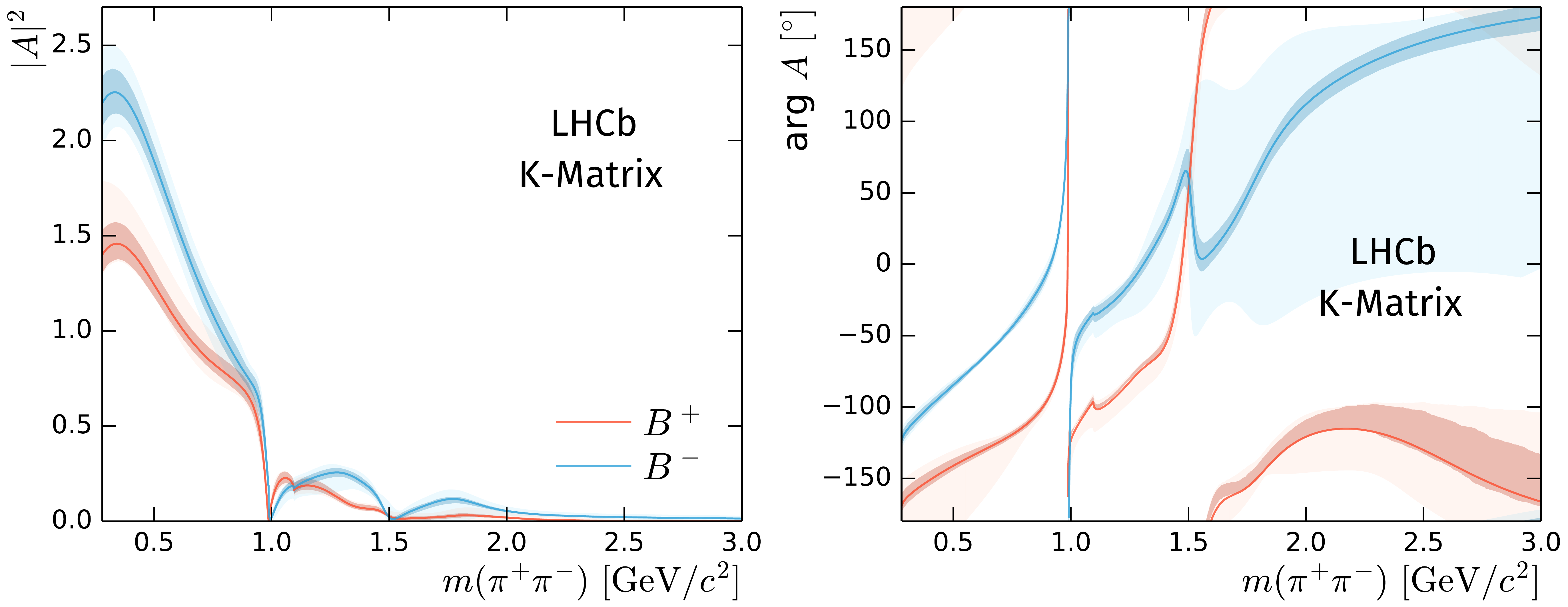}
  \put(-255,150){(c)}
  \put(-30,150){(d)}

  \includegraphics[width=1.0\textwidth]{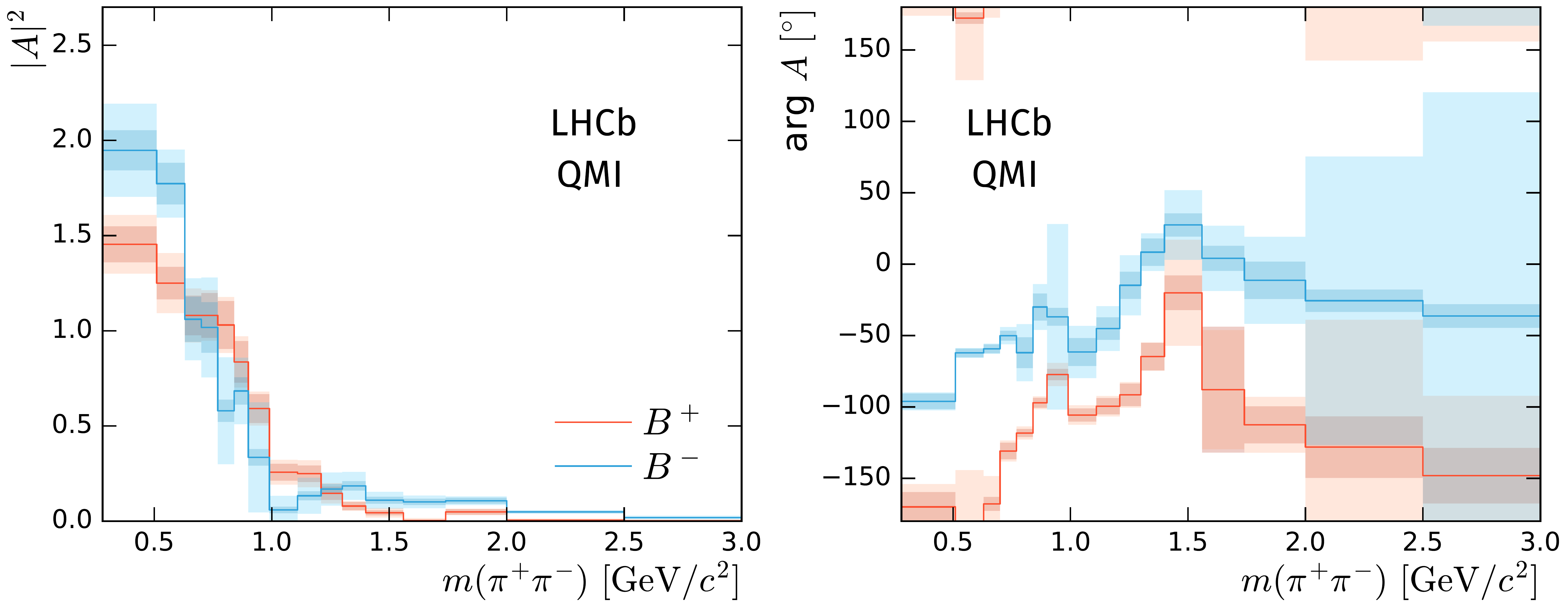}
  \put(-255,150){(e)}
  \put(-30,150){(f)}

  \caption{
    The (top) isobar, (middle) K-matrix and (bottom) QMI S-wave results where (a), (c) and (e) show the magnitude squared while (b), (d) and (f) show the phase motion.
    Discontinuities in the phase motion are due to presentation in the range $[-180^\circ,180^\circ]$.
    Red curves indicate \Bp\ while blue curves represent \Bm\ decays, with the statistical and total uncertainties bounded by the dark and light bands, respectively (incorporating only the dominant systematic uncertainties). Note that the overall scale of the squared magnitude contains no physical meaning, but is simply a manifestation of the different scale factors and conventions adopted by each of the three amplitude analysis approaches.
  }
  \label{fig:Swave}
\end{figure}

\begin{figure}[ht]
  \centering
  \includegraphics[width=1.0\textwidth]{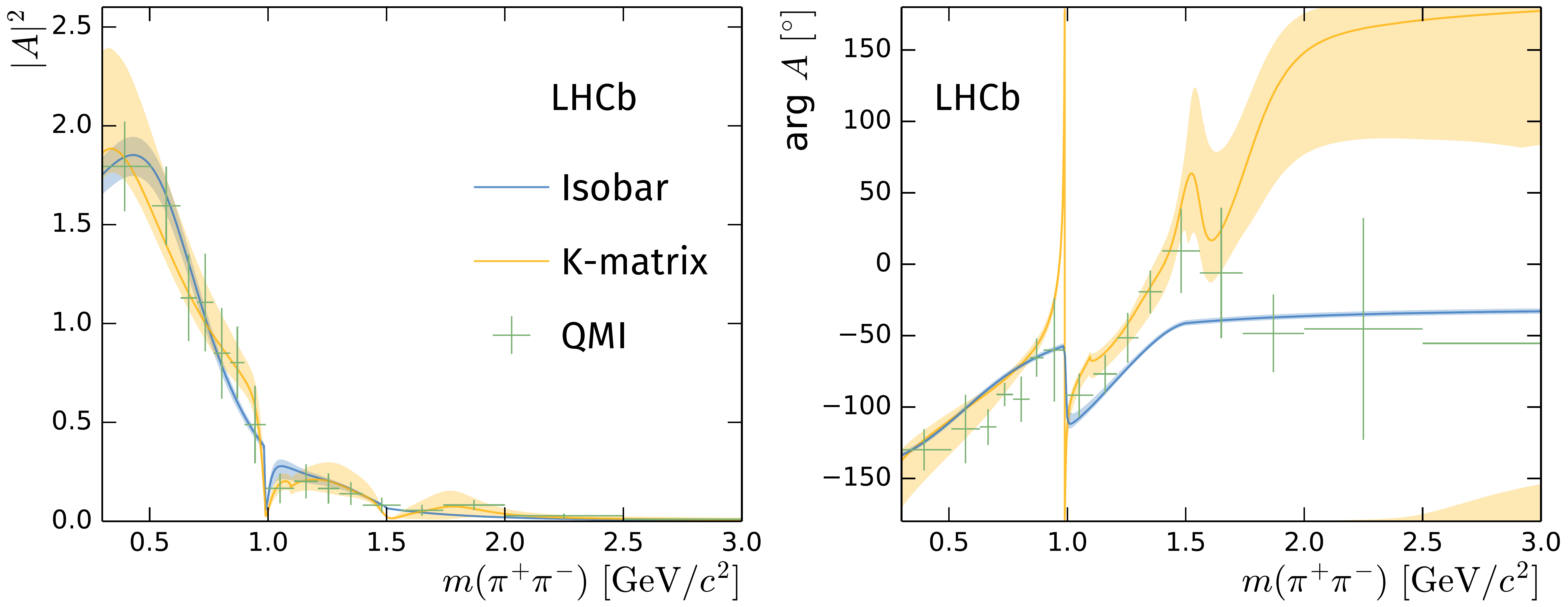}
  \put(-260,40){(a)}
  \put(-30,40){(b)}

  \caption{
    Comparison of results for the \CP-averaged S-wave obtained in the three different approaches, where (a) shows the magnitude squared while (b) shows the phase motion.
    Discontinuities in the phase motion are due to presentation in the range $[-180^\circ,180^\circ]$.
    The blue curve indicates the isobar S-wave, the amber curve indicates the K-matrix S-wave, and the green points with error bars represent the QMI S-wave. The band or error bars in each case represent the total uncertainty, incorporating the dominant systematic uncertainties.
    As the integral of the $|A|^2$ plot in each approach is proportional to its respective S-wave fit fraction, the overall scale of the K-matrix and QMI plots are set relative to the isobar S-wave fit fraction in order to facilitate comparison between the three approaches.
  }
  \label{fig:comparison:Swave}
\end{figure}

The squared amplitude and phase motion of the S-wave models as a function of $m(\pip\pim)$ can be seen in Fig.~\ref{fig:Swave}(a) and~(b) for the isobar approach,
Fig.~\ref{fig:Swave}(c) and~(d) for the K-matrix approach and Fig.~\ref{fig:Swave}(e) and~(f) for the QMI approach.
A comparison of all three models, for the \CP-averaged S-wave projections, can be seen in Fig.~\ref{fig:comparison:Swave}.
The QMI S-wave is recorded in Table~\ref{tab:qmi:Swave}, while the statistical and systematic correlation matrices obtained with this approach are given in the Supplemental Material.

\begin{table}[ht]
  \centering
  \caption{QMI S-wave fit results where the first uncertainty is statistical and the second the quadratic sum of systematic and model sources.}
  \label{tab:qmi:Swave}
  \resizebox{\textwidth}{!}{
    \begin{tabular}
      {@{\hspace{0.5cm}}l@{\hspace{0.25cm}}  @{\hspace{0.25cm}}r@{\hspace{0.25cm}}  @{\hspace{0.25cm}}r@{\hspace{0.25cm}}  @{\hspace{0.25cm}}r@{\hspace{0.25cm}}  @{\hspace{0.25cm}}r@{\hspace{0.25cm}}  @{\hspace{0.25cm}}r@{\hspace{0.5cm}}}
      \hline \hline
      Region, $i$ ($\gevcc$) & \multicolumn{1}{c}{$|A^+_i|^2$ ($10^{-2}$)} & \multicolumn{1}{c}{$\delta^+_i \, (^\circ)$} & \multicolumn{1}{c}{$|A^-_i|^2$ ($10^{-2}$)} & \multicolumn{1}{c}{$\delta^-_i \, (^\circ)$} \\ \hline
       $0.28 \leq m(\pi^+\pi^-) < 0.51$ & $5.87 \pm 0.38 \pm 0.93$ & $-170 \pm 10 \pm \phantom{0}24$ & $7.86 \pm 0.43 \pm 1.35$ & $\phantom{0}{-}96 \pm \phantom{0}6 \pm \phantom{0}10$ \\
      $0.51 \leq m(\pi^+\pi^-) < 0.63$ & $5.06 \pm 0.34 \pm 0.92$ & $+172 \pm \phantom{0}4 \pm \phantom{0}47$ & $7.14 \pm 0.44 \pm 1.08$ & $\phantom{0}{-}63 \pm \phantom{0}3 \pm \phantom{00}6$ \\
      $0.63 \leq m(\pi^+\pi^-) < 0.70$ & $4.34 \pm 0.42 \pm 0.90$ & $-168 \pm \phantom{0}5 \pm \phantom{0}24$ & $4.28 \pm 0.48 \pm 1.26$ & $\phantom{0}{-}59 \pm \phantom{0}3 \pm \phantom{00}6$ \\
      $0.70 \leq m(\pi^+\pi^-) < 0.77$ & $4.37 \pm 0.48 \pm 0.90$ & $-131 \pm \phantom{0}6 \pm \phantom{0}12$ & $4.06 \pm 0.54 \pm 1.51$ & $\phantom{0}{-}50 \pm \phantom{0}4 \pm \phantom{00}9$ \\
      $0.77 \leq m(\pi^+\pi^-) < 0.84$ & $4.15 \pm 0.51 \pm 0.98$ & $-118 \pm \phantom{0}3 \pm \phantom{00}7$ & $2.34 \pm 0.23 \pm 1.35$ & $\phantom{0}{-}61 \pm 11 \pm \phantom{0}29$ \\
      $0.84 \leq m(\pi^+\pi^-) < 0.90$ & $3.46 \pm 0.45 \pm 0.88$ & $\phantom{0}{-}97 \pm \phantom{0}3 \pm \phantom{00}7$ & $2.76 \pm 0.29 \pm 0.95$ & $\phantom{0}{-}31 \pm \phantom{0}9 \pm \phantom{0}24$ \\
      $0.90 \leq m(\pi^+\pi^-) < 0.99$ & $2.36 \pm 0.31 \pm 0.59$ & $\phantom{0}{-}77 \pm \phantom{0}4 \pm \phantom{0}11$ & $1.36 \pm 0.18 \pm 1.33$ & $\phantom{0}{-}36 \pm \phantom{0}6 \pm \phantom{0}71$ \\
      $0.99 \leq m(\pi^+\pi^-) < 1.11$ & $1.04 \pm 0.18 \pm 0.41$ & $-105 \pm \phantom{0}5 \pm \phantom{0}11$ & $0.24 \pm 0.07 \pm 0.37$ & $\phantom{0}{-}61 \pm 10 \pm \phantom{0}26$ \\
      $1.11 \leq m(\pi^+\pi^-) < 1.21$ & $1.00 \pm 0.17 \pm 0.43$ & $\phantom{0}{-}99 \pm \phantom{0}6 \pm \phantom{0}12$ & $0.54 \pm 0.10 \pm 0.47$ & $\phantom{0}{-}45 \pm \phantom{0}8 \pm \phantom{0}22$ \\
      $1.21 \leq m(\pi^+\pi^-) < 1.30$ & $0.58 \pm 0.14 \pm 0.33$ & $\phantom{0}{-}91 \pm \phantom{0}8 \pm \phantom{0}15$ & $0.68 \pm 0.11 \pm 0.45$ & $\phantom{0}{-}15 \pm 10 \pm \phantom{0}29$ \\
      $1.30 \leq m(\pi^+\pi^-) < 1.40$ & $0.33 \pm 0.09 \pm 0.17$ & $\phantom{0}{-}65 \pm 10 \pm \phantom{0}17$ & $0.75 \pm 0.10 \pm 0.38$ & $\phantom{00}{+}8 \pm \phantom{0}9 \pm \phantom{0}21$ \\
      $1.40 \leq m(\pi^+\pi^-) < 1.56$ & $0.17 \pm 0.05 \pm 0.15$ & $\phantom{0}{-}20 \pm 12 \pm \phantom{0}48$ & $0.45 \pm 0.07 \pm 0.24$ & $\phantom{0}{+}28 \pm \phantom{0}8 \pm \phantom{0}32$ \\
      $1.56 \leq m(\pi^+\pi^-) < 1.74$ & $0.01 \pm 0.01 \pm 0.07$ & $\phantom{0}{-}86 \pm 44 \pm \phantom{0}74$ & $0.41 \pm 0.07 \pm 0.19$ & $\phantom{00}{+}5 \pm \phantom{0}9 \pm \phantom{0}30$ \\
      $1.74 \leq m(\pi^+\pi^-) < 2.00$ & $0.20 \pm 0.06 \pm 0.11$ & $-111 \pm 13 \pm \phantom{0}30$ & $0.43 \pm 0.06 \pm 0.14$ & $\phantom{0}{-}11 \pm 13 \pm \phantom{0}42$ \\
      $2.00 \leq m(\pi^+\pi^-) < 2.50$ & $0.02 \pm 0.02 \pm 0.05$ & $-128 \pm 21 \pm 109$ & $0.19 \pm 0.02 \pm 0.06$ & $\phantom{0}{-}25 \pm \phantom{0}8 \pm 109$ \\
      $2.50 \leq m(\pi^+\pi^-) < 3.50$ & $0.01 \pm 0.00 \pm 0.05$ & $-149 \pm 19 \pm \phantom{0}73$ & $0.08 \pm 0.01 \pm 0.05$ & $\phantom{0}{-}36 \pm \phantom{0}8 \pm 165$ \\
      $3.50 \leq m(\pi^+\pi^-) < 5.14$ & $0.00 \pm 0.01 \pm 0.03$ & $+100 \pm 72 \pm 173$ & $0.02 \pm 0.01 \pm 0.07$ & $-144 \pm 15 \pm 121$ \\
      \hline \hline
    \end{tabular}
  }
\end{table}

\subsection{\texorpdfstring{\boldmath{$\rho(770)^0$} mass and width}{rho(770)0 mass and width}}

The $\rho(770)^0$ mass and width are allowed to vary freely in each fit as mentioned in Section~\ref{sec:modelSel}.
The fitted results are consistent with the world-average values, and can be seen in Table~\ref{tab:rhoMW}.

\renewcommand{\arraystretch}{1.35}
\begin{table}[ht]
  \centering
  \caption{The obtained $\rho(770)^0$ mass and width parameters, for each approach, where the uncertainty is statistical.
}
  \label{tab:rhoMW}
  % \resizebox{\textwidth}{!}{
    \begin{tabular}
      {@{\hspace{0.5cm}}l@{\hspace{0.25cm}}  @{\hspace{0.25cm}}r@{\hspace{0.25cm}}  @{\hspace{0.25cm}}r@{\hspace{0.25cm}}}
      \hline \hline

      S-wave approach & \multicolumn{1}{c}{Mass (\mevcc)} & \multicolumn{1}{c}{Width (\mev)} \\
      \hline
      Isobar & $770.8 \pm 1.3$ & $153.4 \pm 3.2$ \\
      K-Matrix & $766.7 \pm 1.4$ & $147.3 \pm 3.1$ \\
      QMI & $766.3 \pm 1.5$ & $148.2 \pm 3.5$ \\
      \hline \hline
    \end{tabular}
  % }
\end{table}

\subsection{Multiple fit solutions}

A search for secondary solutions with negative log-likelihood values worse than, but close to, that of the best fit is performed for each S-wave approach by setting
the initial values of the complex coefficients of the model to random values and repeating the fit to data.
In both the isobar and QMI approaches, no secondary solutions are found within 25 units of $-2\log{\cal L}$.
For the K-matrix approach, however, secondary solutions are found in which some of the pole or SVP amplitude coefficients are rotated in the Argand plane with respect to the best fit result.
Studies using data sampled from the nominal model indicate that these could
potentially be resolved with larger data samples, and further improvements may also be possible by fitting for
the scattering parameters along with the amplitude coefficients.
Isobar coefficients and S-wave projections corresponding to the secondary minimum closest to the most-negative minimum, with a change in log-likelihood of $0.8$, are given in Appendix~\ref{sec:kMatrixTables}.

\section{Discussion}
\label{sec:discussion}

The results and figures presented in Section~\ref{sec:results} show that the three models exhibit good overall agreement with the data and with each other, both in \CP-average projections and in the variation of the asymmetries across the phase space.
In this section the main features observed in the data and in the models are discussed in more detail.

Many of the interference fit fractions are close to zero, as expected, since interference effects between partial waves with even and odd values of the relative angular momentum cancel when integrated over the helicity angle.
The largest interference fit fraction is between the combined $\rho$--$\omega$ component and the $\rho(1450)^0$ resonance; since each of these is spin-$1$, the interference does not vanish when integrated over the Dalitz plot.
No significant interference fit-fraction asymmetries are observed, however this does not preclude sizeable asymmetries in localised regions of the Dalitz plot.

\subsection{\texorpdfstring{The \boldmath{$\rho(770)^0$--$\omega(782)$} region}{The rho(770)0 - omega(782) region}}

The interference between the  spin-1 $\rho(770)^0$ and $\omega(782)$ resonances is well described by the models, as shown in Fig.~\ref{fig:projLowRho}(b).
No significant asymmetry is observed in this region when integrating over $\cos\theta_{\rm hel}$ as shown in Fig.~\ref{fig:projLowRho}(d), and also seen in the $\rho(770)^0$ and $\omega(782)$ quasi-two-body \CP asymmetry parameters in Table~\ref{tab:acp}.
A number of theoretical calculations of these quantities are available in the literature, with some authors~\cite{Beneke:2003zv,Wang:2008rk, Cheng:2009cn,Nogueira:2016mqf} predicting values for $\mathcal{A}_{\CP}\left(\decay{\Bp}{\rho(770)^0\pip}\right)$ that are consistent with the measured result, albeit sometimes with large uncertainties.
Other approaches~\cite{Cheng:2014rfa,Li:2014haa,Zhou:2016jkv,Li:2016tpn} give predictions for this quantity which appear to now be ruled out.
There is also no evident \CP-violation effect associated with $\rho$--$\omega$ mixing, contrary to some theoretical predictions~\cite{Guo:2000uc, Wang:2015ula, Cheng:2016shb}.

A significant \CP\ asymmetry in the $\rho(770)^0$ region can, however, be seen in the $\cos\theta_{\rm hel}$ projections shown in Figs.~\ref{fig:projHelRhoUL}(c) and~(d) when dipion masses below and above the known $\rho(770)^0$ mass are inspected separately.
The same effect can be seen in Fig.~\ref{fig:projHelCut} where the data are separated by the sign of the value of $\cos \theta_{\rm hel}$.
This feature, previously observed through a model-independent analysis~\cite{LHCb-PAPER-2014-044}, is characteristic of \CP violation originating from a sizeable interference between the spin-$1$ $\rho(770)^0$ resonance and the broad spin-$0$ contribution present in this region.
This effect cancels when integrating over the helicity angle, since the interference term is proportional to $\cos\theta_{\rm hel}$.
In addition, the change in the asymmetry below and above the $\rho(770)^0$ peak indicates that the effect is mediated by a strong phase difference dominated by the evolution of the $\rho(770)^0$ Breit--Wigner amplitude phase.
All three approaches to the modelling of the $\pip\pim$ S-wave describe this effect well.

\subsection{\texorpdfstring{The \boldmath{$\pip\pim$} S-wave}{The pi+pi- S-wave}}

A notable feature in Fig.~\ref{fig:projLowRho}(c) is the small but approximately constant asymmetry at $m(\pip\pim)$ values below the $\rho(770)^0$ mass.
This region is dominated by the S-wave component with a small contribution from the $\rho(770)^0$ low-mass tail;
the \CP asymmetry in the S-wave in this region is also seen in all three approaches in Fig.~\ref{fig:Swave}(a),~(c) and~(e).
A \CP asymmetry in the S-wave below the inelastic ($\kaon\Kbar$) threshold cannot be explained via $\pi\pi \leftrightarrow \kaon\Kbar$ rescattering, and therefore has a different origin to effects seen elsewhere in the Dalitz plot.

The combined significance of \CP violation in the S-wave and in the interference between the S- and P-waves is evaluated from the change in log-likelihood between the baseline fit and in fits where all relevant \CP-violation parameters are fixed to be zero.
Since the $\rho$--$\omega$ component serves as the reference amplitude in the baseline fit, this means that all $\delta x$ and $\delta y$ parameters, defined in Eq.~\eqref{eq:cartesian}, associated with the S-wave are fixed to zero.
This is done in fits with each of the approaches to the S-wave model, with the resulting change in log-likelihood converted into a $p$-value, and subsequently into the number of Gaussian standard deviations ($\sigma$), accounting for the number of fixed parameters.
The values obtained are around $30\sigma$ in all cases, despite the very different number of degrees of freedom associated with the S-wave in the different approaches.
While this method can only be considered to give an approximation to the significance, it is sufficient to establish the presence of \CP violation far beyond any reasonable doubt.

In order to separate the effects of \CP violation in the S-wave and in the interference between the S- and P-waves, additional fits are performed in which the reference amplitude is changed to the S-wave.
The $\delta x$ and $\delta y$ parameters associated with the S-wave are then fixed to zero, while those associated with the P-wave are allowed to vary in the fits.
In this case, \CP violation in the interference between the S- and P-waves is allowed, while none is possible in the S-wave itself, and hence the significance of each effect individually can be assessed.
The values obtained are above $10\sigma$ in each of the S-wave modelling approaches, thus establishing that both \CP-violation effects are present.

At low $m(\pip\pim)$ values, the S-wave magnitude and phase motion of the three approaches broadly agree, particularly for the \CP-averaged $|A|^2$, and all models capture similar
behaviour around $1\gevcc$.
However, in the $\kaon\Kbar$ threshold region shown in Fig.~\ref{fig:projHelCut}, the change in sign of the difference
between the number of \Bp and \Bm candidates between positive and negative $\cos\theta_{\rm hel}$ is
captured only by the K-matrix model. It is worth noting that this is the only model with an
explicit $f_0(980)$ term: the isobar model includes only $\pi\pi \leftrightarrow \kaon\Kbar$ rescattering above the $\kaon\Kbar$ threshold and the QMI binning is not sufficiently fine in this region to resolve a narrow structure.

At $1.5\gevcc$, the K-matrix has a clear phase motion, seen in Fig.~\ref{fig:Swave}, that is associated with the $f_0(1500)$ contribution.
Consistent behaviour is seen in the QMI approach, although the uncertainties preclude a definite corroboration of the presence of the $f_0(1500)$ state.
The isobar model does not include this component explicitly, and therefore it is expected that the phase is broadly constant here, continuing to the upper kinematic boundary.
Above $3\gevcc$, the magnitude of the S-wave component in all three approaches is consistent with zero and therefore the phase values are dominated by statistical and systematic uncertainties.

\subsection{\texorpdfstring{The \boldmath{$f_2(1270)$} region}{The f_2(1270) region}}
\label{sec:f2Discussion}

\begin{figure}[tb]
  \begin{center}
    \includegraphics[width=0.46\linewidth]{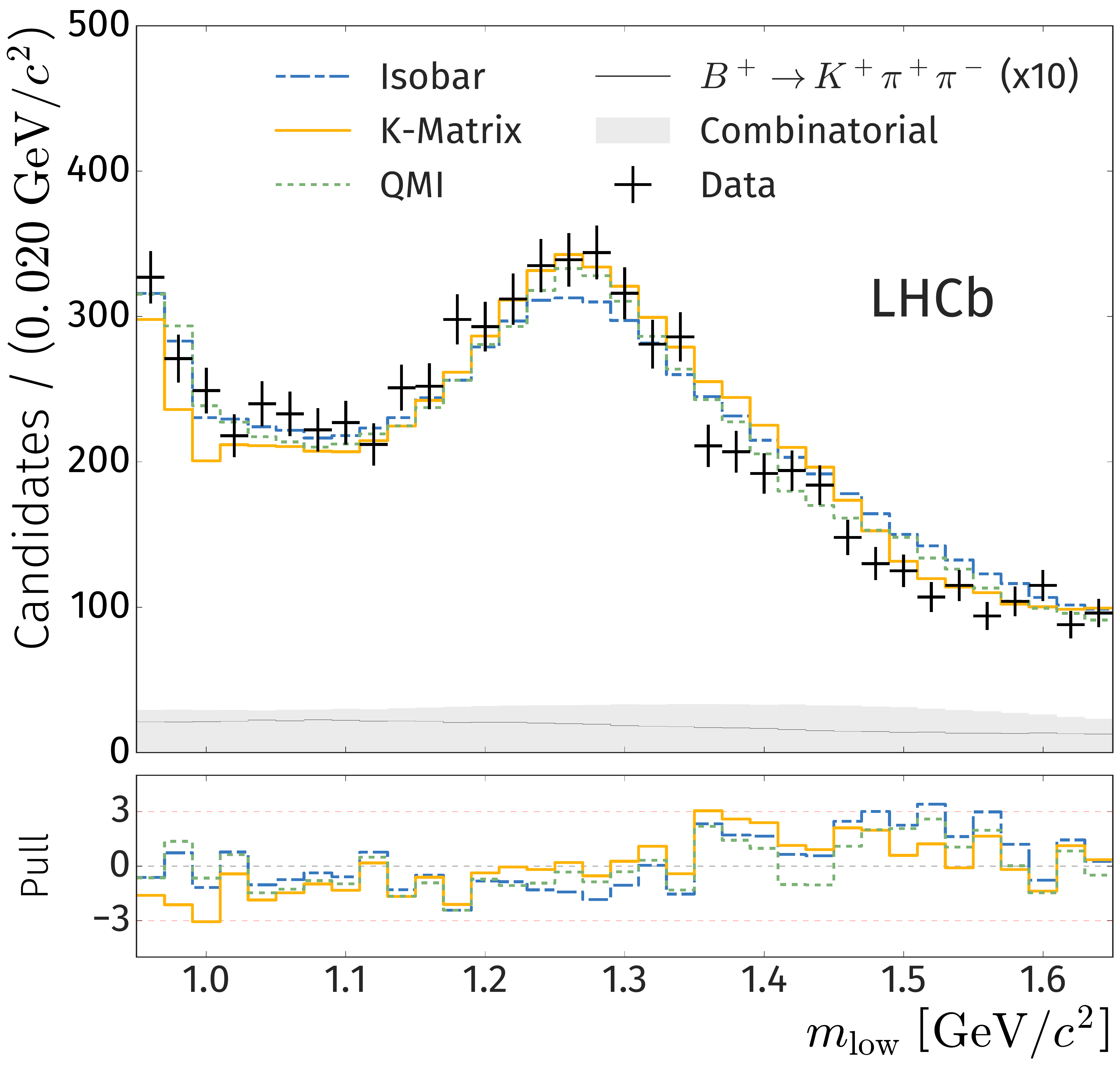}
    \includegraphics[width=0.46\linewidth]{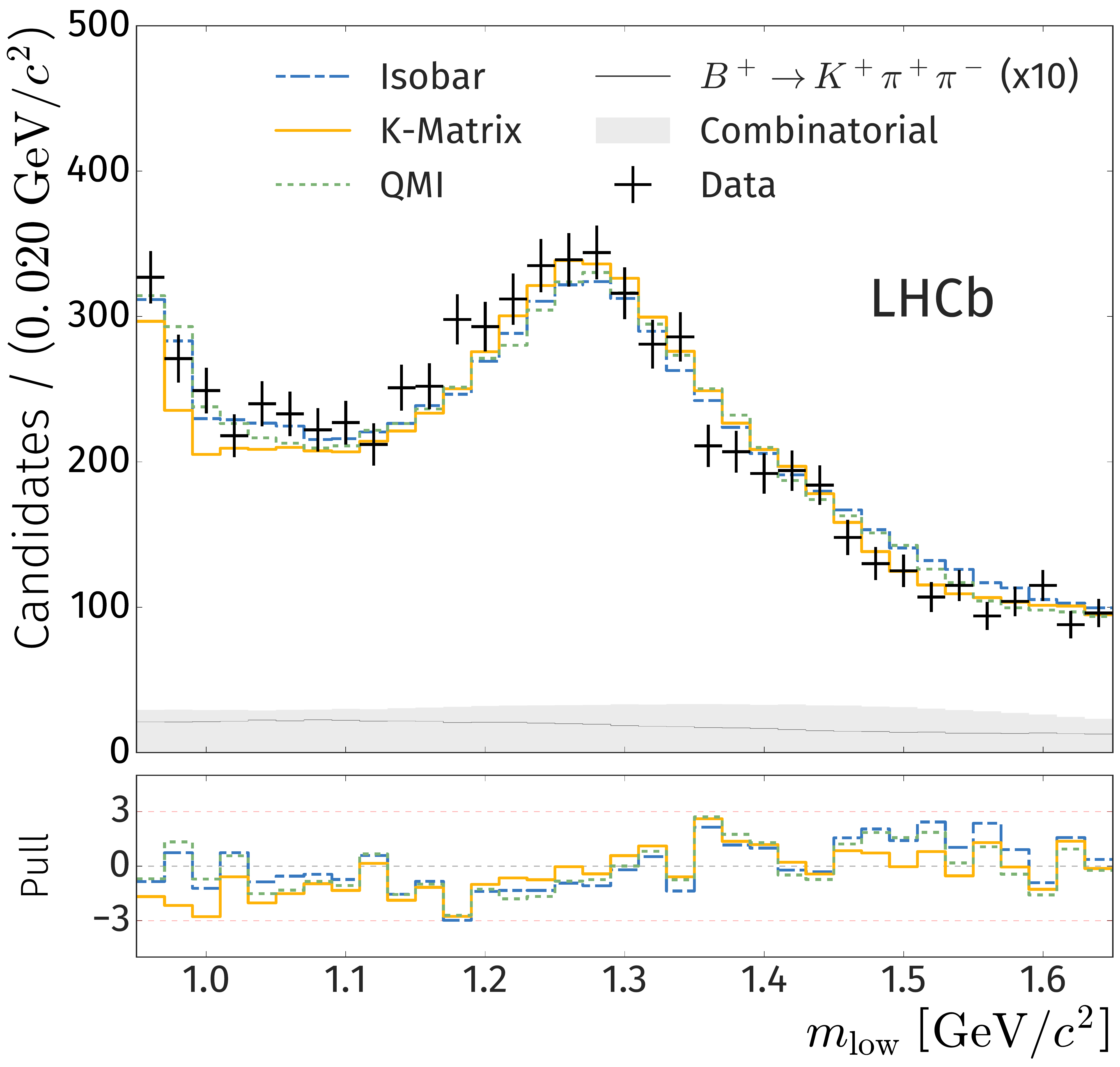}
    \put(-385,70){(a)}
    \put(-170,70){(b)}
  \end{center}
  \caption{
   \small
    Data and fit model projections in the $f_2(1270)$ region with (a)~freely varied $f_2(1270)$ resonance parameters, and (b)~with an additional spin-$2$ component with mass and width parameters determined by the fit.
  }
  \label{fig:f2Variations}
\end{figure}

Despite broad consistency between the three fit models, a clear discrepancy with the data is apparent for all of them in the $f_2(1270)$ region shown in Fig.~\ref{fig:projf2High}(a).
All fit model projections lie under (over) the data below (above) the $f_2(1270)$ peak, which is set to the known value of $1275.5 \pm 0.8 \mevcc$~\cite{PDG2018}.
Better agreement with the data is obtained when the $f_2(1270)$ mass and width are allowed to vary in the fits, as shown in Fig.~\ref{fig:f2Variations}(a).
However, the obtained masses, equal to
$1256 \pm 4\mevcc$, $1252 \pm 4\mevcc$ and $1260 \pm 4\mevcc$ for the isobar, K-matrix and QMI S-wave approaches, respectively, where the uncertainties are statistical only,
are at least four standard deviations away from the world average.
The values obtained for the width are, however, consistent with the world average.
Moreover, if the $f_2(1270)$ mass and width are allowed to vary independently in the \Bm and \Bp subsamples, inconsistent values are obtained.

Alternatively, the discrepancy between the data and the models can be reduced by adding another spin-2 resonance in the $f_2(1270)$ region, as shown in Fig.~\ref{fig:f2Variations}(b).
The established states $f_2^\prime(1525)$, with $m_{f_2^\prime(1525)} = 1525 \pm 5\mevcc$ and
$\Gamma_{f_2^\prime(1525)} = 73\,^{+6}_{-5}\mev$, and $f_2(1565)$, with $m_{f_2(1565)} = 1562 \pm 13\mevcc$
and $\Gamma_{f_2(1565)} = 134 \pm 8\mev$, are too high in mass and too narrow to be likely to induce a significant effect in the $f_2(1270)$ region.
Therefore, an additional spin-2 resonance with mass and width parameters free to vary in the fit is introduced, with initial values corresponding to those of the not-well-established
$f_2(1430)$ resonance.
The fit results for the mass are consistent between each S-wave approach, with $m_{f_2(X)} = 1600 \pm 60$, $1541 \pm 24$ and $1560 \pm 14 \mevcc$ for the isobar, K-matrix and QMI fits, respectively. However, the
obtained values for the width are inconsistent, varying between $\Gamma_{f_2(X)} = 367 \pm 97$, $204 \pm 78$ and $115 \pm 37 \mev$, where the uncertainties are statistical only.
Therefore the addition of a second spin-2 state does not appear in the baseline model, but rather is considered as a source of systematic uncertainty on the model composition.

Consequently, the baseline model includes in the $\pi\pi$ D-wave only the $f_2(1270)$ resonance, with its mass and width fixed to the known values.
Analysis of larger data samples will be required to obtain a more detailed understanding of the $\pi\pi$ D-wave in \decay{\Bp}{\pip\pip\pim} decays.

The effect of additional decay channels on the energy-dependent width of the $f_2(1270)$ is also considered
as another possibility for the baseline fit model discrepancy. A global analysis is performed to
express the known branching fractions of $f_2(1270)$ decays to $\pi\pi$, \kaon\Kbar, $\eta\eta$,
$\pip\pim\piz\piz$, $\pip\pim\pip\pim$ and $\piz\piz\piz\piz$ in terms of their respective couplings in
$f_2(1270)$ decays to these final states.
Subsequent fits accounting for the energy-dependent width in a similar way as in the Flatt\'{e} lineshape~\cite{flatte}
are found to have minimal impact on the model and therefore do not contribute to the systematic uncertainties.

\begin{figure}[ht]
  \centering
  \begin{tabular}{m{0.48\textwidth}@{\hspace{0.0em}}m{0.48\textwidth}}
  \includegraphics[width=0.48\textwidth]{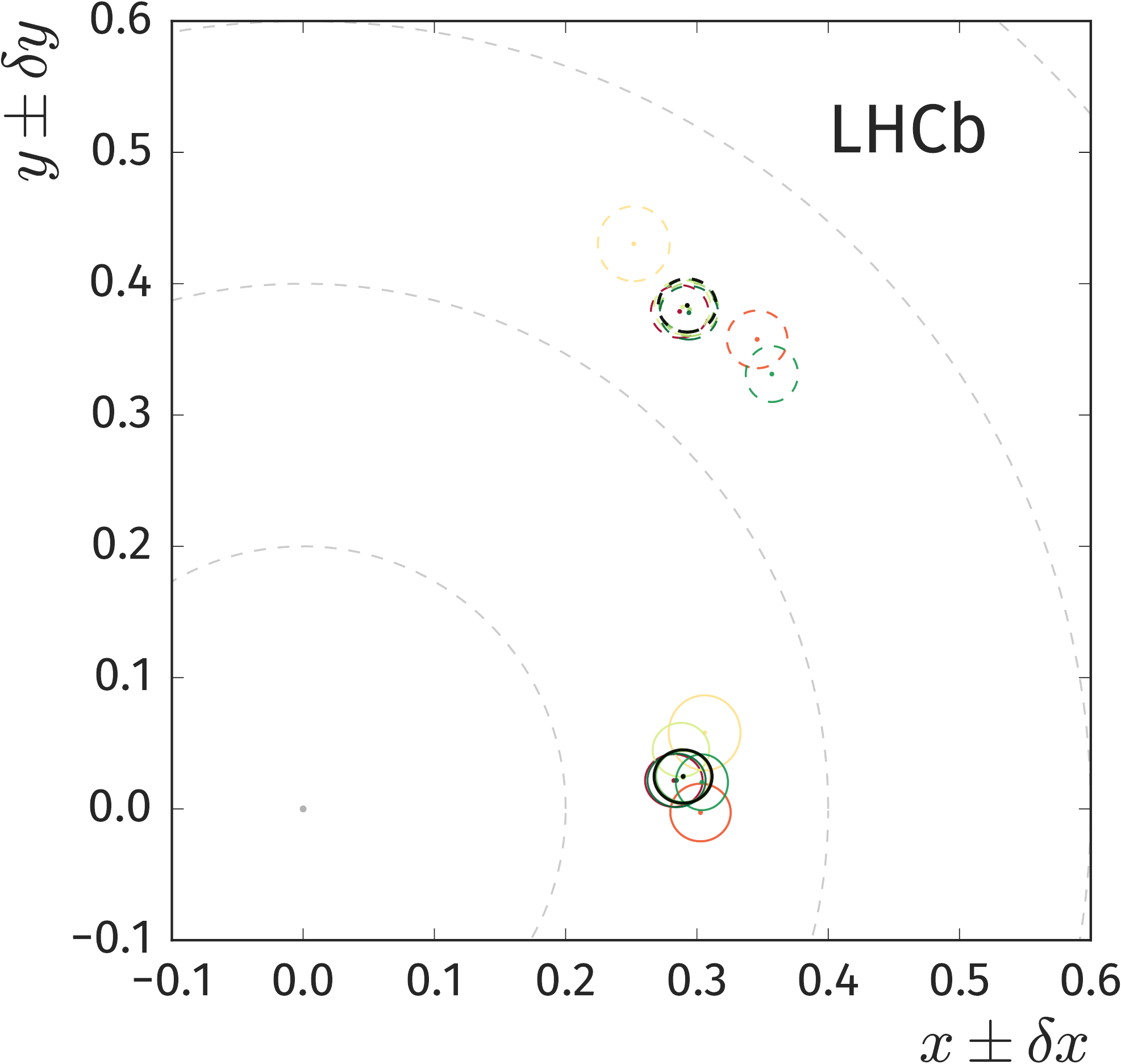}\put(-181,190){(a)} &
  \includegraphics[width=0.48\textwidth]{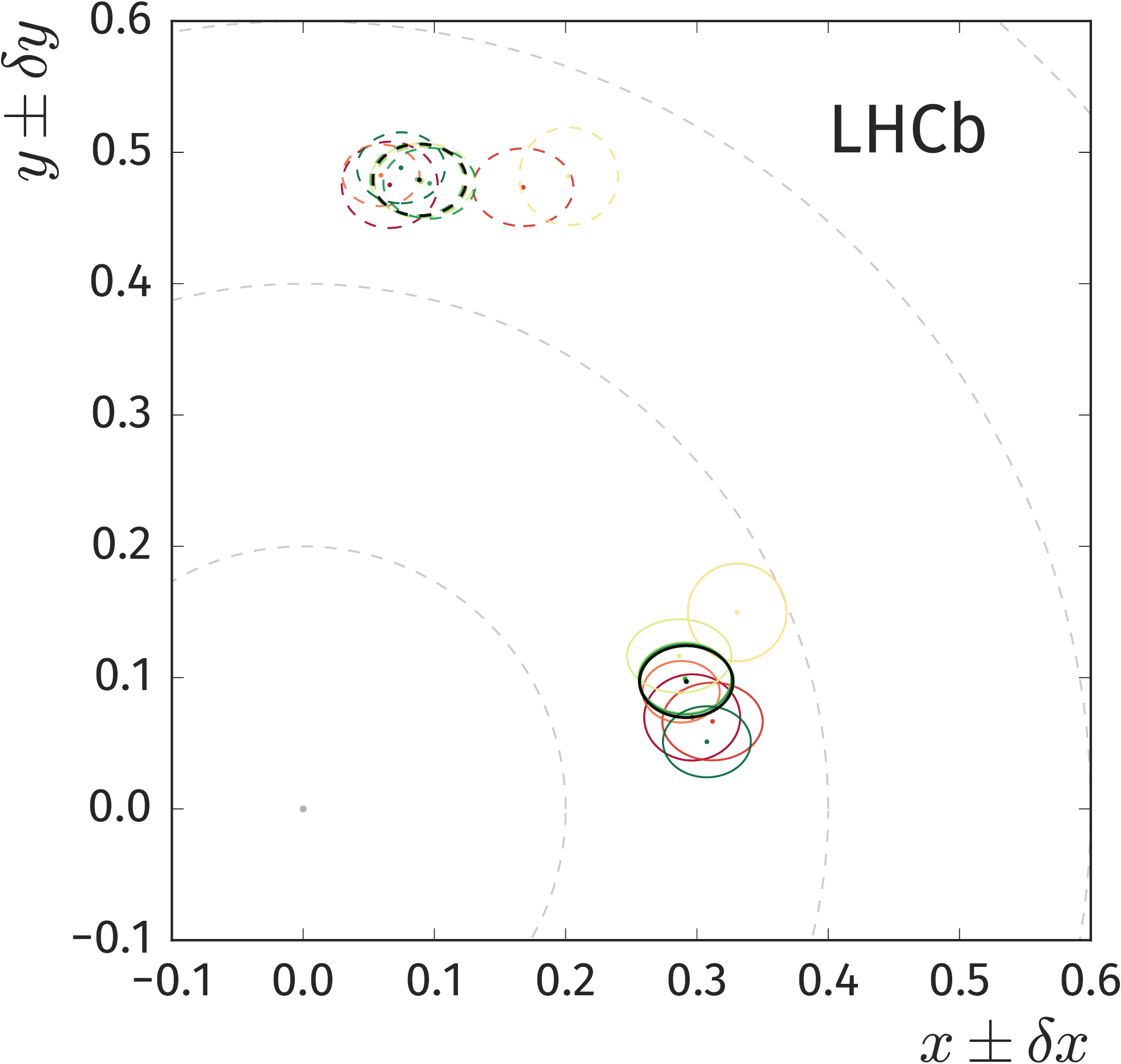}\put(-181,190){(b)} \\
  \includegraphics[width=0.48\textwidth]{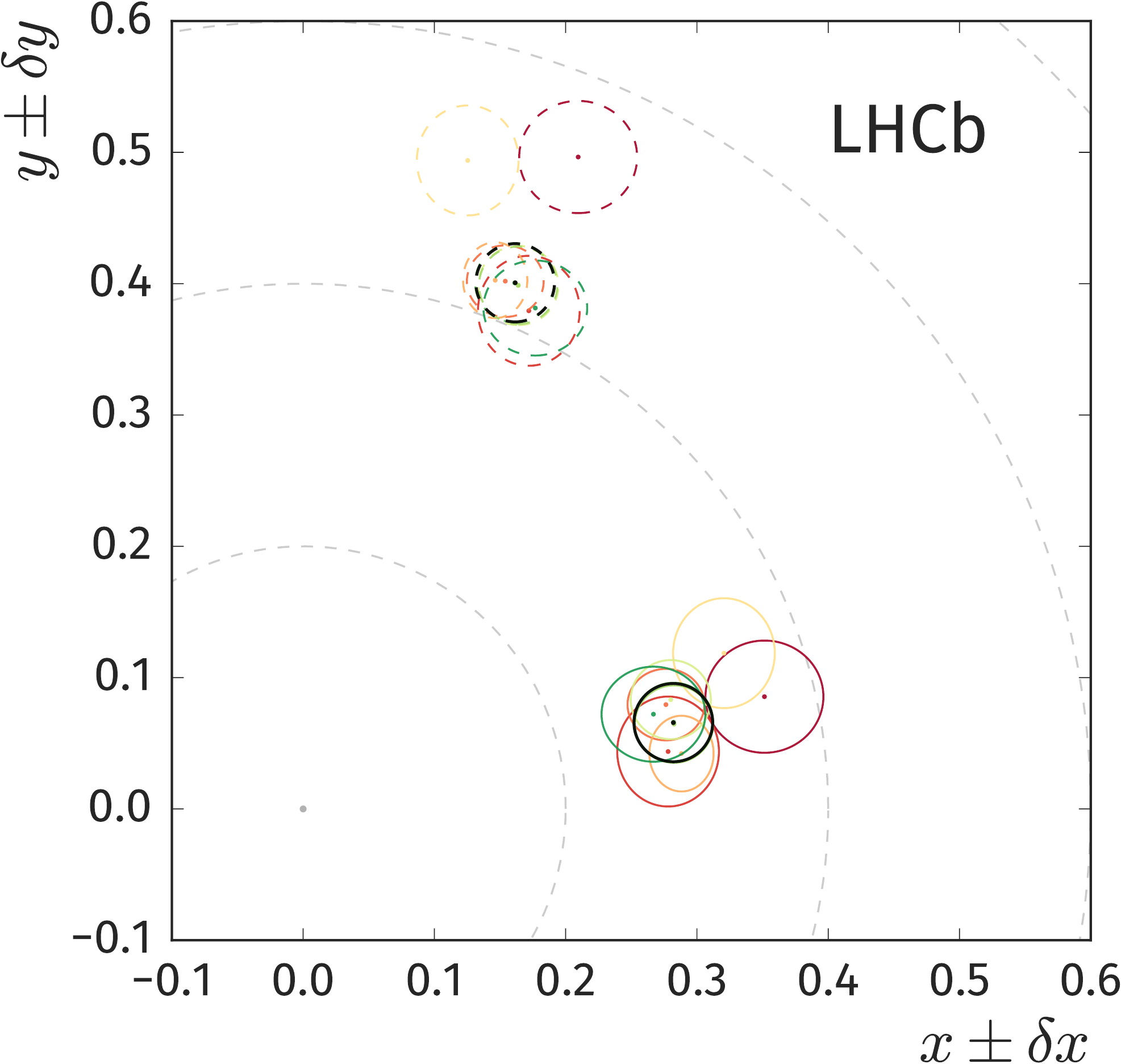}\put(-181,190){(c)} &
  \includegraphics[width=0.48\textwidth]{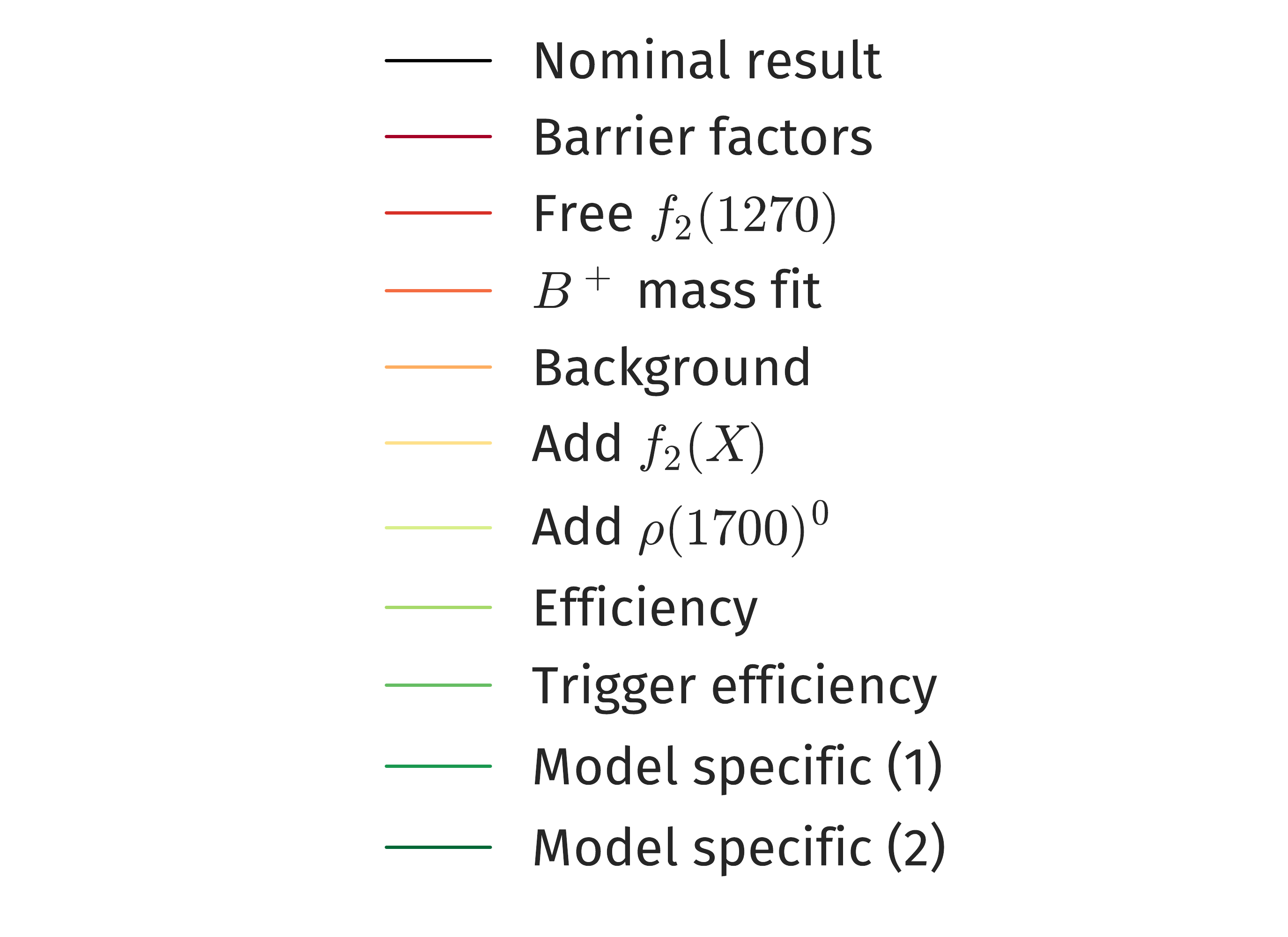} \\
  \end{tabular}

  \caption{Central values (points) and statistical $68\%$ Gaussian confidence regions (ellipses)
    for the complex coefficients associated with the $f_2(1270)$ resonance under various systematic assumptions
    for the \Bp (solid) and \Bm (dashed) decay amplitude models. The nominal result and statistical uncertainty
    is given in black, while the results of the dominant systematic variations to the nominal model (per Section~\ref{sec:systematics}) are given by the coloured ellipses, as noted in the legend, for each of the (a)~isobar, (b)~K-matrix and (c)~QMI S-wave approaches.
    The model specific systematic uncertainties are discussed in Sec.~\ref{sec:systematics}.}
  \label{fig:f2Parameters}
\end{figure}

Despite the considerations outlined above, the \CP violation associated with the $f_2(1270)$ resonance
is robust with respect to the experimental and model systematic uncertainties documented
in Section~\ref{sec:systematics}. This can be seen by comparing the coefficients that describe the
$f_2(1270)$ resonance with those obtained during systematic variations as shown in Fig.~\ref{fig:f2Parameters}.
The fact that the contours for $\Bp$ and $\Bm$ coefficients do not overlap is a visual representation of the significantly non-zero values of the $\delta x$ and $\delta y$ parameters of Eq.~\eqref{eq:cartesian}.
The quasi-two-body \CP asymmetry, defined in Eq.~\eqref{eq:cpAsy}, is related to the difference in the magnitudes of the \Bp and \Bm complex coefficients, \ie the difference in distance from $(0,0)$ to the centres of the corresponding ellipses in Fig.~\ref{fig:f2Parameters}.
In addition to this, \CP violation can also be observed in the difference in the phases relative to the $\rho(770)^0$--$\omega(782)$ reference component in the \Bp and \Bm amplitudes, which would manifest in the Dalitz plot as a difference between \Bp and \Bm decays in the interference pattern between $\rho(770)^0$ and $f_2(1270)$ resonances.
This is indicated by the difference in the polar angle in the Argand plane of Fig.~\ref{fig:f2Parameters}.
It is evident that systematic variations do not significantly modify the distance between the solid and dashed ellipses in the $(x, y)$ plane, meaning that
the significant overall \CP violation associated with the $f_2(1270)$ resonance is robust.

When interpreting Fig.~\ref{fig:f2Parameters}, it should be noted that in the QMI approach the amplitude components are not individually normalised, unlike the case for the isobar and K-matrix approaches.
Since some systematic variations can change the overall scale of various lineshapes, their respective deviations as depicted in this plot cannot be directly interpreted as entirely systematic in origin, and as such naturally manifest as the
largest deviations from the nominal model.

The statistical significance of \CP violation in $\Bp \to f_2(1270) \pip$ is estimated by finding, for
each S-wave method, the difference of twice the log-likelihood between two fits: the nominal one and another
where the \CP-violating parameters of the $f_2(1270)$ are fixed to zero. Since this quantity behaves like a
$\chi^2$ distribution with two degrees of freedom, it is converted into a $p$-value from which confidence intervals are derived.
The significance of \CP violation is found to be $20\sigma$, $15\sigma$ and $14\sigma$ for the isobar, K-matrix and QMI approaches, respectively.
This corresponds to the first observation of \CP violation in \decay{\Bp}{f_2(1270)\pip} decay, which is the first observation of \CP violation in any process with a final state containing a tensor resonance.
The measured central value of $\mathcal{A}_{\CP}\left(\decay{\Bp}{f_2(1270)\pip}\right)$ is consistent with some theoretical predictions~\cite{Li:2018lbd,Cheng:2010yd,Zou:2012td} that, however, have large uncertainties.

\subsection{\texorpdfstring{The \boldmath{$\rho_3(1690)^0$} region}{The rho_3(1690) region}}

The contribution from a spin-$3$ $\rho_3(1690)^0$ component is evident in Fig.~\ref{fig:projHelRho3}, with a dip in intensity at $\cos\theta_{\rm hel} = 0$, characteristic of an odd-spin resonance, as well as multiple ``lobes'' associated with a spin greater than $1$.
The central value of the \CP asymmetry of this component is large and positive, however its systematic uncertainty is also large, mostly driven by ambiguities in the amplitude model.
These uncertainties preclude any conclusive statement about \CP violation in \decay{\Bp}{\rho_3(1690)^0\pip} decays; an analysis with a larger data sample will be necessary to clarify this point.

\section{Conclusions}
\label{sec:conclusions}

An amplitude analysis of the \decay{\Bp}{\pip\pip\pim} decay is performed, using a data sample corresponding to $3\invfb$ collected by the LHCb experiment during Run 1.
Three complementary approaches are used to describe the large S-wave contribution to this decay.
Overall good agreement is found between all three models and the data, although some discrepancies in the region around the $f_2(1270)$ region persist in the baseline models.
Significant \CP\ violation associated with the $f_2(1270)$ resonance is observed, the first observation of \CP\ violation in any process containing a tensor resonance, which is robust with respect to systematic uncertainties related to both experimental effects and to the composition of the amplitude model.

The quasi-two-body \CP asymmetry in the \decay{\Bp}{\rho(770)^0\pip} decay is found to be compatible with zero.
However, \CP-violation effects that are characteristic of interference between the spin-$1$ $\rho(770)^0$ resonance and the spin-$0$ S-wave contribution are observed, and are well described in all three approaches to the S-wave.
This is the first observation of \CP\ violation mediated entirely by interference between hadronic resonances.

All three approaches to the description of the $\pip\pim$ S-wave broadly agree on the variation of its magnitude and phase.
One striking feature is the presence of a significant \CP asymmetry in the S-wave that is not associated with the aforementioned interference effect, of which a substantial component is at low $m(\pip\pim)$.
Further phenomenological and experimental investigations will be required to better understand the underlying dynamics of these and other effects in \decay{\Bp}{\pip\pip\pim} decays, and to elucidate connections with \CP-violation effects observed in \decay{\Bp}{\Kp\Kp\pim} decays~\cite{LHCb-PAPER-2018-051}.

The results of this analysis provide valuable input to phenomenological work on the underlying nature of the remarkably large \CP violation observed in the charmless three-body decays of the charged \B meson, and in \B-meson decays in general.
Furthermore, the robust description of low mass S-wave achieved with the approaches documented here gives insight into the effects of low-energy QCD in \B-meson decays.

% Comment this in for papers
\section*{Acknowledgements}
%
% These Acknowledgements valid from 3-May-2019
%
\noindent We express our gratitude to our colleagues in the CERN
accelerator departments for the excellent performance of the LHC. We
thank the technical and administrative staff at the LHCb
institutes.
We acknowledge support from CERN and from the national agencies:
CAPES, CNPq, FAPERJ and FINEP (Brazil); 
MOST and NSFC (China); 
CNRS/IN2P3 (France); 
BMBF, DFG and MPG (Germany); 
INFN (Italy); 
NWO (Netherlands); 
MNiSW and NCN (Poland); 
MEN/IFA (Romania); 
MSHE (Russia); 
MinECo (Spain); 
SNSF and SER (Switzerland); 
NASU (Ukraine); 
STFC (United Kingdom); 
DOE NP and NSF (USA).
We acknowledge the computing resources that are provided by CERN, IN2P3
(France), KIT and DESY (Germany), INFN (Italy), SURF (Netherlands),
PIC (Spain), GridPP (United Kingdom), RRCKI and Yandex
LLC (Russia), CSCS (Switzerland), IFIN-HH (Romania), CBPF (Brazil),
PL-GRID (Poland) and OSC (USA).
We are indebted to the communities behind the multiple open-source
software packages on which we depend.
Individual groups or members have received support from
AvH Foundation (Germany);
EPLANET, Marie Sk\l{}odowska-Curie Actions and ERC (European Union);
ANR, Labex P2IO and OCEVU, and R\'{e}gion Auvergne-Rh\^{o}ne-Alpes (France);
Key Research Program of Frontier Sciences of CAS, CAS PIFI, and the Thousand Talents Program (China);
RFBR, RSF and Yandex LLC (Russia);
GVA, XuntaGal and GENCAT (Spain);
the Royal Society
and the Leverhulme Trust (United Kingdom).

\clearpage

{\noindent\normalfont\bfseries\Large Appendices}

\appendix

\addcontentsline{toc}{section}{Appendices}
\section{Isobar model tables}
\label{sec:isobarTables}
\begin{figure}
    \centering
    \includegraphics[width=0.49\textwidth]{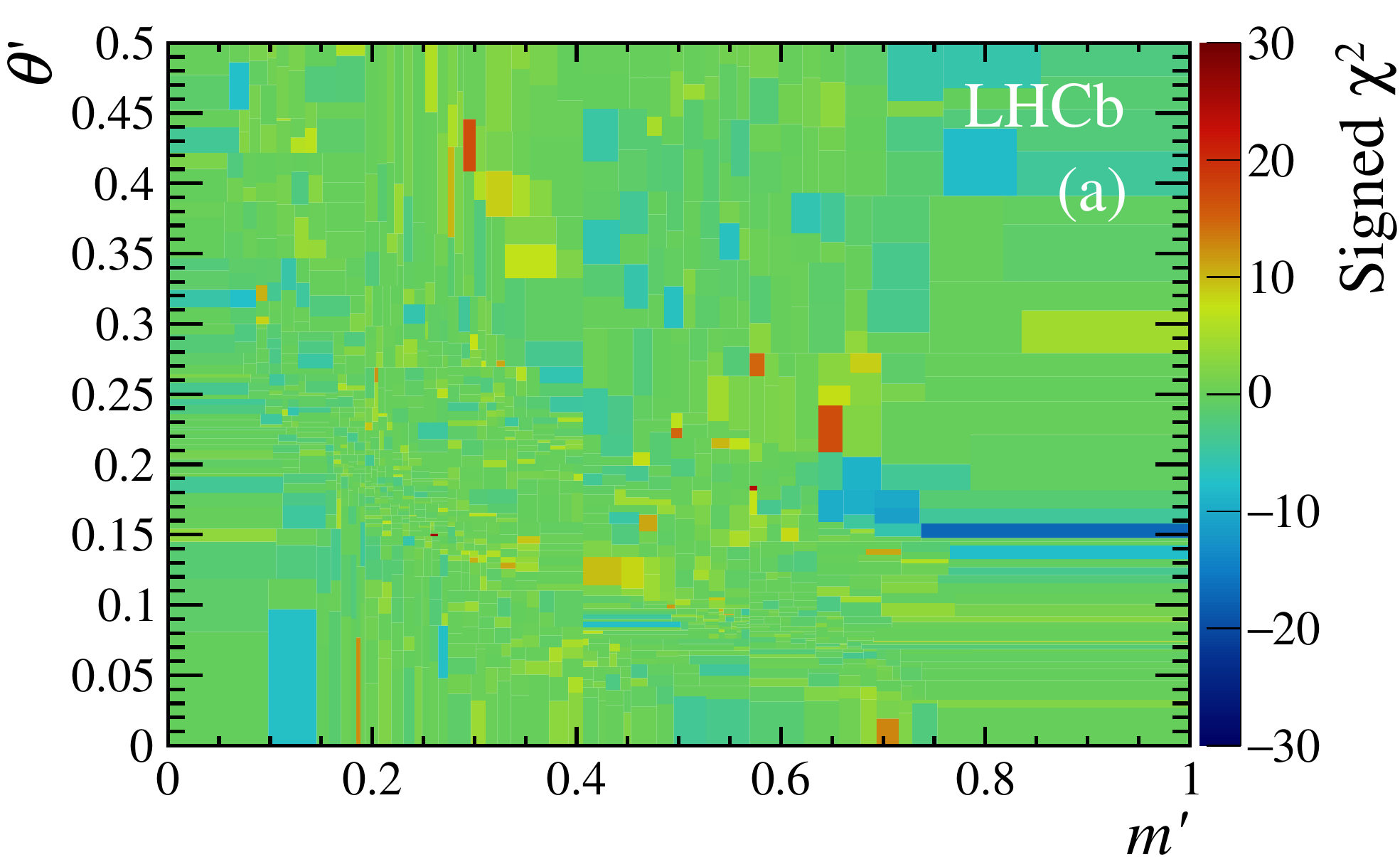}
    \includegraphics[width=0.49\textwidth]{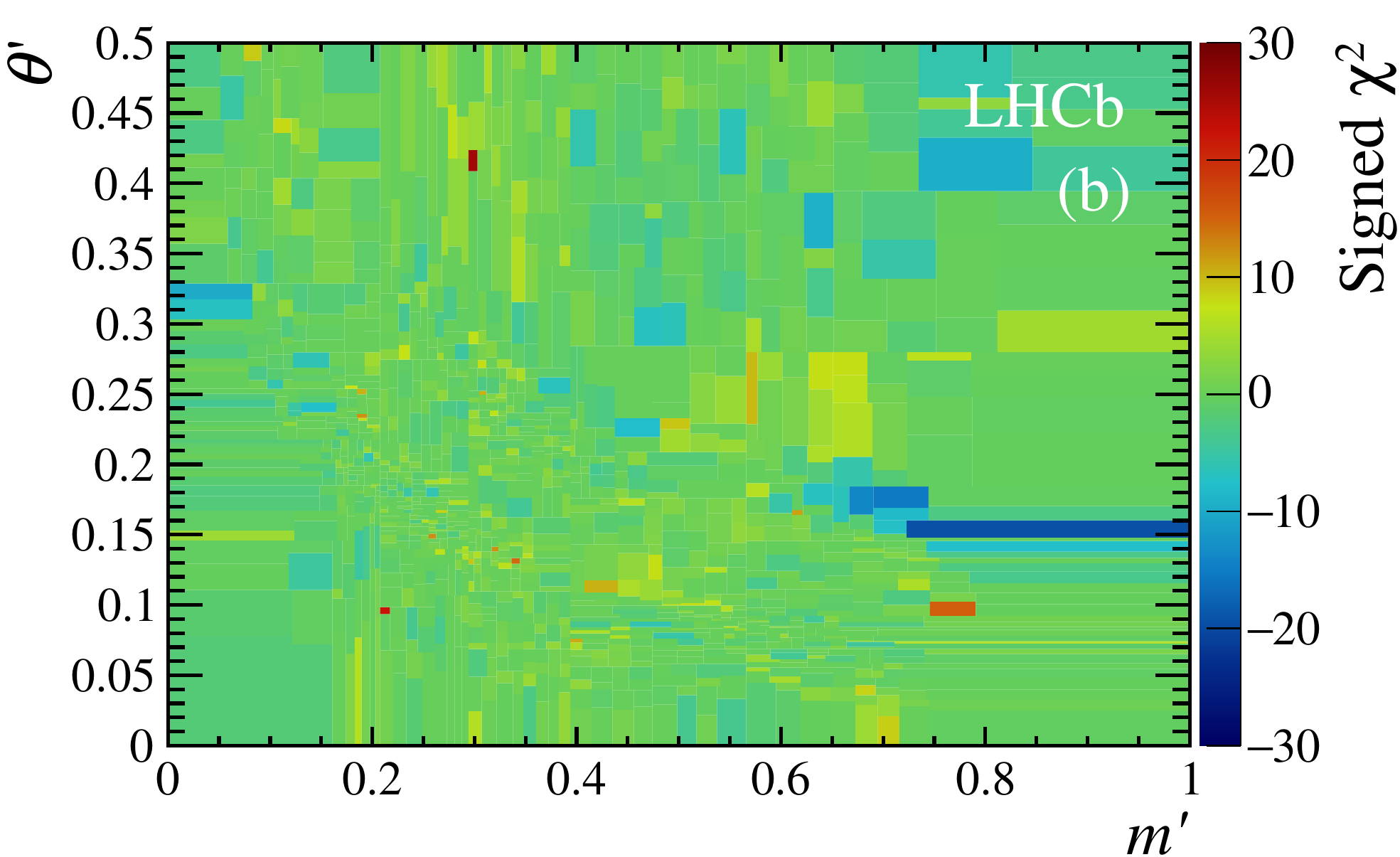}
    \caption{
    Signed $\chi^2$ distributions indicating the agreement between the isobar model fit and the data for (a)~\Bp and (b)~\Bm decays.
    }
    \label{fig:chisqIso}
\end{figure}

The results for the Cartesian coefficients obtained from the fit with the isobar description of the S-wave are reported in Table~\ref{tab:im:isobarParams}, with the fitted pole parameters given in Table~\ref{tab:freeparameter:isobarParams-pole}. 
The correlation matrices for the \CP-averaged fit fractions and quasi-two-body decay \CP asymmetries, corresponding to those presented in Tables~\ref{tab:ff} and~\ref{tab:acp}, can be found in Tables from~\ref{tab:im:isobarParams} to~\ref{tab:iso:covcpsyst}.
As an indication of fit quality, signed $\chi^2$ distributions in the square Dalitz plot, separated by charge, are produced with an adaptive binning procedure requiring at least 15 events per bin. These are shown in Fig.~\ref{fig:chisqIso}.

\renewcommand{\arraystretch}{1.35}
\begin{table}[ht]
  \centering
  \caption{
    Cartesian coefficients, $c_j$, for the components of the isobar model fit.
     }
  \label{tab:im:isobarParams}

  \resizebox{1.0\textwidth}{!}{
   \begin{tabular}{@{\hspace{0.5cm}}lr@{$\,\pm\,$}r@{$\,\pm\,$}lr@{$\,\pm\,$}r@{$\,\pm\,$}lr@{$\,\pm\,$}r@{$\,\pm\,$}lr@{$\,\pm\,$}r@{$\,\pm\,$}l@{\hspace{0.5cm}}}
\hline\hline

Component & \multicolumn{3}{c}{$x$} & \multicolumn{3}{c}{$y$} & \multicolumn{3}{c}{$\delta x$} & \multicolumn{3}{c}{$\delta y$} \\
%\toprule
   \hline
   $\rho(770)^0$              &  \multicolumn{3}{c}{$1.000~({\rm fixed})$} &  \multicolumn{3}{c}{$0.000~({\rm fixed})$}   &  $-0.003$ & $0.000$ & $0.010$ &  \multicolumn{3}{c}{$0.000~({\rm fixed})$} \\
   $\omega(782)$               &  $0.091$ & $0.004$ & $0.000$    &  $-0.007$  &  $0.007$ & $0.010$            &  $0.000$  &  $0.003$ & $0.000$   &  $-0.022$  &  $0.006$ & $0.010$\\
   $f_{2}(1270)$               &  $0.291  $&$ 0.011$& $0.058$   &  $0.204  $&$ 0.009$  & $0.067$             &  $-0.002 $&$ 0.009$ & $0.044$ &  $-0.179 $&$ 0.009$ & $0.044$\\
   $\rho(1450)^0$              &  $-0.223$ & $0.015$ & $0.010$   &  $0.191$  &  $0.010$ & $0.001$             &  $0.031$  &  $0.014$  & $0.020$  &  $0.068$  &  $0.009$ & $0.020$\\
   $\rho_{3}(1690)^0$             &  $0.073  $&$ 0.009$& $0.022$  &  $-0.045 $&$ 0.010$& $0.022$  &  $0.044 $&$ 0.009$ & $0.014$      &  $-0.013 $&$ 0.010$ & $0.014$\\
   Rescattering               &  $0.142  $&$ 0.009$& $0.025$  &  $-0.040 $&$ 0.009$& $0.042$  &  $-0.047 $&$ 0.008$ & $0.031$     &  $-0.027 $&$ 0.008$ & $0.031$\\
   $\sigma$               &  $-0.485 $&$ 0.013$& $0.130$  &  $0.284  $&$ 0.017$& $0.091$  &  $0.231 $&$ 0.011$  & $0.022$     &  $0.270 $&$ 0.015$ & $0.022$\\
   \hline\hline
    \end{tabular}
    }
\end{table}

\begin{table}[ht]
  \centering
  \caption{
    Fitted values of the pole parameters in the isobar model fit.
  }
  \label{tab:freeparameter:isobarParams-pole}
  \begin{tabular}{l|c}
    \hline \hline
    Parameter & Value \\ 
    \hline 
    % $magB$[$\rho^{0}(770)$] & $0.8081 \pm 0.1014$ \\$phiB$[$\rho^{0}(770)$] & $-0.2976 \pm 0.1268$ \\$m$[$\rho^{0}(770)$] & $0.7685 \pm 0.0017$ \\$\Gamma$[$\rho^{0}(770)$] & $0.1437 \pm 0.0042$ \\$magB$[$\rho^{0}(770)$] & $0.9384 \pm 0.0957$ \\$phiB$[$\rho^{0}(770)$] & $0.2436 \pm 0.0931$ \\
    $m_{\sigma}$ & $0.563 \pm 0.010$ \\$\Gamma_{\sigma}$ & $0.350 \pm 0.013$ \\
    \hline \hline
  \end{tabular} 
\end{table}

\begin{table}[ht]
  \centering
  \caption{Statistical correlation matrix for the isobar model \CP-averaged fit fractions.}
  \label{tab:iso:covffstat}
  \resizebox{1.0\textwidth}{!}{
  \begin{tabular}
    {@{\hspace{0.5cm}}c@{\hspace{0.25cm}}  @{\hspace{0.25cm}}|c@{\hspace{0.25cm}}  @{\hspace{0.25cm}}c@{\hspace{0.25cm}}  @{\hspace{0.25cm}}c@{\hspace{0.25cm}}  @{\hspace{0.25cm}}c@{\hspace{0.25cm}}  @{\hspace{0.25cm}}c@{\hspace{0.25cm}}  @{\hspace{0.25cm}}c@{\hspace{0.5cm}} @{\hspace{0.25cm}}c@{\hspace{0.5cm}} }
    \hline \hline

    & $\rho(770)^0$ & $\omega(782)$ & $f_2(1270)$ & $\rho(1450)^0$ & $\rho_3(1690)^0$ & Rescattering & $\sigma$ \\ \hline
    $\rho(770)^0$ & $+1.00$ & $-0.12$ & $-0.31$ & $+0.07$ & $-0.02$ & $-0.11$& $-0.62$ \\
    $\omega(782)$ & & $+1.00$ & $+0.04$ & $+0.05$ & $\phantom{+}0.00$ & $-0.09$& $-0.02$ \\
    $f_2(1270)$ & & & $+1.00$ & $-0.11$ & $+0.04$ & $-0.12$ & $+0.11$\\
    $\rho(1450)^0$ & & & & $+1.00$ & $-0.07$ & $-0.16$ & $-0.41$\\
    $\rho_3(1690)^0$ & & & & & $+1.00$ & $+0.10$ & $-0.05$\\
    Rescattering & & & & & & $+1.00$ & $+0.15$\\
    $\sigma$ & & & & & & & $+1.00$ \\

    \hline \hline
  \end{tabular}
  }
\end{table}

\begin{table}[ht]
  \centering
  \caption{Systematic correlation matrix for the isobar model \CP-averaged fit fractions.}
  \label{tab:iso:covffsyst}
  \resizebox{1.0\textwidth}{!}{
  \begin{tabular}
    {@{\hspace{0.5cm}}c@{\hspace{0.25cm}}  @{\hspace{0.25cm}}|c@{\hspace{0.25cm}}  @{\hspace{0.25cm}}c@{\hspace{0.25cm}}  @{\hspace{0.25cm}}c@{\hspace{0.25cm}}  @{\hspace{0.25cm}}c@{\hspace{0.25cm}}  @{\hspace{0.25cm}}c@{\hspace{0.25cm}}  @{\hspace{0.25cm}}c@{\hspace{0.5cm}} @{\hspace{0.25cm}}c@{\hspace{0.5cm}} }
    \hline \hline

    & $\rho(770)^0$ & $\omega(782)$ & $f_2(1270)$ & $\rho(1450)^0$ & $\rho_3(1690)^0$ & Rescattering & $\sigma$ \\ \hline
    $\rho(770)^0$ & $+1.00$ & $+0.77$ & $+0.58$ & $-0.72$ & $+0.86$ & $-0.53$& $-0.75$ \\
    $\omega(782)$ & & $+1.00$ & $+0.55$ & $-0.46$ & $+0.73$ & $-0.25$& $-0.49$ \\
    $f_2(1270)$ & & & $+1.00$ & $-0.34$ & $+0.71$ & $-0.78$ & $-0.47$\\
    $\rho(1450)^0$ & & & & $+1.00$ & $-0.53$ & $+0.57$ & $+0.68$\\
    $\rho_3(1690)^0$ & & & & & $+1.00$ & $-0.71$ & $-0.89$\\
    Rescattering & & & & & & $+1.00$ & $+0.77$\\
    $\sigma$ & & & & & & & $+1.00$ \\

    \hline \hline
  \end{tabular}
  }
\end{table}

\begin{table}[ht]
  \centering
  \caption{Statistical correlation matrix for the isobar model quasi-two-body decay \CP asymmetries.}
  \label{tab:iso:covcpstat}
  \resizebox{1.0\textwidth}{!}{
  \begin{tabular}
    {@{\hspace{0.5cm}}c@{\hspace{0.25cm}}  @{\hspace{0.25cm}}|c@{\hspace{0.25cm}}  @{\hspace{0.25cm}}c@{\hspace{0.25cm}}  @{\hspace{0.25cm}}c@{\hspace{0.25cm}}  @{\hspace{0.25cm}}c@{\hspace{0.25cm}}  @{\hspace{0.25cm}}c@{\hspace{0.25cm}}  @{\hspace{0.25cm}}c@{\hspace{0.5cm}} @{\hspace{0.25cm}}c@{\hspace{0.5cm}} }
    \hline \hline

    & $\rho(770)^0$ & $\omega(782)$ & $f_2(1270)$ & $\rho(1450)^0$ & $\rho_3(1690)^0$ & Rescattering & $\sigma$ \\ \hline
    $\rho(770)^0$ & $+1.00$ & $-0.07$ & $-0.25$ & $\phantom{+}0.00$ & $-0.02$ & $-0.21$& $-0.55$ \\
    $\omega(782)$ & & $+1.00$ & $-0.02$ & $-0.07$ & $-0.02$ & $+0.01$& $-0.08$ \\
    $f_2(1270)$ & & & $+1.00$ & $-0.01$ & $+0.23$ & $-0.03$ & $-0.02$\\
    $\rho(1450)^0$ & & & & $+1.00$ & $-0.03$ & $-0.26$ & $-0.28$\\
    $\rho_3(1690)^0$ & & & & & $+1.00$ & $\pm0.00$ & $-0.17$\\
    Rescattering & & & & & & $+1.00$ & $+0.15$\\
    $\sigma$ & & & & & & & $+1.00$ \\

    \hline \hline
  \end{tabular}
  }
\end{table}

\begin{table}[ht]
  \centering
  \caption{Systematic correlation matrix for the isobar model quasi-two-body decay \CP asymmetries.}
  \label{tab:iso:covcpsyst}
  \resizebox{1.0\textwidth}{!}{
  \begin{tabular}
    {@{\hspace{0.5cm}}c@{\hspace{0.25cm}}  @{\hspace{0.25cm}}|c@{\hspace{0.25cm}}  @{\hspace{0.25cm}}c@{\hspace{0.25cm}}  @{\hspace{0.25cm}}c@{\hspace{0.25cm}}  @{\hspace{0.25cm}}c@{\hspace{0.25cm}}  @{\hspace{0.25cm}}c@{\hspace{0.25cm}}  @{\hspace{0.25cm}}c@{\hspace{0.5cm}} @{\hspace{0.25cm}}c@{\hspace{0.5cm}} }
    \hline \hline

    & $\rho(770)^0$ & $\omega(782)$ & $f_2(1270)$ & $\rho(1450)^0$ & $\rho_3(1690)^0$ & Rescattering & $\sigma$ \\ \hline
    $\rho(770)^0$ & $+1.00$ & $+0.58$ & $+0.50$ & $-0.46$ & $+0.39$ & $+0.06$& $+0.27$ \\
    $\omega(782)$ & & $+1.00$ & $+0.28$ & $-0.20$ & $+0.71$ & $+0.08$& $+0.53$ \\
    $f_2(1270)$ & & & $+1.00$ & $+0.38$ & $-0.28$ & $+0.82$ & $+0.57$\\
    $\rho(1450)^0$ & & & & $+1.00$ & $-0.46$ & $+0.77$ & $+0.40$\\
    $\rho_3(1690)^0$ & & & & & $+1.00$ & $-0.49$ & $+0.02$\\
    Rescattering & & & & & & $+1.00$ & $+0.60$\\
    $\sigma$ & & & & & & & $+1.00$ \\

    \hline \hline
  \end{tabular}
  }
\end{table}

\clearpage
\section{K-matrix model tables}
\label{sec:kMatrixTables}

The results for the Cartesian coefficients, and other fitted parameters, obtained with the K-matrix S-wave approach can be found in Table~\ref{tab:km:isobarParams}.
Correlation matrices for these parameters are reported in Ref.~\cite{supplemental}.
Here, as the reference amplitude is the $\rho$--$\omega$ mixing component, rather than a component only representing the $\rho(770)^0$ resonance, the magnitude of the positive isobar coefficient describing the $\rho(770)^0$ resonance is not unity, but is calculated accounting for the small $\omega(782)$ contribution.

\renewcommand{\arraystretch}{1.35}
\begin{table}[t]
  \centering
  \caption{Cartesian coefficients, $c_j$, for the components of the K-matrix model fit. For the K-matrix model, the $\beta_{\alpha}$ and $f^{\rm prod}_{v}$ parameters describe the relative contributions of the production pole $\alpha$ and production slowly varying part corresponding to channel $v$, respectively. In the absence of \CP\ violation, $\delta x = \delta y = 0$.}
  \label{tab:km:isobarParams}

 \resizebox{1.0\textwidth}{!}{
   \begin{tabular}{@{\hspace{0.5cm}}lr@{$\,\pm\,$}r@{$\,\pm\,$}lr@{$\,\pm\,$}r@{$\,\pm\,$}lr@{$\,\pm\,$}r@{$\,\pm\,$}lr@{$\,\pm\,$}r@{$\,\pm\,$}l@{\hspace{0.5cm}}}
\hline\hline

Component & \multicolumn{3}{c}{$x$} & \multicolumn{3}{c}{$y$} & \multicolumn{3}{c}{$\delta x$} & \multicolumn{3}{c}{$\delta y$} \\
\hline
$\rho(770)^0$  &  $1.006$ & $0.007$ & $ 0.001$ &  \multicolumn{3}{c}{$0$ (fixed)}  &  $-0.021$ & $0.008$ & $ 0.065$ &  \multicolumn{3}{c}{$0$ (fixed)} \\
$\omega(782)$& $ 0.054$ & $ 0.033 $ & $ 0.000 $ & $ -0.016$ & $ 0.045$ & $ 0.000 $ & $0.025$ & $ 0.031$ & $ 0.000 $ & $0.068 $ & $ 0.048$ & $ 0.000 $ \\
$f_2(1270)$& $ 0.190$ & $ 0.017 $ & $ 0.107$ & $ 0.289$ & $ 0.012$ & $ 0.045$ & $0.102$ & $ 0.016$ & $ 0.053$ & $-0.190 $ & $ 0.012$ & $ 0.049$\\
$\rho(1450)^0$& $ -0.408$ & $ 0.016 $ & $ 0.129$ & $ 0.034$ & $ 0.021$ & $ 0.148$& $0.032$ & $ 0.017$ & $ 0.100$& $0.141 $ & $ 0.020$ & $ 0.143$\\
$\rho_3(1690)^0$& $ 0.155$ & $ 0.009 $ & $ 0.034$ & $ 0.037$ & $ 0.018$ & $ 0.083$ & $0.024$ & $ 0.009$ & $ 0.036$ & $0.026 $ & $ 0.016$ & $ 0.056$ \\
\midrule
$\beta_{1}$& $ -0.210$ & $ 0.033 $ & $ 0.187$ & $ -0.303$ & $ 0.038$ & $ 0.275$ & $-0.139$ & $ 0.032$ & $ 0.159$ & $0.102 $ & $ 0.042$ & $ 0.243$ \\
$\beta_{2}$& $ -0.236$ & $ 0.056 $ & $ 0.058$ & $ -0.065$ & $ 0.055$ & $ 0.247$ & $0.122$ & $ 0.062$ & $ 0.114$ & $0.187 $ & $ 0.063$ & $ 0.286$ \\
$\beta_{3}$& $ 0.055$ & $ 0.057 $ & $ 0.169$ & $ 0.072$ & $ 0.074$ & $ 0.293$ & $-0.121$ & $ 0.067$ & $ 0.152$ & $-0.027 $ & $ 0.071$ & $ 0.286$ \\
$\beta_{4}$& $ 0.072$ & $ 0.060 $ & $ 0.125$ & $ 0.087$ & $ 0.065$ & $ 0.297$ & $-0.124$ & $ 0.068$ & $ 0.137$ & $-0.034 $ & $ 0.065$ & $ 0.272$ \\
$\beta_{5}$& $ -0.038$ & $ 0.076 $ & $ 0.205$ & $ -0.040$ & $ 0.097$ & $ 0.446$ & $0.148$ & $ 0.086$ & $ 0.164$ & $0.234 $ & $ 0.092$ & $ 0.487$ \\
$f^{\rm prod}_{1}$ & $ -0.329$ & $ 0.057 $ & $ 0.215$ & $ -0.047$ & $ 0.072$ & $ 0.288$ & $0.148$ & $ 0.058$ & $ 0.200$ & $0.147 $ & $ 0.067$ & $ 0.360$ \\
$f^{\rm prod}_{2}$ & $ -0.190$ & $ 0.052 $ & $ 0.178$ & $ -0.022$ & $ 0.056$ & $ 0.197$ & $-0.057$ & $ 0.056$ & $ 0.163$ & $0.231 $ & $ 0.064$ & $ 0.250$ \\
$f^{\rm prod}_{3}$ & $ -0.017$ & $ 0.097 $ & $ 0.349$ & $ 0.139$ & $ 0.066$ & $ 0.196$ & $0.082$ & $ 0.089$ & $ 0.409$ & $-0.095 $ & $ 0.075$ & $ 0.155$ \\
$f^{\rm prod}_{4}$ & $ 0.033$ & $ 0.036 $ & $ 0.138$ & $ 0.142$ & $ 0.040$ & $ 0.085$ & $0.068$ & $ 0.032$ & $ 0.052$ & $0.037 $ & $ 0.040$ & $ 0.094$ \\

\hline\hline
\end{tabular}
}
\end{table}

\begin{figure}
    \centering
    \includegraphics[width=0.49\textwidth]{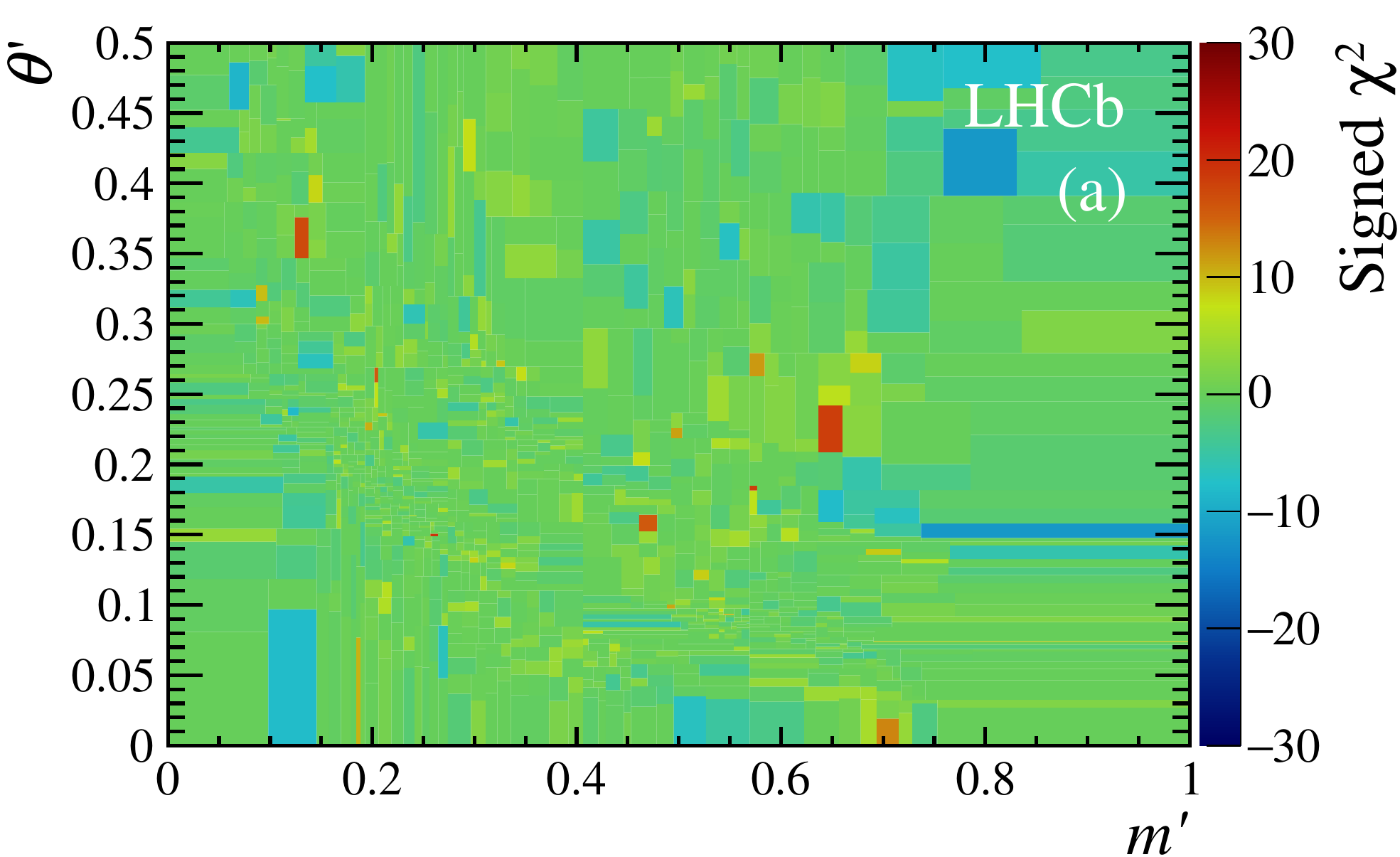}
    \includegraphics[width=0.49\textwidth]{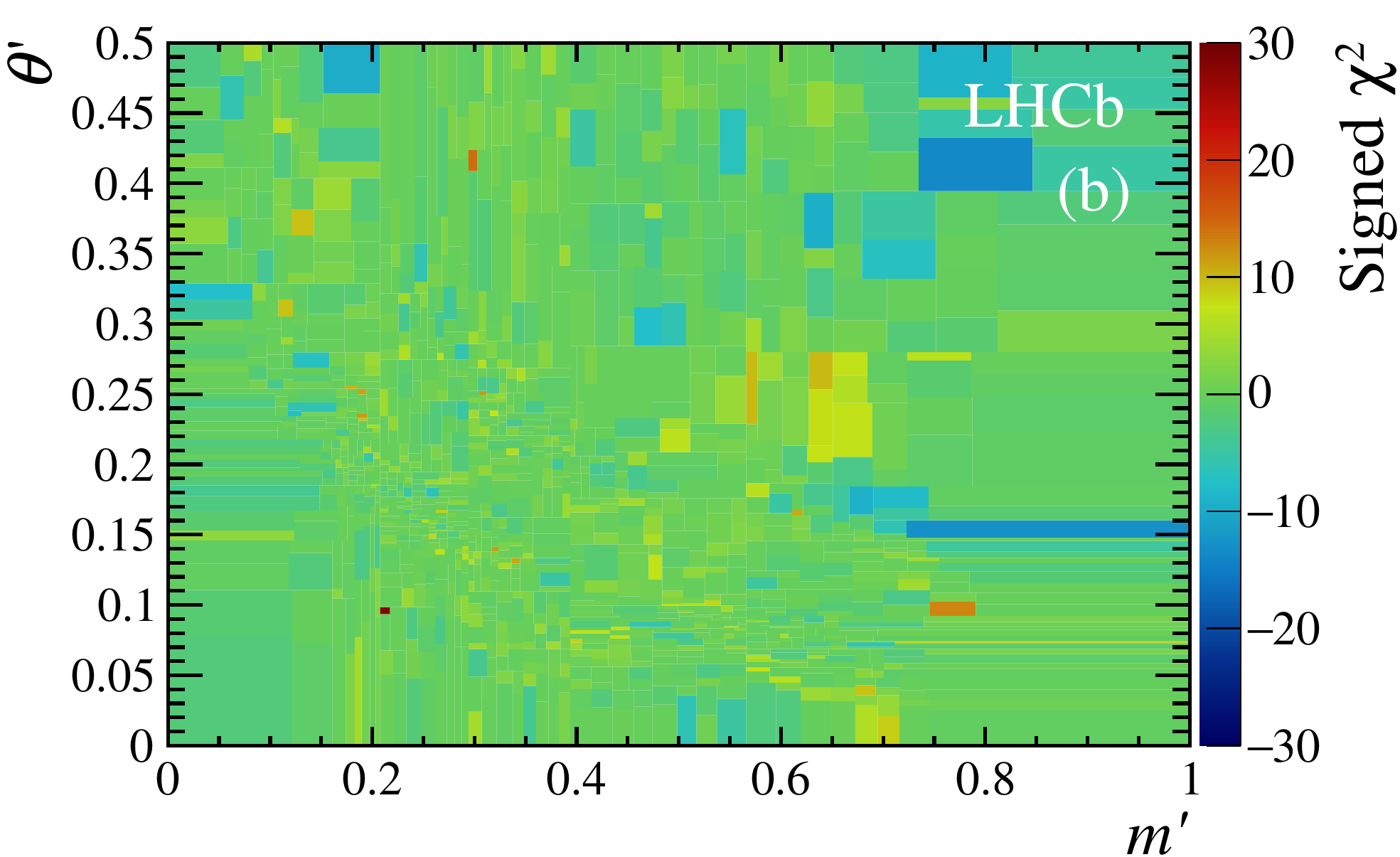}
    \caption{
    Signed $\chi^2$ distributions indicating the agreement between the K-matrix model fit and the data for (a)~\Bp and (b)~\Bm decays.
    }
    \label{fig:chisqKM}
\end{figure}

Furthermore, the statistical and systematic uncertainty correlation matrices for the K-matrix fit \CP-averaged fit fractions and quasi-two-body decay \CP asymmetries, corresponding to those presented in Tables~\ref{tab:ff} and~\ref{tab:acp}, can be found in Tables~\ref{tab:km:covffstat} and~\ref{tab:km:covffsyst}, and~\ref{tab:km:covcpstat} and~\ref{tab:km:covcpsyst}, respectively.
As an indication of fit quality, signed $\chi^2$ distributions in the square Dalitz plot, separated by charge, are produced with an adaptive binning procedure requiring at least 15 events per bin. These are shown in Fig.~\ref{fig:chisqKM}.

\begin{table}[ht]
  \centering
  \caption{Statistical correlation matrix for the K-matrix \CP-averaged fit fractions.}
  \label{tab:km:covffstat}
  \begin{tabular}
    {@{\hspace{0.5cm}}c@{\hspace{0.25cm}}  @{\hspace{0.25cm}}|c@{\hspace{0.25cm}}  @{\hspace{0.25cm}}c@{\hspace{0.25cm}}  @{\hspace{0.25cm}}c@{\hspace{0.25cm}}  @{\hspace{0.25cm}}c@{\hspace{0.25cm}}  @{\hspace{0.25cm}}c@{\hspace{0.25cm}}  @{\hspace{0.25cm}}c@{\hspace{0.5cm}}}
    \hline \hline

    & $\rho(770)^0$ & $\omega(782)$ & $f_2(1270)$ & $\rho(1450)^0$ & $\rho_3(1690)^0$ & S-wave \\ \hline
    $\rho(770)^0$ & $+1.00$ & $+0.06$ & $-0.02$ & $+0.13$ & $+0.17$ & $-0.47$\\
    $\omega(782)$ & & $+1.00$ & $+0.06$ & $+0.14$ & $-0.03$ & $-0.12$\\
    $f_2(1270)$ & & & $+1.00$ & $-0.08$ & $+0.08$ & $-0.53$\\
    $\rho(1450)^0$ & & & & $+1.00$ & $+0.03$ & $+0.09$\\
    $\rho_3(1690)^0$ & & & & & $+1.00$ & $-0.30$\\
    S-wave & & & & & & $+1.00$\\

    \hline \hline
  \end{tabular}
\end{table}

\begin{table}[ht]
  \centering
  \caption{Systematic correlation matrix for the K-matrix \CP-averaged fit fractions.}
  \label{tab:km:covffsyst}
  \begin{tabular}
    {@{\hspace{0.5cm}}c@{\hspace{0.25cm}}  @{\hspace{0.25cm}}|c@{\hspace{0.25cm}}  @{\hspace{0.25cm}}c@{\hspace{0.25cm}}  @{\hspace{0.25cm}}c@{\hspace{0.25cm}}  @{\hspace{0.25cm}}c@{\hspace{0.25cm}}  @{\hspace{0.25cm}}c@{\hspace{0.25cm}}  @{\hspace{0.25cm}}c@{\hspace{0.5cm}}}
    \hline \hline

    & $\rho(770)^0$ & $\omega(782)$ & $f_2(1270)$ & $\rho(1450)^0$ & $\rho_3(1690)^0$ & S-wave \\ \hline
    $\rho(770)^0$ & $+1.00$ & $+0.69$ & $-0.46$ & $+0.67$ & $+0.73$ & $+0.66$\\
    $\omega(782)$ & & $+1.00$ & $+0.05$ & $+0.25$ & $+0.28$ & $+0.66$\\
    $f_2(1270)$ & & & $+1.00$ & $-0.80$ & $-0.85$ & $+0.08$\\
    $\rho(1450)^0$ & & & & $+1.00$ & $+0.95$ & $+0.23$\\
    $\rho_3(1690)^0$ & & & & & $+1.00$ & $+0.27$\\
    S-wave & & & & & & $+1.00$\\

    \hline \hline
  \end{tabular}
\end{table}

\begin{table}[ht]
  \centering
  \caption{Statistical correlation matrix for the K-matrix quasi-two-body decay \CP asymmetries.}
  \label{tab:km:covcpstat}
  \begin{tabular}
    {@{\hspace{0.5cm}}c@{\hspace{0.25cm}}  @{\hspace{0.25cm}}|c@{\hspace{0.25cm}}  @{\hspace{0.25cm}}c@{\hspace{0.25cm}}  @{\hspace{0.25cm}}c@{\hspace{0.25cm}}  @{\hspace{0.25cm}}c@{\hspace{0.25cm}}  @{\hspace{0.25cm}}c@{\hspace{0.25cm}}  @{\hspace{0.25cm}}c@{\hspace{0.5cm}}}
    \hline \hline

    & $\rho(770)^0$ & $\omega(782)$ & $f_2(1270)$ & $\rho(1450)^0$ & $\rho_3(1690)^0$ & S-wave \\ \hline
    $\rho(770)^0$ & $+1.00$ & $+0.08$ & $+0.09$ & $+0.24$ & $-0.04$ & $-0.37$\\
    $\omega(782)$ & & $+1.00$ & $+0.05$ & $+0.17$ & $+0.15$ & $\phantom{+}0.00$\\
    $f_2(1270)$ & & & $+1.00$ & $-0.02$ & $+0.09$ & $-0.41$\\
    $\rho(1450)^0$ & & & & $+1.00$ & $-0.02$ & $\phantom{+}0.00$\\
    $\rho_3(1690)^0$ & & & & & $+1.00$ & $-0.14$\\
    S-wave & & & & & & $+1.00$\\
    \hline \hline
  \end{tabular}
\end{table}

\begin{table}[ht]
  \centering
  \caption{Systematic correlation matrix for the K-matrix quasi-two-body decay \CP asymmetries.}
  \label{tab:km:covcpsyst}
  \begin{tabular}
    {@{\hspace{0.5cm}}c@{\hspace{0.25cm}}  @{\hspace{0.25cm}}|c@{\hspace{0.25cm}}  @{\hspace{0.25cm}}c@{\hspace{0.25cm}}  @{\hspace{0.25cm}}c@{\hspace{0.25cm}}  @{\hspace{0.25cm}}c@{\hspace{0.25cm}}  @{\hspace{0.25cm}}c@{\hspace{0.25cm}}  @{\hspace{0.25cm}}c@{\hspace{0.5cm}}}
    \hline \hline

    & $\rho(770)^0$ & $\omega(782)$ & $f_2(1270)$ & $\rho(1450)^0$ & $\rho_3(1690)^0$ & S-wave \\ \hline
    $\rho(770)^0$ & $+1.00$ & $+0.64$ & $+0.79$ & $+0.71$ & $+0.63$ & $+0.30$ \\
    $\omega(782)$ & & $+1.00$ & $+0.39$ & $+0.43$ & $+0.30$ & $+0.14$\\
    $f_2(1270)$ & & & $+1.00$ & $+0.92$ & $+0.94$ & $+0.75$\\
    $\rho(1450)^0$ & & & & $+1.00$ & $+0.97$ & $+0.67$\\
    $\rho_3(1690)^0$ & & & & & $+1.00$ & $+0.78$\\
    S-wave & & & & & & $+1.00$\\
    \hline \hline
  \end{tabular}
\end{table}

\subsection{Secondary minimum}
\label{sec:kMatrixSecondary}

A secondary minimum is observed in the maximum likelihood fit of the model with the K-matrix S-wave approach, located approximately $0.8$ units of negative-log-likelihood away from the primary minimum. This minimum results in fit results that are statistically consistent with the best minimum, except for in the parameters of the individual K-matrix components (fit fractions and overall \CP-violation parameters are otherwise consistent).

\begin{figure}[ht]
  \centering
  \includegraphics[width=1.0\textwidth]{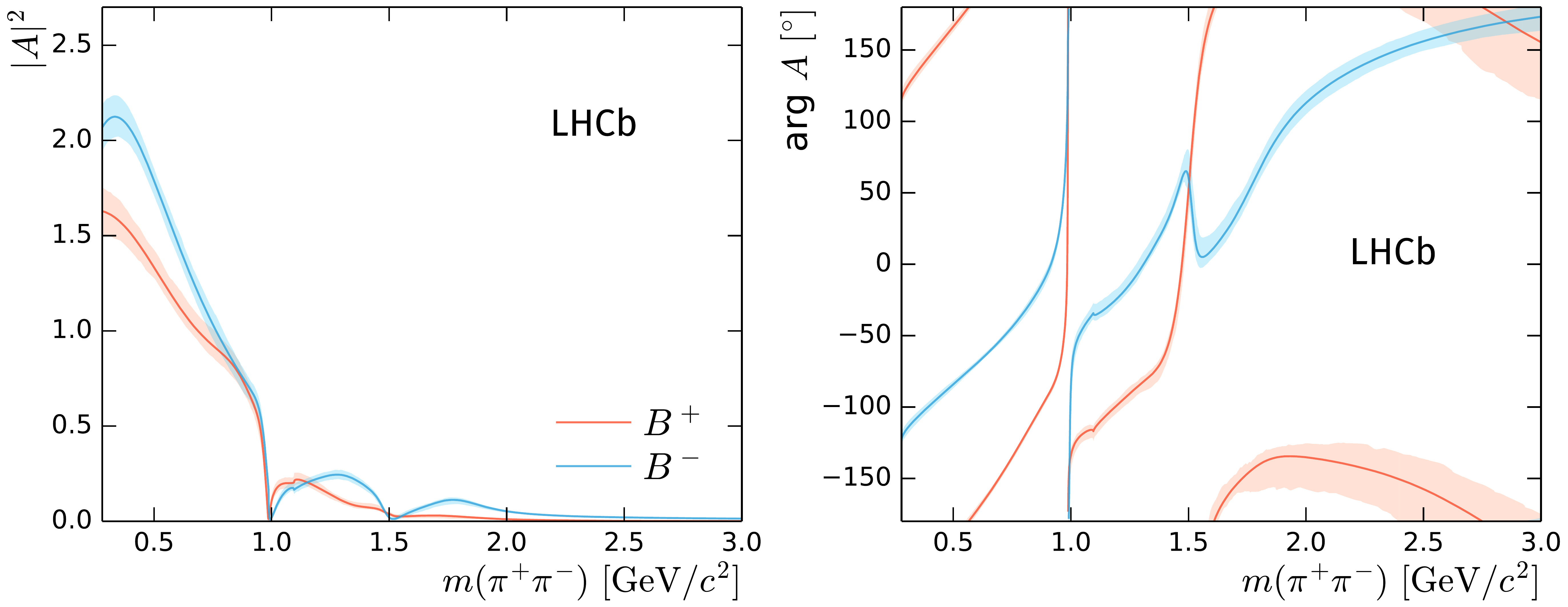}
  \put(-265,70){(a)}
  \put(-35,70){(b)}

  \caption{
    The K-matrix S-wave projections for the secondary solution, where (a) shows the magnitude-squared while (b) shows the phase motion.
    The red curve indicates \Bp, while the blue curve represents \Bm\ decays. The light bands represent the $68\%$ confidence interval around the central values, including statistical uncertainties only.
  }
  \label{fig:km:swaveProjSecondary}
\end{figure}

The parameters obtained from this secondary solution can be seen in Table~\ref{tab:km:isobarParamsSecondary} (to be compared to the nominal results in Table~\ref{tab:km:isobarParams}). Projections of the S-wave amplitude on $m(\pip\pim)$ can be seen in Fig.~\ref{fig:km:swaveProjSecondary}.

\renewcommand{\arraystretch}{1.35}
\begin{table}[ht]
  \centering
  \caption{Isobar coefficients, $c_j$, for the components of the second solution of the K-matrix model fit, where uncertainties are statistical only. For the K-matrix model, the $\beta_{\alpha}$ and $f^{\rm prod}_{v}$ parameters describe the relative contributions of the production pole $\alpha$ and production slowly varying part corresponding to channel $v$, respectively. In the absence of \CP\ violation, $\delta x = \delta y = 0$.}
  \label{tab:km:isobarParamsSecondary}

\begin{tabular}{@{\hspace{0.5cm}}lr@{$\,\pm\,$}lr@{$\,\pm\,$}lr@{$\,\pm\,$}lr@{$\,\pm\,$}l@{\hspace{0.5cm}}}
  \hline\hline

Component & \multicolumn{2}{c}{$x$} & \multicolumn{2}{c}{$y$} & \multicolumn{2}{c}{$\delta x$} & \multicolumn{2}{c}{$\delta y$} \\
\hline
$\rho(770)^0$  &  $1.015$ & $0.008$ & \multicolumn{2}{c}{$0$ (fixed)}  &  $-0.030$ & $0.008$ &  \multicolumn{2}{c}{$0$ (fixed)} \\
$\omega(782)$& $ 0.069$ & $ 0.031 $ & $ 0.006$ & $ 0.046$ & $0.007$ & $ 0.034$ & $-0.063 $ & $ 0.047$ \\
$f_2(1270)$& $ 0.196$ & $ 0.017 $ & $ 0.278$ & $ 0.012$ & $0.106$ & $ 0.017$ & $-0.207 $ & $ 0.014$ \\
$\rho(1450)^0$& $ -0.395$ & $ 0.019 $ & $ 0.051$ & $ 0.023$ & $0.052$ & $ 0.020$ & $0.164 $ & $ 0.025$ \\
$\rho_3(1690)^0$& $ 0.162$ & $ 0.010 $ & $ 0.026$ & $ 0.021$ & $0.029$ & $ 0.010$ & $0.015 $ & $ 0.021$ \\
\midrule

$\beta_{1}$ & $ -0.031$ & $ 0.040 $ & $ -0.030$ & $ 0.037$ & $0.036$ & $ 0.038$ & $0.386 $ & $ 0.035$ \\
$\beta_{2}$ & $ -0.294$ & $ 0.066 $ & $ 0.153$ & $ 0.062$ & $0.052$ & $ 0.057$ & $0.401 $ & $ 0.059$ \\
$\beta_{3}$ & $ 0.045$ & $ 0.061 $ & $ -0.113$ & $ 0.062$ & $-0.116$ & $ 0.052$ & $-0.202 $ & $ 0.083$ \\
$\beta_{4}$ & $ 0.048$ & $ 0.065 $ & $ -0.156$ & $ 0.056$ & $-0.135$ & $ 0.055$ & $-0.268 $ & $ 0.074$ \\
$\beta_{5}$ & $ -0.025$ & $ 0.080 $ & $ 0.195$ & $ 0.082$ & $0.134$ & $ 0.065$ & $0.458 $ & $ 0.109$ \\

$f^{\rm prod}_{1}$ & $ -0.325$ & $ 0.053 $ & $ 0.144$ & $ 0.066$ & $0.140$ & $ 0.049$ & $0.337 $ & $ 0.075$ \\
$f^{\rm prod}_{2}$ & $ -0.315$ & $ 0.063 $ & $ -0.211$ & $ 0.064$ & $-0.166$ & $ 0.058$ & $0.033 $ & $ 0.054$ \\
$f^{\rm prod}_{3}$ & $ 0.218$ & $ 0.081 $ & $ 0.013$ & $ 0.074$ & $0.308$ & $ 0.101$ & $-0.199 $ & $ 0.065$ \\
$f^{\rm prod}_{4}$ & $ -0.078$ & $ 0.036 $ & $ 0.025$ & $ 0.038$ & $-0.042$ & $ 0.036$ & $-0.084 $ & $ 0.036$ \\

\hline\hline
\end{tabular}
\end{table}

\clearpage
\section{QMI model tables}
\label{sec:qmiTables}

\begin{figure}
    \centering
    \includegraphics[width=0.49\textwidth]{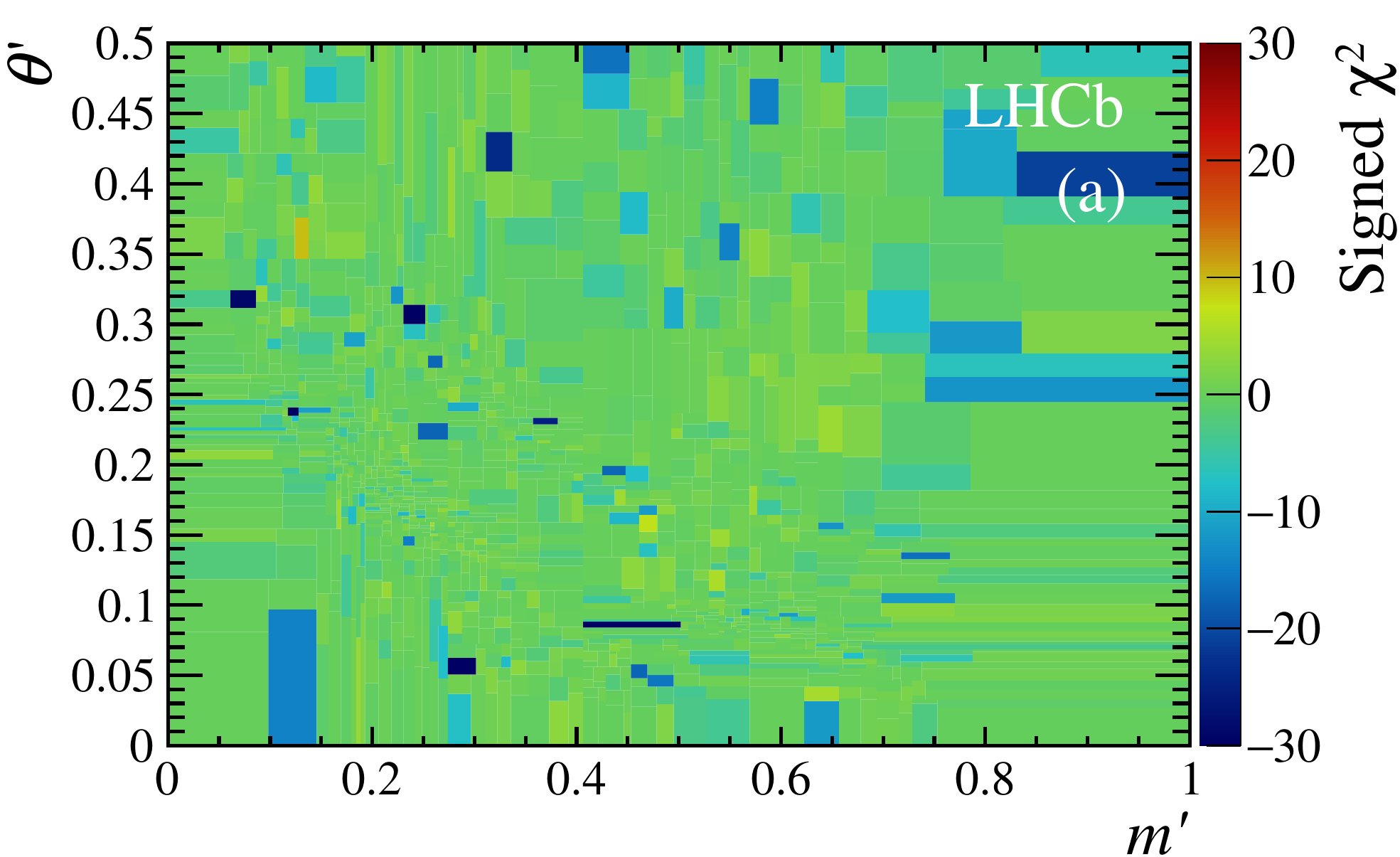}
    \includegraphics[width=0.49\textwidth]{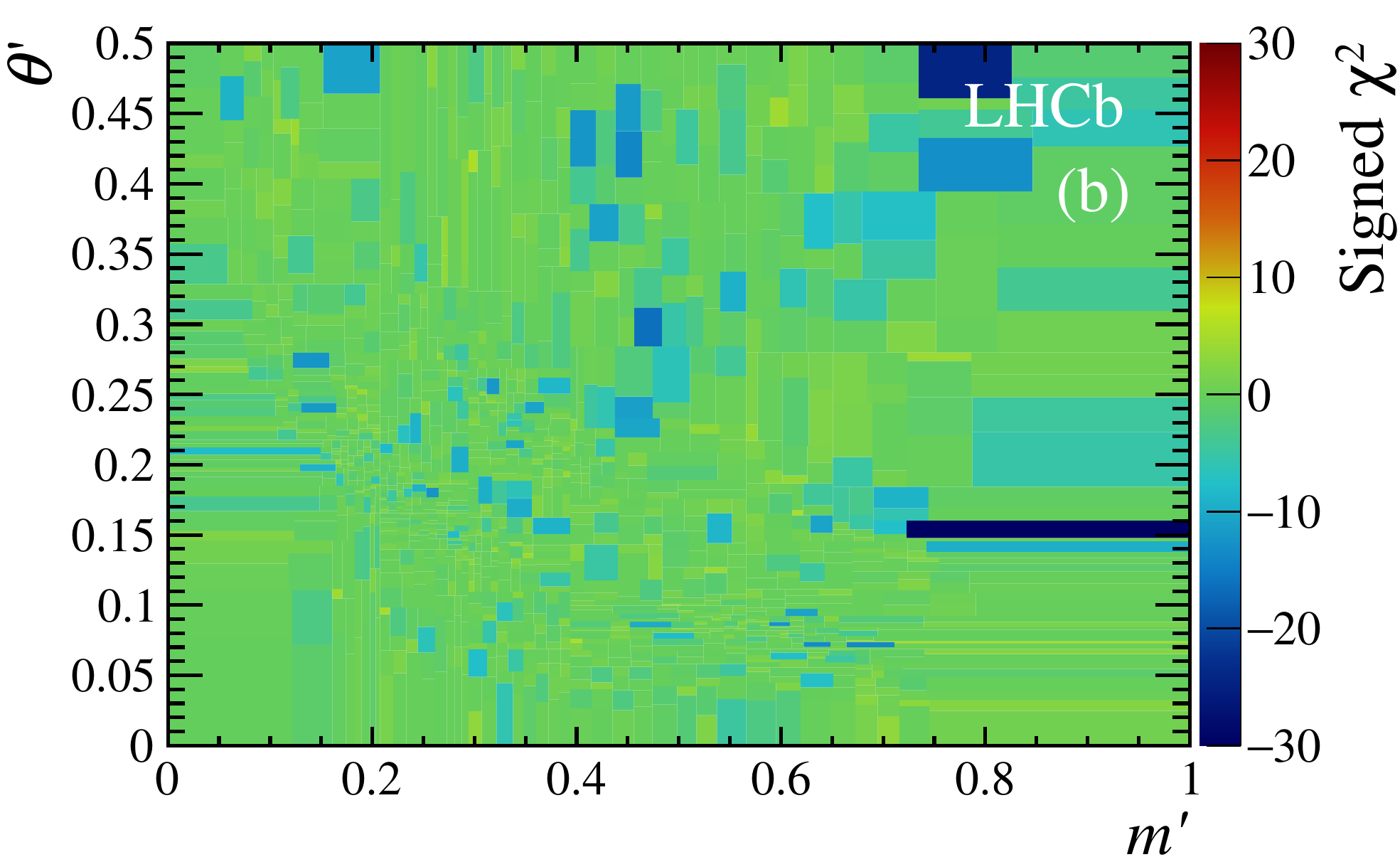}
    \caption{
    Signed $\chi^2$ distributions indicating the agreement between the QMI model fit and the data for (a)~\Bp and (b)~\Bm decays.
    }
    \label{fig:chisqQMI}
\end{figure}

The results for the Cartesian coefficients from the fit with the QMI S-wave can be found in Table~\ref{tab:pwa:covffstat0}.
Correlation matrices for these parameters are reported in Ref.~\cite{supplemental}.
The statisical and systematic correlation matrices for the \CP-averaged fit fractions are given in Tables~\ref{tab:pwa:covffstat1} and~\ref{tab:pwa:covffstat2}, respectively,
while those for the quasi-two-body decay \CP asymmetries can be found in Tables~\ref{tab:pwa:covffstat3} and~\ref{tab:pwa:covffstat4}.
As an indication of fit quality, signed $\chi^2$ distributions in the square Dalitz plot, separated by charge, are produced with an adaptive binning procedure requiring at least 15 events per bin. These are shown in Fig.~\ref{fig:chisqQMI}.

\begin{table}[ht]
  \centering
  \caption{Cartesian coefficients obtained with the QMI model. Only the statistical uncertainties are shown as some systematic variations change the overall scale of various lineshapes at this level.}
  \label{tab:pwa:covffstat0}
  \begin{tabular}
    {@{\hspace{0.5cm}}c@{\hspace{0.25cm}}  @{\hspace{0.25cm}}c@{\hspace{0.25cm}}  @{\hspace{0.25cm}}c@{\hspace{0.25cm}}  @{\hspace{0.25cm}}c@{\hspace{0.25cm}}  @{\hspace{0.25cm}}c@{\hspace{0.5cm}}}
    \hline \hline

    Component & $x$ & $y$ & $\delta x$ & $\delta y$ \\ \hline
    $\rho(770)^0$ & 1 (fixed) & 0 (fixed) & $-0.022 \pm 0.009$ & 0 (fixed) \\
    $\omega(782)$ & $+0.822 \pm 0.069$ & $-0.213 \pm 0.065$ & $+0.010 \pm 0.069$ & $-0.179 \pm 0.065$ \\
    $f_2(1270)$ & $+0.737 \pm 0.047$ & $+0.779 \pm 0.044$ & $+0.205 \pm 0.046$ & $-0.556 \pm 0.049$ \\
    $\rho(1450)^0$ & $-0.679 \pm 0.040$ & $+0.189 \pm 0.042$ & $+0.006 \pm 0.043$ & $+0.251 \pm 0.048$ \\
    $\rho_3(1690)^0$ & $+0.279 \pm 0.023$ & $+0.059 \pm 0.046$ & $+0.205 \pm 0.023$ & $+0.005 \pm 0.046$ \\

    \hline \hline
  \end{tabular}
\end{table}

\begin{table}[ht]
  \centering
  \caption{Statistical correlation matrix for the QMI \CP-averaged fit fractions.}
  \label{tab:pwa:covffstat1}
  \begin{tabular}
    {@{\hspace{0.5cm}}c@{\hspace{0.25cm}}  @{\hspace{0.25cm}}|c@{\hspace{0.25cm}}  @{\hspace{0.25cm}}c@{\hspace{0.25cm}}  @{\hspace{0.25cm}}c@{\hspace{0.25cm}}  @{\hspace{0.25cm}}c@{\hspace{0.25cm}}  @{\hspace{0.25cm}}c@{\hspace{0.25cm}}  @{\hspace{0.25cm}}c@{\hspace{0.5cm}}}
    \hline \hline

& $\rho(770)^0$ & $\omega(782)$ & $f_2(1270)$ & $\rho(1450)^0$ & $\rho_3(1690)^0$ & S-wave \\ \hline
$\rho(770)^0$ & $+1.00$ & $-0.06$ & $+0.01$ & $+0.11$ & $-0.35$ & $-0.63$\\
$\omega(892)$ & & $+1.00$ & $-0.11$ & $+0.16$ & $+0.09$ & $+0.01$\\
$f_2(1270)$ & & & $+1.00$ & $-0.20$ & $+0.10$ & $-0.42$\\
$\rho(1450)^0$ & & & & $+1.00$ & $+0.13$ & $-0.01$\\
$\rho_3(1690)^0$ & & & & & $+1.00$ & $+0.16$\\
S-wave & & & & & & $+1.00$\\
    \hline \hline
  \end{tabular}
\end{table}

\begin{table}[ht]
  \centering
  \caption{Correlation matrix corresponding to the quadratic sum of systematic and model uncertainties for the QMI fit \CP-averaged fractions.}
  \label{tab:pwa:covffstat2}
  \begin{tabular}
    {@{\hspace{0.5cm}}c@{\hspace{0.25cm}}  @{\hspace{0.25cm}}|c@{\hspace{0.25cm}}  @{\hspace{0.25cm}}c@{\hspace{0.25cm}}  @{\hspace{0.25cm}}c@{\hspace{0.25cm}}  @{\hspace{0.25cm}}c@{\hspace{0.25cm}}  @{\hspace{0.25cm}}c@{\hspace{0.25cm}}  @{\hspace{0.25cm}}c@{\hspace{0.5cm}}}
    \hline \hline

& $\rho(770)^0$ & $\omega(782)$ & $f_2(1270)$ & $\rho(1450)^0$ & $\rho_3(1690)^0$ & S-wave \\ \hline
$\rho(770)^0$ & +1.00 & +0.10 & +0.10 & $-$0.23 & $-$0.21 & $-$0.43\\
$\omega(782)$ & & +1.00 & +0.04 & $-$0.48 & $-$0.46 & $-$0.40\\
$f_2(1270)$ & & & +1.00 & $-$0.15 & +0.05 & $-$0.32\\
$\rho(1450)^0$ & & & & +1.00 & +0.82 & +0.32\\
$\rho_3(1690)^0$ & & & & & +1.00 & +0.29\\
S-wave & & & & & & +1.00\\
    \hline \hline
  \end{tabular}
\end{table}

\begin{table}[ht]
  \centering
  \caption{Statistical correlation matrix for the QMI quasi-two-body decay \CP asymmetries.}
  \label{tab:pwa:covffstat3}
  \begin{tabular}
    {@{\hspace{0.5cm}}c@{\hspace{0.25cm}}  @{\hspace{0.25cm}}|c@{\hspace{0.25cm}}  @{\hspace{0.25cm}}c@{\hspace{0.25cm}}  @{\hspace{0.25cm}}c@{\hspace{0.25cm}}  @{\hspace{0.25cm}}c@{\hspace{0.25cm}}  @{\hspace{0.25cm}}c@{\hspace{0.25cm}}  @{\hspace{0.25cm}}c@{\hspace{0.5cm}}}
    \hline \hline

& $\rho(770)^0$ & $\omega(782)$ & $f_2(1270)$ & $\rho(1450)^0$ & $\rho_3(1690)^0$ & S-wave \\ \hline
$\rho(770)^0$ & $+1.00$ & $-0.10$ & $-0.06$ & $+0.23$ & $-0.24$ & $-0.52$\\
$\omega(892)$ & & $+1.00$ & $-0.02$ & $+0.06$ & $+0.10$ & $+0.07$\\
$f_2(1270)$ & & & $+1.00$ & $-0.15$ & $+0.08$ & $-0.32$\\
$\rho(1450)^0$ & & & & $+1.00$ & $-0.11$ & $+0.02$\\
$\rho_3(1690)^0$ & & & & & $+1.00$ & $+0.08$\\
S-wave & & & & & & $+1.00$\\
    \hline \hline
  \end{tabular}
\end{table}

\begin{table}[ht]
  \centering
  \caption{Correlation matrix corresponding to the quadratic sum of systematic and model uncertainties for the QMI quasi-two-body decay \CP asymmetries.}
  \label{tab:pwa:covffstat4}
  \begin{tabular}
    {@{\hspace{0.5cm}}c@{\hspace{0.25cm}}  @{\hspace{0.25cm}}|c@{\hspace{0.25cm}}  @{\hspace{0.25cm}}c@{\hspace{0.25cm}}  @{\hspace{0.25cm}}c@{\hspace{0.25cm}}  @{\hspace{0.25cm}}c@{\hspace{0.25cm}}  @{\hspace{0.25cm}}c@{\hspace{0.25cm}}  @{\hspace{0.25cm}}c@{\hspace{0.5cm}}}
    \hline \hline

& $\rho(770)^0$ & $\omega(782)$ & $f_2(1270)$ & $\rho(1450)^0$ & $\rho_3(1690)^0$ & S-wave \\ \hline
$\rho(770)^0$ & +1.00 & +0.05 & $-$0.06 & +0.04 & $-$0.13 & $-$0.19\\
$\omega(782)$ & & +1.00 & $-$0.13 & $-$0.23 & $-$0.20 & +0.22\\
$f_2(1270)$ & & & +1.00 & +0.40 & +0.46 & $-$0.61\\
$\rho(1450)^0$ & & & & +1.00 & +0.77 & $-$0.54\\
$\rho_3(1690)^0$ & & & & & +1.00 & $-$0.42\\
S-wave & & & & & & +1.00\\

    \hline \hline
  \end{tabular}
\end{table}

\clearpage
\section{Results with S-wave model variation included as systematic uncertainty}
\label{sec:extraApp}

Results are presented throughout this paper for each of three different approaches to the modelling of the $\pi\pi$ S-wave: the isobar, K-matrix and QMI models.  
As discussed in Sections~\ref{sec:results} and~\ref{sec:discussion}, all three give good descriptions of the data, with each describing some regions of the Dalitz plot better than the others.  
Therefore, it is not possible to conclude that one is preferred to the others.

Nonetheless, it is anticipated that for some purposes it will be more useful to have a single set of results rather than three sets.  
Therefore, Table~\ref{tab:extraTab} provides such a presentation.
The central values, statistical and experimental systematic uncertainties are taken from the results with the isobar model, while the largest deviation in the central value between the isobar model and the other two S-wave approaches is combined in quadrature with the other sources of model uncertainty.

\renewcommand{\arraystretch}{1.35}
\begin{table}[ht]
  \centering
  \caption{Results with S-wave model variation included as a source of systematic uncertainty. The first uncertainty is statistical, the second is experimental systematic and the third is the adjusted model systematic uncertainty.}
  \label{tab:extraTab}
  \resizebox{\textwidth}{!}{
    \begin{tabular}
      {@{\hspace{0.5cm}}l@{\hspace{0.25cm}}  @{\hspace{0.25cm}}r@{\hspace{0.25cm}}  @{\hspace{0.25cm}}r@{\hspace{0.25cm}}}
      \hline \hline

      Component & \multicolumn{1}{c}{\CP-averaged fit fractions ($10^{-2}$)} & \multicolumn{1}{c}{Quasi-two-body \CP asymmetries ($10^{-2}$)} \\
      \hline
      $\rho(770)^0$ & $55.5\phantom{0} \pm 0.6\phantom{0} \pm 0.4\phantom{0} \pm 2.7\phantom{0}$ &
      $+0.7 \pm \phantom{0}1.1 \pm \phantom{0}0.6 \pm \phantom{0}4.0$\hspace{7.5ex} \\

      $\omega(782)$ & $0.50 \pm 0.03 \pm 0.01 \pm 0.08$ & 
      $-4.8 \pm \phantom{0}6.5\pm \phantom{0}1.3 \pm \phantom{0}4.7$\hspace{7.5ex} \\

      $f_2(1270)$ & $9.0\phantom{0} \pm 0.3\phantom{0} \pm 0.7\phantom{0} \pm 1.5\phantom{0}$ &
      $+46.8 \pm \phantom{0}6.1 \pm \phantom{0}1.5 \pm 10.2$\hspace{7.5ex} \\

      $\rho(1450)^0$ & $5.2\phantom{0} \pm 0.3\phantom{0} \pm 0.2\phantom{0} \pm 5.6\phantom{0}$ &
      $-12.9 \pm \phantom{0}3.3 \pm \phantom{0}3.6 \pm 41.9$\hspace{7.5ex} \\

      $\rho_3(1690)^0$ & $0.5\phantom{0} \pm 0.1\phantom{0} \pm 0.1\phantom{0} \pm 1.0\phantom{0}$ &
      $-80.1 \pm 11.4 \pm \phantom{0}7.8 \pm 50.5$\hspace{7.5ex} \\

      S-wave & $25.4\phantom{0} \pm 0.5\phantom{0} \pm 0.5\phantom{0} \pm 3.9\phantom{0}$ &
      $+14.4 \pm \phantom{0}1.8 \pm \phantom{0}1.0 \pm \phantom{0}2.4$\hspace{7.5ex} \\

      \hline \hline
    \end{tabular}
  }
\end{table}

\clearpage
\section{Phase comparison}
\label{sec:phaseComp}

The presentation of the complex coefficients $c_j$ in the Cartesian convention makes it difficult to compare the relative phases of the components in the different models.
To facilitate this, the relevant information is presented in Table~\ref{tab:phaseComp}.

\begin{table}[ht]
  \centering
  \caption{Phase comparison in degrees for (top) \Bp and (bottom) \Bm between the three S-wave approaches where the first uncertainty is statistical, the second systematic and the third from the model. Note that the phase of the $\rho(770)^0$ component of the $\rho$--$\omega$ lineshape is fixed to zero as it is selected to be the reference contribution.}
  \label{tab:phaseComp}
    \begin{tabular}
      {@{\hspace{0.5cm}}l@{\hspace{0.25cm}}  @{\hspace{0.25cm}}|r@{\hspace{0.25cm}}  @{\hspace{0.25cm}}r@{\hspace{0.25cm}}  @{\hspace{0.25cm}}r@{\hspace{0.5cm}}}
      \hline \hline

      Component & \multicolumn{1}{c}{Isobar} & \multicolumn{1}{c}{K-matrix} & \multicolumn{1}{c}{QMI} \\
      \hline
      $\omega(782)$ & $-19 \pm 6 \pm \phantom{0}1$ &
      $-15 \pm 6 \pm \phantom{0}4$ &
      $-25 \pm \phantom{0}6 \pm \phantom{0}27$\\

      $f_2(1270)$ & $+5 \pm 3 \pm 12$ &
      $+19 \pm 4 \pm 18$ &
      $+13 \pm \phantom{0}5 \pm \phantom{0}21$\\

      $\rho(1450)^0$ & $+127 \pm 4 \pm 21$ &
      $+155 \pm 5 \pm 29$ &
      $+147 \pm \phantom{0}7 \pm 152$\\

      $\rho_3(1690)^0$ & $-26 \pm 7 \pm 14$ &
      $+19 \pm 8 \pm 34$ &
      $+8 \pm 10 \pm \phantom{0}24$\\

      \hline \hline

      Component & \multicolumn{1}{c}{Isobar} & \multicolumn{1}{c}{K-matrix} & \multicolumn{1}{c}{QMI} \\
      \hline
      $\omega(782)$ & $+8 \pm 6 \pm \phantom{0}1$ &
      $+8 \pm 7 \pm \phantom{00}4$ &
      $-2 \pm \phantom{0}7 \pm \phantom{0}11$\\

      $f_2(1270)$ & $+53 \pm \phantom{0}2 \pm 12$ &
      $+80 \pm 3 \pm \phantom{0}17$ &
      $+68 \pm \phantom{0}3 \pm \phantom{0}66$\\

      $\rho(1450)^0$ & $+154 \pm 4 \pm \phantom{0}6$ &
      $-166 \pm 4 \pm \phantom{0}51$ &
      $-175 \pm \phantom{0}5 \pm 171$\\

      $\rho_3(1690)^0$ & $-47 \pm 18 \pm 25$ &
      $+5 \pm 8 \pm \phantom{0}46$ &
      $+36 \pm 26 \pm \phantom{0}46$\\

      \hline \hline
    \end{tabular}
\end{table}

\clearpage
\section{Isobar model component projections}
\label{sec:isobarProj}

\begin{figure}[tbh]
\centering
\includegraphics[width=0.49\linewidth]{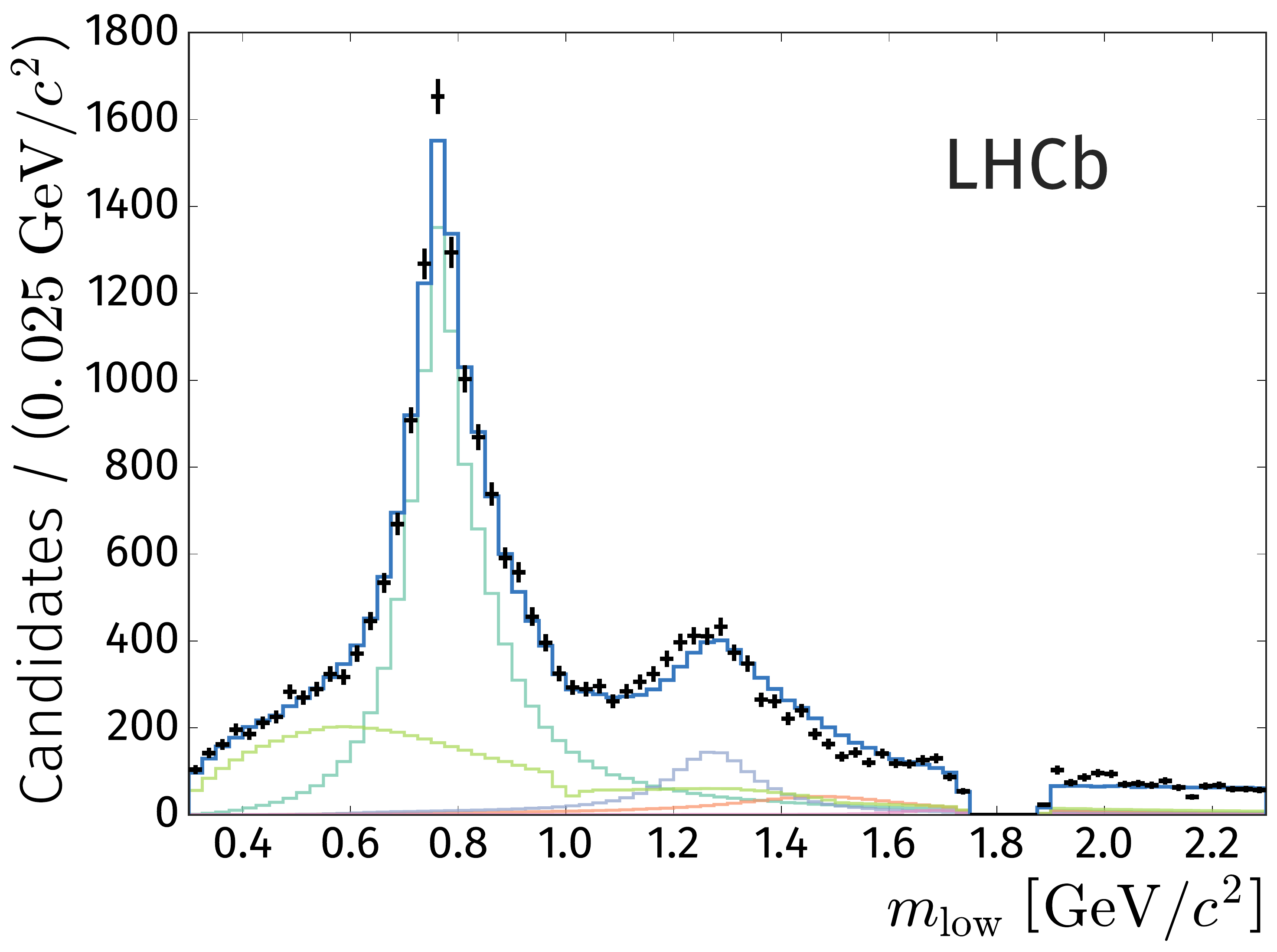}
\includegraphics[width=0.49\linewidth]{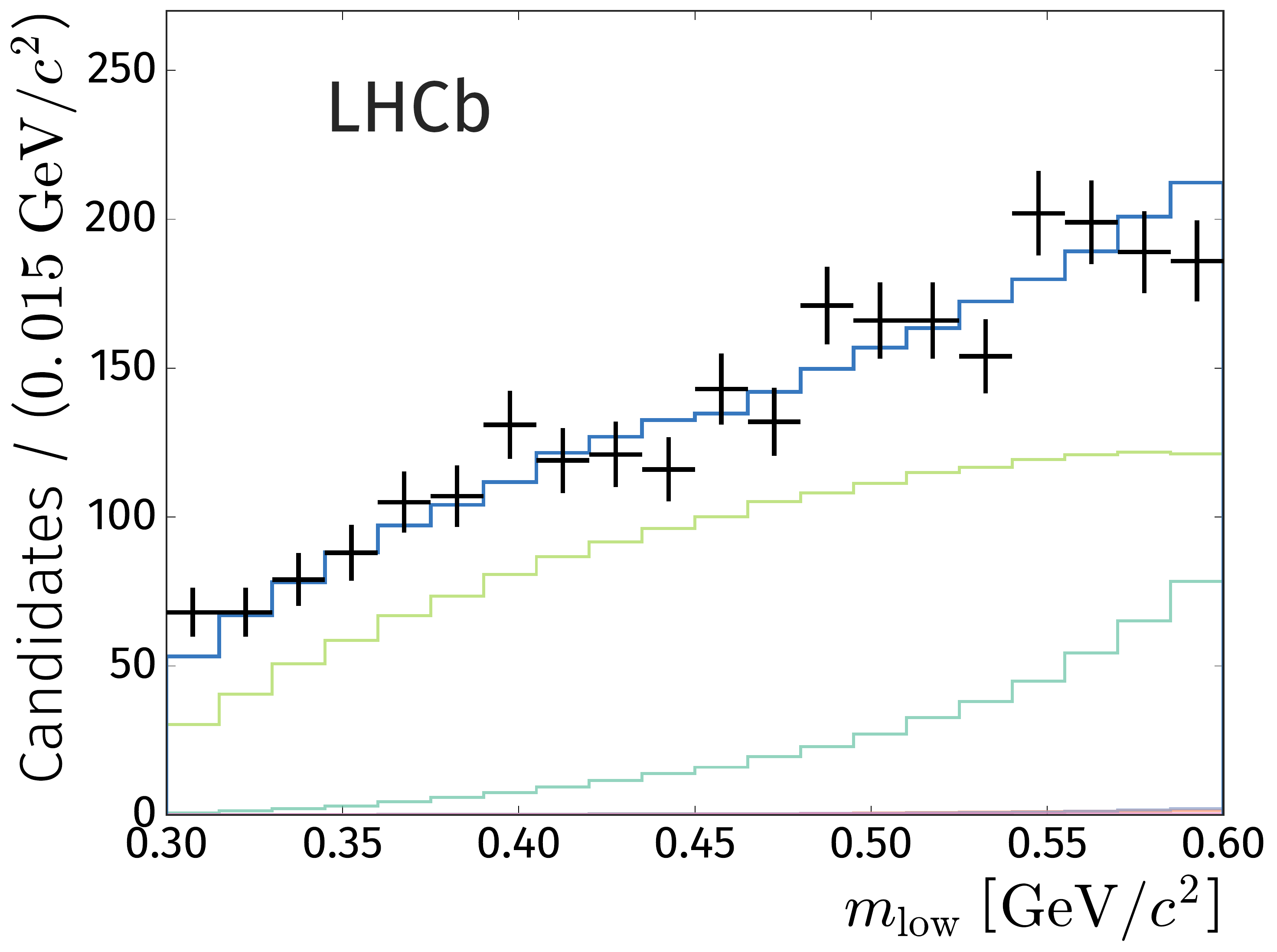}
\put(-405,27){(a)}
\put(-180,27){(b)}

\includegraphics[width=0.49\linewidth]{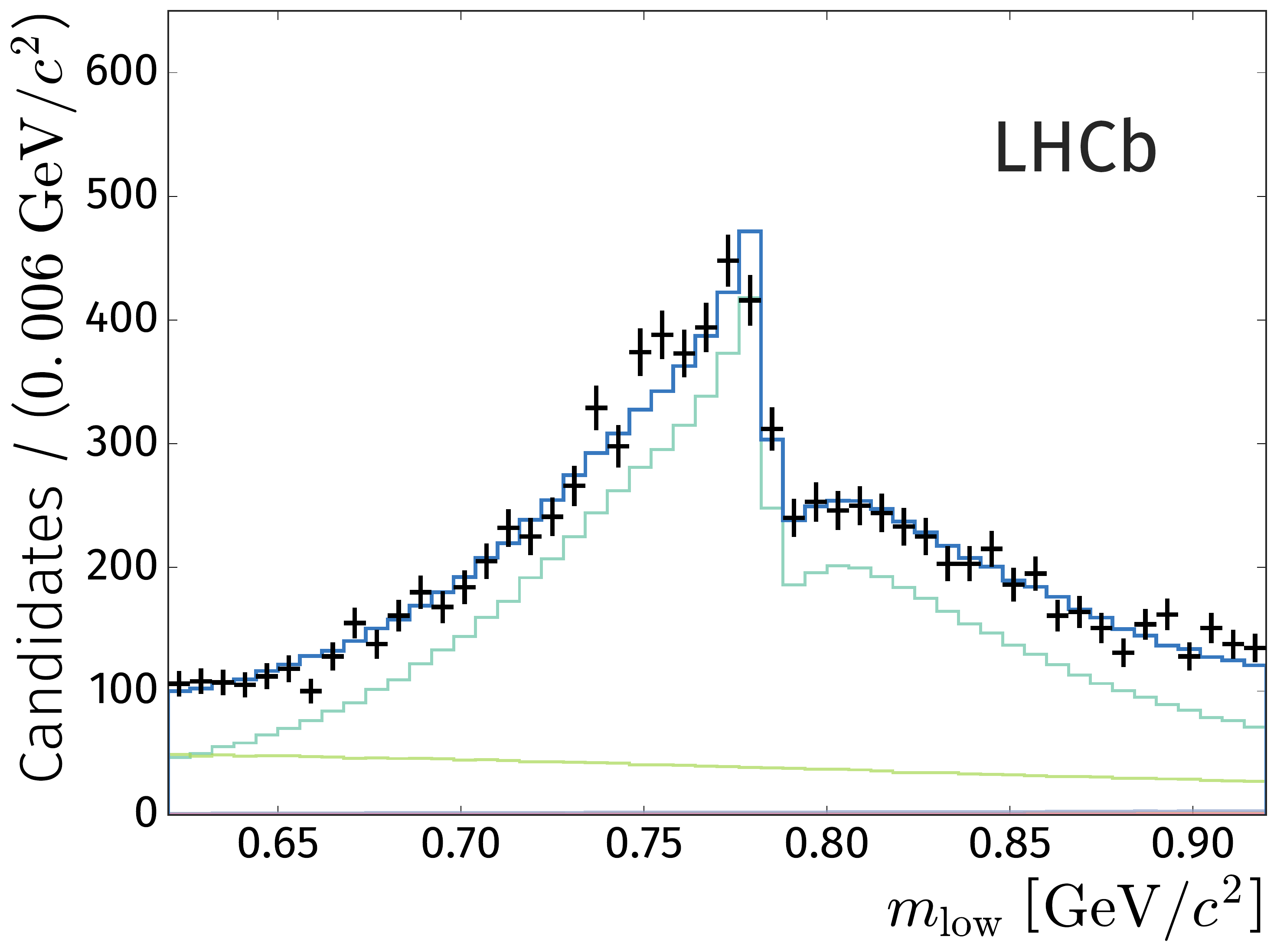}
\includegraphics[width=0.49\linewidth]{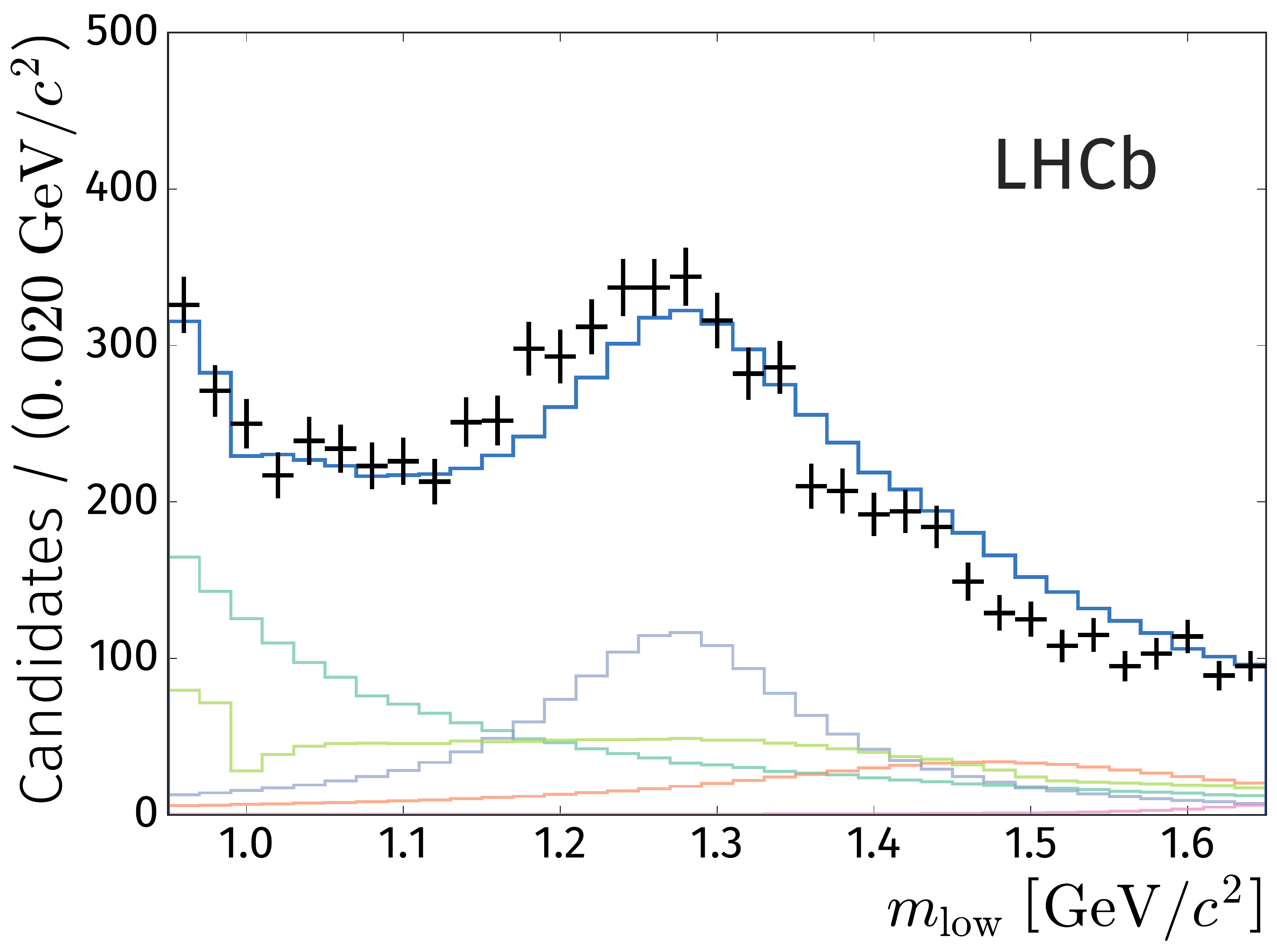}
\put(-405,30){(c)}
\put(-180,30){(d)}

\caption{Fit projections on $m_{\rm low}$ of the result with the isobar S-wave model (a)~in the low $m_{\rm low}$ region, (b)~below the $\rho(770)^0$ region, (c)~in the $\rho(770)^0$ region, and (d)~in the $f_2(1270)$ region. The thick blue curve represents the total model, and the coloured curves represent the contributions of individual model components (not including interference effects), as per the legend in Fig.~\ref{fig:projComponentsI3}.}
\label{fig:projComponentsI1}
\end{figure}

\begin{figure}[tbh]
\centering
\includegraphics[width=0.49\linewidth]{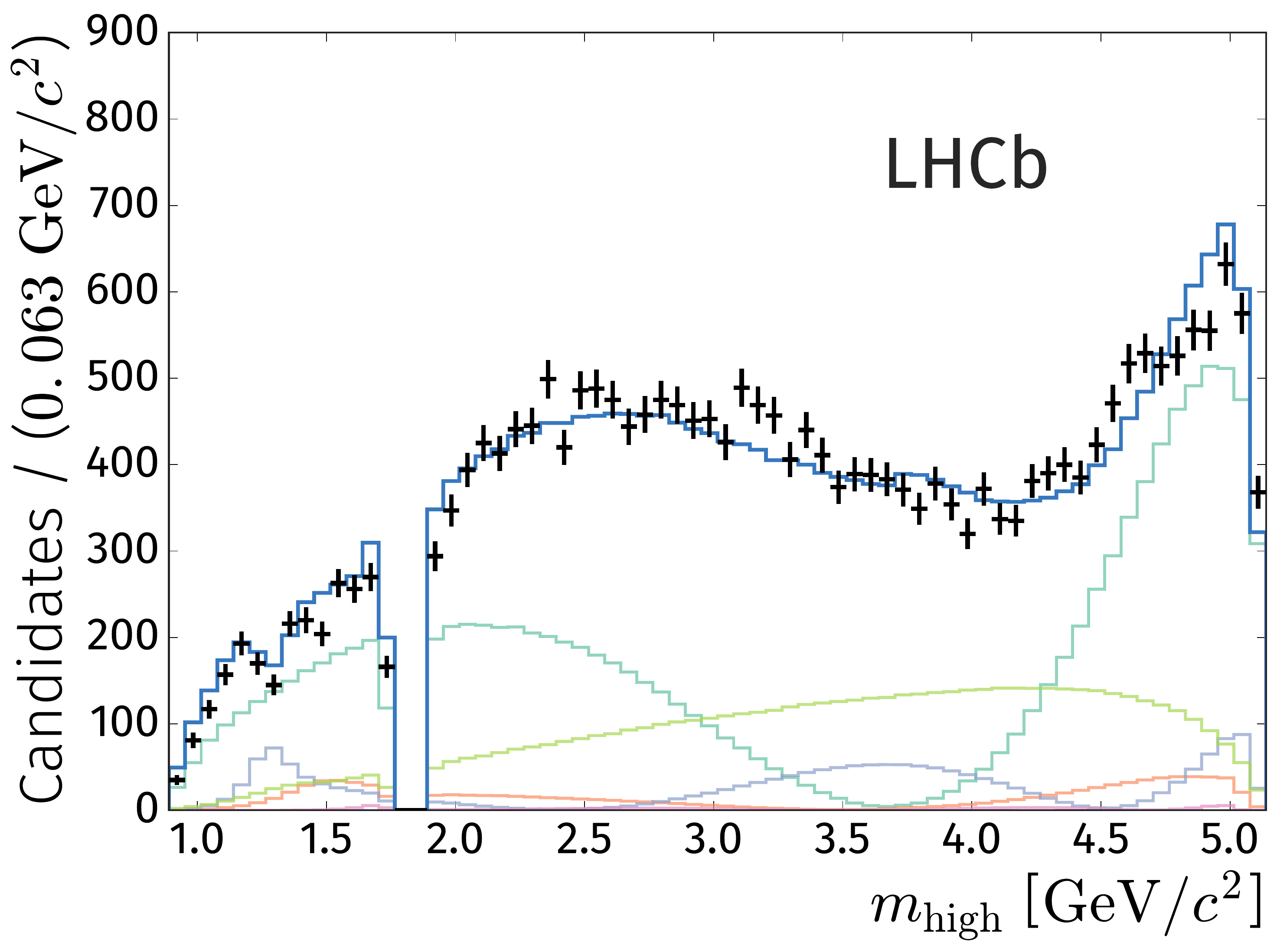}
\includegraphics[width=0.49\linewidth]{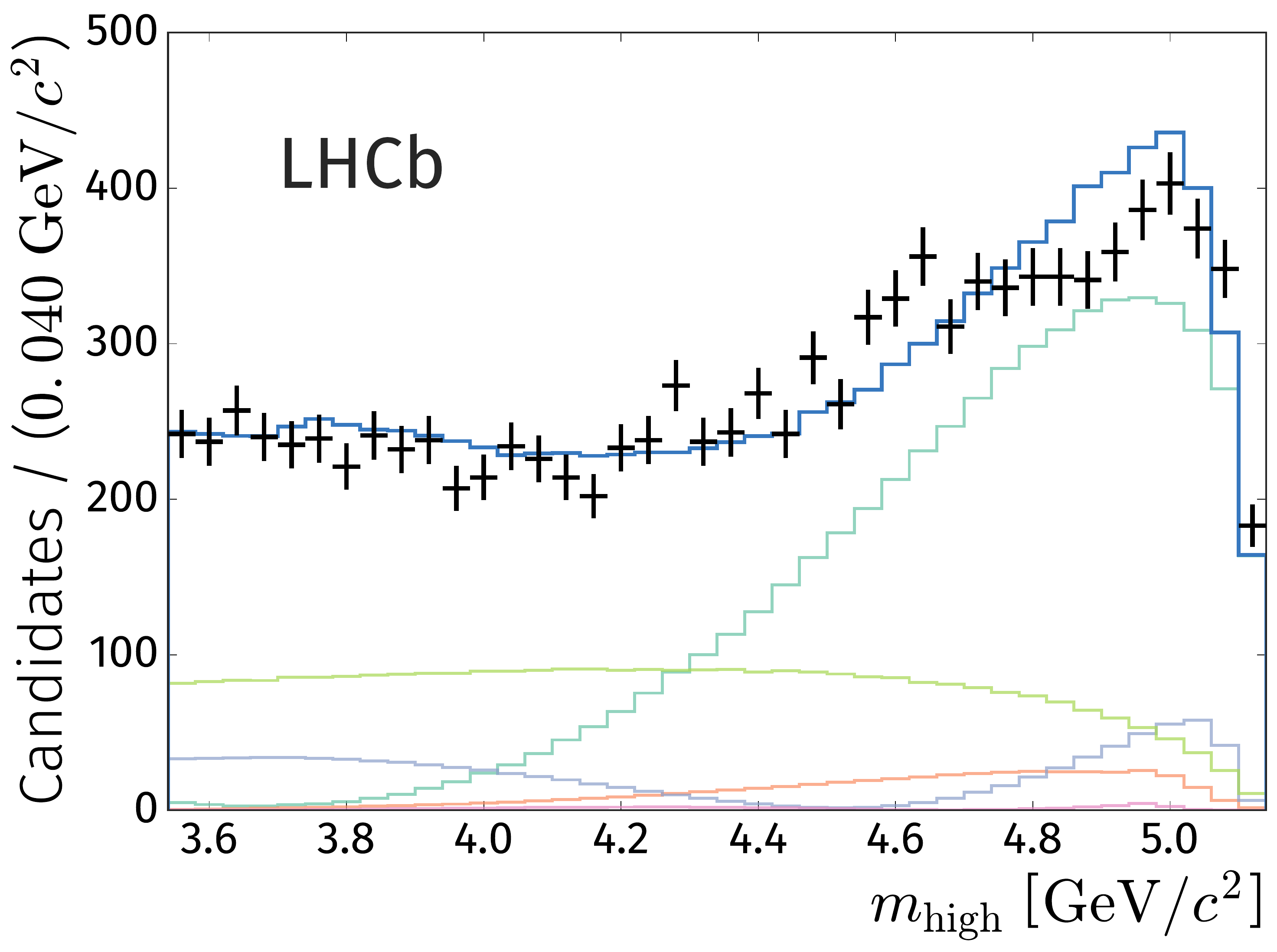}
\put(-405,27){(a)}
\put(-180,27){(b)}

\includegraphics[width=0.49\linewidth]{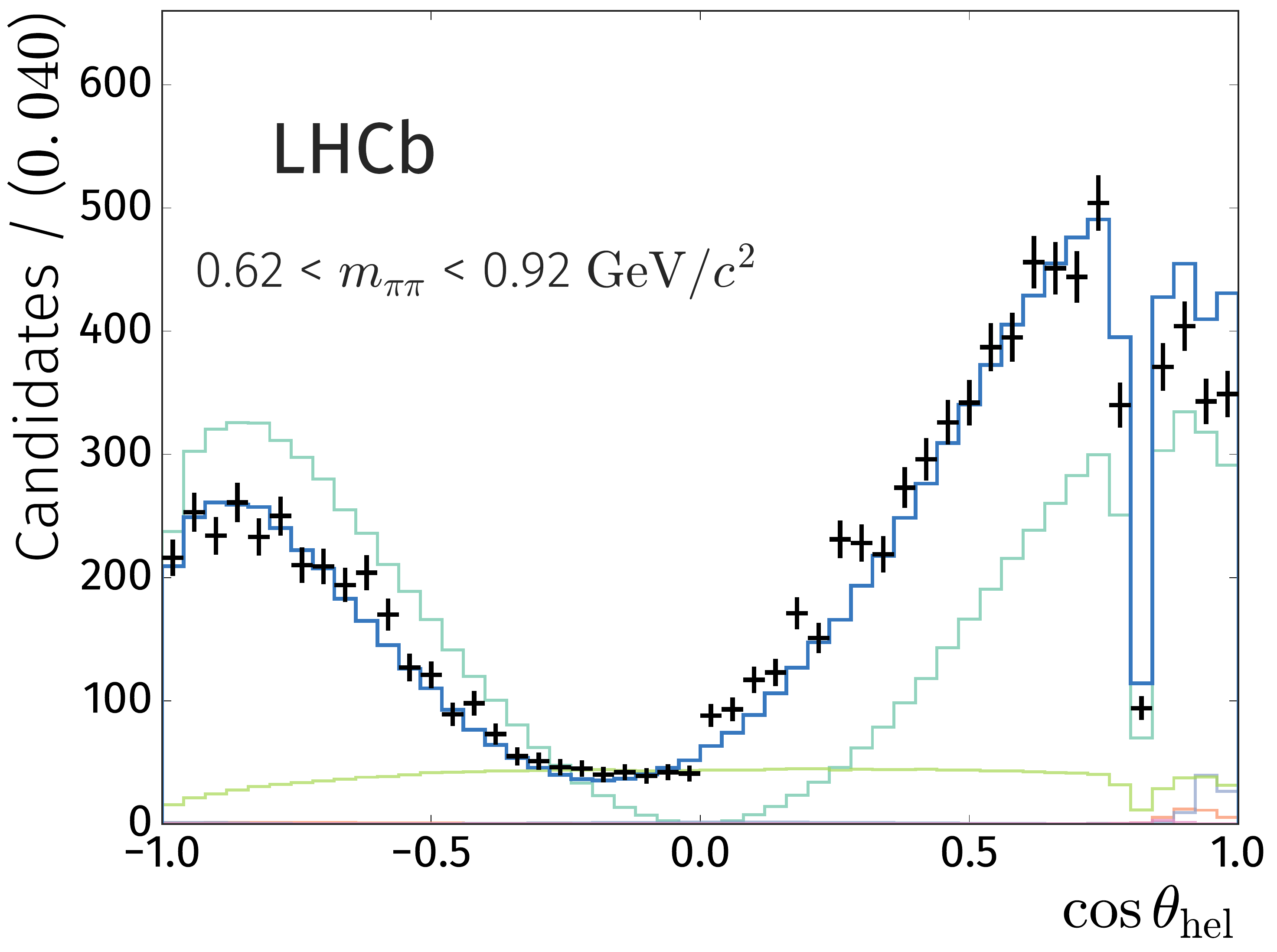}
\includegraphics[width=0.49\linewidth]{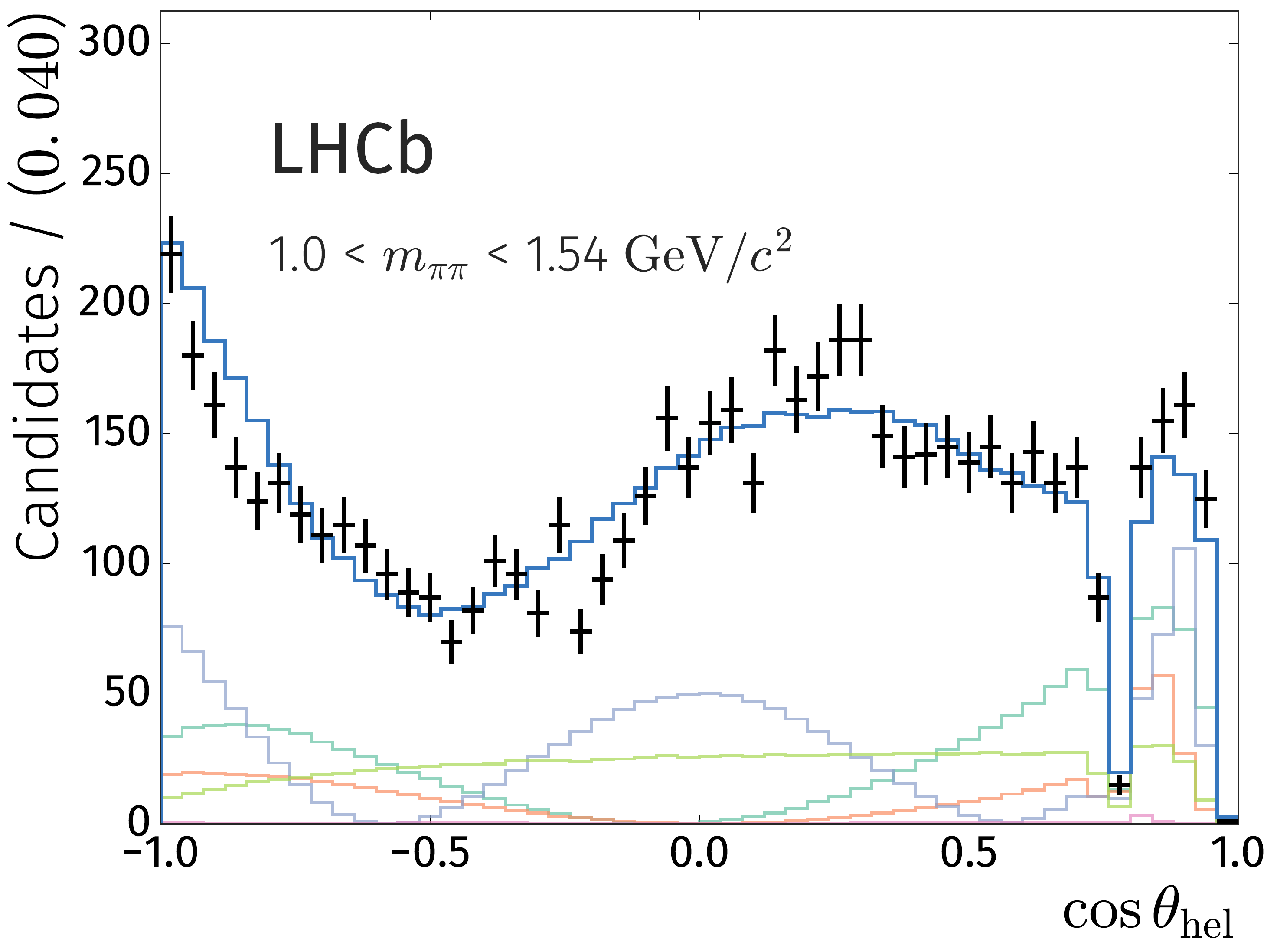}
\put(-405,30){(c)}
\put(-180,30){(d)}

\caption{Fit projections on $m_{\rm high}$ of the result with the isobar S-wave model (a)~in the full $m_{\rm high}$ range, (b)~in the high $m_{\rm high}$ region, and on $\cos\theta_{\rm hel}$ (c)~in the $\rho(770)^0$ region and (d)~in the $f_2(1270)$ region. The thick blue curve represents the total model, and the coloured curves represent the contributions of individual model components (not including interference effects), as per the legend in Fig.~\ref{fig:projComponentsI3}.}
\label{fig:projComponentsI2}
\end{figure}

\begin{figure}[tbh]
\centering
\includegraphics[width=0.49\linewidth]{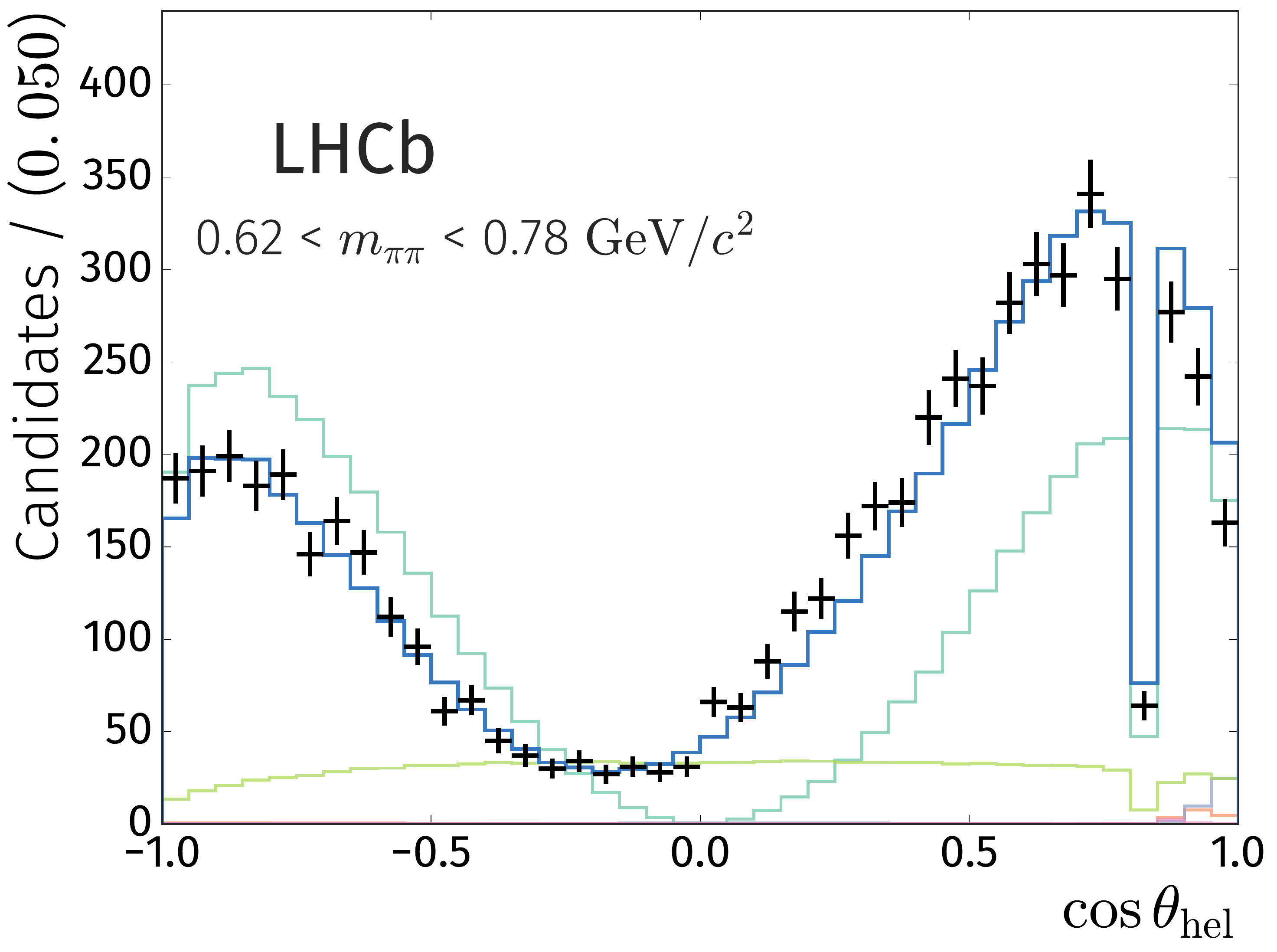}
\includegraphics[width=0.49\linewidth]{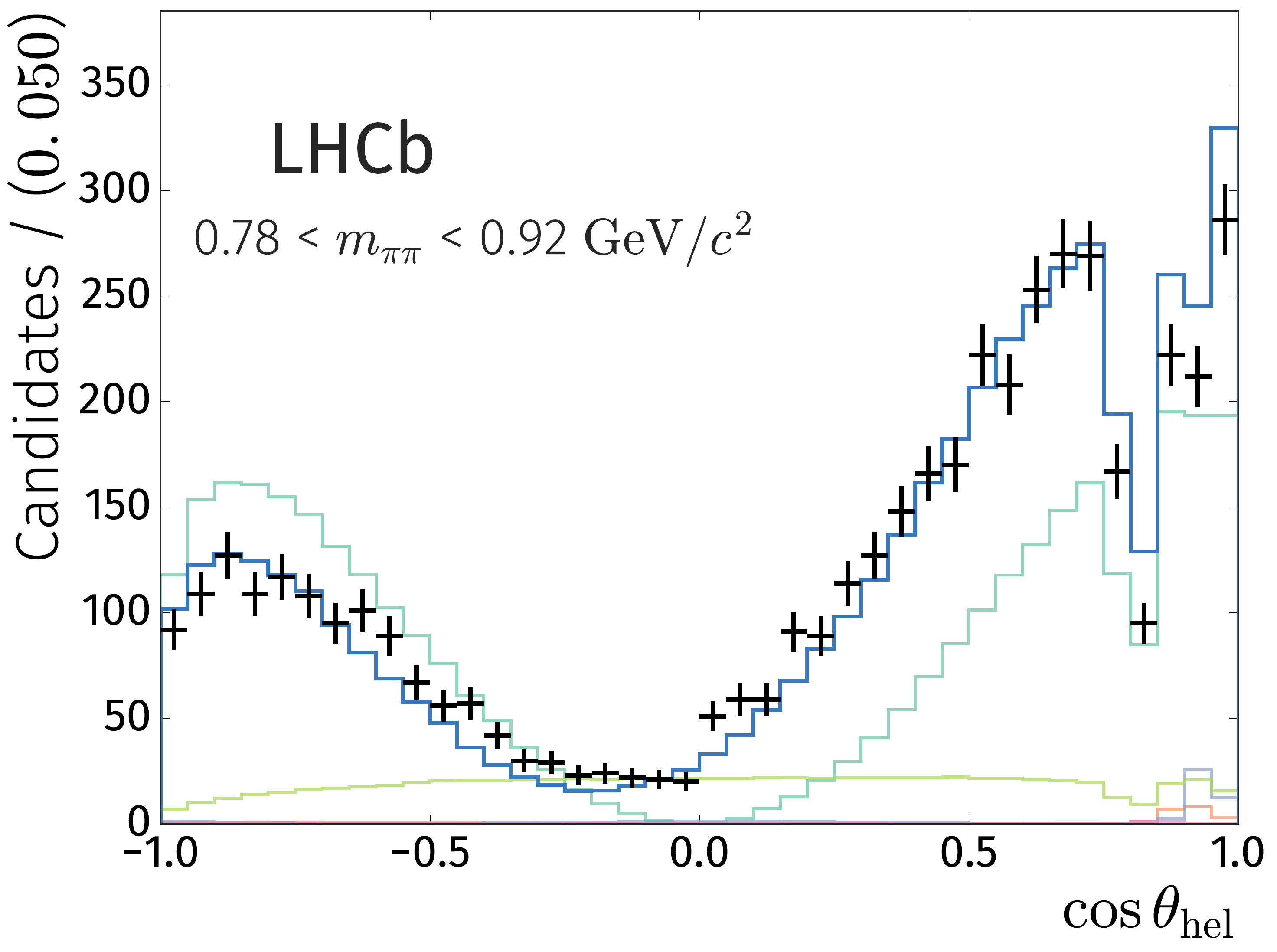}
\put(-405,27){(a)}
\put(-180,27){(b)}

\includegraphics[width=0.49\linewidth]{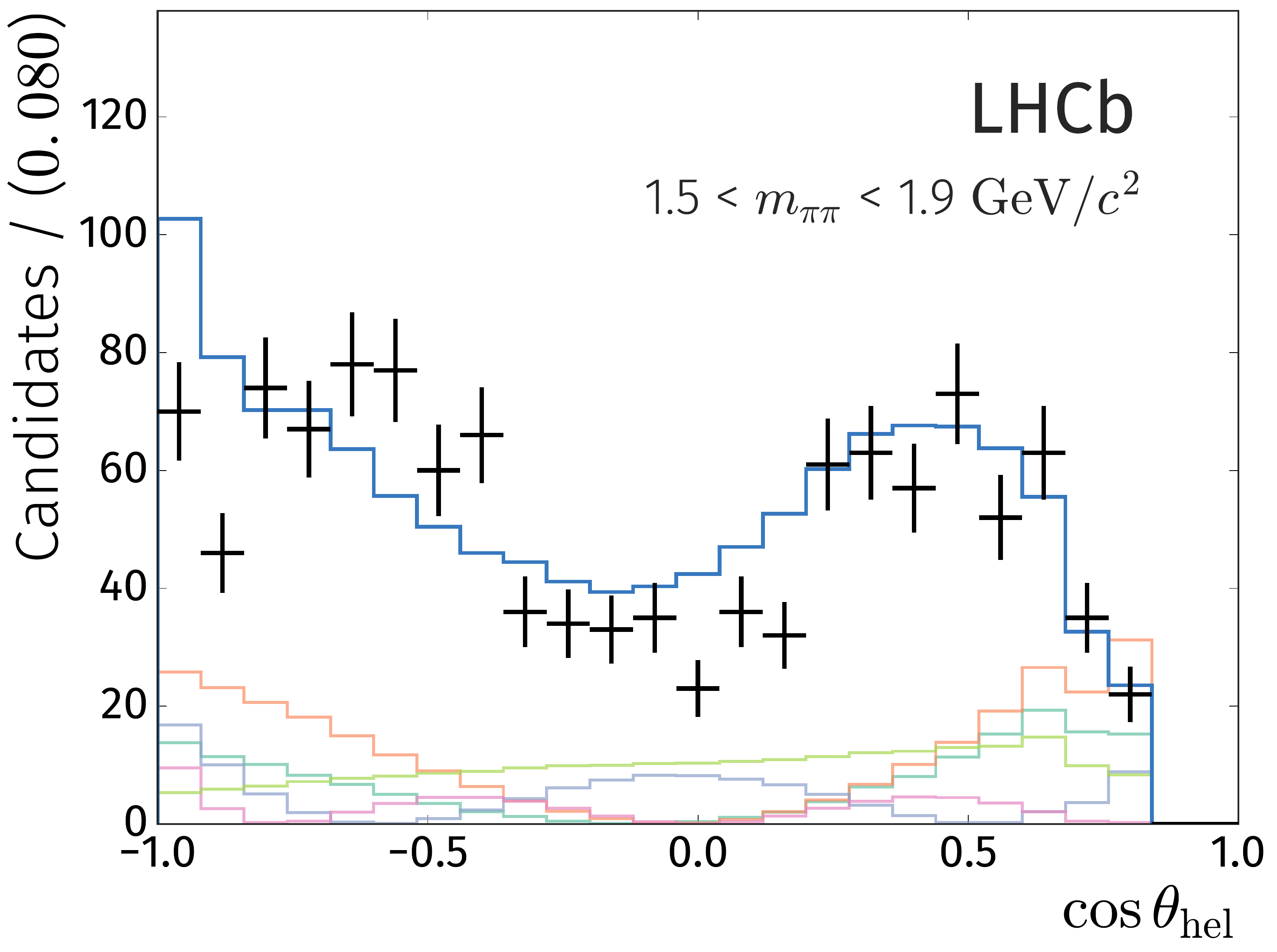}
\includegraphics[width=0.49\linewidth]{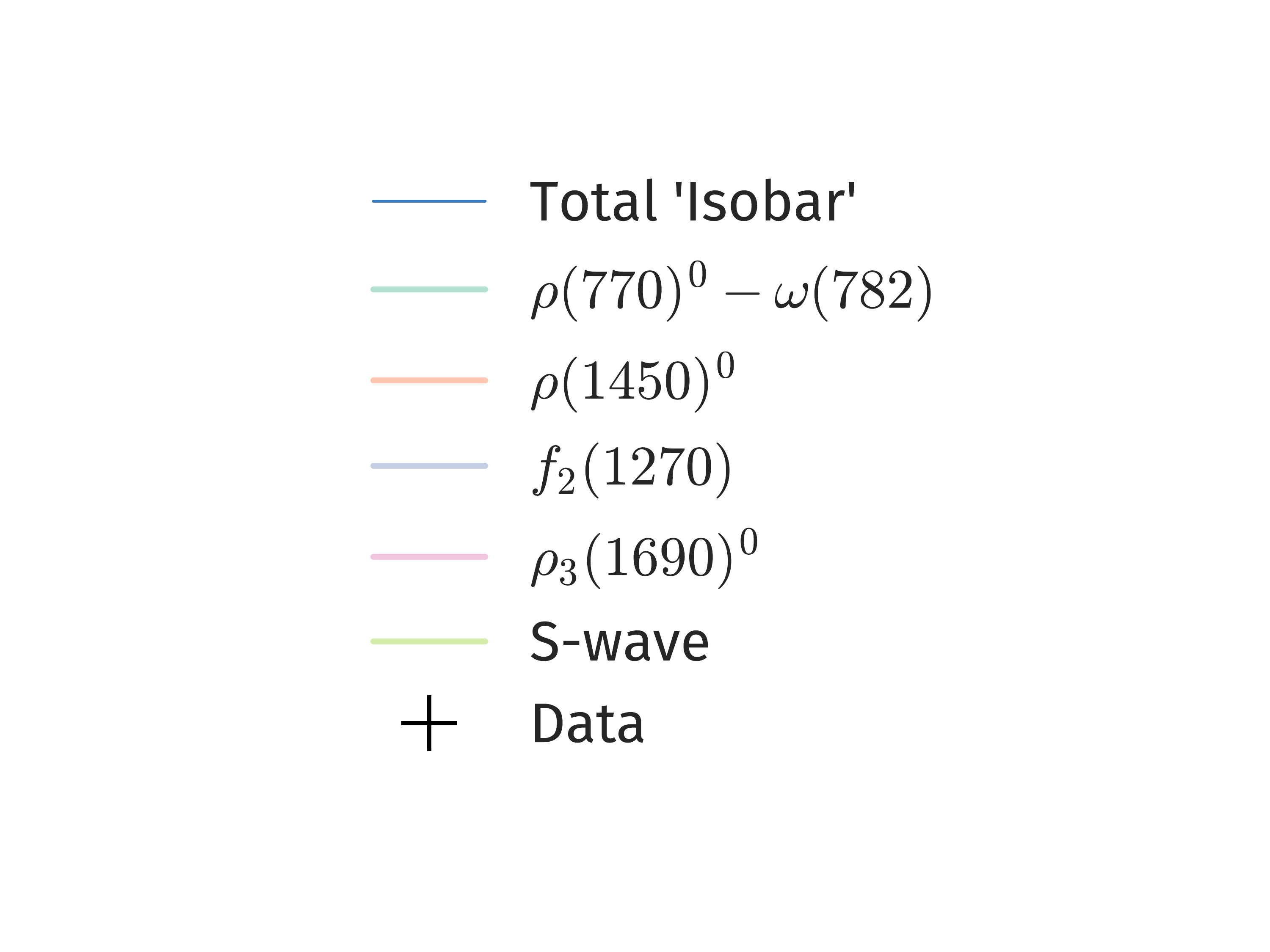}
\put(-405,30){(c)}

\caption{Fit projections on $\cos\theta_{\rm hel}$ of the result with the isobar S-wave model in the region (a)~below and (b)~above the $\rho(770)^0$ mass, and (c)~in the $\rho_3(1690)^0$ region. The thick blue curve represents the total model, and the coloured curves represent the contributions of individual model components (not including interference effects), as per the legend.}
\label{fig:projComponentsI3}
\end{figure}

Various projections of the data and the result of fit with the isobar description of the S-wave are shown in Figs.~\ref{fig:projComponentsI1}--\ref{fig:projComponentsI3}. The colour legend for each contribution is given in the final figure.

\section{K-matrix model component projections}
\label{sec:kMatrixProj}

\begin{figure}[tbh]
\centering
\includegraphics[width=0.49\linewidth]{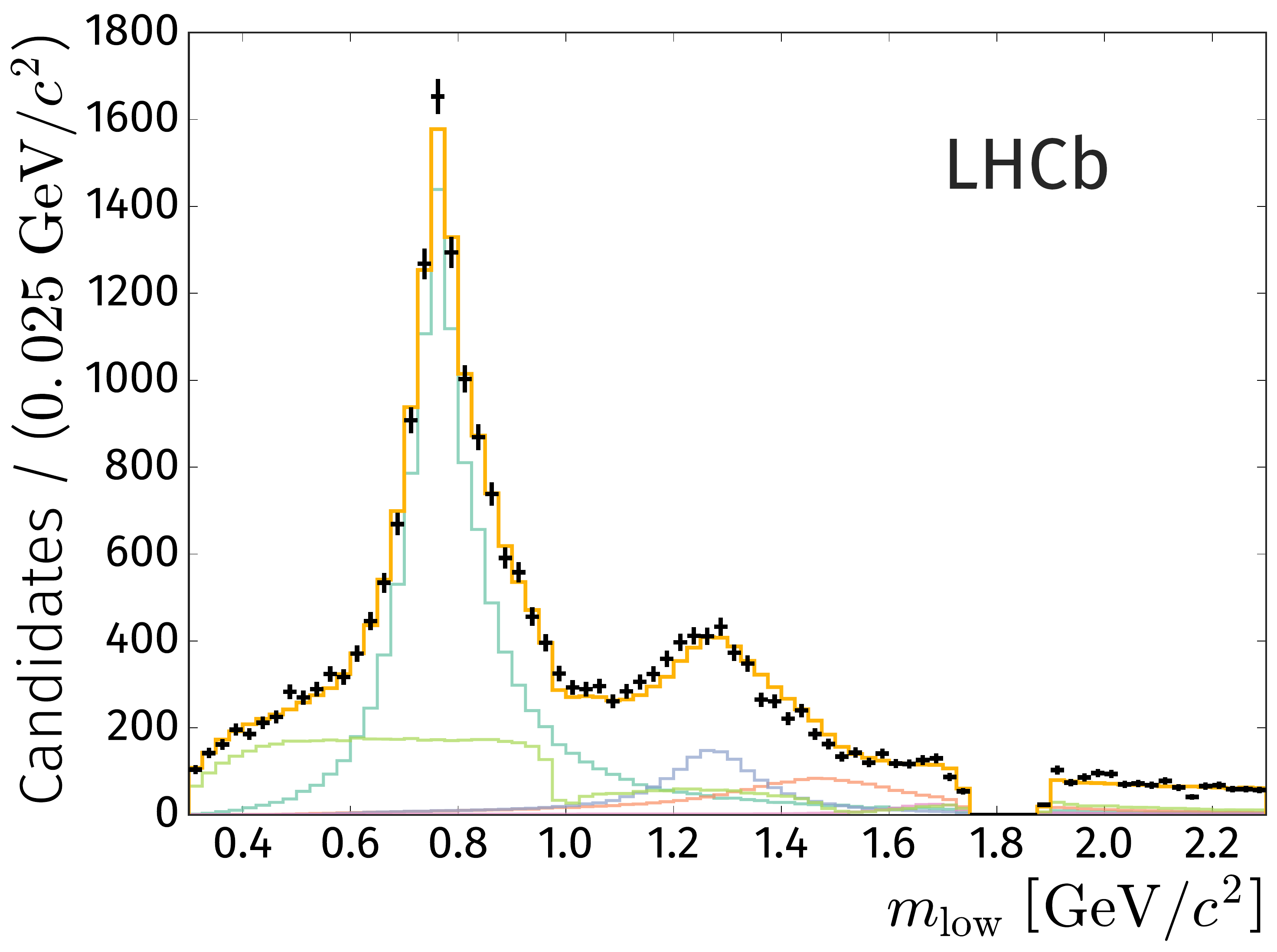}
\includegraphics[width=0.49\linewidth]{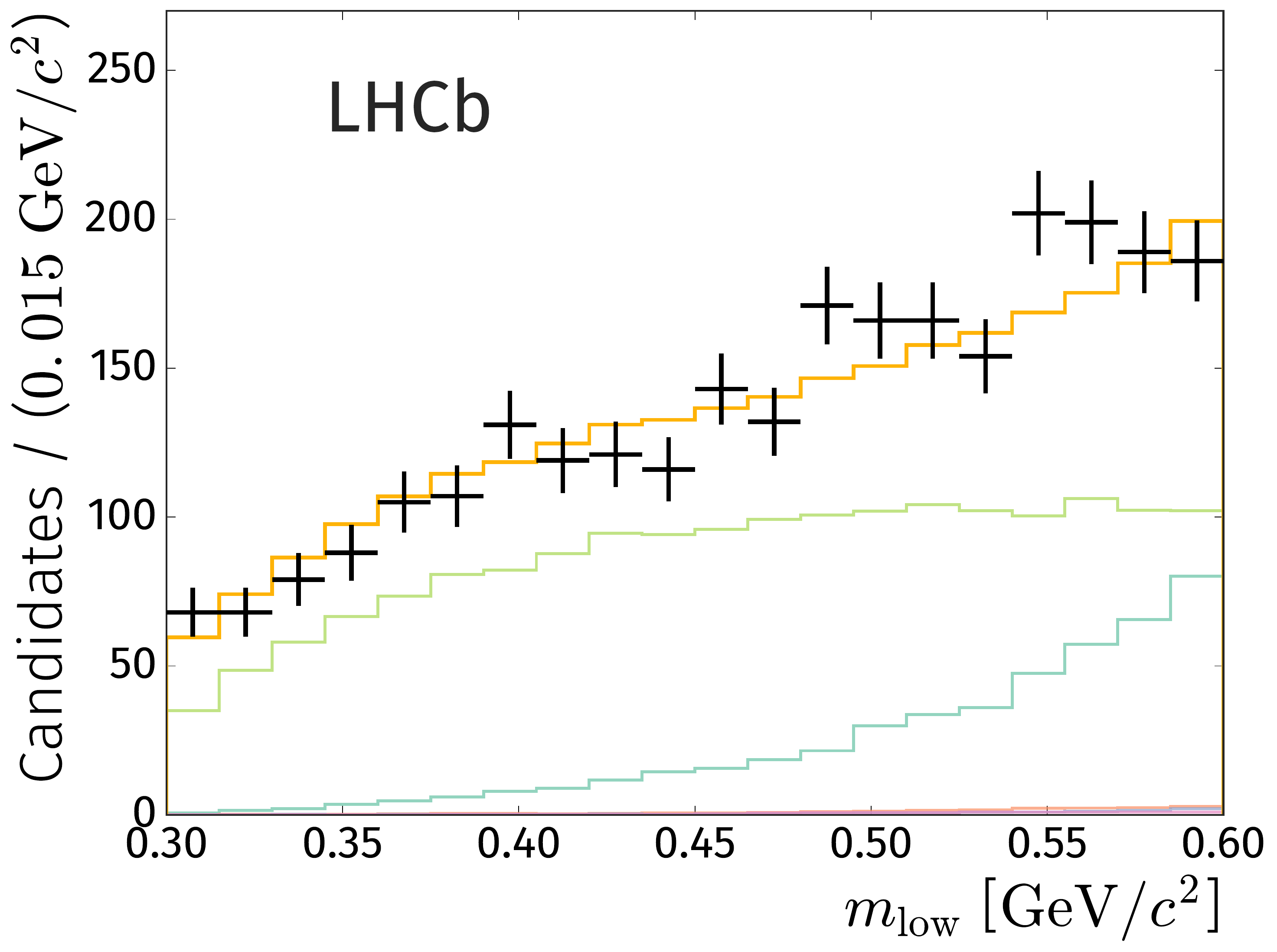}
\put(-405,27){(a)}
\put(-180,27){(b)}

\includegraphics[width=0.49\linewidth]{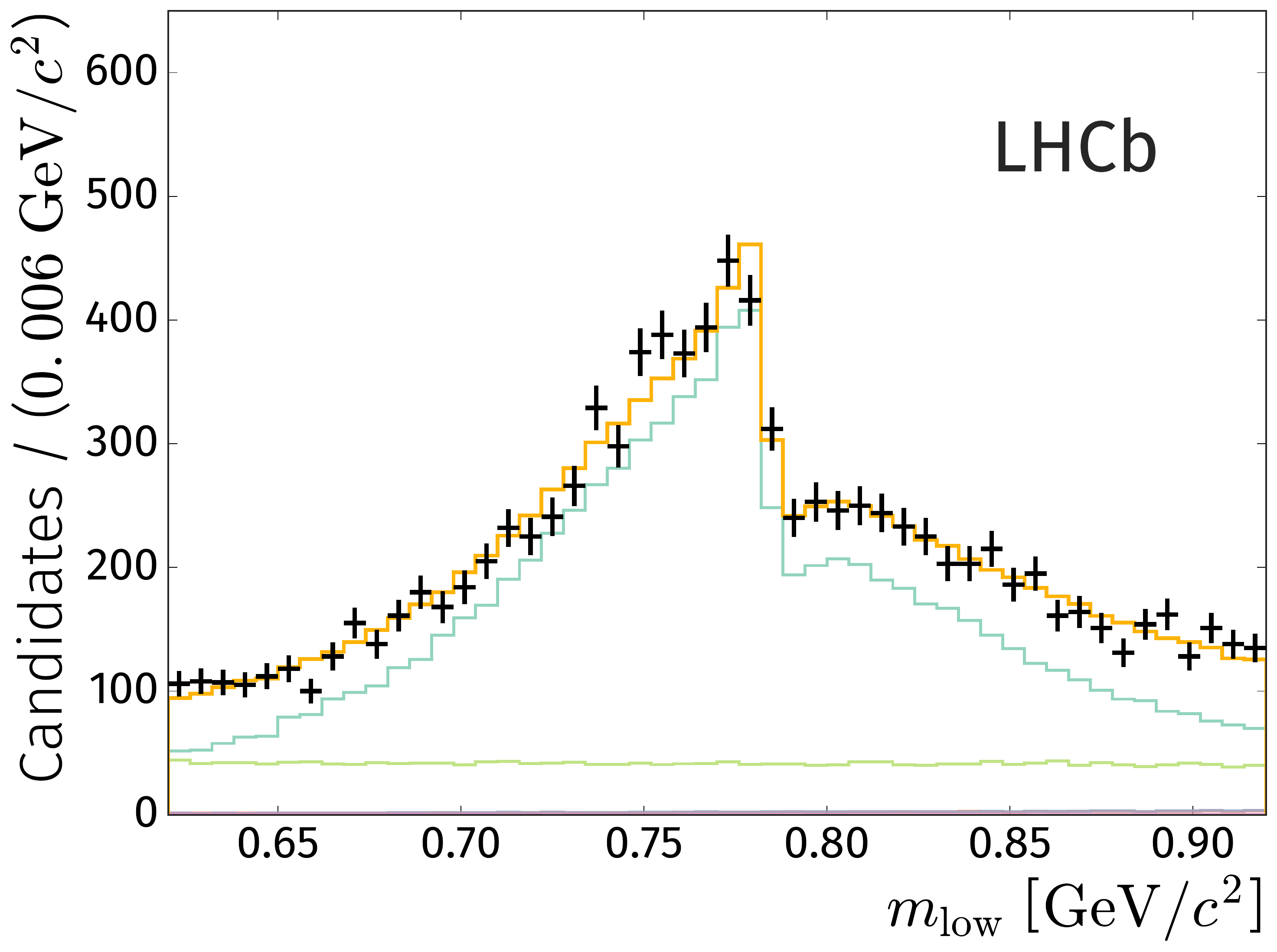}
\includegraphics[width=0.49\linewidth]{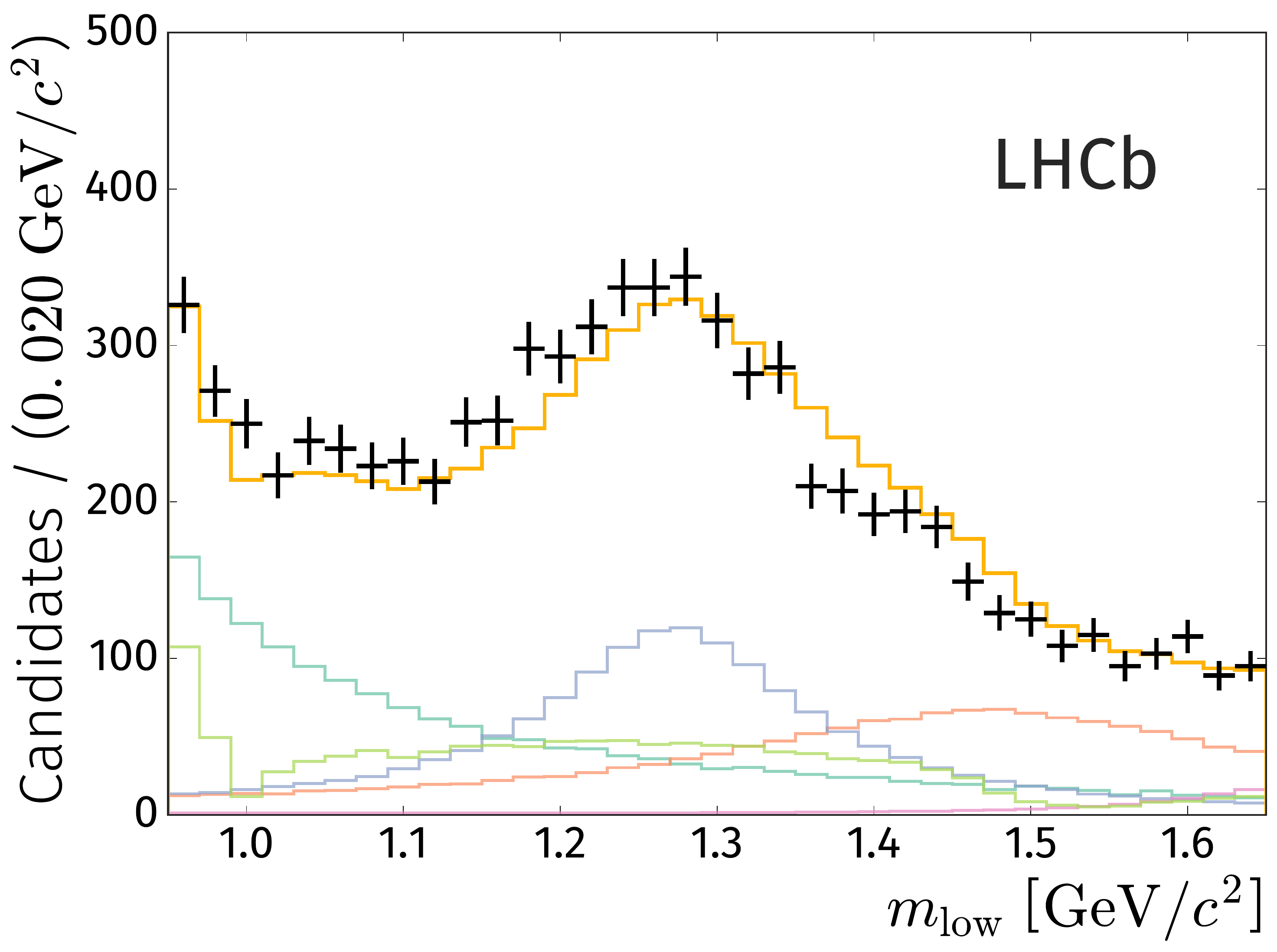}
\put(-405,30){(c)}
\put(-180,30){(d)}

\caption{Fit projections on $m_{\rm low}$ of the result with the K-matrix S-wave model (a)~in the low $m_{\rm low}$ region, (b)~below the $\rho(770)^0$ region, (c)~in the $\rho(770)^0$ region, and (d)~in the $f_2(1270)$ region. The thick amber curve represents the total model, and the coloured curves represent the contributions of individual model components (not including interference effects), as per the legend in Fig.~\ref{fig:projComponentsKM3}.}
\label{fig:projComponentsKM1}
\end{figure}

\begin{figure}[tbh]
\centering
\includegraphics[width=0.49\linewidth]{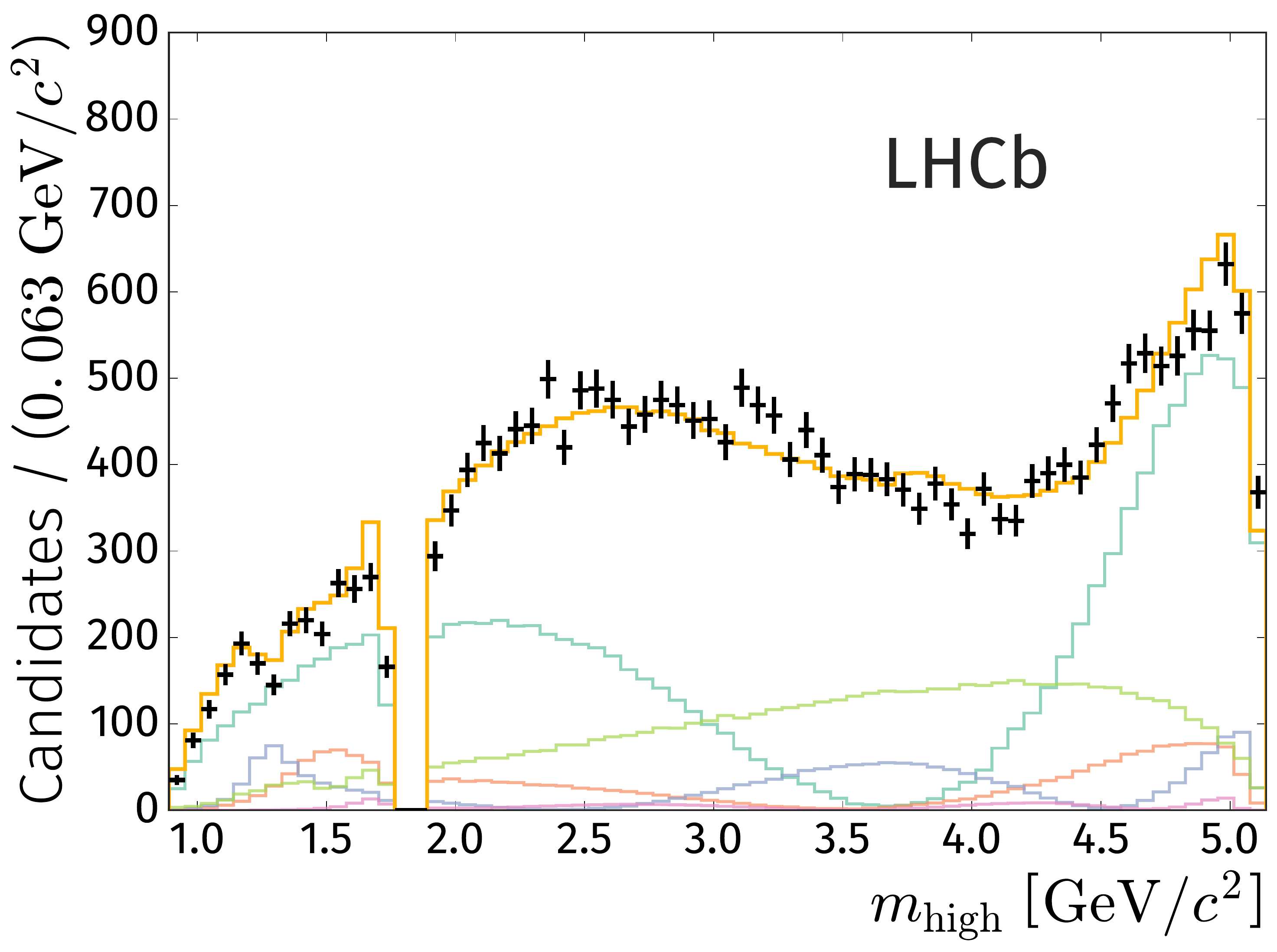}
\includegraphics[width=0.49\linewidth]{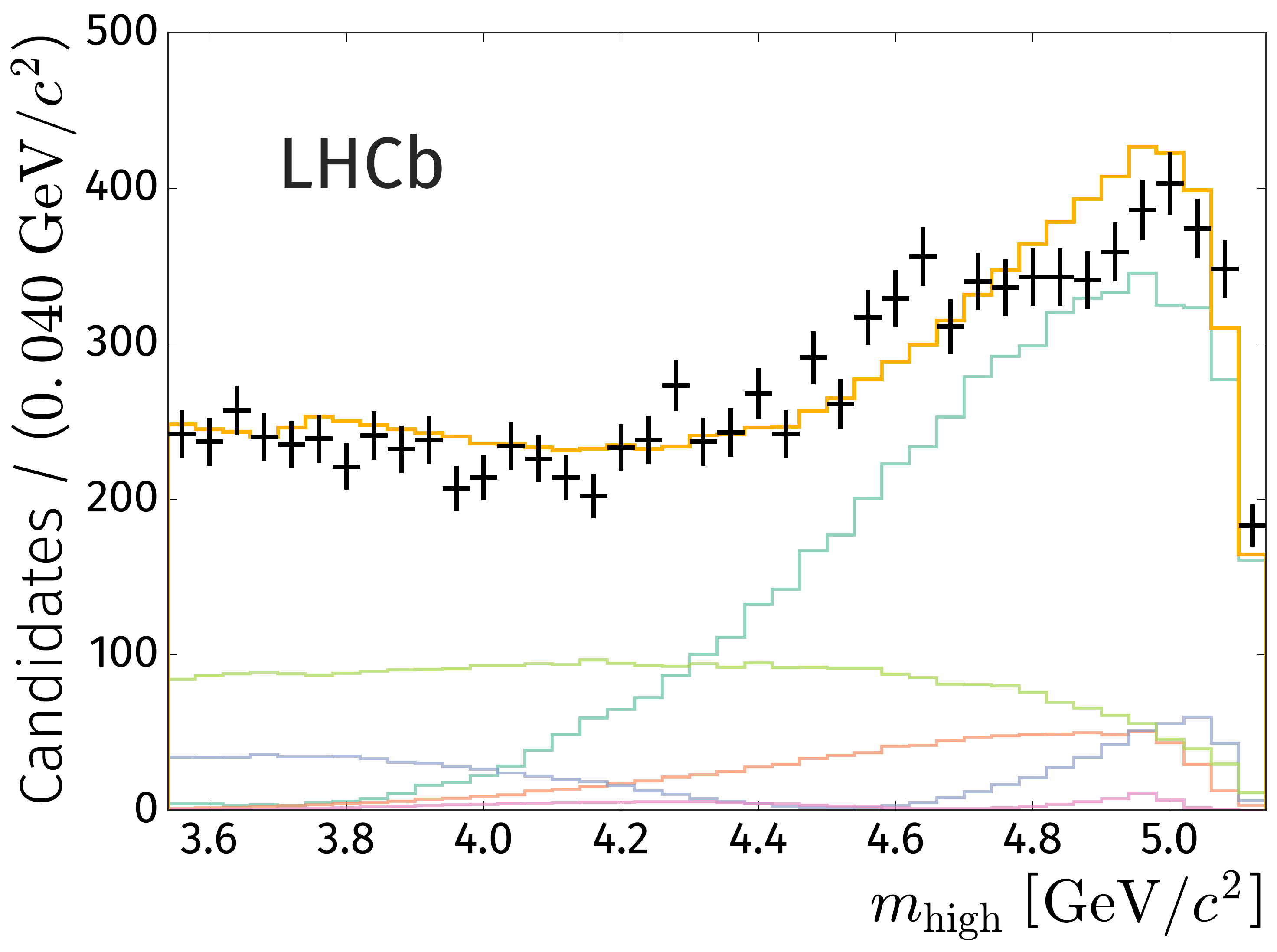}
\put(-405,27){(a)}
\put(-180,27){(b)}

\includegraphics[width=0.49\linewidth]{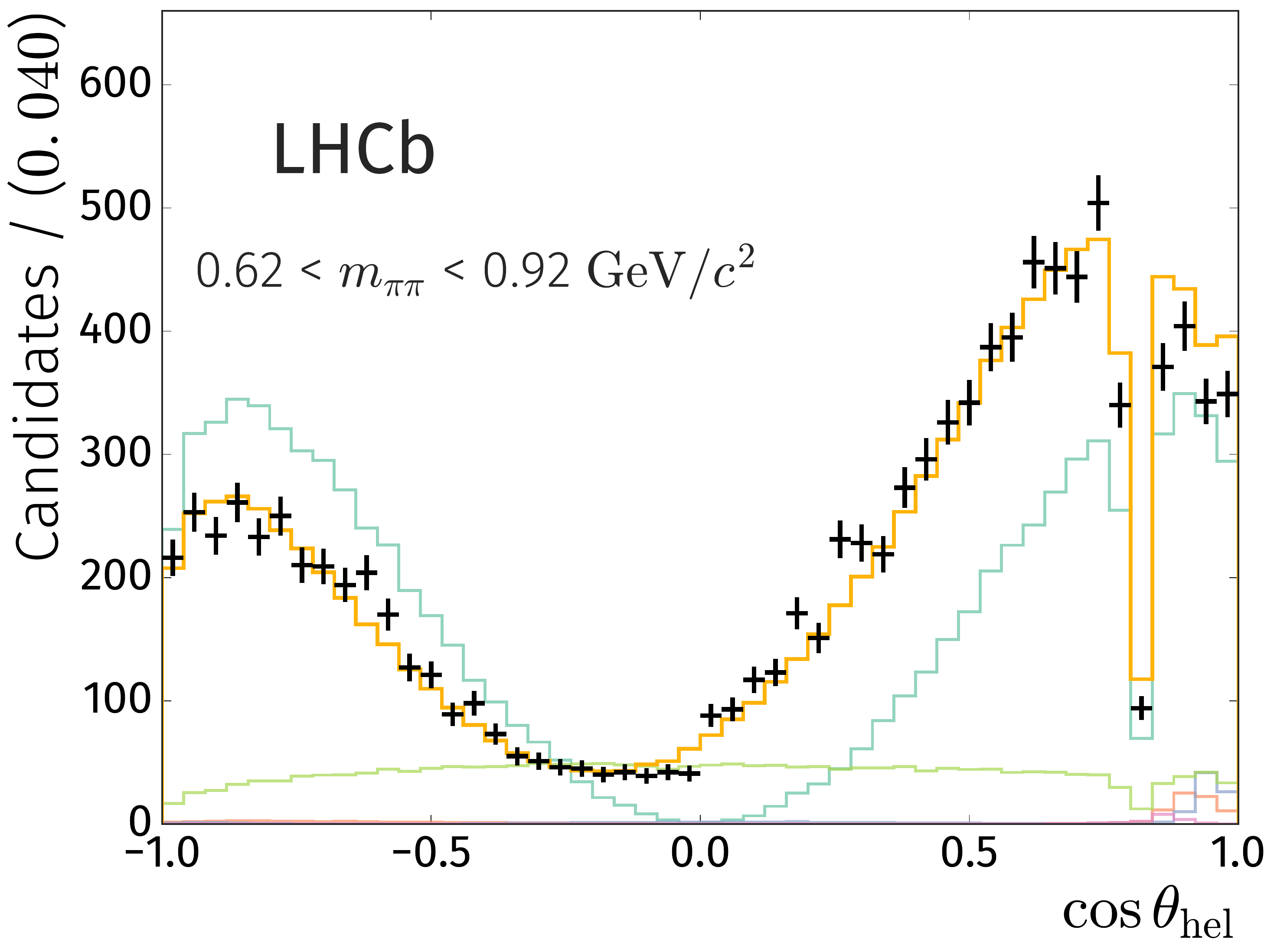}
\includegraphics[width=0.49\linewidth]{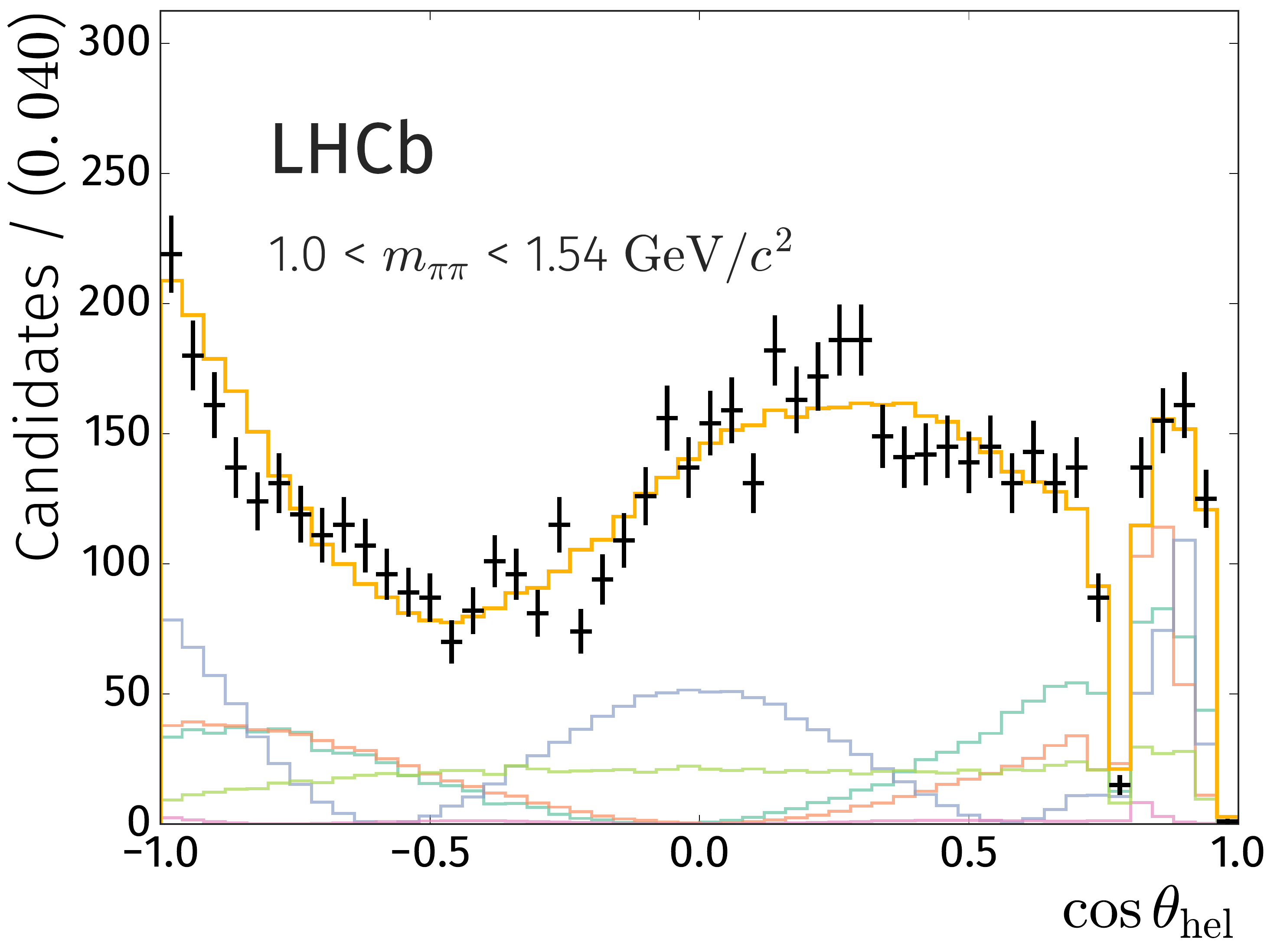}
\put(-405,30){(c)}
\put(-180,30){(d)}

\caption{Fit projections on $m_{\rm high}$ of the result with the K-matrix S-wave model (a)~in the full $m_{\rm high}$ range, (b)~in the high $m_{\rm high}$ region, and on $\cos\theta_{\rm hel}$ (c)~in the $\rho(770)^0$ region), and (d)~in the $f_2(1270)$ region. The thick amber curve represents the total model, and the coloured curves represent the contributions of individual model components (not including interference effects), as per the legend in Fig.~\ref{fig:projComponentsKM3}.}
\label{fig:projComponentsKM2}
\end{figure}

\begin{figure}[tbh]
\centering
\includegraphics[width=0.49\linewidth]{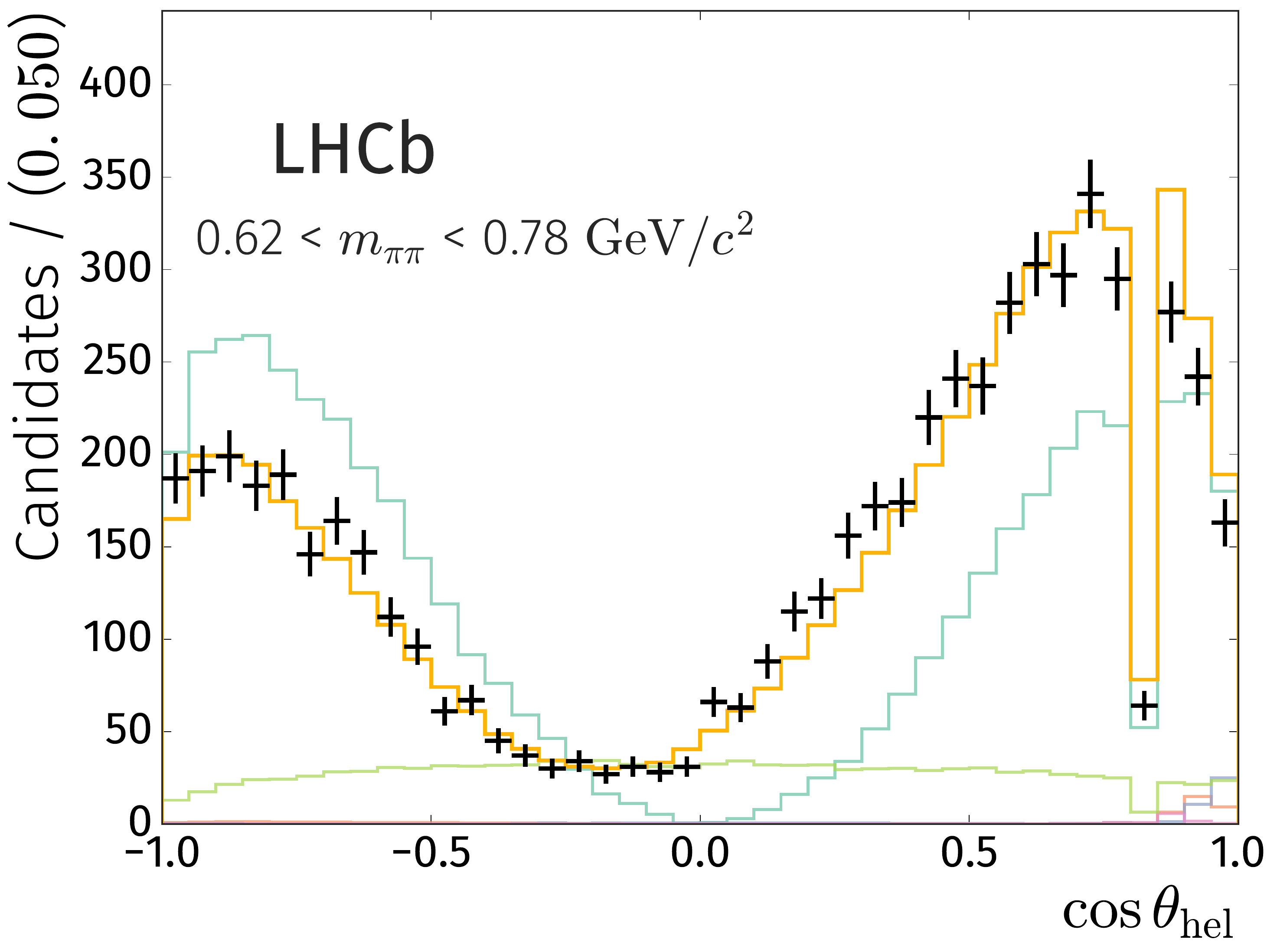}
\includegraphics[width=0.49\linewidth]{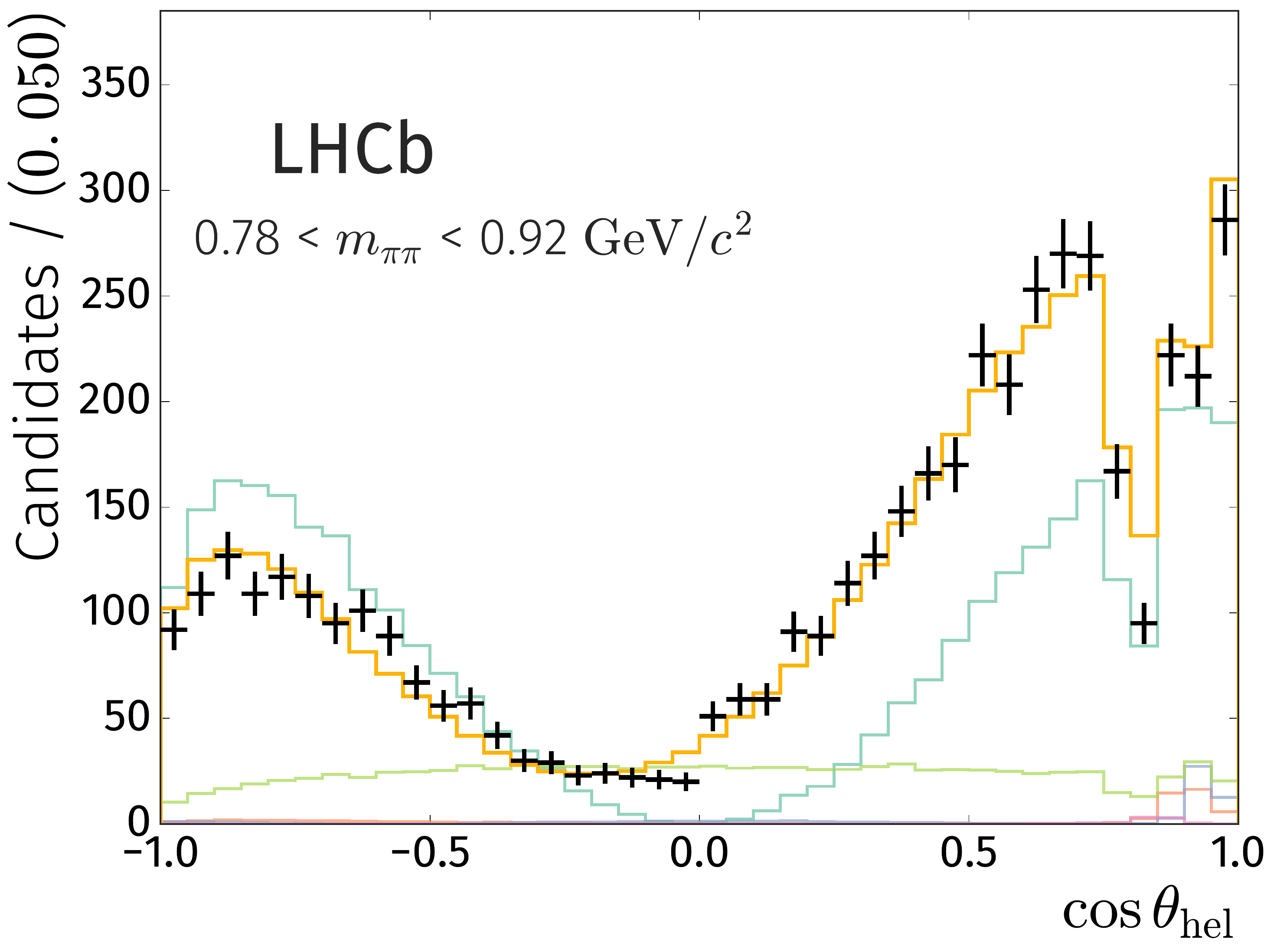}
\put(-405,27){(a)}
\put(-180,27){(b)}

\includegraphics[width=0.49\linewidth]{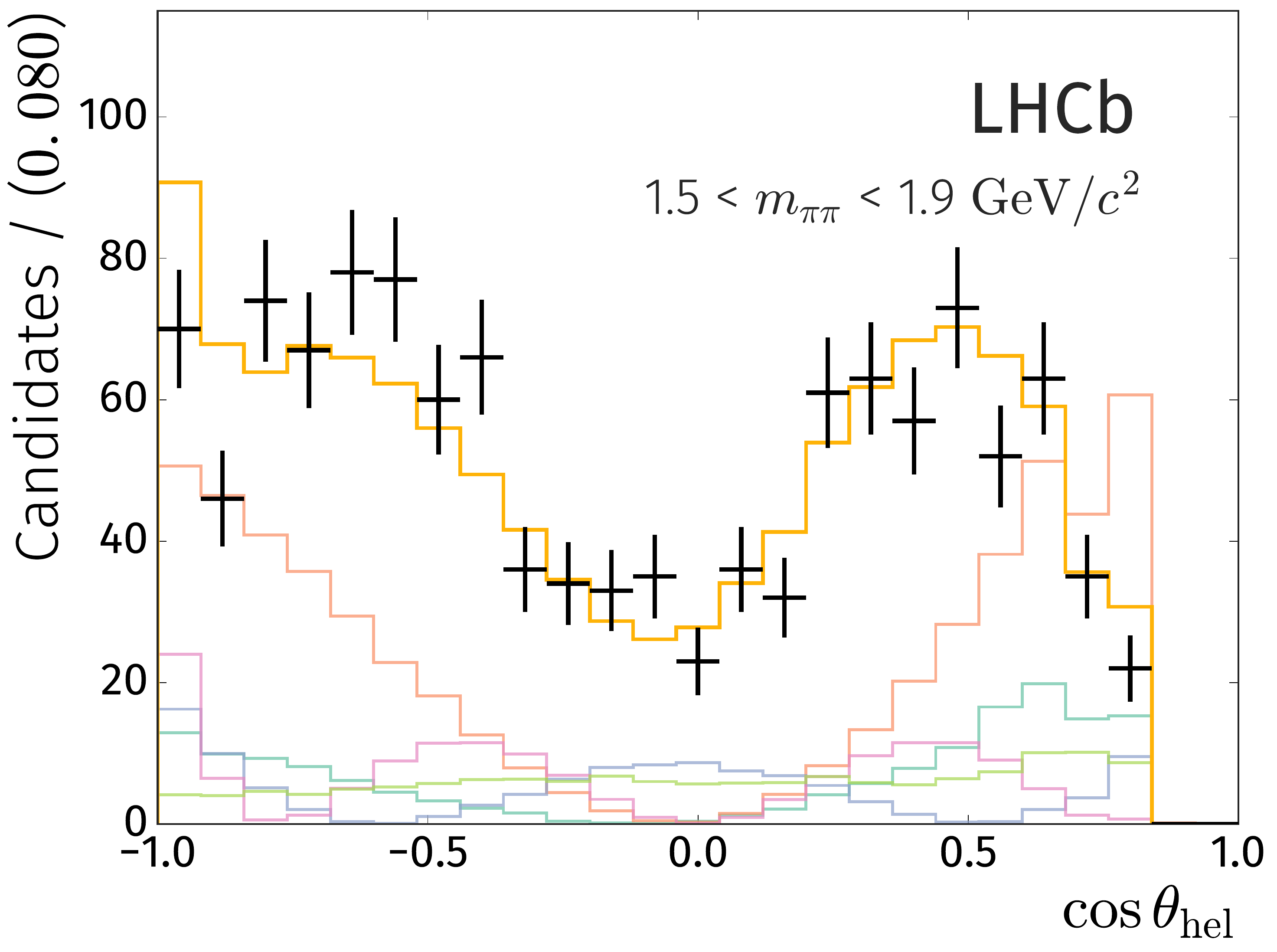}
\includegraphics[width=0.49\linewidth]{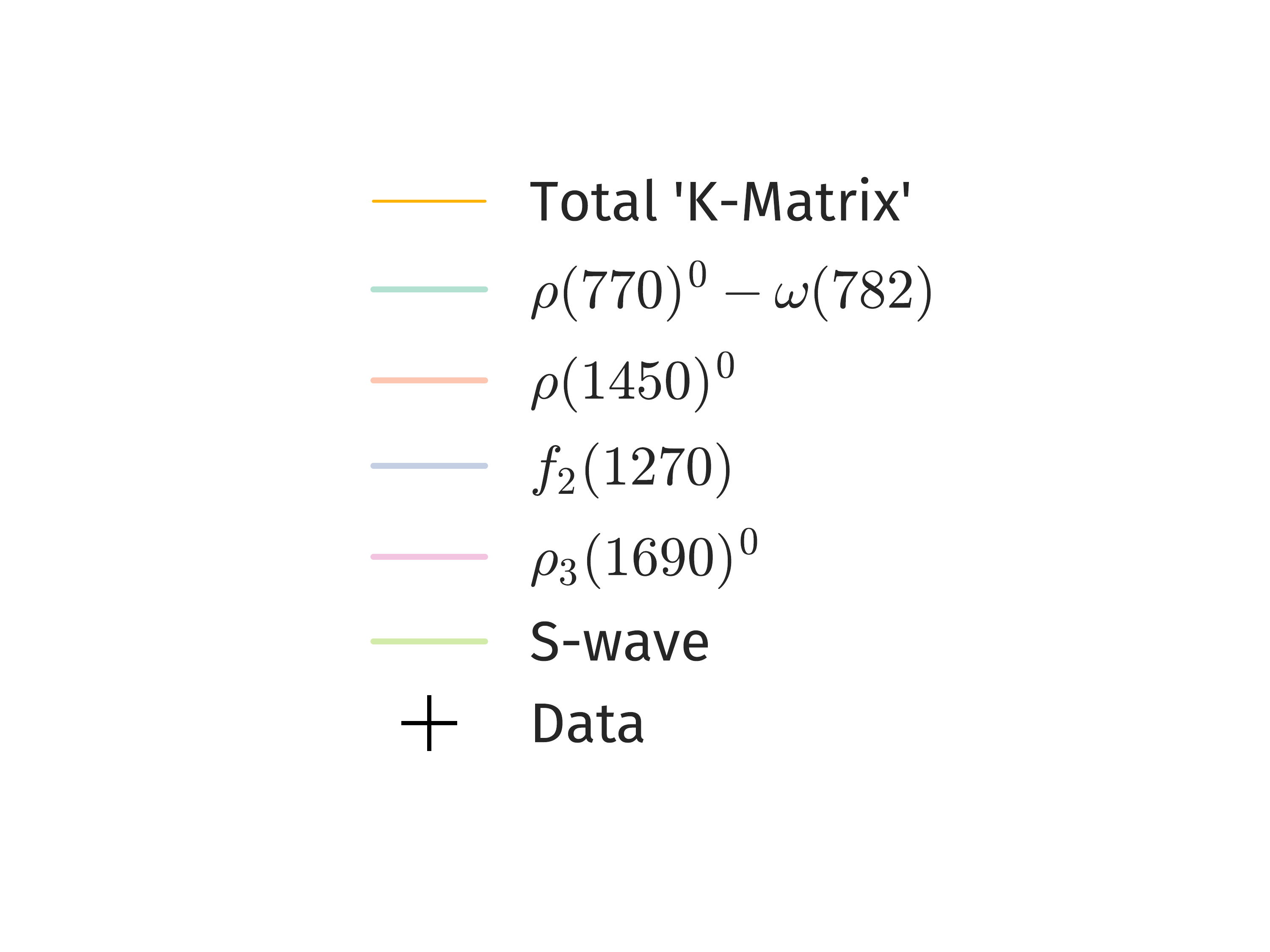}
\put(-405,30){(c)}

\caption{Fit projections on $\cos\theta_{\rm hel}$ of the result with the K-matrix S-wave model in the region (a)~below and (b)~above the $\rho(770)^0$ mass, and (c)~in the $\rho_3(1690)^0$ region. The thick amber curve represents the total model, and the coloured curves represent the contributions of individual model components (not including interference effects), as per the legend.}
\label{fig:projComponentsKM3}
\end{figure}

Various projections of the data and the result of fit with the K-matrix description of the S-wave are shown in Figs.~\ref{fig:projComponentsKM1}--\ref{fig:projComponentsKM3}. The colour legend for each contribution is given in the final figure.

\clearpage

\section{QMI model component projections}
\label{sec:qmiProj}

\begin{figure}[tbh]
\centering
\includegraphics[width=0.49\linewidth]{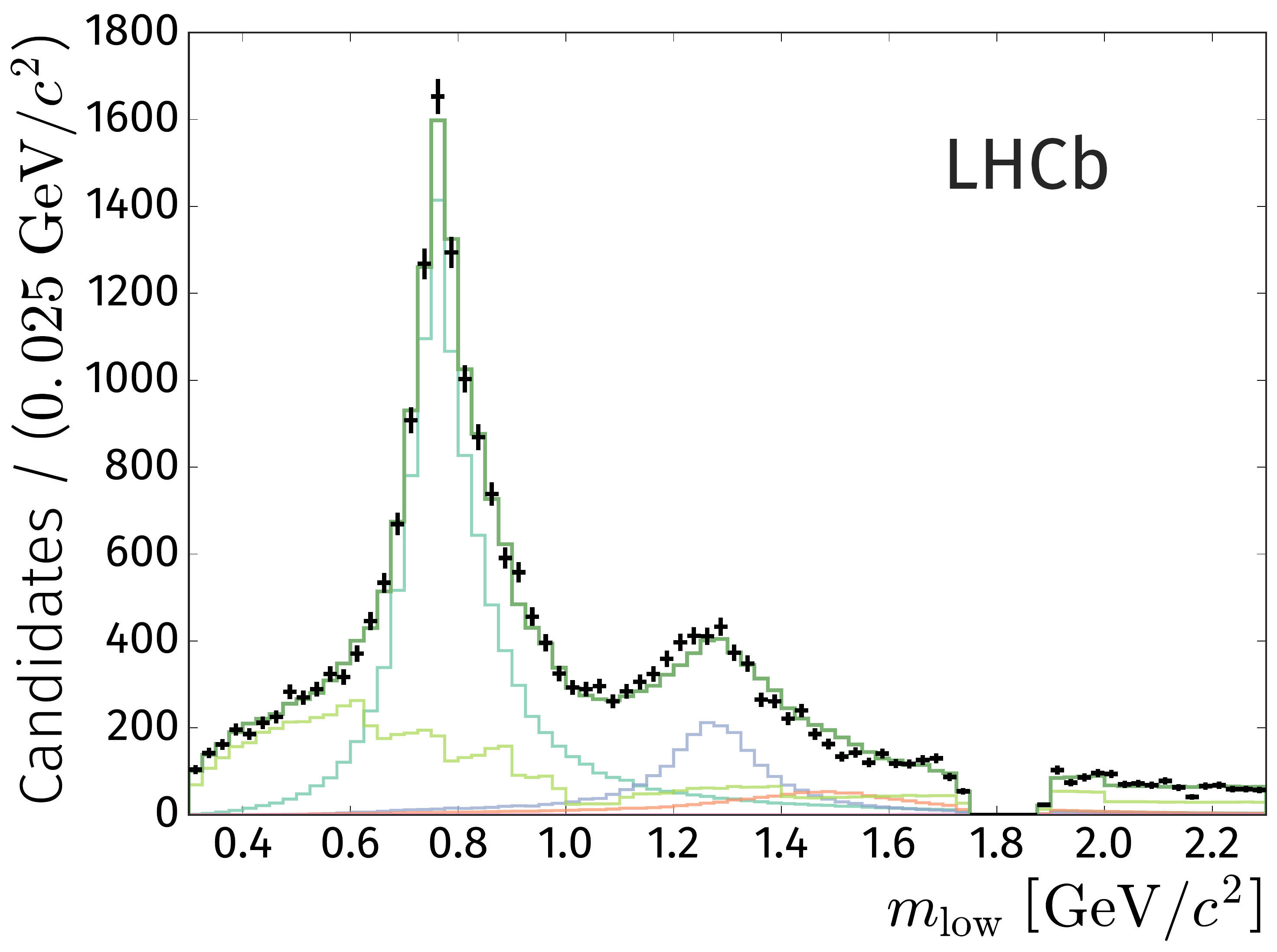}
\includegraphics[width=0.49\linewidth]{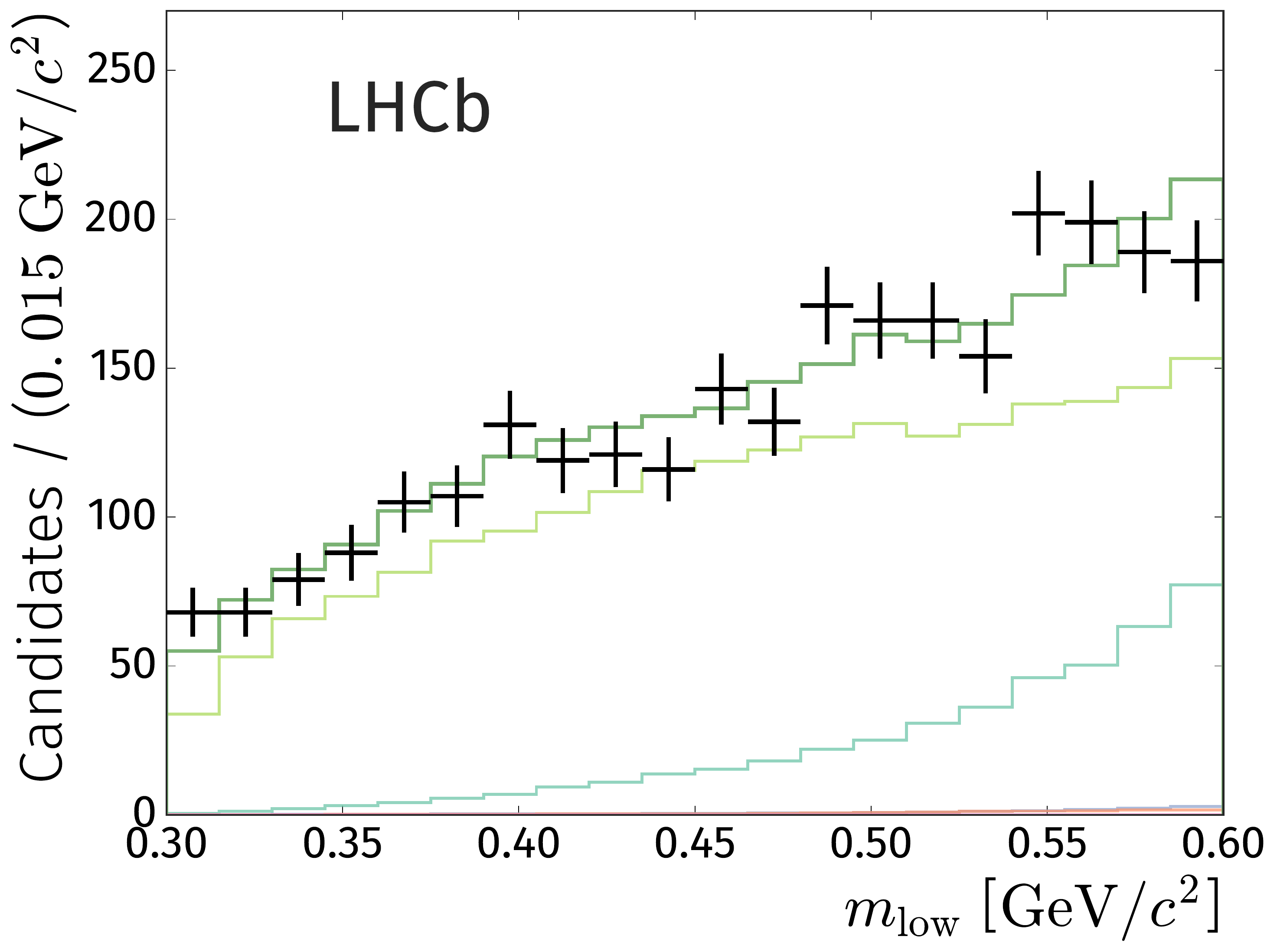}
\put(-405,27){(a)}
\put(-180,27){(b)}

\includegraphics[width=0.49\linewidth]{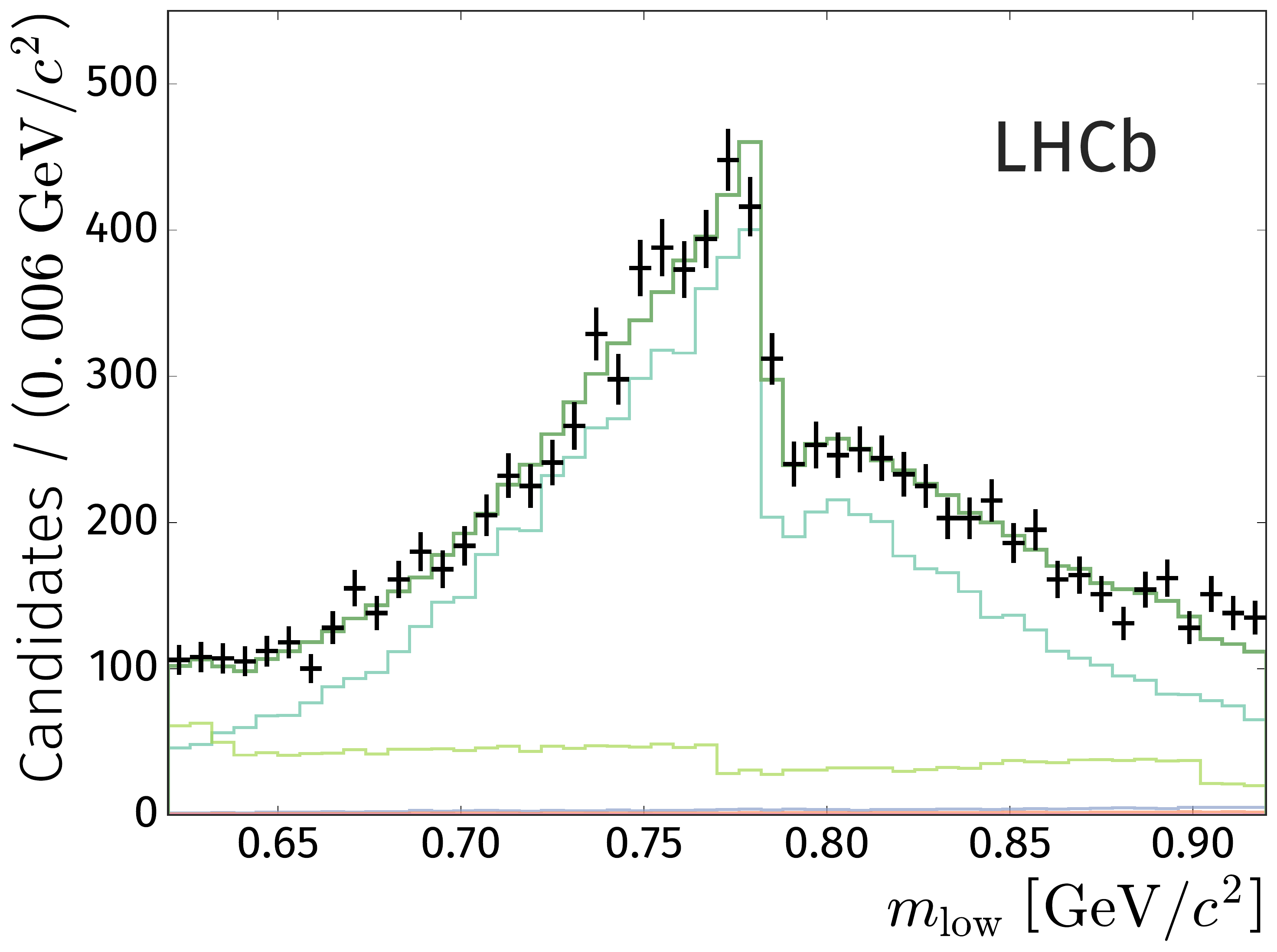}
\includegraphics[width=0.49\linewidth]{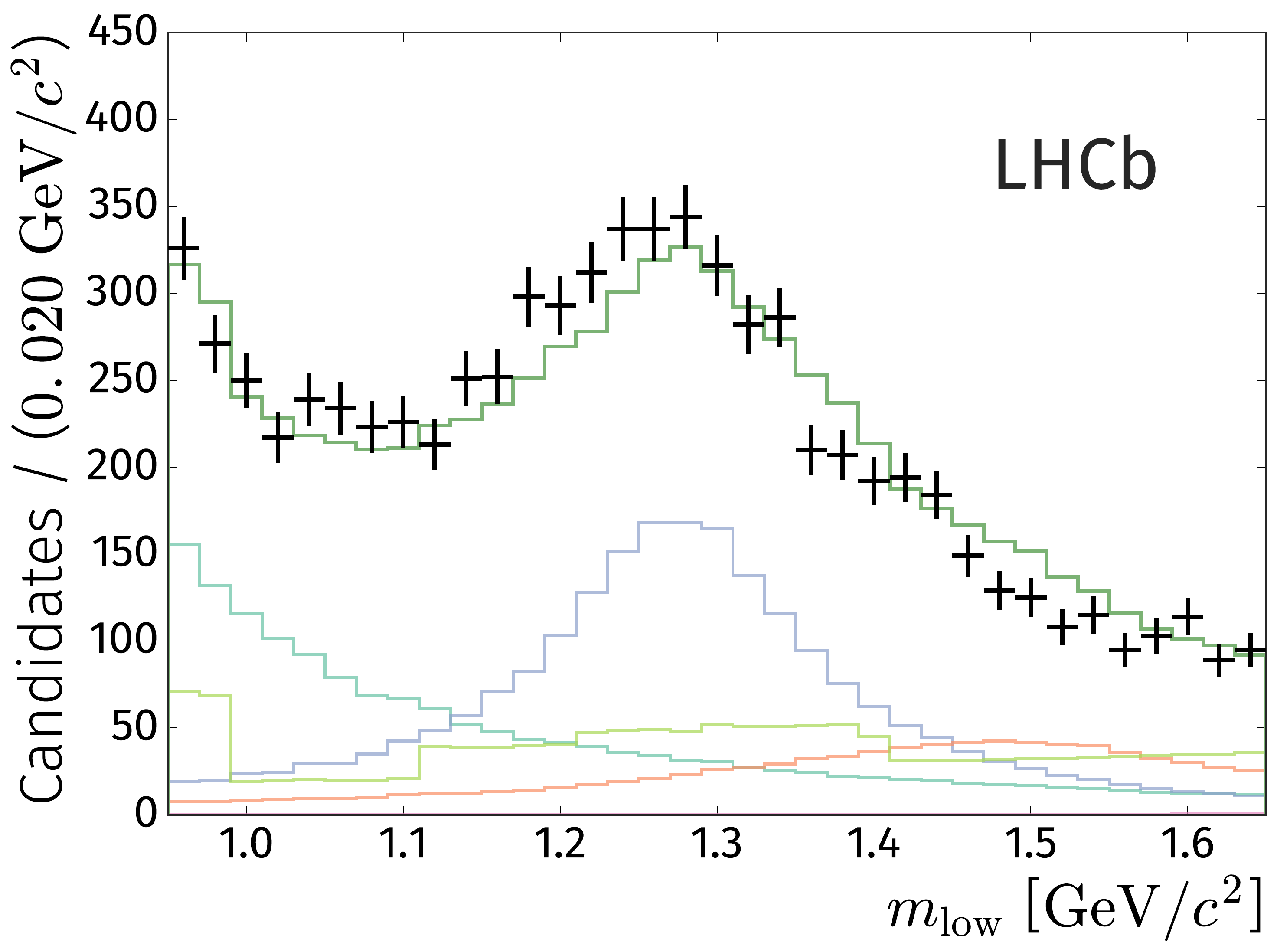}
\put(-405,30){(c)}
\put(-180,30){(d)}

\caption{Fit projections on $m_{\rm low}$ of the result with the QMI S-wave model (a)~in the low $m_{\rm low}$ region, (b)~below the $\rho(770)^0$ region, (c)~in the $\rho(770)^0$ region, and (d)~in the $f_2(1270)$ region. The thick dark-green curve represents the total model, and the coloured curves represent the contributions of individual model components (not including interference effects), as per the legend in Fig.~\ref{fig:projComponentsQMI3}.}
\label{fig:projComponentsQMI1}
\end{figure}

\begin{figure}[tbh]
\centering
\includegraphics[width=0.49\linewidth]{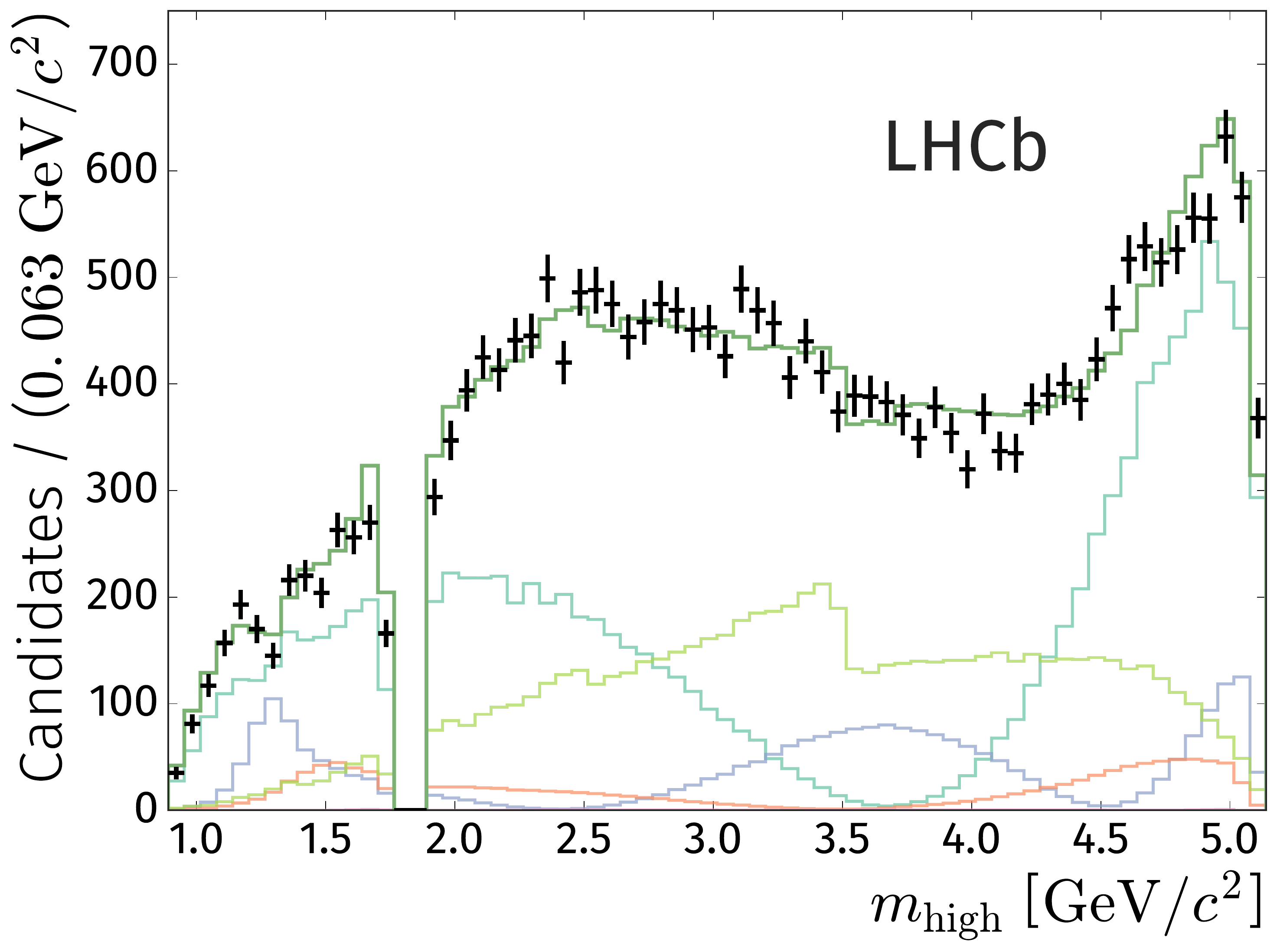}
\includegraphics[width=0.49\linewidth]{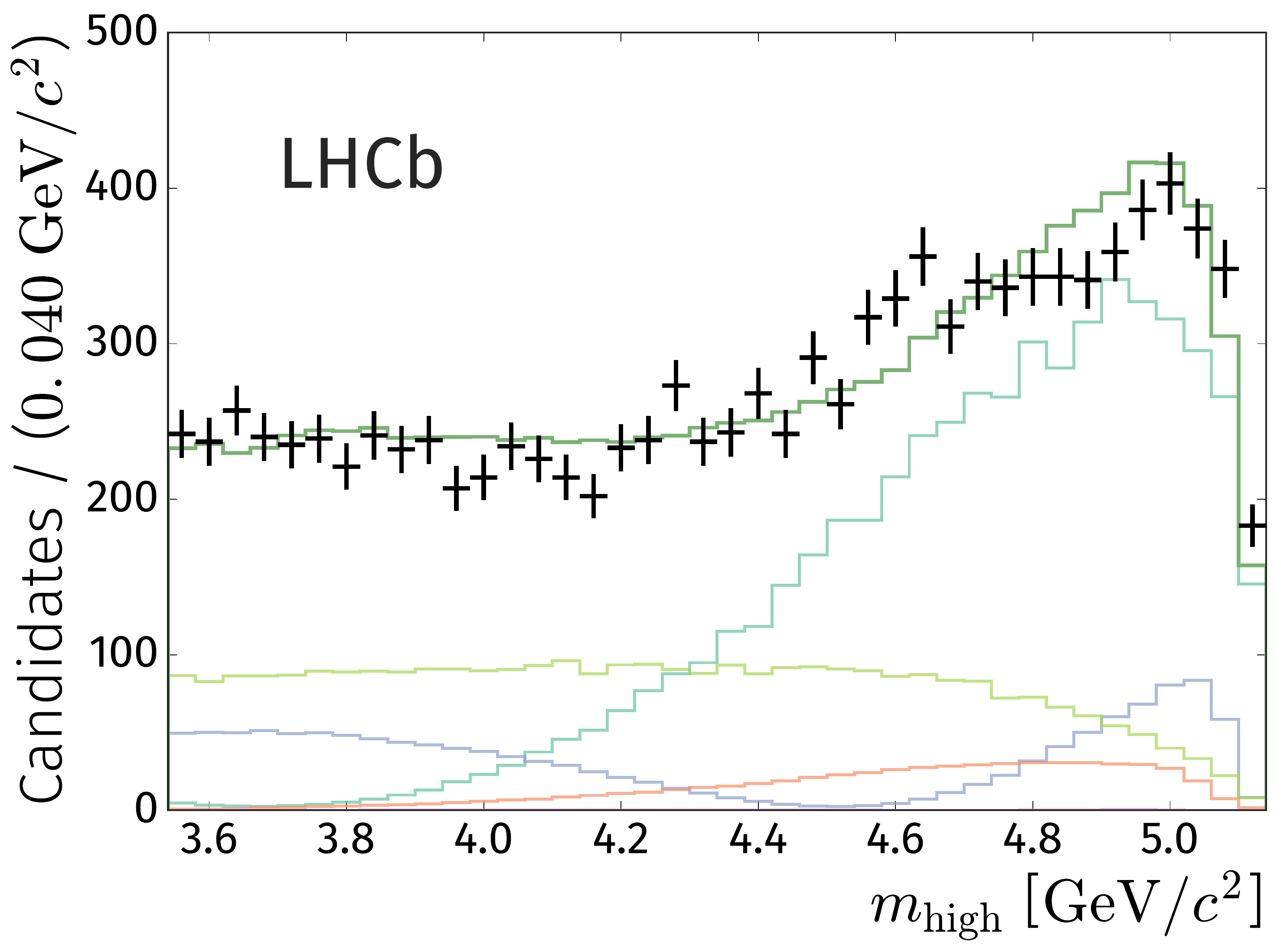}
\put(-405,27){(a)}
\put(-180,27){(b)}

\includegraphics[width=0.49\linewidth]{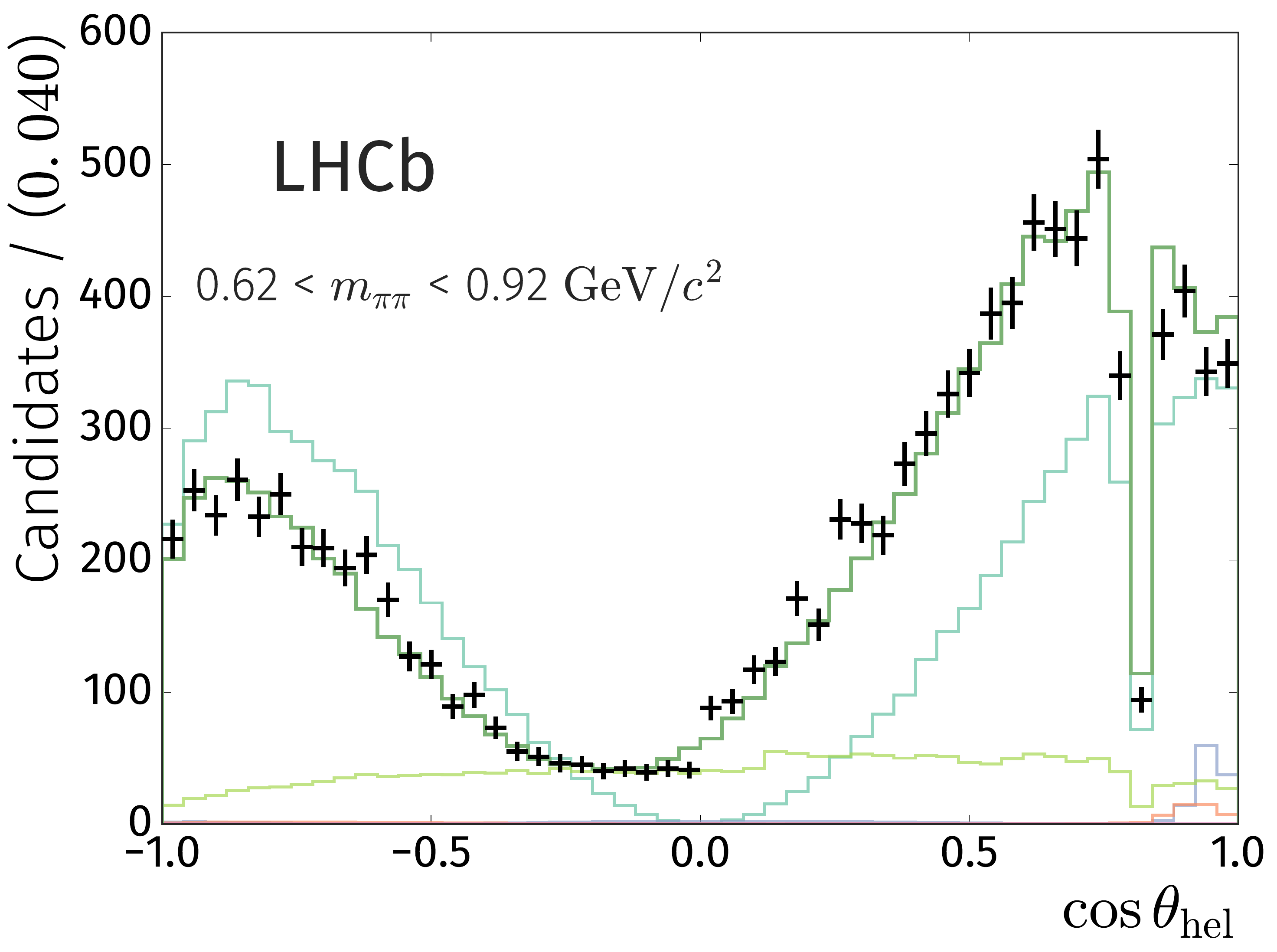}
\includegraphics[width=0.49\linewidth]{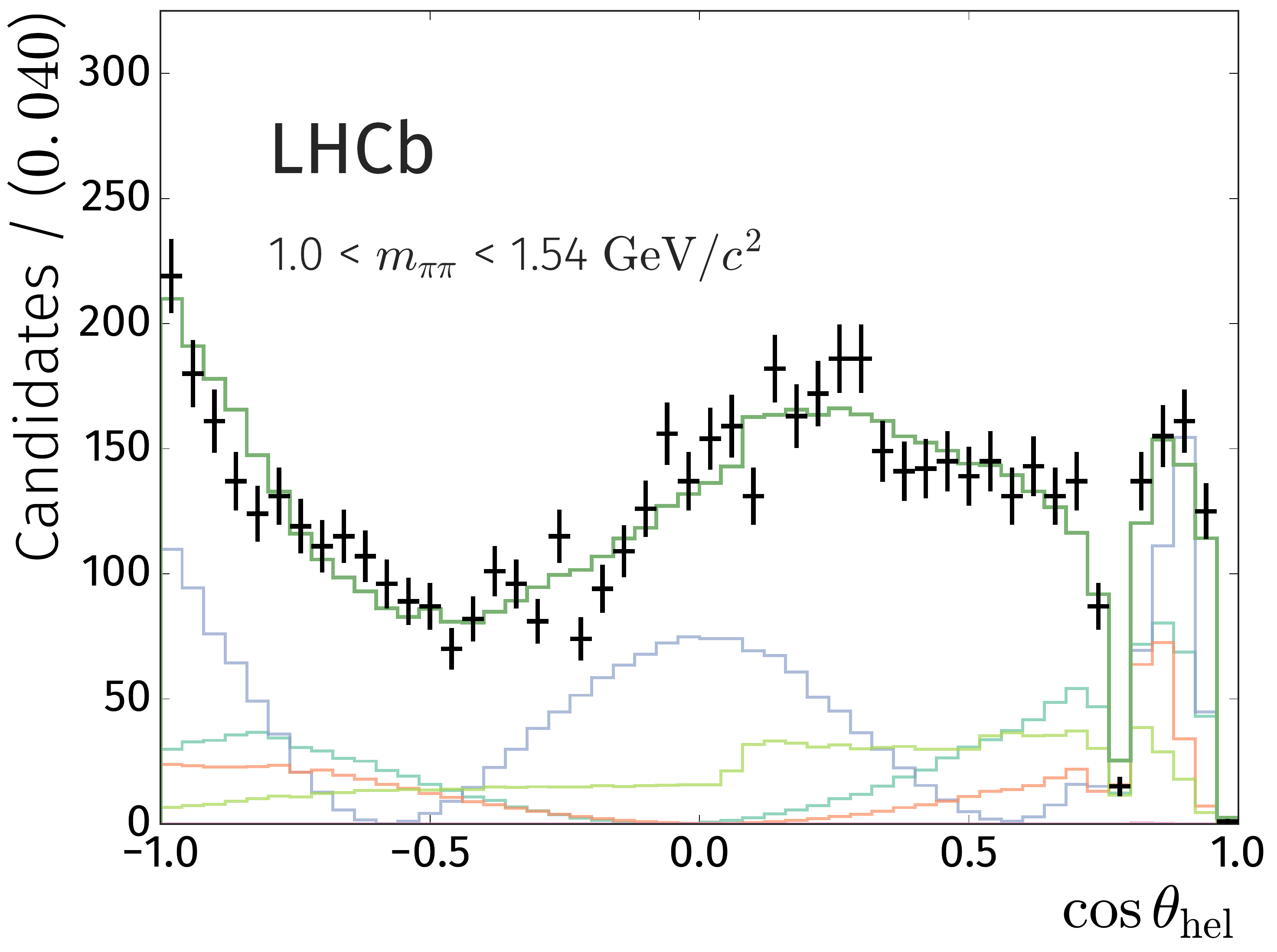}
\put(-405,30){(c)}
\put(-180,30){(d)}

\caption{Fit projections on $m_{\rm high}$ of the result with the QMI S-wave model (a)~in the full $m_{\rm high}$ range, (b)~in the high $m_{\rm high}$ region, and on $\cos\theta_{\rm hel}$ in (c)~the $\rho(770)^0$ region, and (d)~in the $f_2(1270)$ region. The thick dark-green curve represents the total model, and the coloured curves represent the contributions of individual model components (not including interference effects), as per the legend in Fig.~\ref{fig:projComponentsQMI3}.}
\label{fig:projComponentsQMI2}
\end{figure}

\begin{figure}[tbh]
\centering
\includegraphics[width=0.49\linewidth]{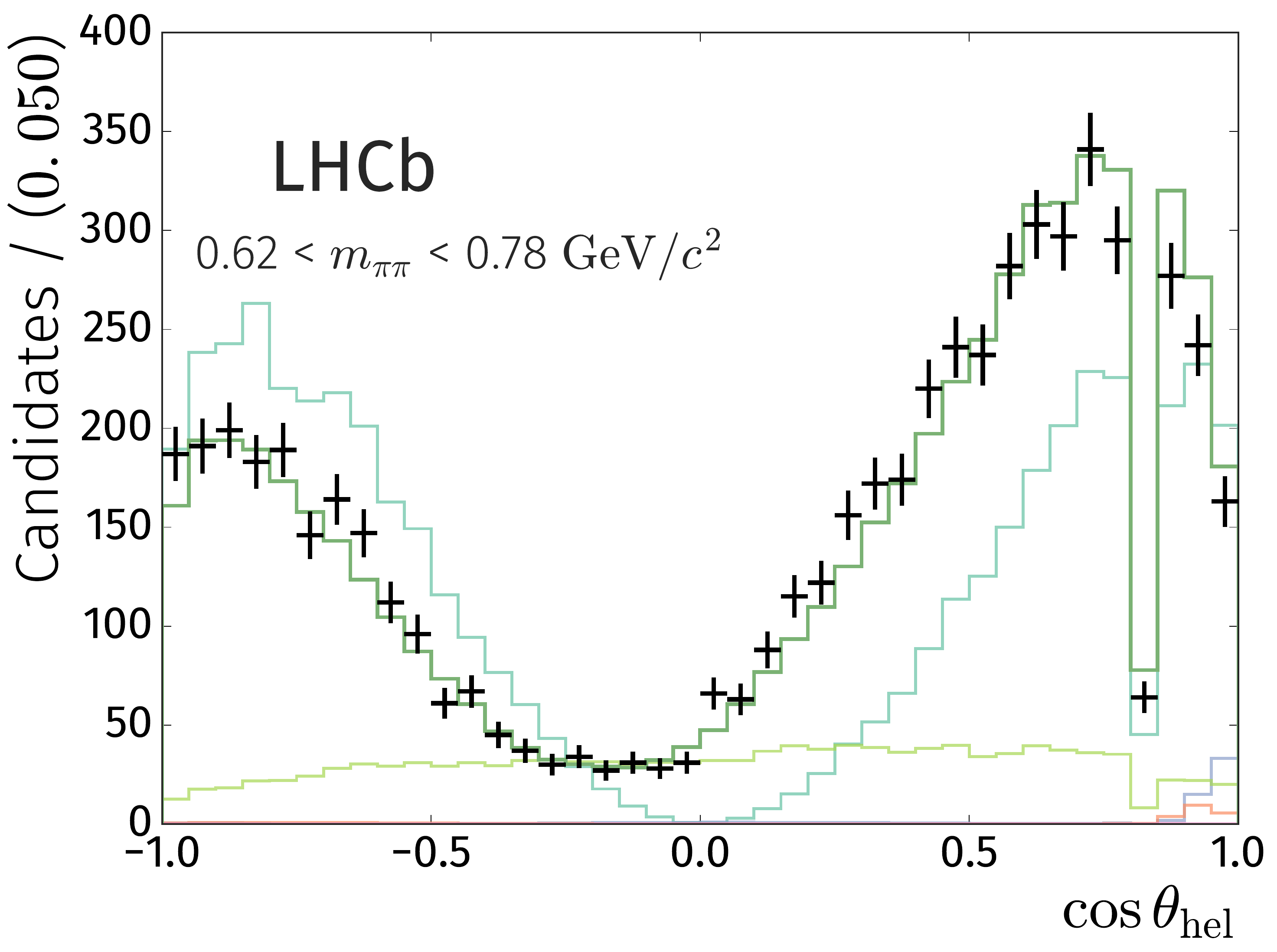}
\includegraphics[width=0.49\linewidth]{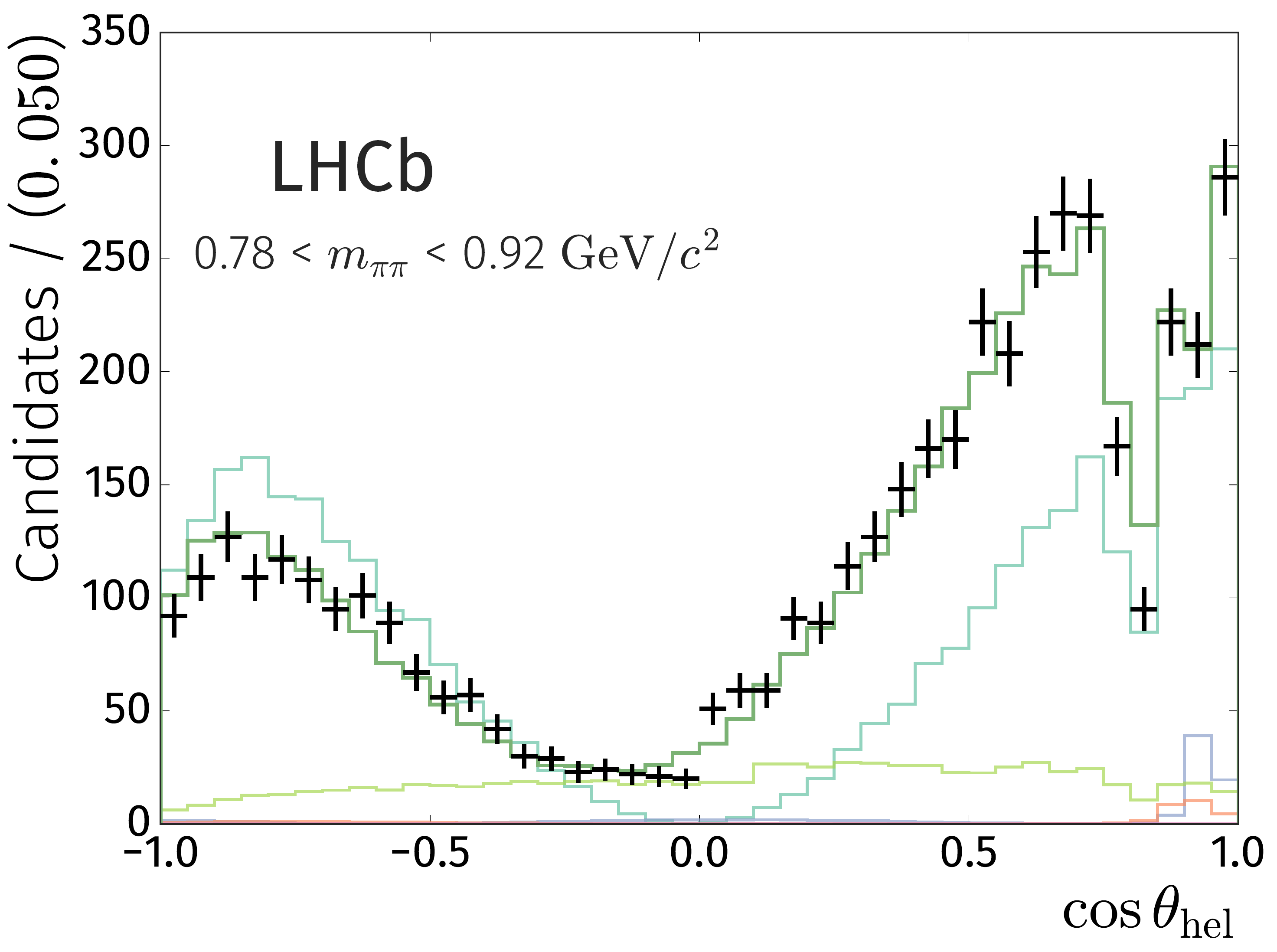}
\put(-405,27){(a)}
\put(-180,27){(b)}

\includegraphics[width=0.49\linewidth]{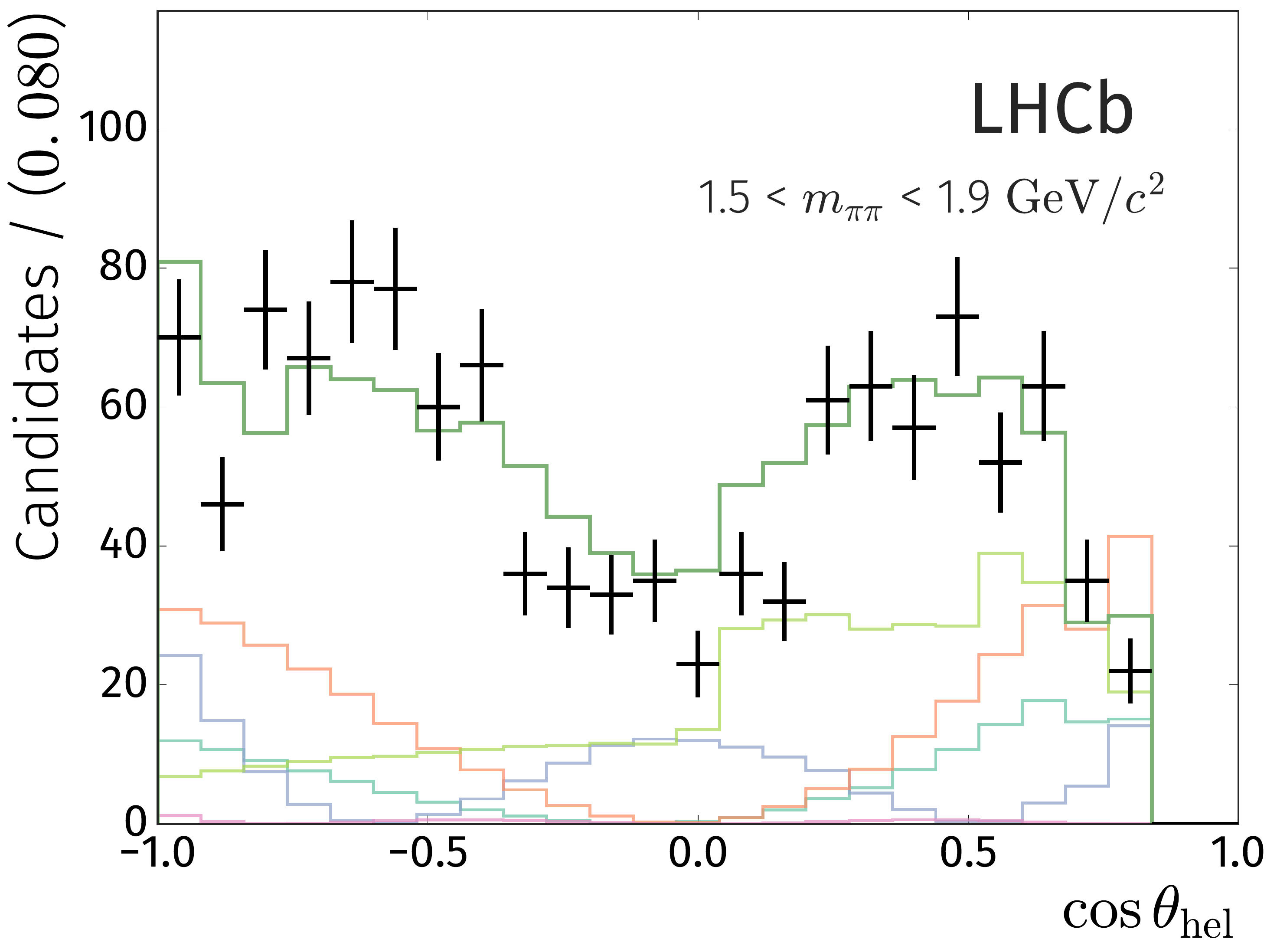}
\includegraphics[width=0.49\linewidth]{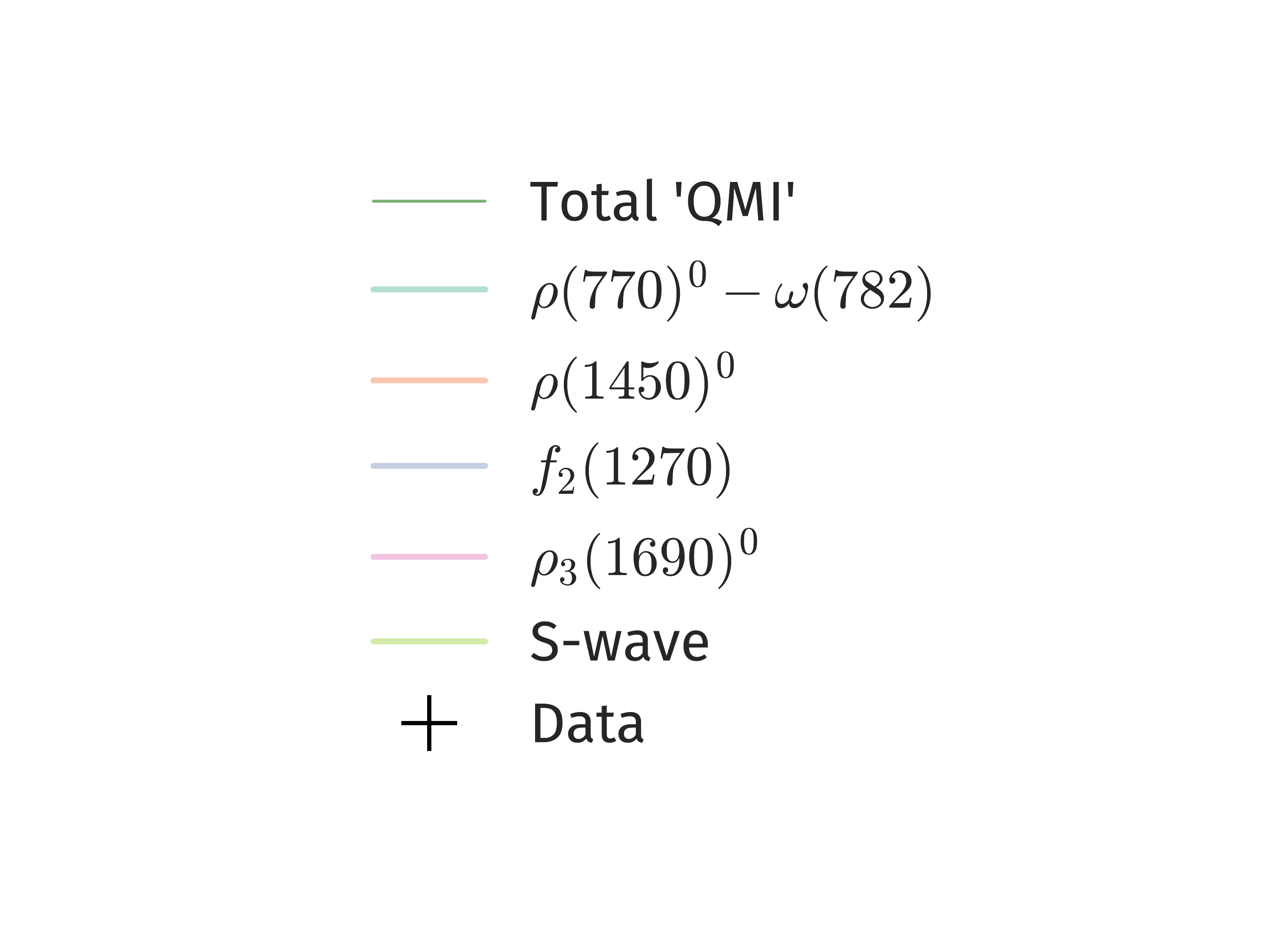}
\put(-405,30){(c)}

\caption{Fit projections on $\cos\theta_{\rm hel}$ of the result with the QMI S-wave model in the region (a)~below and (b)~above the $\rho(770)^0$ mass, and (c)~in the $\rho_3(1690)^0$ region. The thick dark-green curve represents the total model, and the coloured curves represent the contributions of individual model components (not including interference effects), as per the legend.}
\label{fig:projComponentsQMI3}
\end{figure}

Various projections of the data and the result of fit with the QMI description of the S-wave are shown in Figs.~\ref{fig:projComponentsQMI1}--\ref{fig:projComponentsQMI3}. The colour legend for each contribution is given in the final figure.

\clearpage

\addcontentsline{toc}{section}{Supplemental material}
\section*{Supplemental material}
\label{sec:Supplemental}

In addition to the results presented in the main body of the paper, we provide a supplementary collection of files containing correlation matrices that are impractical to publish either in the main text or appendices. Files are provided in the JSON (JavaScript Object Notation) format, which is both machine and human readable, and contain parameter names, two-dimensional arrays corresponding to the correlations between parameters in the order given in the parameter list, and maps from strings describing entries in the correlation matrix and the associated value.

Three files obtained from the QMI approach, related to tables presented in Section~\ref{sec:results}, are given here. These are comprised of statistical and systematic correlation matrices of the parameters given in Tables~\ref{tab:qmi:ffp} and~\ref{tab:qmi:ffm} in file \verb|qmi_ff_corr.json|, statistical uncertainties on the isobar coefficients in file \verb|qmi_params_corr.json|, along with statistical and systematic correlation matrices for the parameters in Table~\ref{tab:qmi:Swave} in file \verb|qmi_SWaveParams_corr.json|.

For the K-matrix fit model results, statistical correlation matrices for the values given in Tables~\ref{tab:ff},~\ref{tab:km:ffp},~\ref{tab:km:ffm}, and~\ref{tab:acp}, along with for the isobar coefficients listed in Table~\ref{tab:km:isobarParams}, can be found in file \verb|kMatrix_stat_matrices.json|.

\addcontentsline{toc}{section}{References}
\setboolean{inbibliography}{true}
\bibliographystyle{LHCb}
\bibliography{main,LHCb-PAPER,LHCb-CONF,LHCb-DP,LHCb-TDR}

\newpage

% LHCb collaboration author list
% Data extracted on July 29th, 2019 at 10:36am for reference date 12-Mar-2019
\centerline
{\large\bf LHCb collaboration}
\begin
{flushleft}
\small
R.~Aaij$^{30}$,
C.~Abell{\'a}n~Beteta$^{47}$,
B.~Adeva$^{44}$,
M.~Adinolfi$^{51}$,
C.A.~Aidala$^{78}$,
Z.~Ajaltouni$^{8}$,
S.~Akar$^{62}$,
P.~Albicocco$^{21}$,
J.~Albrecht$^{13}$,
F.~Alessio$^{45}$
% ...

{\footnotesize \it

% ...
$ ^{8}$Universit{\'e} Clermont Auvergne, CNRS/IN2P3, LPC, Clermont-Ferrand, France\\
% ...
$ ^{13}$Fakult{\"a}t Physik, Technische Universit{\"a}t Dortmund, Dortmund, Germany\\
% ...
$ ^{21}$INFN Laboratori Nazionali di Frascati, Frascati, Italy\\
% ...
$ ^{30}$Nikhef National Institute for Subatomic Physics, Amsterdam, Netherlands\\
% ...
$ ^{44}$Instituto Galego de F{\'\i}sica de Altas Enerx{\'\i}as (IGFAE), Universidade de Santiago de Compostela, Santiago de Compostela, Spain\\
$ ^{45}$European Organization for Nuclear Research (CERN), Geneva, Switzerland\\
% ...
$ ^{47}$Physik-Institut, Universit{\"a}t Z{\"u}rich, Z{\"u}rich, Switzerland\\
% ...
$ ^{51}$H.H. Wills Physics Laboratory, University of Bristol, Bristol, United Kingdom\\
% ...
$ ^{62}$University of Cincinnati, Cincinnati, OH, United States\\
% ...
$ ^{65}$Syracuse University, Syracuse, NY, United States\\
% ...
$ ^{78}$University of Michigan, Ann Arbor, United States, associated to $^{65}$\\
\bigskip
% ...
}
\end{flushleft}

\end{document}